\documentclass[times, twocolumn, iop, preprint, 10pt]{aastex63}

\usepackage[]{natbib}
\usepackage{graphicx}
\usepackage{subfigure}
\usepackage{float}
\usepackage{booktabs}
\usepackage{amsmath}
\usepackage{fancyref}
\definecolor{orange}{RGB}{255,69,0}
\definecolor{green}{RGB}{0,255,0}
\definecolor{darkred}{RGB}{139,0,0}
\usepackage{amssymb}
\usepackage{lineno}
\usepackage{nameref}
\usepackage{array,multirow,makecell}
\usepackage{tabularx}
\usepackage{longtable}

\usepackage{morefloats}
\usepackage{gensymb}
\usepackage{enumitem}

\usepackage{array,longtable}
\renewcommand*{\arraystretch}{1.5}

\begin{document}

\title
{Gamma-Ray Flares in Long Term light curve of 3C 454.3}
\author{Avik Kumar Das$^{1}$, Raj Prince$^{1}$, Nayantara Gupta}
\affiliation{Raman Research Institute, C.V. Raman Avenue, Sadashivanagar, Bangalore 560080, India\\
}
\email{avikdas@rri.res.in}
\begin{abstract}
3C 454.3 is frequently observed in flaring state. The long term light curve of this source has been analysed with 9 year (August 2008 - July 2017) data from Fermi LAT detector. 
We have identified five flares and one quiescent state. The flares have sub-structures with many peaks during flaring phase. We have estimated the rise and decay time of the flares
 and compared with flares of other similar sources. The modeling of gamma ray spectral energy distributions shows in most cases Log-parabola function gives the best fit to the data.
 We have done time dependent leptonic modeling of two of the flares, for which simultaneous multi-wavelength data are available. 
 These two long lasting flares Flare-2A and Flare-2D continued for 95 days and 133 days respectively. We have used the average values of Doppler factor, injected luminosity in electrons, size of the emission region and the magnetic field in the emission region in modeling these flares. The emission region is assumed to be in the broad line region in our single zone model. The energy losses (synchrotron, synchrotron self-Compton, external Compton) and escape of electrons from the emission region have been included while doing the modelling. Although, the total jet powers required to model these flares with leptonic model are higher compared to other sources,  they are always found to be lower than the Eddington's luminosity of 3C 454.3.  We also select some flaring peaks and show that time variation of the Doppler factor or the injected luminosity in electrons over short time scales can explain their light curves.

\end{abstract}


\section{Introduction}
 The FSRQ (Flat Spectrum Radio Quasar) 3C 454.3, located at redshift  0.859, is frequently monitored due to its high flux variability. During an intense flare in 1992 it was observed by EGRET (\citep{Hartman et al. (1992)}, \citep{Hartman et al. (1993)}), when its flux varied in the range of (0.4-1.4)$\times 10^{-6}$ photons cm$^{-2}$ sec$^{-1}$. Subsequently, 3C 454.3 remained a source of interest for multi-wavelength observations due to its variable nature. 
 This source was active in 2000 and even more in 2005. The 2005 outburst was recorded in optical and X-ray frequencies \citep{Giommi et al. (2006)}. 
 The high activity of  3C 454.3 in autumn 2007 was observed by  Whole Earth Blazar  Telescope (WEBT) in radio to optical frequencies. The gamma-ray satellite Astro-rivelatore Gamma a Immagini  LEggero (AGILE) detected this source in late July and November-December of 2007 \citep{Raiteri et al. (2008)}.
 
 The AGILE 2007 November campaign was reported by \citet{Vercellone et al. (2009)}. AGILE, International Gamma-ray Astrophysics Laboratory (INTEGRAL), Swift, WEBT consortium and the optical-NIR  telescope Rapid Eye Mount (REM) observed 3C 454.3 during the campaign. During three weeks of observation period the average gamma-ray flux above 100 MeV was 1.7$\times 10^{-6}$ photons cm$^{-2}$ sec$^{-1}$. The source was extremely variable in optical band. The gamma-ray emission was found to be correlated with optical emission.
 AGILE 2007 December campaign \citep{Donnarumma et al. (2009)} observed this source with average flux 2.5$\times 10^{-6}$ photons cm$^{-2}$ sec$^{-1}$  above 100 MeV and the delay between gamma-ray and optical emissions was found to be 12 hours.
 
Fermi-LAT is regularly monitoring this source since July 2008. An intense flare was observed during July 7 to Oct 6 , 2008 and the average flux above 100 MeV was found to be 3$\times 10^{-6}$ photons cm$^{-2}$ sec$^{-1}$. Strong, distinct and symmetric flares were observed with increase in flux by several factors within 3 days \citep{Abdo et al. (2009)}.

 A multi-wavelength study was carried out to find out correlation between emissions in different wavelengths (IR, optical, UV, X-ray and gamma-ray) during August to December of 2008 \citep{Bonning et al. (2009)}. They found correlation of less than a day between light curves in different frequencies except in X-rays. The X-ray flux is not correlated with fluxes in gamma-ray or longer wavelengths.

 Similar result was also reported for the high state in 2009, November-December \citep{Gaur et al. (2011)}. They found strong correlation between optical and gamma-ray emission with a time lag of four days but the X-ray emission is not correlated to any of them.
  
  The strong flare of 3C 454.3 in 2009 during December 3-12 in gamma-rays, X-rays  and optical/near-infrared bands was studied by  \citet{Gupta et al. (2017)}. Optical polarisation
  measurements showed dramatic changes during flare with a strong anti-correlation between optical flux and degree of polarisation during the decay phase of the flare. They used one zone model with variations in magnetic field, spectral break energies and normalisation to fit the spectral energy distributions at different times.

\citet{Raiteri et al. (2011)} studied the multi-wavelength light curves in 18 bands to analyse the flux variability for the period  April 2008  to March 2010. The X-ray flux variation appeared 
 to follow the  gamma-ray and optical ones by about 0.5 and 1 day respectively. They speculated that there is a slight variable misalignment  between the synchrotron and  Comptonisation zones, which can explain the  increases in gamma-ray and X-ray flux levels in 2009-2010, and also the change in gamma-ray to optical flux ratio at the peaks of the outbursts.
 
 During high gamma-ray states of 3C 454.3 in December 2009, April 2010 and November 2010 the parsec-scale jet was highly active. Superluminal radio knots K09 and K10 were found to be associated with autumn 2009 and 2010 outbursts \citep{Jorstad et al. (2013)}. It was argued that gamma-ray outbursts of as short as 3 hours duration can occur on parsec scales if flares take place in localised regions such as turbulent cells.
  
   Multi-wavelength  variations of 3C 454.3 during the 2010 November to 2011 January outburst were studied before \citep{Wehrle et al. (2012)}. Their discrete correlation analysis of 
  the milli-meter, far-infrared and gamma-ray light curves showed simultaneous variations indicating their common origin. They located the site of outburst in  parsec scale ``core". In their model the turbulent plasma crosses a conical shock in the parsec scale region of the jet. The seed photons for inverse Compton emission are produced in nonthermal radiation by a Mach disk, thermal emission from hot dust, or synchrotron emission from moving plasma. Extremely high polarisation in the 2010 outbursts was reported by \citet{Sasada et al. (2014)}.

  Long term and rapid radio variability of  3C 454.3 was studied on the RATAN-600  radio telescope of the  Special  Astrophysical Observatory  at 4.6, 8.2, 11.2  and  21.7 GHz and on the 32-m Zelenchuk and Badary radio telescopes \citep{Gorshkov et al. (2018)}. Two flares were observed in the long term light curve in 2010 and in 2015-2017. The delay in the maximum of the first flare at 4.85 GHz  relative to the maximum at 21.7 GHz was six months. Intraday variability was detected at 8.57 GHz on the 32-m telescopes in 30 of 61 successful observations and  it was found to be correlated with the maxima of the flares. The characteristic time scale for this variability was found to be two to ten hours.
  
  Multi-wavelength temporal variability of 3C 454.3 for the gamma-ray high state during May to December 2014 was studied by  \citet{Kushwaha et al. (2017)}. Their correlation study showed that no lag between infrared (IR) and gamma-ray, optical and gamma-ray, optical and IR, the source went to a state where gamma-ray lags the optical/IR by 3 days.
  
  Fermi LAT observations of the 2014 May-July outburst was studied by \citet{Britto et al. (2016)}. The average flux during the highest state from 7-29 June, 2014 was found to be $7.2\times 10^{-6}$ photons cm$^{-2}$ sec$^{-1}$. Several photons above 20 GeV were detected, including one above 45 GeV on MJD  56827. The emission region was speculated to be near the outer boundary of BLR. Temporal correlation between the optical and  gamma-ray flux  variations in the blazar 3C 454.3 has been studied with 9 years of Fermi LAT data \citep{Rajput et al. (2019)}. Out of four epochs of intense optical flares, in two epochs the gamma-ray and optical flares are found to be correlated. In the other two epochs gamma-rays are weak or absent.
  
   The long term optical spectroscopic variations of blazar 3C 454.3 have recently been investigated with 10 years of data from the Steward Observatory \citep{Nalewajko et al. (2019)}. The data revealed that the line flux from the broad line region (BLR) changed dramatically with the blazar activity from a very high state in 2010 to a significantly low state in 
  2012. Inverse Compton emission of relativistic electrons by the seed photons from BLR is the well established scenario for explaining gamma-ray emission from FSRQs. Due to this reason the radius of the BLR is a crucial input parameter in modelling of multi-wavelength emission from FSRQs. They have obtained the lower bound on the radius of the BLR to be 0.28 pc.
  
  The long term variability for the period between February 2008 and April 2016 in radio, IR and optical bands have been analysed recently by \citet{Sarkar et al. (2019)}. This source showed significant multi-wavelength variability with the time scale of variability  in the range of  months to  years. The variations in radio band has been observed to be lagging behind the variations in optical/IR bands by 15 to 100 days. Strong correlation in optical/IR bands indicates their co-spatial origin. They inferred from their analysis that the emission regions change their orientation with our line of sight as the time lag between radio and optical/IR emission varies over the years.  
  
  Recently,  \citet{Weaver et al. (2019)} analysed the uniquely structured  multifrequency  outburst of  2016 June. 
  This outburst was monitored in optical R-band by several  ground based  telescopes in photometric and polarimetric modes, and also by the Fermi LAT gamma-ray detector.
  Intra-day variability continued throughout the outburst.
  They constrained the Doppler factor and the size of the emission region from the observed minimum variability timescale.

  Leptonic and lepto-hadronic models have been used previously to  model the multi-wavelength spectral energy distributions. 
  In MHD jet launching  models a large scale poloidal magnetic field at the jet base  extends to a helical magnetic field downstream along the jet. A large scale ordered helical magnetic field at a distance of hundreds of parsecs was used by \citet{Zamaninasab et al. (2013)} to explain the radio emission of 3C 454.3.
 Several theoretical models were proposed to explain the spectral energy distributions of 3C 454.3 (\citet{Finke and Dermer (2010)}; \citet{Cerruti et al. (2013)}; \citet{Hunger and Reimer (2016)}).
  The flare observed during Nov 2010 is well explained by one zone lepto-hadronic model by \citet{Diltz and Bottcher (2016)}. Another flare in August 2015 was observed with simultaneous data in optical, UV, X-ray and gamma-ray energy  \citep{Shah et al. (2017)}. They suggested that X-ray and gamma-ray emission of 3C 454.3 cannot be attributed to a single emission zone and both SSC (synchrotron self Compton) and EC (external Compton) mechanisms are required to explain the data. They further suggested that the flare region lies beyond the broad line region (BLR) of this source.

   Motivated by the earlier studies we have analysed the Fermi LAT data from August, 2008 to July, 2017 to identify the flares of 3C 454.3 and study their characteristics. 
   In section 2 we have discussed about Fermi LAT and Swift XRT/UVOT data analysis. In section 3 the flaring states of 3C 454.3 are identified from the 9-year gamma ray light curve.
   The flares are studied in section 4, their sub-structures and peaks are identified. The variability time in gamma ray is calculated by scanning the light curves.
The spectral energy distributions of the flares in gamma rays are studied in section 5.    
In section 6 we have discussed about the multi-wavelength modelling of two flares. In section 7 we have discussed how time dependent Doppler factor or injected luminosity in electrons can explain the flare peaks. Our results are discussed in section 8 and conclusions are drawn in section 9.

\section{DATA ANALYSIS}

\subsection{FERMI-LAT ANALYSIS}
 Fermi-LAT (Large Area Telescope) is a imaging pair conversion telescope, which covers $\gamma$-ray energy range from 20 MeV to $>$300 GeV with energy 
 resolution $<15$\% at energy $>$100 Gev \citep{Atwood et al. (2009)}. The detailed description of LAT-characteristics have been provided in Fermi Webpage
 \footnote{https://fermi.gsfc.nasa.gov/ssc/data/analysis/software/}. 
 Fermi typically scans the entire sky in survey mode with a time period of $\sim$ 3.2 hour. We have extracted the data of blazar 3C 454.3 source from FSSC's 
 website data server\footnote{https://fermi.gsfc.nasa.gov/cgi-bin/ssc/LAT/LATDataQuery.cgi} over the period of 9 years (August, 2008 - July, 2017) and analyzed it with the help of Fermi science tool software package version- 1.0.10, which includes galactic diffuse emission model (\texttt{gll\_iem\_v06.fits}) and extragalactic isotropic diffuse emission model (\texttt{iso\_P8R2\_SOURCE\_V6\_v06.txt}). 
 The ``unbinned Likelihood analysis" (using python) method has been used to analyze the Fermi-LAT Pass8 data with appropriate selections and recommended cuts. The photon-like events are classified as 
 ‘evclass=128, evtype=3’ with energies range from 100 MeV to 300 GeV. We have extracted the photons from a radius (Region of interest or ROI) of 10$^\circ$ 
 around the source and used maximum zenith angle value of 90$^\circ$, which is the standard value provided by the LAT-instrument team, in order to avoid the $\gamma$-ray detection from the earth's limb. Filter expression ``(\texttt{DATA\_QUAL}$>$0)\&(\texttt{LAT\_CONFIG}==1)" is implemented to select the good 
 time interval data, which is recommended by the LAT team. The live-time, exposure-map and diffuse response of the instrument have been computed subsequently for 
 each event with the latest instrument response function (IRF)- \texttt{"P8R2\_SOURCE\_V6"}. To localize the source detection 
 a quantity called `Test Statistic' (TS) is computed, which is defined as
\begin{equation} 
TS=-2\log(\frac{L_0}{L_1})
\end{equation}
where, $L_0$ \& $L_1$ are the maximum likelihood values for a given model without (null hypothesis) and with the point like source at the position of source. We have 
always maintained the criterion to choose the sources with TS$\geqslant$25 (corresponds to $\sim$$(TS)^{1/2}$$\sigma$ or 5$\sigma$ detection level) for 
each data sets. To generate the light curve we have fixed the model parameters of all the sources within the ROI excluding our source of interest from third 
fermi catalog (3FGL; \citet{Acero et al. (2015)}). In our work, we have studied the light curve of three different time bins: 7-day, 1-day \& 6-hour. Apart 
from this, we have also generated the spectral data points for different periods of activity in the energy range 0.1$\leqslant$E$\leqslant$300 GeV.

\subsection{SWIFT-XRT/UVOT}
 We have analyzed the archival data from the Swift-XRT/UVOT for the source 3C 454.3 during the time period of April, 2009 - April, 2011 ($\sim$ 2 years.), 
 which has been retrieved from HEASARC webpage\footnote{https://heasarc.gsfc.nasa.gov/cgi-bin/W3Browse/swift.pl}. Total 203 observations were made in this 
 time span. A task `$xrtpipeline$' (version of 0.13.2.) has been used to process the XRT-data (\citet{Burrows et al. (2005)}, 0.2-10 KeV) files for each 
 observation sets. The latest calibration files (CALDB version of 20160609) and standard screening criteria have been implemented in this process. We have 
 chosen a circular radius of 20 arc seconds around the source to analyze the XRT-data. Background region is also chosen of same radius (20 arc seconds) but 
 far away from the source region. A tool `$xselect$' has been used to extract the X-ray light curve and spectra. The tool called `$xrtmkarf$' \& 
 `$grppha$' have been used to create the ancillary response file and group the spectra of 30 counts per bin respectively. Subsequently, the grouped spectra have been
 Modelled in XSPEC (version of 12.10.0) with the `$tbabs*log parabola$' model and with the fixed neutral hydrogen column density of $n_H = 1.34\times10^{21} 
 cm^{-2}$ \citep{Villata et al (2006)}.

The source 3C 454.3 was also observed by the Swift Ultraviolet/Optical telescope (UVOT, \citet{Roming et al. (2005)}) in all the six filters: U, V, B, W1, M2, 
\& W2. The source region has been extracted from 10 arc seconds circular region around the source and the background region has also been chosen with a radius of 
25 arc seconds away from the source. The source magnitudes have been extracted by the task `$uvotsource$' and corrected for galactic extinction 
\citep{Schlafly and Finkbeiner (2011)}. Subsequently, these magnitudes have been converted into flux by using the zero points \citep{Breeveld et al. (2011)} 
and conversion factors \citep{Larionov et al. (2016)}.
 
\section{FLARING STATE OF 3C 454.3}
Seven day binning gamma-ray light curve of 3C 454.3 has been shown in Figure-1, which is observed by Fermi-LAT from MJD 54686 (August,2008) to MJD 57959 (July, 2017). 
From this 9-year light curve history we have clearly identified (shown by broken green line) five major flaring states and one quiescent state. As 
alluded to previously \citep{Prince et al. (2017)}, we have defined these states as Flare-1, Flare-2, Flare-3, Flare-4 and Flare-5 with time span from 
MJD 54683-54928, MJD 54928-55650, MJD 56744-57169, MJD 57169-57508 and MJD 57508-57933 respectively. The quiescent state has time duration of almost about 3
years (MJD 55650-56744). In our work, we are more interested in flaring states and hence further analysis has been carried out on these states only. We have studied these 
flares in detail for one-day binning (where the sub-structures are not clearly visible) 
and then six-hour binning to identify the various sub-structures properly.

In the 6-hour binning study we have found several sub-structures for each flaring state. Flare-1 has only one sub-structure, we labeled that as Flare-1A. 
Four sub-structures were noticed in Flare-2 , defined as Flare-2A, Flare-2B, Flare-2C \& Flare-2D respectively. Flare-3A and Flare-3B  are two sub-structures of Flare-3. Similarly, Flare-4 and 
Flare-5 have four (Flare-4A, Flare-4B, Flare-4C \& Flare-4D) and two (Flare-5A \& Flare-5B) sub-structures respectively. There are two sub-substructures (Flare-1A \& Flare-2A) which are well observed in one-day binning but we are unable to study them in 6-hour binning due to large error in the photon flux. 

\section{GAMMA-RAY light curve HISTORY OF FLARES \& VARIABILITY}
We have studied each sub-structure separately and observed different states of activity (e.g. pre-flare, flare, post-flare etc.) as shown in 6-hour binning light curve. There are various ways in which one can define the different states of the source. One of these methods is to estimate the average flux for each time period (pre-flare, flare, etc.) and compare their values. The flare period can be defined as the period when the average flux is more than 3-4 times of its average flux during the pre-flare period. The other way is to estimate the fractional variability in each period. The flux is high and more variable during the flaring period while during preflare or post flare the fractional variability is less and also the flux will be constant for a long period of time (e.g. \cite{Prince et al. (2018)}). In our case we have used both these methods to identify the various states of the source and our result is consistent with both these methods. 

We have fitted only the flaring state of each sub-structure with sum of exponential function to show the temporal 
evolution. These fitted flares have characteristic rising and decay times for different peaks (P1, P2 etc.). The functional form of the sum of exponential function is given by \citet{Abdo et al. (2010b)}
\begin{equation} \label{eq:2}
F(t)=2F_0[\exp(\frac{t_0-t}{T_r})+\exp(\frac{t-t_0}{T_d})]^{-1} 
\end{equation}
where, $t_0$ is the peak time and $F_0$ is the flux observed at time $t_0$. $T_r$ and $T_d$ represent the rising and decay time respectively. For few flares we are able to show the constant state (shown by horizontal grey line). All reported gamma-ray fluxes throughout the paper are mentioned in unit of 10$^{-6}$ ph cm$^{-2}$ s$^{-1}$.

\subsection{FLARE-1}
A 6 hour binning has been carried out for Flare-1 during MJD 54683-54928. We have found only one sub-structure (defined as Flare-1A) in this period. But we 
are unable to identify the peaks in this binning due to rapid fluctuation and large error in photon counts. For this reason we have shown the sub-structure in 1-day binning in Figure-2.

Flare-1A (MJD 54712-54783) has two distinct states of activity, these are defined as Flare and Post-flare. There are several peaks in the Flare epoch 
(shown in Figure-3), but we have considered only three prominent major peaks which are labeled as P1, P2 \& P3 with the fluxes of 5.45$\pm$0.42, 4.31$\pm$0.35
\& 3.66$\pm$0.28 at time MJD 54719.1, 54729.1 and 54738.1 respectively. The details of the modelling parameters 
($T_r$ \& $T_d$) have been elucidated in Table-1. Post-flare epoch (MJD 54759 - 54783) follows immediately after Flare epoch with the time span of 24 days, which has small variations in flux and the average flux is found to be 1.27$\pm$0.04.

\subsection{FLARE-2}
We have performed 6-hour binning of the light curve of Flare-2 during MJD 54928-55650 and identified four sub-structures (Flare-2A, Flare-2B, Flare-2C \& Flare-2D). As Flare-1A we are unable to study the temporal evolution of Flare-2A in 6 hour binning due to large error in flux. Here one day binning light curve of Flare-2A are considered for further study which is shown in Figure-4. The six hour binning light curve of Flare-2B, Flare-2C \& Flare-2D are presented in Figure-6, Figure-9 and Figure-12 respectively.

Flare-2A shows two different phases during MJD 55045 - 55140 which are labeled as Pre-flare and Flare. Pre-flare epoch has time span of 19 days (MJD 55045 - 55064) with average flux of 1.22$\pm$0.04. After that source enters into flaring state with time duration of MJD 55064 - 55140. Figure-5 shows the fitted light curve of flaring state in 1 day binning which has four prominent peaks (P1, P2, P3 \& P4) with fluxes of 3.32$\pm$0.29, 3.31$\pm$0.36, 5.95$\pm$0.52, 4.08$\pm$0.36 at MJD 55070.5, 55077.5, 55091.5 and 55103.5 respectively. The details of the parameters have been described in Table-2.    

Flare-2B (MJD 55140-55201) also shows two different states of activity regions: Pre-flare and Flare. Pre-flare has been considered from MJD 55140-55152, during 
which flux does not vary much. Rest of the region of the light curve is considered as Flare (MJD 55152-55201). Figure-7 and 
Figure-8 represent the fitted light curve of Flaring state in two different parts, as we are unable to fit the entire Flare in a single plot.
In the 1st part of the Flare (Figure 7, MJD 55152-55177), six major peaks (P1, P2, P3, P4, P5 and P6) are observed at MJD 55154.9, 55163.1, 55165.1 55167.9, 55170.4, 55172.1 with fluxes($F_0$) of 7.48$\pm$1.24, 9.69$\pm$1.41, 9.69$\pm$0.99, 22.86$\pm$1.48, 18.70$\pm$1.24 and 14.56$\pm$1.21 respectively. A small hump 
kind of structure has been observed in the beginning of light curve during MJD 55152.0-55153.9 (Figure 7), but we have not considered it as a distinct peak due to low flux value. Similarly, five different peaks (P1, P2, P3, P4 and P5)  have been noticed in the 2nd part of Flare (Figure 8, MJD 55177-55201). The Flux values
($F_0$) of these peaks are: 8.73$\pm$1.04, 7.85$\pm$0.95, 8.48$\pm$0.98, 7.68$\pm$0.87 and 8.52$\pm$0.83 at MJD 55178.4, 55180.4, 55182.6, 55185.1 and 55195.1 
respectively. The values of the fitted parameters are given in Table-3.

There are four different phases of activity (Pre-flare, Flare-I, Flare-II and Post-flare) in Flare-2C during MJD 55250-55356, which are shown in Figure-9.
Pre-flare phase has small variation in counts with average flux of 2.68$\pm$0.06 and then the source goes to Flare-I and Flare-II state with the time span of 36 days
\& 18 days respectively. Fitted light curve of Flare-I phase (shown in Figure-10) shows five distinguishable major peaks which are labeled as P1, P2, P3, P4 \& P5 respectively. After peak P5 flux counts gradually decrease with small variation
and at the end of Flare-I epoch (During MJD 55312.2-55314.7) a sudden increase in flux has been observed, although we have not considered it as 
a distinct peak since it is far away from the main peaks. Flare-II (shown in Figure-11) phase also shows 5 distinctive major peaks (defined as P1, P2, P3, P4 \& P5)
with fluxes of 9.71$\pm$0.94, 10.05$\pm$0.95, 7.79$\pm$0.90, 9.12$\pm$1.26 and 5.96$\pm$0.78 at MJD 55320.6, 55321.6, 55322.6, 55327.1 \& 55329.4 respectively. 
After Flare-II photon flux starts to decay slowly and the source comes back to its quiescent state, which we have identified as Post-flare phase in Figure-9. 
The detailed description of the modelling parameters have been elucidated in Table-4.

Flare-2D (MJD 55467-55600) is observed to be the most violent sub-structure in the whole 9-years of light curve history with six different phases (shown in Figure-12) of
activity: Pre-flare, Plateau-I, Flare-I, Flare-II, Plateau-II and Post-flare. There is no rapid fluctuation in flux during 
MJD 55467-55480, this phase is considered as Pre-flare phase. After that (MJD 55467) the flux starts to rise slowly up to MJD 55511, which is labeled as Plateau-I
phase, with the average flux of 6.26$\pm$0.07. We have identified three major peaks (P1, P2, P3) from the fitted light curve (see Figure-13) of Flare-I phase 
with time duration of 25 days, which has peak fluxes($F_0$) 53.51$\pm$2.08, 65.66$\pm$2.34, 80.41$\pm$5.92 at MJD 55517.6, 55518.6 and 55519.9 respectively.
Peak P3 corresponds to the highest observed flux in our analysis. A small variation compared to peaks P1, P2, and P3 has been noticed in flux after peak P3 
in the Flare-I phase, but no major peak has been identified. 
Flare-II state is observed immediately after Flare-I during MJD 55536 - 55572. Large variation in flux is seen during this period and six major peaks are observed (see Figure-14). After Flare-II (see Figure-12.) the source went into a state of steady diminution of flux defined as Plateau-II, which eventually
ends up into a Post-flare state having almost constant temporal flux distribution. The details of the  values of the parameter are displayed in Table-5.

\subsection{FLARE-3}
Following the similar procedure executed for Flare-2, a 6 hour binning light curve analysis has also been carried out for Flare-3 and two sub-structures
(Flare-3A and Flare-3B) of moderate time duration (51 \& 30 days respectively) have been found in our study.

Four different epochs of flaring phases are identified in Flare-3A (shown in Figure-15). The time span of pre-flare is about 14 days. After the pre-flare two 
 flaring states (Flare-I and Flare-II) of similar time durations have been identified, both of which have 5 prominent peaks and shown in Figure-16 and Figure-17 respectively. 
The values of the modelling parameters for these state has been elucidated in Table-6. Small fluctuations in photon flux are noticed during MJD 56838-56850 
with the average flux of 3.56$\pm$0.12, which is defined as post-flare phase (Figure-15).

Flare-3B has the least complicated substructure with three clear states shown in Figure-18. A pre-flare phase has been 
 identified from MJD 56799 - 57002. In the Flare region, the source shows only two major peaks at MJD 57006.1, 57008.4 with fluxes of 4.95$\pm$0.69 and 
7.90$\pm$0.90 respectively (Figure-19). After spending around 10 days in flaring state, it comes back again to the constant flux state, which is labeled as 
Post-flare. The values of the fitted parameters have been displayed in Table-7.

\subsection{FLARE-4}
Six hour binning of the light curve of Flare-4 shows four distinct sub-structures, defined as Flare-4A, Flare-4B, Flare-4C and Flare-4D (Figure-20, 
Figure-22, Figure-24 and Figure-26). In this period, we are able to fit the light curve by showing the constant flux state (shown by horizontal grey line) 
for Flare-4A, Flare-4B and Flare-4D, which are shown in Figure-21, Figure-23 and Figure-27 respectively. 

A Pre-flare phase has been noticed in Flare-4A during MJD 57178 to MJD 57194 with small-scale variation in photon flux and the average flux is observed to be
1.57$\pm$0.08. After that, the source enters into the Flaring state (shown in Figure-21) with time span of 19 days (MJD 57164 - 57213), which has 5 well defined peaks (labeled as P1, P2, P3, P4 and P5).
The values of the peak fluxes ($F_0$) at time $t_0$ and the fitted parameters have been given in Table-8. Post-flare region promptly follows after this with
a duration of 19 days and having an almost constant flux throughout this period. 

Similarly, Flare-4B also shows three phases (see Figure-22): Pre-flare, Flare and Post-flare. Pre-flare and Post-flare epochs have almost constant flux
with the average fluxes of 2.50$\pm$0.12 and 1.91$\pm$0.10 respectively. Two distinct major peaks (P1 and P2) are observed during the Flare phase (see Figure-23), which have
peak fluxes of 11.43$\pm$0.48 and 12.00$\pm$0.49 at MJD 57254.1 \& 57256.1 respectively. The details of the fitted parameters are given in Table-9. 

Flare-4C (Figure 24) has much more error in flux compared to other sub-structures and three different phases (pre-flare, flare, and post-flare) are observed.
Pre-flare and Post-flare states have time span of 8 days and 11 days before and after the flare phase respectively. During the flare phase 4 major peaks have been clearly identified with fluxes of 5.15$\pm$0.83, 5.49$\pm$0.80, 7.44$\pm$0.94 \& 5.44$\pm$0.83 at MJD 57401.4, 57402.9, 57407.1 and 57408.9 respectively, which are shown in Figure-25. The values of the fitted parameters are given in Table-10.

Flare-4D (shown in Figure-26) has three phases similar to Flare-4A \& Flare-4B. The Pre-flare phase shows small variation in the flux and the average flux is
observed to be 1.31$\pm$0.11, which lasts from MJD 57450-57456.
The Flare phase has two sharp peaks labeled as P1 \& P2 with fluxes of 3.49$\pm$0.69, 9.99$\pm$0.45 at MJD 57457.4 \& 57460.1 respectively, which are shown in Figure-27.
We have identified the Post-flare region during MJD 57461-57468.  Table-11 displays the values of the fitted parameters.

\subsection{FLARE-5}
Similarly, a 6 hour binning of Flare-5 has also been carried out, and two sub-structures have been found. One during June-July, 2016 (MJD 57542 - 57576) and another in 
December,2016 (MJD 57727 - 57752) with time span of 34 days and 25 days respectively. Both the sub-structures (defined as Flare-5A and Flare-5B) have the simplest time profile,
where the three phases Pre-flare, Flare \& Post-flare can be clearly identified.

The Pre-flare phase of Flare-5A has almost constant flux during MJD 57542-57549 (see Figure-28). Figure-29 shows the fitted light curve
of the Flare phase with time duration (MJD 57549 - 57568) of 19 days and five major peaks have been identified. A small fluctuation is noticed in flux during the Post-flare phase (MJD 57568 - 57576) 
and the average flux is estimated to be 1.84$\pm$0.09. The values of the modelling parameters have been given in Table-12.

Figure-30 shows the three different states of Flare-5B. Pre-flare phase has been considered during MJD 57727-57737. After that a Flare having two distinct major peaks have been identified, which is shown in Figure-31. Small variation in flux has been noticed in flare phase during MJD 57737.1 to MJD 57741.9, which is also fitted with the sum of exponential function (equation \ref{eq:2}). However, we have not considered any peak in this time interval due to low count of photons. Post-flare region has time duration of around six days with an average flux of 1.96$\pm$0.14. The fitted parameters values have been described in Table-13.

Constant flux value in the steady state (shown by constant grey line) for Flare-4A, Flare-4B, Flare-4D and Flare-5A have been shown in Table-14.

\subsection{VARIABILITY}
Variability time ($t_{var}$) is the timescale of variation in flux during flare. This can be computed by scanning the 6 hour binned $\gamma$-ray light curve with the following equation -

\begin{equation}
 F(t_2) = F(t_1)2^\frac{(t_2-t_1)}{\tau_{d/h}}   
\end{equation}

where, $F(t_1)$ \& $F(t_2)$ are the fluxes at consecutive time instants $t_1$ \& $t_2$ respectively. Doubling/Halving (indicated by `$+$' \& `$-$' sign 
respectively) timescale is indicated by $\tau_{d/h}$. We have used the following two criteria while scanning the light curve \citep{Prince et al. (2017)}. They
are:
\begin{itemize}
  \item Flux should be half or double between two successive instants of time.
  \item The condition TS $>$ 25 (corresponds to $\sim$5$\sigma$ detection) on flux must always be fulfilled for these two consecutive time instants.
\end{itemize}
The value of $\tau_{d/h}$ for each sub-structure has been shown in Table-15 \& Table-16. The minimum value of {$\mid\tau_{d/h}\mid$} is defined as the variability
time ($t_{var}$) in our work.

In our 9-year light curve study we have found the shortest time as $\tau_{d/h}$ (or $t_{var}$) = $1.70\pm0.38$ hour during MJD 56815.625-56815.875
(Flare-3A), which is consistent with previously calculated hour scale variability time for other FSRQs e.g. PKS 1510-089 and CTA 102 (\cite{Prince et al. (2017)}, \cite{Prince et al. (2018)}).

\section{GAMMA-RAY SPECTRAL ENERGY DISTRIBUTION (SED) OF FLARING STATES}
We have fitted the SEDs of different epochs with three different spectral models \citep{Abdo et al. (2010a)}. These are 
\begin{enumerate}[label=(\roman*)]
  \item A powerlaw model(PL), whose functional form is
  \begin{equation}
   \frac{dN}{dE} = N_0(\frac{E}{E_0})^{-\Gamma}
   \end{equation}
   where, $N_0$ and $\Gamma$ are the prefactor \& spectral index respectively. We have kept fixed the value of $E_0$ (Scaling factor) to 100 MeV for all the SEDs.
  \item A log parabola model(LP), whose functional form is
  \begin{equation}
   \frac{dN}{dE} = N_0(\frac{E}{E_0})^{-(\alpha+\beta\log(E/E_0))}
  \end{equation}
  where, $\alpha$ \& $\beta$ are the photon index \& curvature index respectively. Scaling factor ($E_0$) is kept fixed to 300 MeV, near the low energy part 
  of the spectrum( "ln" is the natural logarithm). 
  \item A broken-powerlaw model(BPL), whose functional form is
  \begin{equation}
    \frac{dN}{dE} = N_0
    \begin{cases}
      (\frac{E}{E_b})^{-\Gamma_1}, & \text{for}\ E < E_b \\
      (\frac{E}{E_b})^{-\Gamma_2}, & \text{otherwise}
    \end{cases}
  \end{equation}
  where $E_b$ is the break energy.
\end{enumerate}  

The values of the fitted parameters for these spectral models (PL, LP \& BPL) have been elucidated in Table-17 - Table-29. We have also mentioned the 
log(Likelihood) value for all the epochs and calculated $\Delta$log(Likelihood) value from that, which is defined by the difference between the log(Likelihood) 
value for logparabola/broken-powerlaw model and simple powerlaw model.

Figure-32 shows the SEDs of the sub-structure of Flare-1A for two different phases: Flare \& Post-flare. Here cyan, black \& magenta color indicate the fitting of 
spectral points with the Powerlaw (PL), Log-parabola (LP) and Broken-powerlaw model (BPL) respectively. The values of the fitted parameters for the different periods of activity for these models (PL, LP, BPL) have been given in Table-17.

The SEDs of the flaring epochs for all the three sub-structures (Flare-2A, Flare-2B, Flare-2C \& Flare-2D) of Flare-2 have been illustrated in Figure-33, Figure-34, Figure-35 and Figure-36 respectively. All of these sub-structure except Flare-2A show the spectral hardening with increasing flux. Spectral index ($\Gamma$) is nearly constant (for PL model) with changing flux in Flare-2A (shown in Table-18). For Flare-2D, in Pre-flare phase index $\Gamma$=2.41$\pm$0.01, then it changes to 2.33$\pm$0.01 for Plateau-I phase, to 2.27$\pm$0.00 and 2.29$\pm$0.00 for Flare-I \& Flare-II phases respectively, which have been described in Table-21. The values of the fitted parameters for Flare-2B and Flare-2C have been displayed in Table-19 \& Table-20 respectively.

A significant spectral hardening is observed during Flare-3A when the source transits from Pre-flare to Flare-I \& Flare-II phase, whereas during Flare-3B the 
spectral hardening is not much significant. The SEDs of these substructures have been shown in Figure-37 and Figure-38 and the corresponding values of the parameters have been given in Table-22 \& Table-23 respectively.

Flare-4A shows the spectral softening when source travels from preflare to flare epoch and spectral index changes from $\Gamma$=2.27$\pm$0.01 to 2.32$\pm$0.00 which is described in Table-24. Two (Flare-4B \& Flare-4D) out of four sub-structures show significant spectral hardening when the source transits from low flux state to high flux state which have been described in Table-25 \& Table-27. The SEDs of different epoch of Flare-4A, Flare-4B, Flare-4C and Flare-4D have been illustrated in Figure-39, Figure-40, Figure-41 and Figure-42 respectively. Table-26 describe the modelling parameter values of SEDs of different period for Flare-4C. 

 A clear indication of spectral hardening is also seen in both sub-structures (Flare-5A \& Flare-5B) of Flare-5. In Flare-5A, a significant change in $\Gamma$ (2.60$\pm$0.01 to 2.11$\pm$0.00) has been noticed during Pre-flare to Flare epoch. The SEDs of different periods of activity of these sub-structures have been given in Figure-43 \& Figure-44 respectively. The values of the fitted parameters have been elucidated in Table-28 \& Table-29. 

From the above $\gamma$-ray SED analysis of 3C 454.3 source, we find spectral hardening is an important feature of this source.
This has been noticed before by \citet{Britto et al. (2016)} during MJD 56570 - 56863. Only one sub-structure (Flare-4A) shows spectral softening 
 during change of state from Pre-flare to Flare. The values of the reduced $\chi^{2}$ for the different spectral models (PL, LP, BPL) have been provided in Table-30, which show that LP is the best fitted model for most of the flaring states.

\section{MULTI-WAVELENGTH STUDY OF 3C 454.3}
This section is dedicated to multi-wavelength study of blazar 3C 454.3. We have chosen the brightest flaring state (Flare-2) of 3C 454.3 from whole 9 year
$\gamma$-ray light curve (shown in Figure-1). We have also collected the simultaneous multi-wavelength data from other instruments and analyzed it. The simultaneous data from other wavebands are: X-ray, Ultraviolet (UV) \& Optical from Swift-XRT and UVOT telescope. We have divided this Flare-2 state in four regions labeled as Flare-2A, Flare-2B, Flare-2C \& Flare-2D with time duration of MJD 55045-55140, MJD 55140-55201, MJD 55250-55356 \& MJD 55467-55600 respectively which are shown in Figure-45 based on gamma-ray flux as mentoined in sub section-4.2. Flare-2A and Flare-2D has the simultaneous observation in gamma-ray, X-ray, optical, and UV wavebands and hence for further study we have concentrated on Flare-2A and Flare-2D. Multi-wavelength light curve of Flare-2 has been shown in Figure-45.

\subsection{MULTI-WAVELENGTH LIGHT CURVE}  

Figure-46 shows the multi-wavelength light curve of Flare-2A with time span of 95 days (MJD 55045 - 55140). In the upper most panel six hour binning of $\gamma$-ray data has been shown and corresponding X-ray, Optical \& UV data have been shown in 2nd, 3rd \& 4th rows respectively. In the $\gamma$-ray light curve flux 
started rising slowly with small fluctuation. The maximum flux was recorded as 6.69$\pm$0.79 at MJD 55091.375 and then the flux decayed slowly with small variation. The average flux of decay period during MJD 55095.6 - 55140.125 is 2.94$\pm$0.05. It is also observed that when the flux was still increasing in gamma-ray, the source started flaring in X-ray, optical and UV bands. In Swift-XRT dataset maximum peak was observed at MJD 55070.37 with flux of (8.66$\pm$0.96)$\times10^{-11}$ $erg$ $cm^{-2}$ $s^{-1}$. During MJD 55094 - 55140 data is not available in XRT photon counting (PC) mode. Similarly Optical \& UV data have been also plotted and brightest peak was found at MJD 55069.91 with fluxes of 3.47$\pm$0.13, 3.07$\pm$0.08, 2.71$\pm$0.11, 2.03$\pm$0.09, 2.04$\pm$0.08 \& 1.74$\pm$0.06 in V, B, U, W1, M2 \& W2 band respectively, which are in units of $10^{-11}$ $erg$ $cm^{-2}$ $s^{-1}$. In V, U, W1, M2 the second brightest peak was observed at MJD 55091.18 which coincided with the first $\gamma$-ray brightest peak with time lag of $\sim$ 5 hour, while in B band maximum peak was noticed at MJD 55090.92 with time lag of $\sim$ 11 hour.

The multi-wavelength light curve of Flare-2D has been shown in Figure-47 which has a time duration of 133 days (MJD 55467.125 - 55600.125). The highest flux
was recorded at MJD 55519.875 with flux of 80.41$\pm$5.93 in six hour bin $\gamma$-ray waveband. After this the flux started decreasing slowly with small
variation during MJD 55536.6 - 55590.1 and the average flux was 9.80$\pm$0.06 . We are unable to observe any peak in XRT PC mode due to unavailability 
of the simultaneous data during MJD 55504 - 55554.
All the peaks in optical \& UV band nearly coincide with the peaks observed in $\gamma$-ray band. Interestingly, the peaks in optical \& UV band 
during MJD 55510.1-55511.4, have no $\gamma$-ray counterpart which has been reported earlier in several cases (\citet{Vercellone et al. (2011)}, \citet{Rajput et al. (2019)}).

\subsection{MULTI-WAVELENGTH SED MODELLING}  
We have Modelled the multi-wavelength SEDs with the time dependent `GAMERA' (\citet{Hahn (2015)}) code, which is publicly available in github 
webpage\footnote{https://github.com/libgamera/GAMERA}. This code solves the time dependent continuity equation, calculates the evolved electron 
spectrum N(E,t) and then computes the synchrotron \& inverse Compton emission for that N(E,t). The continuity equation is given by -
\begin{equation}\label{eq:7}
 \frac{\partial N(E,t)}{\partial t} = Q(E,t)-\frac{\partial }{\partial E}(b(E,t)N(E,t))-\frac{N(E,t)}{\tau_{esc}(E,t)}   
\end{equation}

where Q(E,t) is the injected electrons spectrum. The energy loss rate is denoted by b(E) = $(\frac{dE}{dt})$ and $\tau_{esc}(E,t)$ represents the escape 
time. 
Log-Parabolic model gives the best fitted parameters for most 
of the sub-structures in the $\gamma$-ray SED, which have been described in Table-30.
 The radiative losses of LP electron spectrum produce LP photon spectrum. 
\citep{Massaro et al. (2004)} gave a general formalism to show that if the efficiency of acceleration decreases with increasing energy, the resulting shock electron spectrum follows LP distribution. Due to this reason we have assumed LP form of Q(E,t). Functional form of Q(E,t) is defined by -
  \begin{equation}\label{eq:8}
   Q(E) = l_0(\frac{E}{E_{ref}})^{-(\alpha+\beta\log(E/E_{ref}))}
  \end{equation}
 where $l_0$ is normalization constant \& $E_{ref}$ is the reference energy which is set at 90.0 Mev.
`GAMERA' uses the full Klein-Nishina cross section to compute the 
inverse-Compton (IC) emission \citet{Blumenthal and Gould (1970)}.

\subsection{PHYSICAL CONSTRAINTS}
Here we discuss about the constraints on the model parameters that we have used in our modelling
\begin{enumerate}[label=(\roman*)]
\item To calculate the EC emission by the relativistic electrons the CMB (Cosmic Microwave Background) and BLR (Broad Line Region) photons are taken into account as target photons. The standard value of CMB photon density (0.25 eV$cm^{-3}$, \citet{Longair (1974)}) has been used. The energy density of BLR photons is computed (in the comoving frame) with the following equation 

\begin{equation}
 U^\prime_{BLR} = \frac{\Gamma^2\zeta_{BLR}L_{Disk}}{4\pi cR^2_{BLR}}   
\end{equation}

where $\Gamma$ is the bulk Lorentz factor of the emitting blob whose value is assumed to be 20 \citep{Vercellone et al. (2011)}. The BLR photon energy density
is only a fraction of 10\% ($\zeta_{BLR} \sim 0.1$) of the accretion disk photon energy density. The value of the disk luminosity $L_{Disk} = 6.75\times10^{46}
erg/sec$ is taken from \citealt{Bonnoli et al. (2010)}. We have computed the radius of the BLR region by the scaling relation $R_{BLR} = 10^{17}L^{1/2}_{d,45}$ where 
$L_{d,45}$ \citep{Ghisellini and Tavecchio (2009)} is the disk luminosity in units of $10^{45} erg/sec$.

\item We have also included emission from accretion disk component in the code to compute the EC emission. We constrain the disk energy density in the comoving jet frame by the following equation (\citet{Dermer and Menon (2009)}) 
  
\begin{equation}\label{eq:10}
 U^\prime_{Disk} = \frac{0.207 R_{g}L_{Disk}}{\pi c Z^3 \Gamma^2}   
\end{equation}

We chose the mass of the central engine or Black Hole ($M_{BH}$) as $5\times10^8 M_\odot$ (\citet{Bonnoli et al. (2010)}) in order to estimate the 
gravitational radius $R_g$ = 1.48$\times10^{14}$ cm. Distance of the emission region from the black hole is represented by 'Z'. The upper limit of this quantity is estimated by the given equation \citep{Paliya et al. (2015)} -

\begin{equation}
 Z \leqslant \frac{2 \Gamma^{2} c t_{var}}{1+z}   
\end{equation}

Where $z$ is the redshift of the source. The variability time estimated during Flare-2 is found to be $t_{var}$ = 1.93 hour during MJD 55068.125 - 55068.375,
(corresponds to Flare-2A) which has been used to estimate `Z'. The value of 'Z' is estimated as $Z$ $\sim$ 1.0$\times10^{17}$ cm.

\item We can estimate the upper limit  on the size of the emission region $R$ with the following relation - 
  
\begin{equation}\label{eq:12}
 R \leqslant \frac {c t_{var} \delta}{1+z}.   
\end{equation}

We have used $t_{var}$ = 1.93 hour \& Doppler factor $\delta$ = 27.5 (comparable to \citet{Bonnoli et al. (2010)}) for Flare-2 which give the value of 
$R$ = $3.08\times10^{15}$ cm.  But it is noted that equation ($\ref{eq:12}$) is just an approximation and there are several effects that may introduce large 
error in determining $R$ \citep{Protheroe (2002)}. Moreover the value of $R$ calculated in this way for $\gamma$-ray wavelength does not give a good fit to the data in our SED modelling. In our work we have chosen $R = 3.0\times10^{16}$ cm which is comparable to the value 5$\times10^{16}$ cm given by \citealt{Gupta et al. (2017)}.

\item We have used typical values of BLR temperature ($T^\prime_{BLR}$) \& Disk temperature ($T^\prime_{Disk}$) in our model, which are 2.0$\times10^{4}$ K
and 1.0$\times10^{6}$ K respectively.

\end{enumerate}

\subsection{MODELLING THE SEDs}
After constraining the above model parameters we have simulated the multi-wavelength SED using the code `GAMERA'. We have included the escape term
($-\frac{N(E,t)}{\tau_{esc}(E,t)}$) for electrons in the continuity equation ($\ref{eq:7}$) and considered two different cases -
\begin{itemize}

\item Case 1: In this case we have studied the model with constant escape time which is $\tau_{esc} \sim R/c$, where $R$ is the size of the 
emission region.
\item Case 2: Next we consider energy dependent escape time which is given by $\tau_{esc} = \eta E^{-0.5}$ \citep{Sinha et al. (2016)}. We have chosen the following values
$\eta$ $\sim$ 387.0 \& 155.0 sec MeV$^{1/2}$ for Flare-2A and Flare-2D respectively, so that at low energy the escape time is comparable to the cooling time of electrons. 
\end{itemize}
SED modelling has been done for the above two cases for both the flares (Flare-2A \& Flare-2D), which have been illustrated in Figure-48 and Figure-49. We have shown the non-simultaneous archival data for both the flares in cyan colour represented by plus symbol, which are taken from \citet{Abdo et al. (2010c)}. There are no simultaneous archival data available for Flare-2A.  However, quasi-simultaneous data from MJD 55515 - 55524 \citep{Vercellone et al. (2011)} and for MJD 55519 \citep{Jorstad et al. (2013)} are available for Flare-2D and they are shown with black triangle and green star symbol in Figure 48 \& 49. In our work the SED is averaged over the whole flaring period, i.e. 133 days from MJD  55467 - 55600. However, the SED data points shown in black and green colour are for the peak of the flare, which lasted for very short period compared to our period (133 days) and hence our SED data points differ from them.

In our study we have adjusted the values of the following parameters to obtain the best fitted model: Magnetic field in the emission region 
($B$), minimum \& maximum Lorentz factor of the injected relativistic electrons ($\gamma_{min} \& \gamma_{max}$) and their spectral index ($\alpha$) \& curvature 
index ($\beta$). We have obtained the values of $B$ = 3.80 \& 2.30 G for Flare-2A and Flare-2D respectively, by fitting the synchrotron emission of the relativistic electrons to the optical data. The value of spectral index ($\alpha$) is 2.00 \& 2.18 for Flare-2A and Flare-2D respectively. For Flare-2A the value of minimum Lorentz factor ($\gamma_{min}$) is 55 \& 45 in case-1 and case-2 respectively. For Flare-2A \& Flare-2D there is no significant difference in the values of the the maximum Lorentz factor of the injected electrons ($\gamma_{max}$), however curvature index ($\beta$) varies significantly for Flare-2D ($\beta$ = 0.09 for case-1 and $\beta$ = 0.14 for Case-2) whereas it remains similar for Flare-2A. The detailed results of the multi-wavelength SED modelling have been described in Table-31.

We have also calculated the total required jet power by using the following equation -
\begin{equation}\label{eq:13}
 P_{tot} = \pi R^2 \Gamma^2 c (U^\prime_e+U^\prime_B+U^\prime_p)   
\end{equation}

where $U^\prime_e$, $U^\prime_B$ \& $U^\prime_p$ are the energy density of electrons (and positrons), magnetic field \& cold protons respectively in the
comoving jet frame. The power carries by the injected electrons in the jet is given by-
\begin{equation}\label{eq:14}
 P_e = \frac{3\Gamma^{2}c}{4R}\int_{E_{min}}^{E_{max}} E Q(E) dE   
\end{equation}
where $Q(E)$ is the injected electron spectrum as defined in equation (\ref{eq:8}).
To compute $U^\prime_p$ we have assumed the ratio of electron-positron pair to proton number in the emission region is 10:1. From 
equation ($\ref{eq:13}$) we have calculated the maximum required jet power ($P_{tot}$) in our model which is found to be 3.04$\times10^{46}$ erg/sec. This value
is lower than the estimated range of Eddington's luminosity ($L_{Edd}$) - $(0.6-5)\times10^{47}$ erg/sec (\citet{Bonnoli et al. (2010)}, \citet{Gu et al. (2001)},
\citet{Khangulyan et al. (2013)}).

\section{MODELLING THE LIGHT CURVE}

Our SEDs represent the time averaged flux over a very long time period. This is why the average values of the model parameters (Doppler factor, magnetic field, luminosity in injected electrons, blob size, viewing angle) are used in modelling the SEDs of Flare 2A and Flare 2D. Light curves represent the photon fluxes at different time epochs. The time variation of photon fluxes, representing complicated structures, reflects instantaneous perturbations in the emission zone. 
Time dependent modelling of blazars (\citet{Saito et al. (2015)}; \citet{Potter (2018)}) has been used earlier to simulate the photon fluxes with time. 
Simulated profiles of flares of PKS 1510-089  were analysed in optical, X-ray, high energy and very high energy gamma-ray for timescale of hours \citep{Saito et al. (2015)}.
Simultaneous multi-wavelength data is not available at different frequencies to test their model predictions.
The time variation of some of the parameters involved in modelling may generate time dependent photon flux to mimic the flare peaks in the light curves over short time intervals.

Here we discuss about the modelling of the $\gamma$-ray light curves using multi-wavelength SED parameters. We have chosen short duration flare peaks of the following three types,  $T_r > T_d$ , $T_r < T_d$ , and $T_r \sim T_d$, since the long duration peaks have much more complex time dependent structures. These peaks are: Peak P5 ($T_r > T_d$), P1 ($T_r < T_d$) \& P3 ($T_r \sim T_d$). Since Flare 2B (Figure-7, MJD 55152 - 55177) has many peaks which include all the three types of peaks $T_r=T_d$, $T_r>T_d$ and $T_r<T_d$, we have chosen three different types of peaks of this flare. We have modelled the light curves by varying separately the Doppler factor ($\delta$) and the normalisation constant of the injected electron flux ($l_0$). While doing this we fixed the other SED model parameters ($E_{min}$, $E_{max}$, B, R, etc.) to their average values as used for SED modelling of Flare-2A since Multi-wavelength data of Flare 2B is not available for SED modelling.
\begin{enumerate}[label=(\roman*)]
\item Case 1: In this case we calculate the light curve by varying the Doppler factor as a function of time which goes as broken-powerlaw 
  \begin{equation}\label{eq:15}
    \delta =
    \begin{cases}
      kt^{a_1}, & \text{for}\ t < t_c \\
      k{t_c}^{(a_1-a_2)}t^{a_2}, & \text{otherwise}
    \end{cases}
  \end{equation}
where $t_c$ is peak time, $k$ is normalization constant and $a_1, a_2$ are the indices of the broken-powerlaw.

 The blob is boosted to a higher Doppler factor which causes the rise in photon flux and then slows down during the decay phase. Due to poor photon statistics a detailed modelling of the time variation of the Doppler factor is not possible at this stage. We have calculated the integrated $\gamma$-ray flux in the Fermi LAT energy range (0.1$\leqslant$E$\leqslant$300 GeV) in each time step from the multi-wavelength SED model and fitted to the light curve data points.
Our results are shown in Figure-50, Figure-51 \& Figure-52 for the three types of flare peaks. The best fitted model parameters in \ref{eq:15} along with the range of values of the Doppler factor ($\delta$) have been displayed in Table-32.
 
\item Case 2: In this case we fix the Doppler factor to its average value of Flare-2A but vary the normalisation constant ($l_0$) of the injected electron flux (equation \ref{eq:8}) with a functional form similar to $\delta$, which is defined by-
 \begin{equation}\label{eq:16}
   l_0 =
   \begin{cases}
    kt^{a_3}, & \text{for}\ t < t_c \\
    k{t_c}^{(a_3-a_4)}t^{a_4}, & \text{otherwise}
   \end{cases}
 \end{equation}
where $t_c$ is peak time, $k$ is normalization constant and $a_3, a_4$ are the indices of the broken-powerlaw.

The normalisation constant of the injected electron flux in the emission region increases which causes the peak in the light curve and subsequently it decreases when the photon flux diminishes. Similar to Case 1, it is not possible to get more accurate result on time variation of normalisation constant $l_0$ due to poor photon statistics.  We have calculated the integrated $\gamma$-ray flux from our SED model in each time step to obtain the simulated light curve as before. 
Our simulated light curves of these peaks have been shown in Figure-53, Figure-54 and Figure-55 respectively. The best fitted values of the parameters in equation \ref{eq:16} along with the ranges in the values of the normalisation constant ($l_0$) and injected power in electrons ($P_e$) for each flare peak have been given in Table-33.

\end{enumerate}

Thus we show that the light curves can be  approximately generated by varying the Doppler factor ($\delta$) or the normalisation constant ($l_0$).

\section{DISCUSSION}
3C 454.3 is one of the most violent source in Fermi 3FGL catalog. We have analyzed the light curve of this source in $\gamma$-ray for 7 day time bin during 
August 2008 - July 2017, which consists of five major flares as shown in Figure-1. Each major flare comprises of several sub-structures (or sub-flares) which are identified in 
1 day \& 6 hour binning analysis. All the sub-structures show different phases of activity (e.g. Pre-flare, Flare, Plateau, Post-flare). Flare regions
of each sub-structure consist of several distinctive peaks (labeled as P1, P2 etc.) of different photon counts. 
Only one sub-structure Flare 1A has been identified in Flare-1. The light curve of Flare-1  is shown in Figure-2 for 1 day binning, which shows Flare and Post-flare phases.
The peaks P1, P2 and P3 are identified in Figure-3 for the flare phase of Flare-1A. The gamma-ray SED data points are fitted with log-parabola, broken-powerlaw and powerlaw functions to find which function gives the best fit to the data. The same procedure has been carried out for all the flares subsequently for 6 hour binning except Flare-2A where we have used same binning as Flare-1A.
Table-30 shows that in most cases the gamma-ray SEDs of flares are well represented by log-parabola function. The scanning of 6 hour binning light curve is done to estimate the variability time scale in gamma-ray emission. The results are displayed in Table-15 \& Table-16. 
The shortest variability time is found to be hour scale (1.70$\pm$0.38).
The rise and decay timescales of flares are studied to see whether they follow any trend.
Characteristic rising \& decay timescales ($T_r$ \& $T_d$) have been computed for each peak which are shown in Table 1 - Table 13. We have found that the values 
of $T_r$ and $T_d$ vary between hour to day scale for different peaks. To compare these two timescales ($T_r$ \& $T_d$) we define a quantity $K$, which is given by 
\citep{Abdo et al. (2010b)} 

\begin{equation}\label{eq:17}
 K = \frac{T_d-T_r}{T_d+T_r}   
\end{equation}  
Depending on the value of $K$ there may be three different possibilities as discussed below,

\begin{itemize}
\item Rising timescale is greater than decay timescale ($T_r > T_d$) when $K < -0.3$. This may happen when injection rate is slower than the cooling rate of
electrons into the emission region. The electrons can lose energy through IC \& synchrotron cooling.  
\item Decay timescale is greater than rising timescale ($T_r < T_d$) when $K > 0.3$. This could be due to longer cooling timescale of electrons. 
\item Nearly equal rising and decay timescale ($T_r \sim T_d$) or symmetric temporal evolution when $-0.3 \leqslant K \leqslant 0.3$. This property can be 
explained by perturbation in the jet or a dense plasma blob passing through a standing shock front in the jet region \citep{Blandford and Konigl (1979)}.
\end{itemize} 
In our study we have found that out of total 69 prominent peaks 16 peaks have $T_r > T_d$, 20 peaks have $T_r < T_d$ and 33 peaks have $T_r \sim T_d$. 
Earlier,  a similar study was done for PKS 1510-089 with 8 years of Fermi LAT data \citet{Prince et al. (2017)}. The rise and decay times were presented in Tables 1-5 and plotted 
in Figure-27 of  \citet{Prince et al. (2017)}. For most of the peaks the decay time was found to be shorter than the rise time.
CTA102 , another flaring FSRQ was studied for a much shorter period Sep, 2016 - March, 2017 \citet{Prince et al. (2018)}. During its flaring state 14 peaks were identified. Out of these 5 peaks had nearly equal decay and rise time, 5 peaks had slower rise time than decay time, and 4 peaks had slower decay time than rise time.
These results suggest that the decay and rise time of peaks do not have any specific trend for flaring FSRQs.

The SED modelling has been done for the two flares Flare-2A and Flare-2D for which multi-wavelength data are available. The modelling has been done with the time dependent code GAMERA, which solves the transport equation for electrons including their energy losses by synchrotron and inverse Compton emission (SSC and EC), and escape. We have considered two cases for the escape timescale (i) constant escape time $R/c=10^6$ sec and (ii) energy dependent escape time, which goes as $E^{-0.5}$.
We note that the cooling timescale of the electrons in case (i) is much shorter than $R/c$ in our case. In case (ii) the escape time is comparable to the cooling time for low energy electrons but for high energy electrons the cooling time decreases faster than the escape time as it goes as $E^{-1}$. Table-31 shows the results of our SED modelling. The results for the two cases are comparable for both the flares Flare-2A and Flare-2D. Magnetic field is slightly higher for Flare-2A. The jet power in relativistic electrons and positrons is higher  for Flare-2D compared to Flare-2A. Figure-48 and Figure-49 show the results of our SED modelling. If we divide the duration of a flare into four equal time intervals the SED calculated for each time interval overlaps with each other. The electron spectrum becomes steady in a short time compared to the durations of Flare-2A and Flare-2D, as a result their radiated photon spectrum also becomes steady. Due to this reason it is not possible to see the time evolution in Figure-48 and Figure-49.

3C 454.3 being highly variable in gamma rays is often monitored.  The data from July 7-October 6 of 2008 was analysed to study the flaring activity during this period \citet{Abdo et al. (2009)}. They observed nearly symmetric flares with rise and decay timescales of 3.5 days. They obtained a lower bound of 8 on the value of Doppler factor. Their gamma-ray SED is best represented by a broken powerlaw with a break near 2 GeV. They suggested this break may be due to an intrinsic break in the electron spectrum.
\citet{Finke and Dermer (2010)} suggested a combination of the Compton scattered disk and BLR radiation to explain the spectral break and also fit 
the quasi-simultaneous  radio, optical, X-ray and gamma-ray data of 2008 flare. \citet{Hunger and Reimer (2016)} used a particle distribution with 
a break to model the flare emission with Compton-scattered BLR radiation alone and also in combination with Compton-scattered disc emission. 
\citet{Kohler and Nalewajko (2015)} studied many short bright flares of blazars including 3C 454.3. They concluded that  the average Fermi-LAT 
spectrum is a superposition of many short lived components where each one having different spectral curvature.
 In our work in many cases (see Table-30) Log-parabola function well represents the gamma-ray SEDs of flares.
  
 While modelling the two flares Flare-2A and Flare-2D we have assumed the emission region to be in the BLR region, which is commonly assumed
  in single zone SED modelling. However, for many of the flares a more complicated and realistic scenario may be required to explain the temporal and 
  spectral features. 
 
 Due to its variable nature this source should be monitored for high energy neutrino emission during its flaring states. High energy neutrinos can escape from the jets even if they are produced in the inner regions of jets. In this case high energy neutrinos may be detected without counterparts in high energy gamma-rays.  
  More simultaneous multi-wavelength data and constraint on neutrino flux from IceCube detector would be useful to model the flares, constrain their hadronic jet power and locate the emission regions of the flares.
  
Below we clarify some important points on our analysis.

The `sum of exponentials' (equation \ref{eq:2}) is the function that the `Blazar community' uses to model the peaks observed in a light curve. 
We have performed the fitting in python with `curvefit' package.
The number of exponentials is chosen based on the number of peaks observed in a particular light curve. The first exponential function is used to fit the rising part of the peak and this gives the rising time. Similarly, the second exponential is used to fit the decaying part of the peak which gives the decay time. The rise and decay time of the peak play an important role for calculating the variability time which is used to do the SED modeling. 
In equation \ref{eq:2}, we have four parameters but among them two parameters peak flux ($F_0$) and corresponding time $t_0$ are fixed from observation and we have varied $T_r$ and $T_d$ to get the best fit value. 
However, the fitting of light curves with the `sum of exponentials'  does not always give very good result. There could be many reasons if the fit is not good. It may be due to low statistics and large error bars on the data points. There is also the possibility that the flux is changing so fast that it is impossible to catch that flux value with any smooth function. The high value of reduced chi-square could also be because of rapid variations in flux (small peaks) which have not been included during the fitting.

A statistical `mixture model' decision process can be used to choose the number of components from the fitted light curve for more sophisticated analysis. But in our case the peaks can be clearly identified from eye inspection. Moreover, we mostly use the brighter peak (where the flux is a few times higher than its initial value in a short duration of time) to estimate the variability time so it does not matter if we leave out some small peaks in our fitting, which will of course increase the chi-square value.

One can also use nonparametric density estimation approach, smoothing the time series with a (Gaussian) kernel or locally fitting with polynomials (e.g. splines).

In our analysis the binning of light curve is not arbitrary as it is based on how good the data is (TS value of each data points). If the source is very bright during a flare and the flux is very high, in this case there is a chance of having good statistics and hence we can bin the light curve up to a minute time scale. This has been done for many flares of various sources previously (see \cite{Shukla et al. (2018)}). In our case we have focussed on 6-hour binning because for this binning the data has good statistics (TS $\sim$ 25; $\sim5\sigma$ significance) and also each and every peak can be clearly identified.

\section{CONCLUSION}
We have identified five flares in the 9 year gamma-ray light curve of 3C 454.3. After scanning the light curve the shortest variability timescale is found to be of hour scale, which is similar to other flaring FSRQs e.g. PKS 1510-089. The gamma ray spectral energy distributions of the flares are in most cases best fitted with Log-parabola function. Similar result was also found earlier for PKS 1510-089. The rise and decay times of flares do not follow any particular trend, in some cases they are equal, but in some other cases they are not. Flare-2D (MJD 55467-55600) is found to be the most violent sub-structure in the 9 year light curve history of this source with six different phases of activity: Pre-flare, Plateau-I, Flare-I, Flare-II, Plateau-II and Post-flare. The most basic sub-structures are having only three phases of activity: Pre-flare, Flare and Post-flare. We have done time dependent leptonic  modelling of Flare-2A and Flare-2D with multi-wavelength data. The magnetic field required to model these flares are 3.8 Gauss and 2.3 Gauss respectively, which are comparable to the magnetic fields found from SED modelling of other blazars e.g. PKS 1510-089 and CTA 102. The jet powers required to model these flares are below the Eddington's luminosity of 3C 454.3.
 In future simultaneous multi-wavelength observations and constraint on neutrino flux from IceCube detector would be useful to understand the composition of the jets and the location of flare.
 
\textbf{Acknowledgements :} 
We thank the referee for helpful comments to improve this paper.
This work has made use of publicly available Fermi LAT data obtained from FSSC's website data server and provided by NASA Goddard Space Flight Center. This work has also made use of data, software/tools obtained from NASA High Energy Astrophysics Science Archive Research Center (HEASARC) developed by Smithsonian astrophysical Observatory (SAO) and the XRT Data Analysis Software (XRTDAS) developed by ASI Science Data Center, Italy. N. Gupta thanks C. S. Stalin and S. Sahaynathan for helpful discussions.
A.K. Das thanks T. Ghosh for proof reading.

\software{Fermitools \href{https://fermi.gsfc.nasa.gov/ssc/data/analysis/software/}{(https://fermi.gsfc.nasa.gov/ssc/data/analysis/)}}\\
GAMERA \href{https://github.com/libgamera/GAMERA}{(https://github.com/libgamera/GAMERA)}
HEASARC \href{https://heasarc.gsfc.nasa.gov/docs/software/heasoft/download.html}{(https://heasarc.gsfc.nasa.gov/docs/software/heasoft/)}
XSPEC \href{https://heasarc.gsfc.nasa.gov/xanadu/xspec/}{(https://heasarc.gsfc.nasa.gov/xanadu/xspec/)}

XSELECT \href{https://heasarc.gsfc.nasa.gov/ftools/xselect/}{(https://heasarc.gsfc.nasa.gov/ftools/xselect/)}

\begin{figure*}[h]
\centering
\includegraphics[height=2.6in,width=5.3in]{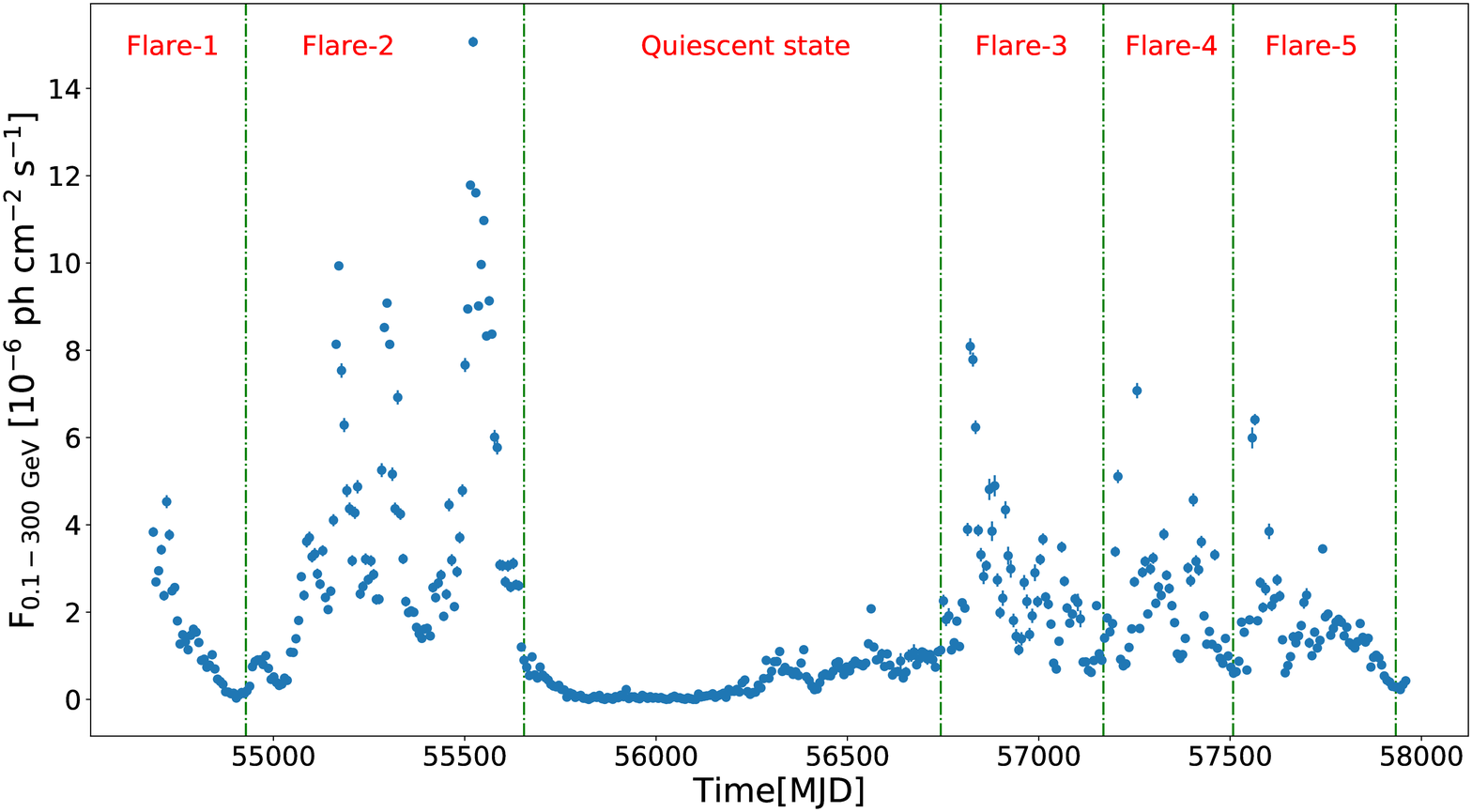}
\caption[optional]{Seven-day binning light curve of 3C 454.3 (MJD 54686-57959). We have identified five major flares (shown by broken green line).}
\includegraphics[height=2.6in,width=5.3in]{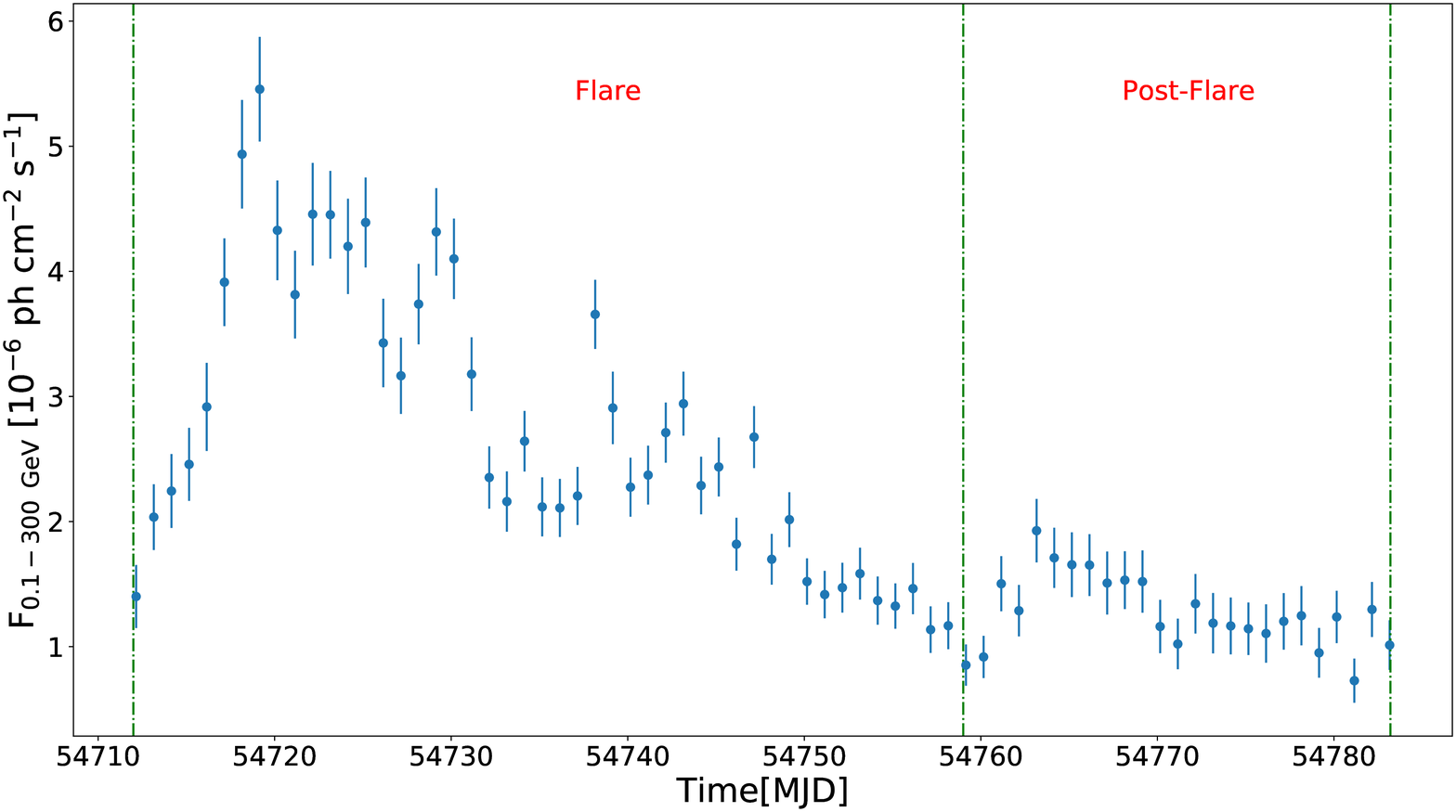}
\caption[optional]{One day binning light curve for Flare-1A. Time durations of all the different periods of activities (shown by broken green line) are: MJD 54712-54759 (Flare), MJD 54759-54783 (Post-Flare).}
\includegraphics[height=2.6in,width=5.3in]{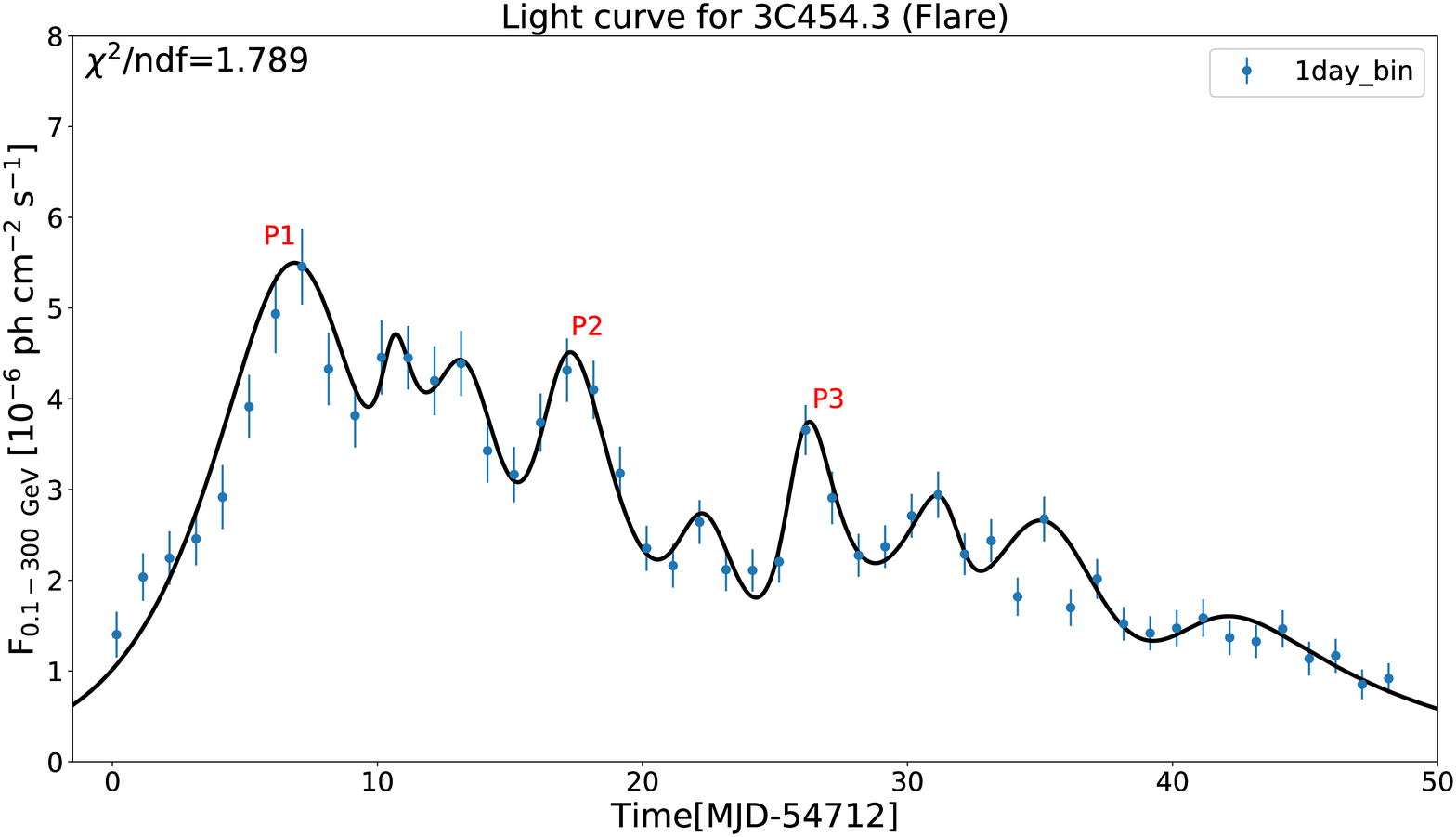}
\caption[optional]{Fitted light curve (fitted by the sum of exponential function) of Flare-1A of Flare (MJD 54712-54759) epoch.}
\label{fig:steps}
\end{figure*}

\begin{figure*}[h]
\centering

\includegraphics[height=2.6in,width=5.3in]{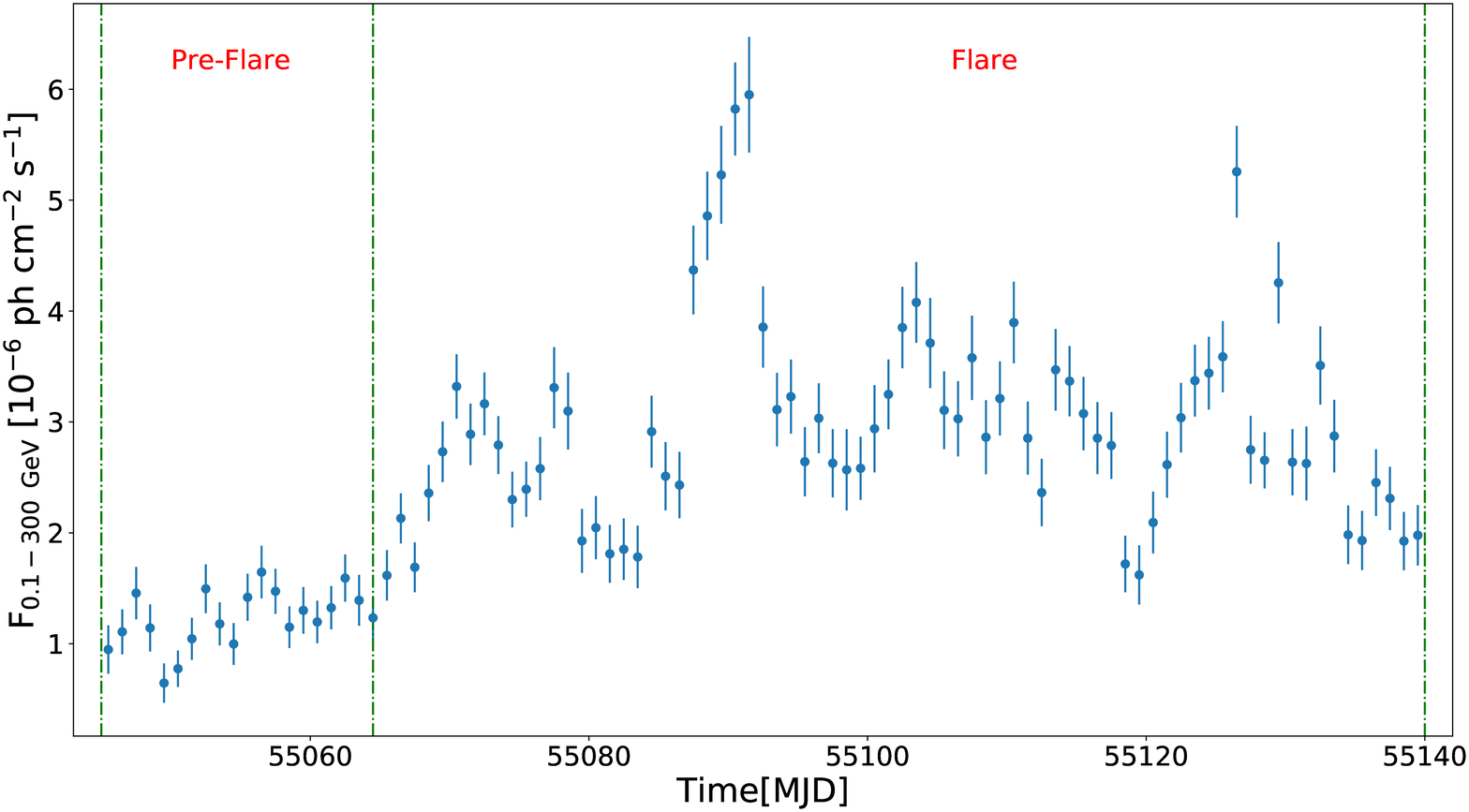}
\caption[optional]{One day binning light curve for Flare-2A. Time durations of all the different periods of activities (shown by broken green line) are: MJD 55045-55064 (Pre-Flare), MJD 55064-55140 (Flare).}

\includegraphics[height=2.6in,width=5.3in]{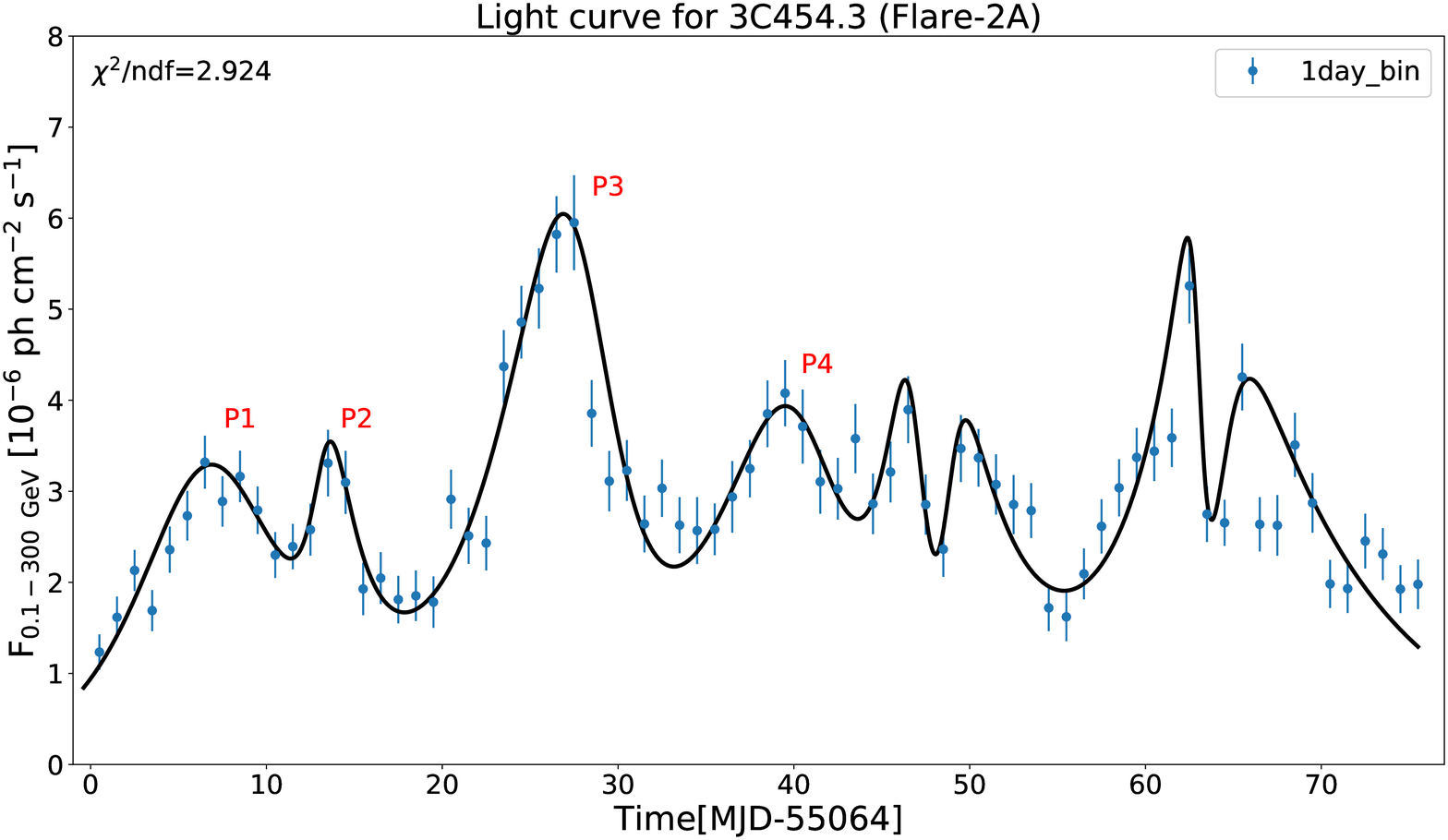}
\caption[optional]{Fitted light curve (fitted by the sum of exponential function) of Flare-2A of Flare (MJD 55064-55140) epoch.}

\includegraphics[height=2.8in,width=5.4in]{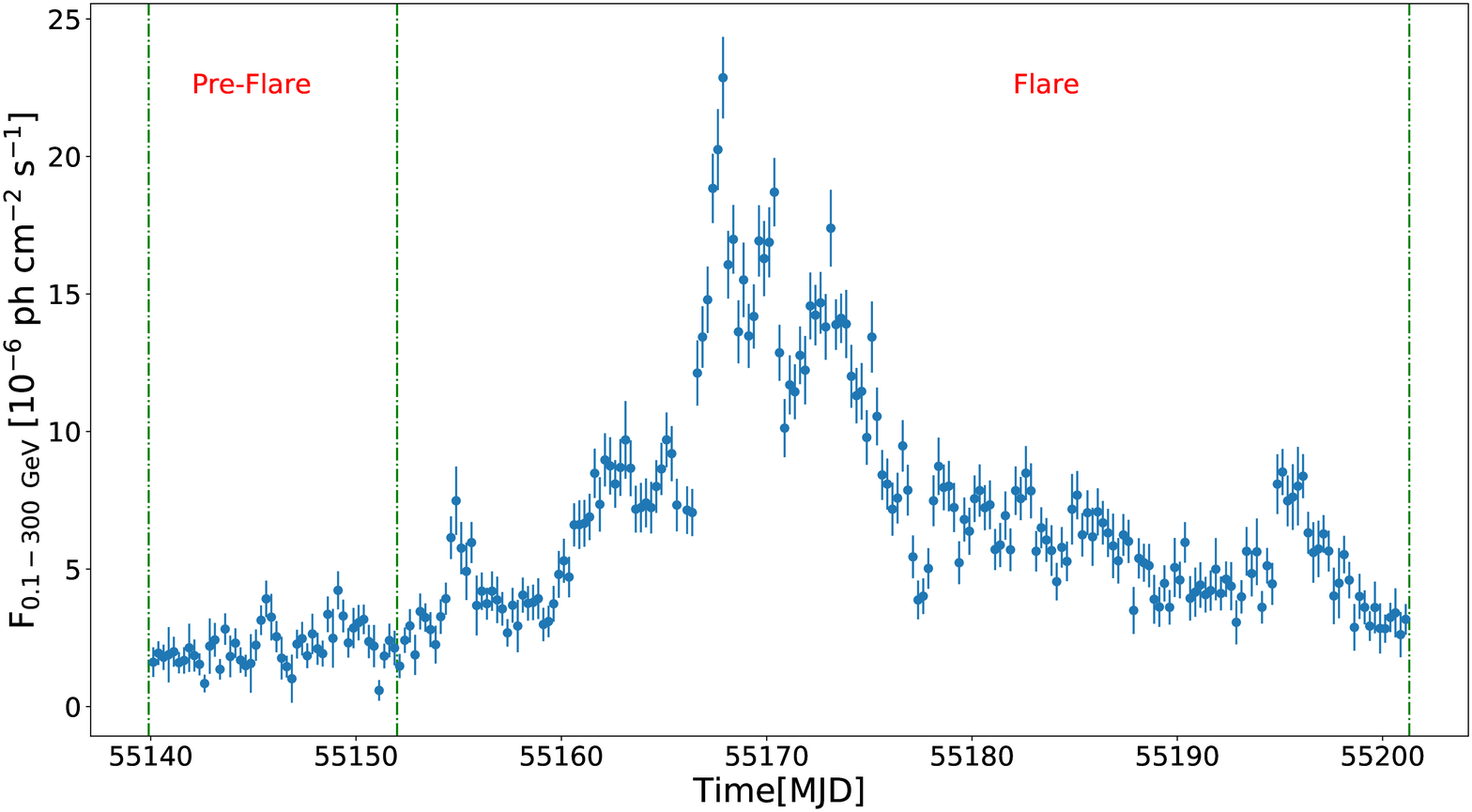}
\caption[optional]{Six-hour binning light curve for Flare-2B. Time durations of all the different periods of activities (shown by broken green line) are: MJD 55140-55152 (Pre-flare), MJD 55152-55201 (Flare).}
\end{figure*}

\begin{figure*}[h]
\centering

\includegraphics[height=2.8in,width=5.4in]{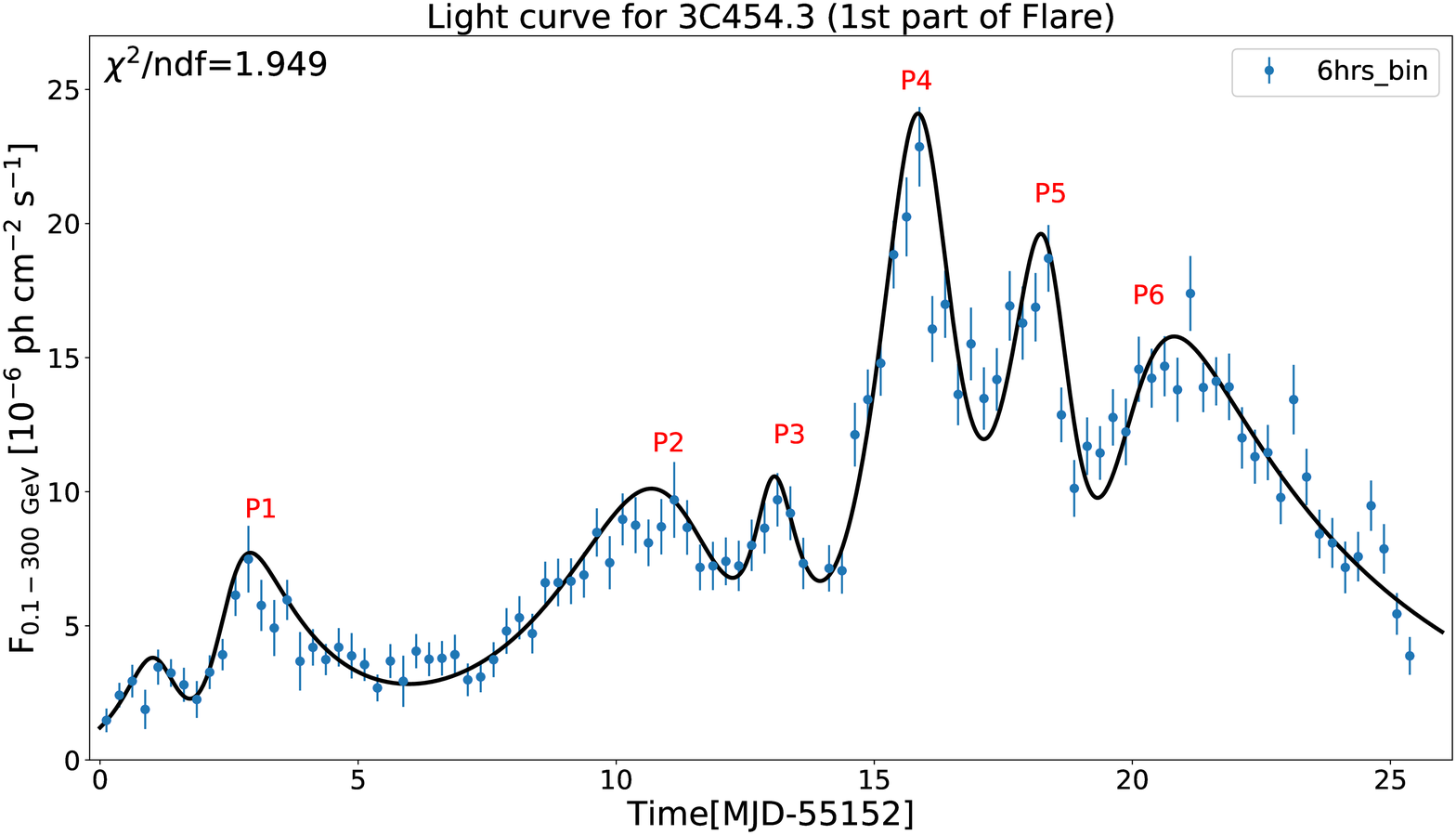}
\caption[optional]{Fitted light curve (fitted by the sum of exponential function) of Flare-2B for 1st part of Flare (MJD 55152-55177) epoch.}

\includegraphics[height=2.8in,width=5.4in]{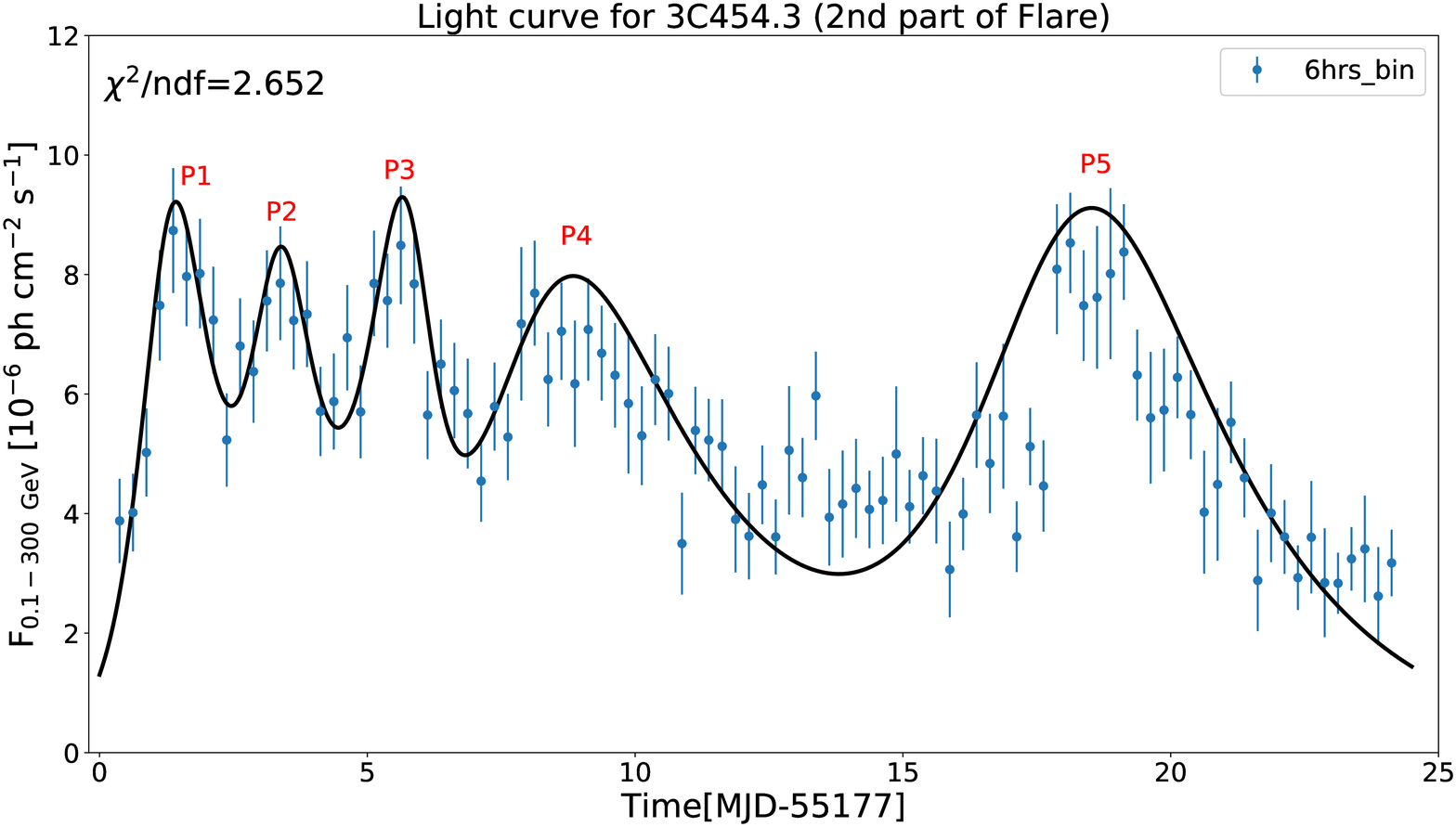}
\caption[optional]{Fitted light curve (fitted by the sum of exponential function) of Flare-2B for 2nd part of Flare (MJD 55177-55201) epoch.}

\includegraphics[height=2.8in,width=5.4in]{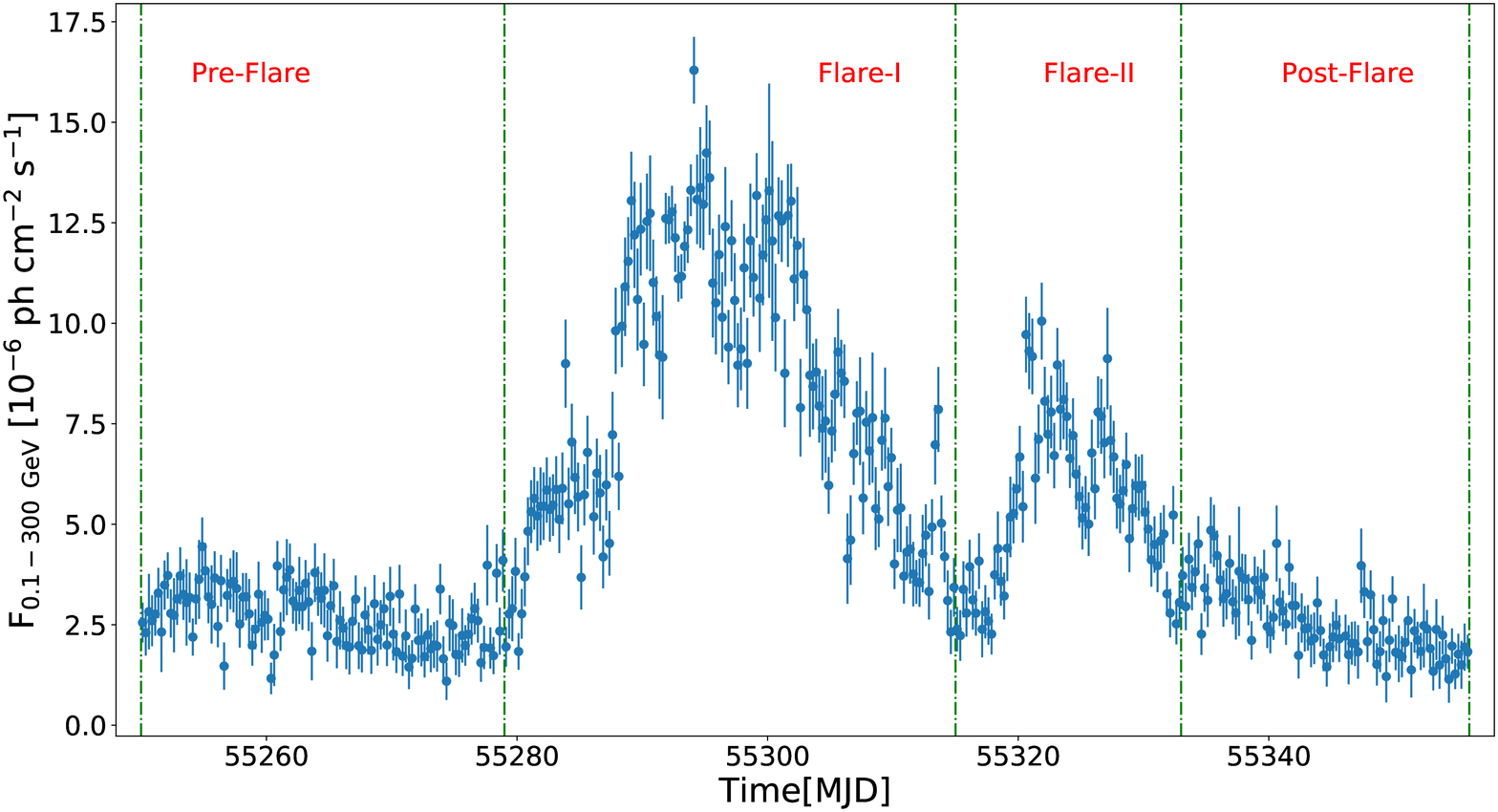}
\caption[optional]{Six-hour binning light curve for Flare-2C. Time durations of all the different periods of activities (shown by broken green line) are: MJD 55250-55279 (Pre-flare), MJD 55279-55315 (Flare-I),MJD 55315-55333 (Flare-II) and MJD 55333-55356 (Post-flare).}

\end{figure*}

\begin{figure*}[h]
\centering

\includegraphics[height=2.8in,width=5.4in]{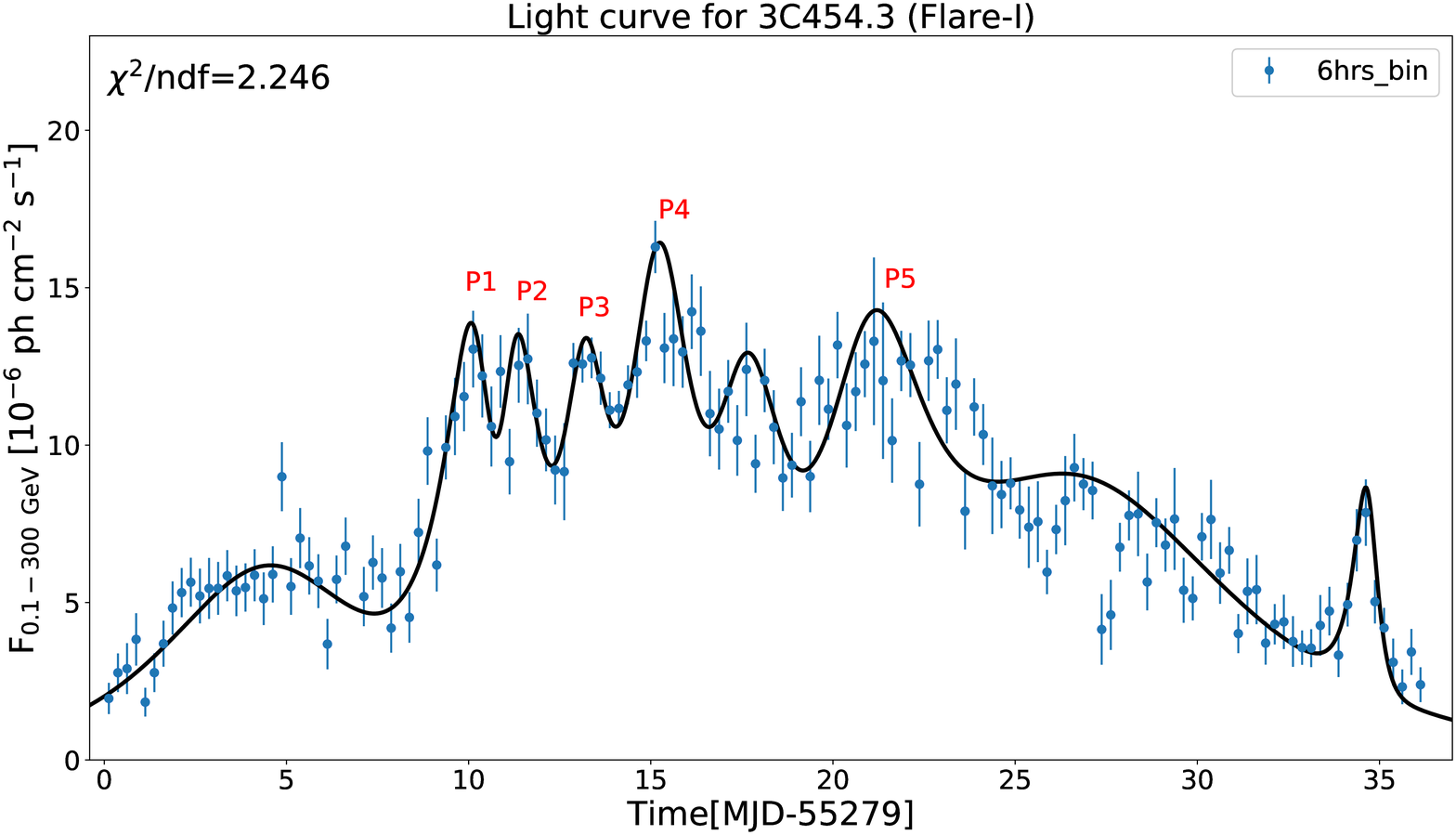}
\caption[optional]{Fitted light curve (fitted by the sum of exponential function) of Flare-2C for Flare-I (MJD 55279-55315) epoch.}

\includegraphics[height=2.8in,width=5.4in]{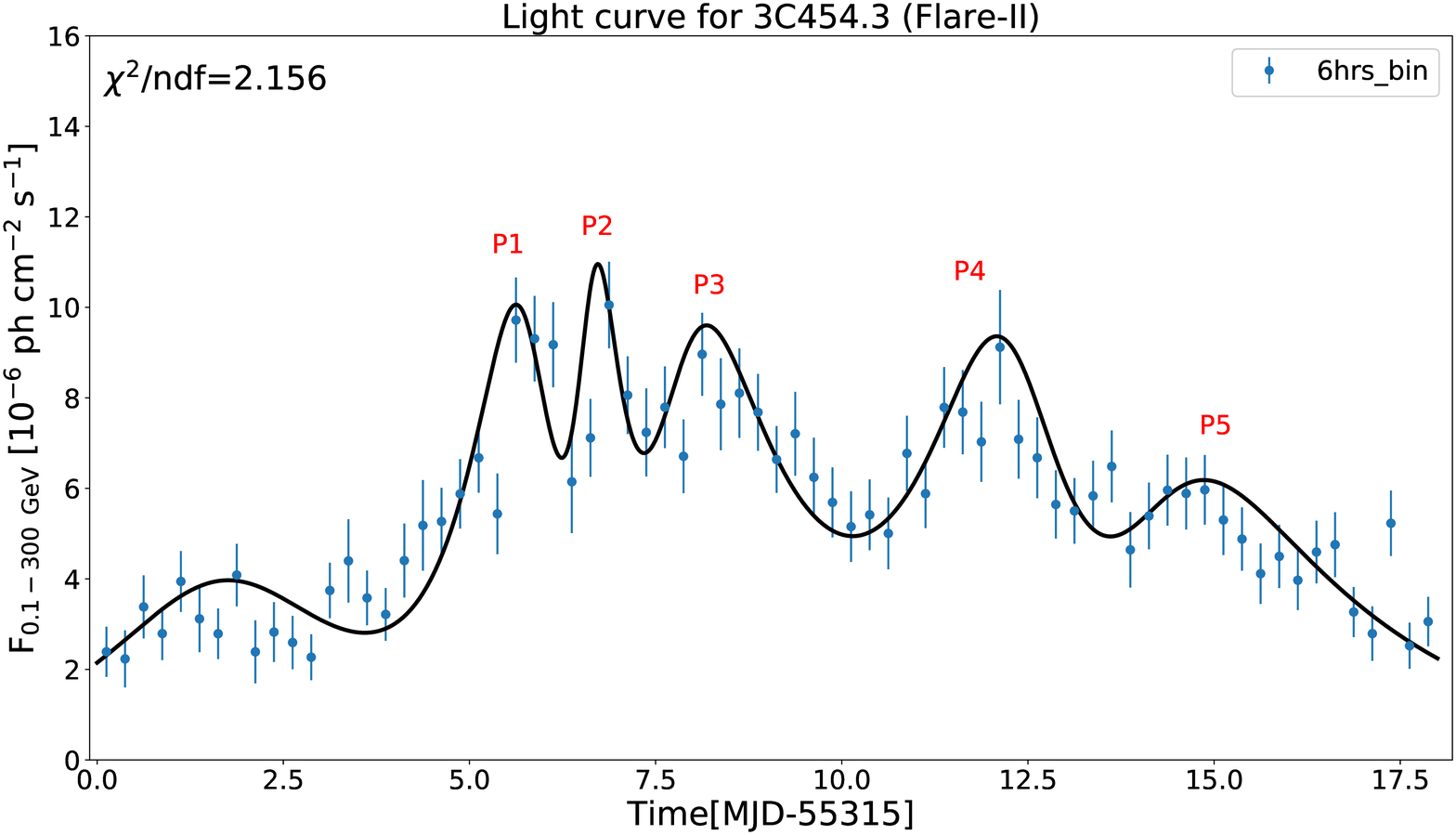}
\caption[optional]{Fitted light curve (fitted by the sum of exponential function) of Flare-2C for Flare-II (MJD 55315-55333) epoch.}

\includegraphics[height=2.8in,width=5.4in]{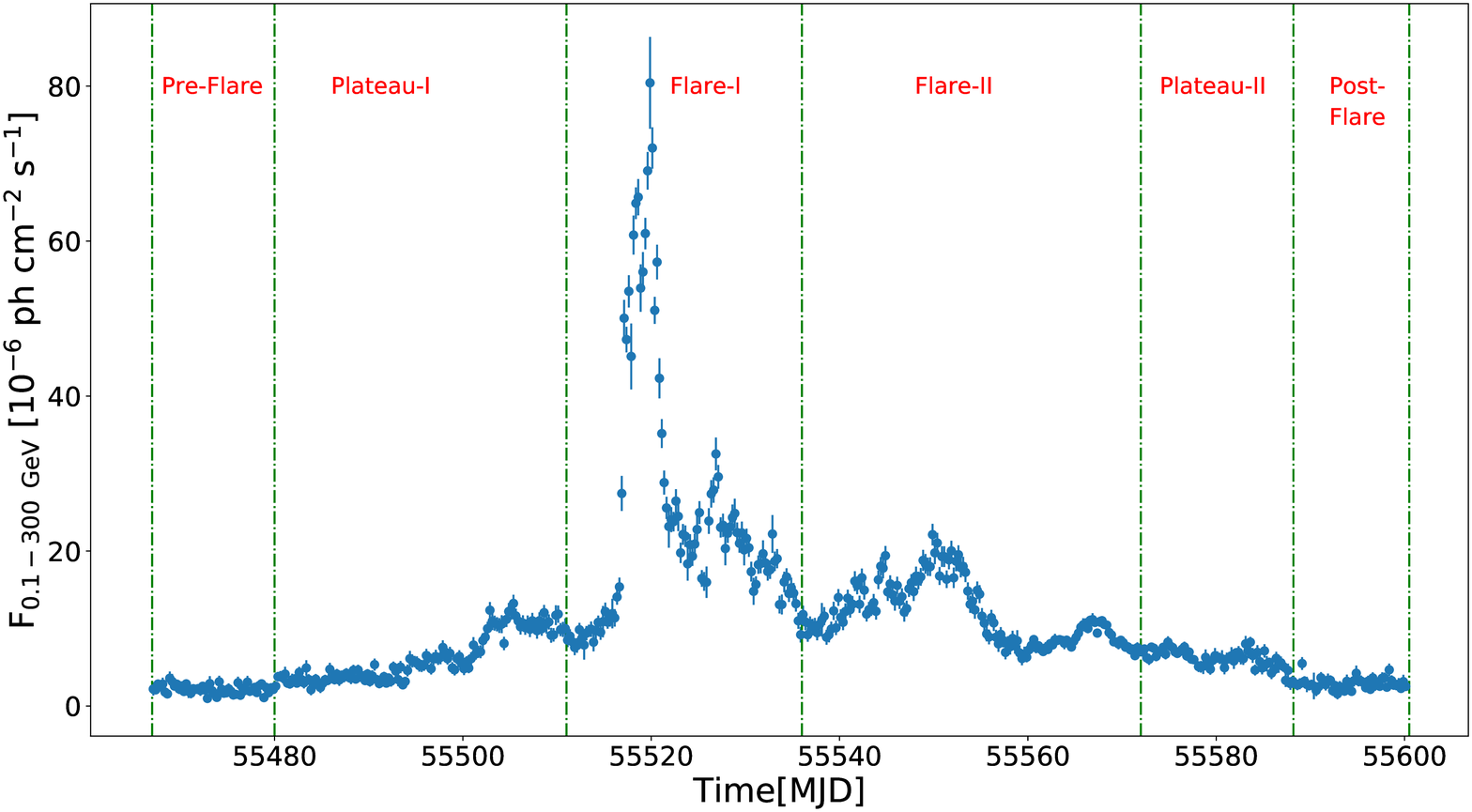}
\caption[optional]{Six-hour binning light curve for Flare-2D. Time durations of all the different periods of activities (shown by broken green line) are: MJD 55467-55480 (Pre-flare), MJD 55480-55511 (Plateau-I), MJD 55511-55536 (Flare-I) and MJD 55536-55572 (Flare-II), 55572-55588 (Plateau-II), 55590-55600 (Post-flare).}

\end{figure*}

\begin{figure*}[h]
\centering

\includegraphics[height=2.8in,width=5.4in]{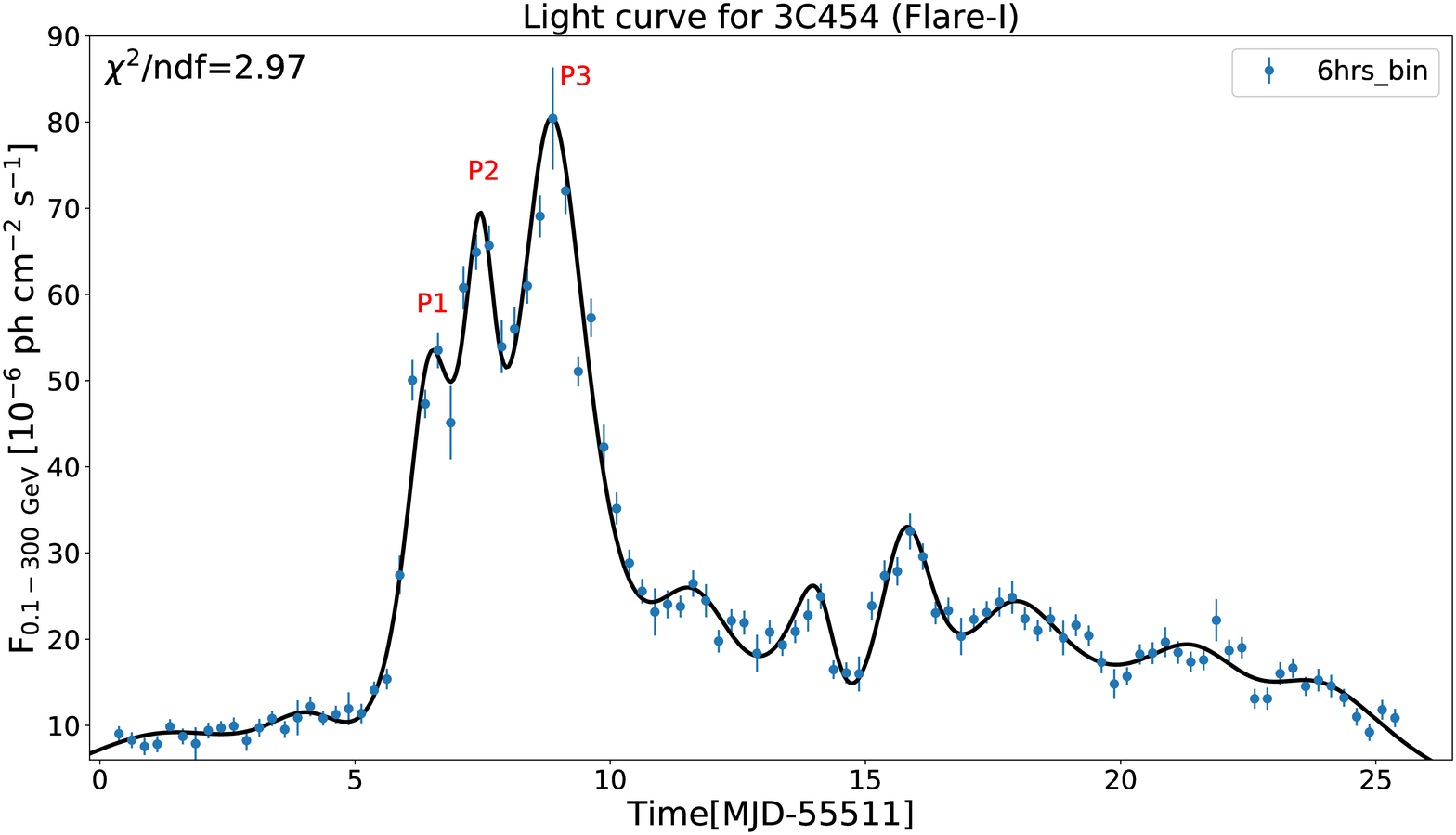}
\caption[optional]{Fitted light curve (fitted by the sum of exponential function) of Flare-2D for Flare-I (MJD 55511-55536) epoch.}

\includegraphics[height=2.8in,width=5.4in]{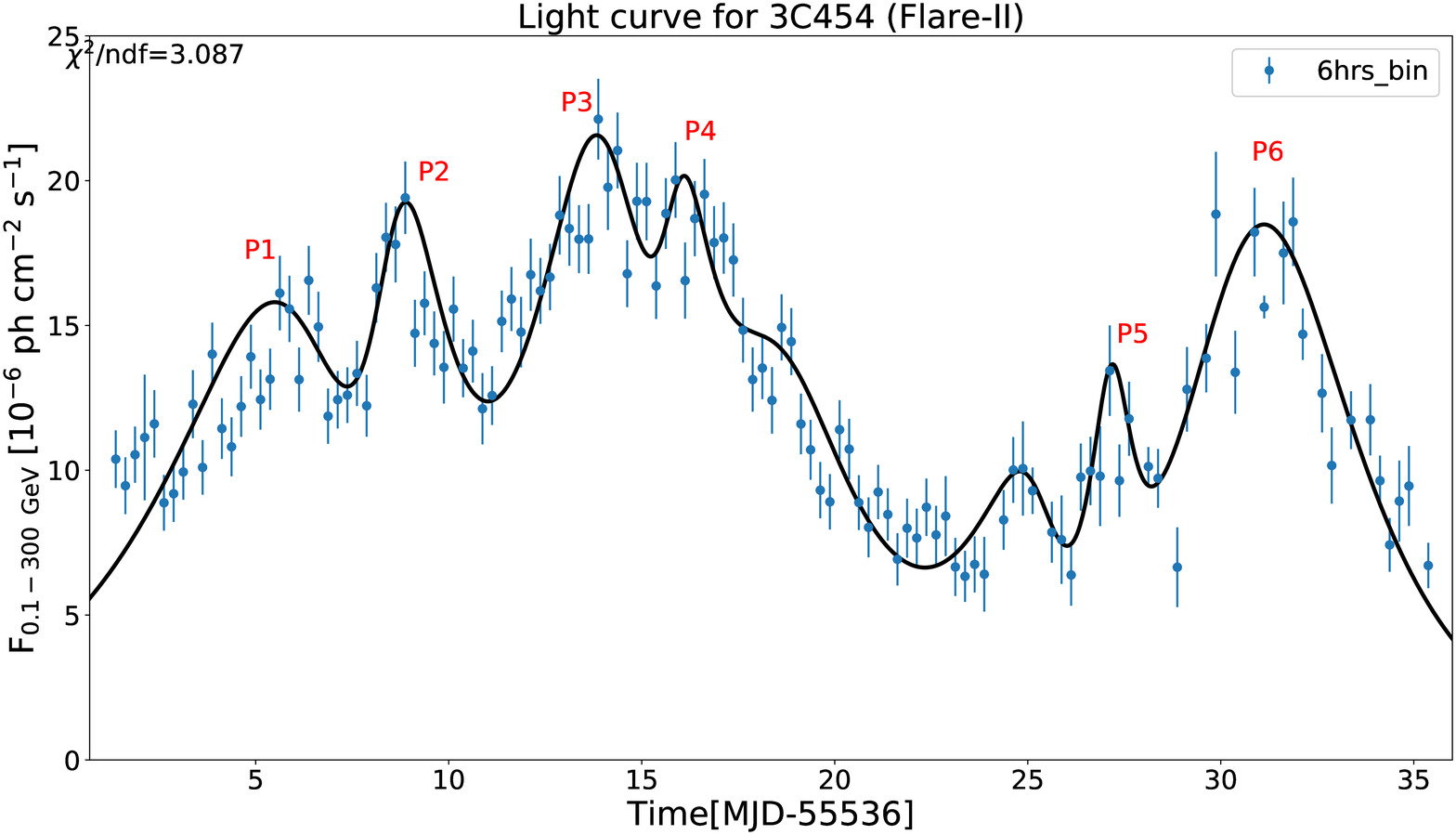}
\caption[optional]{Fitted light curve (fitted by the sum of exponential function) of Flare-2D for Flare-II (MJD 55536-55572) epoch.}

\includegraphics[height=2.6in,width=5.3in]{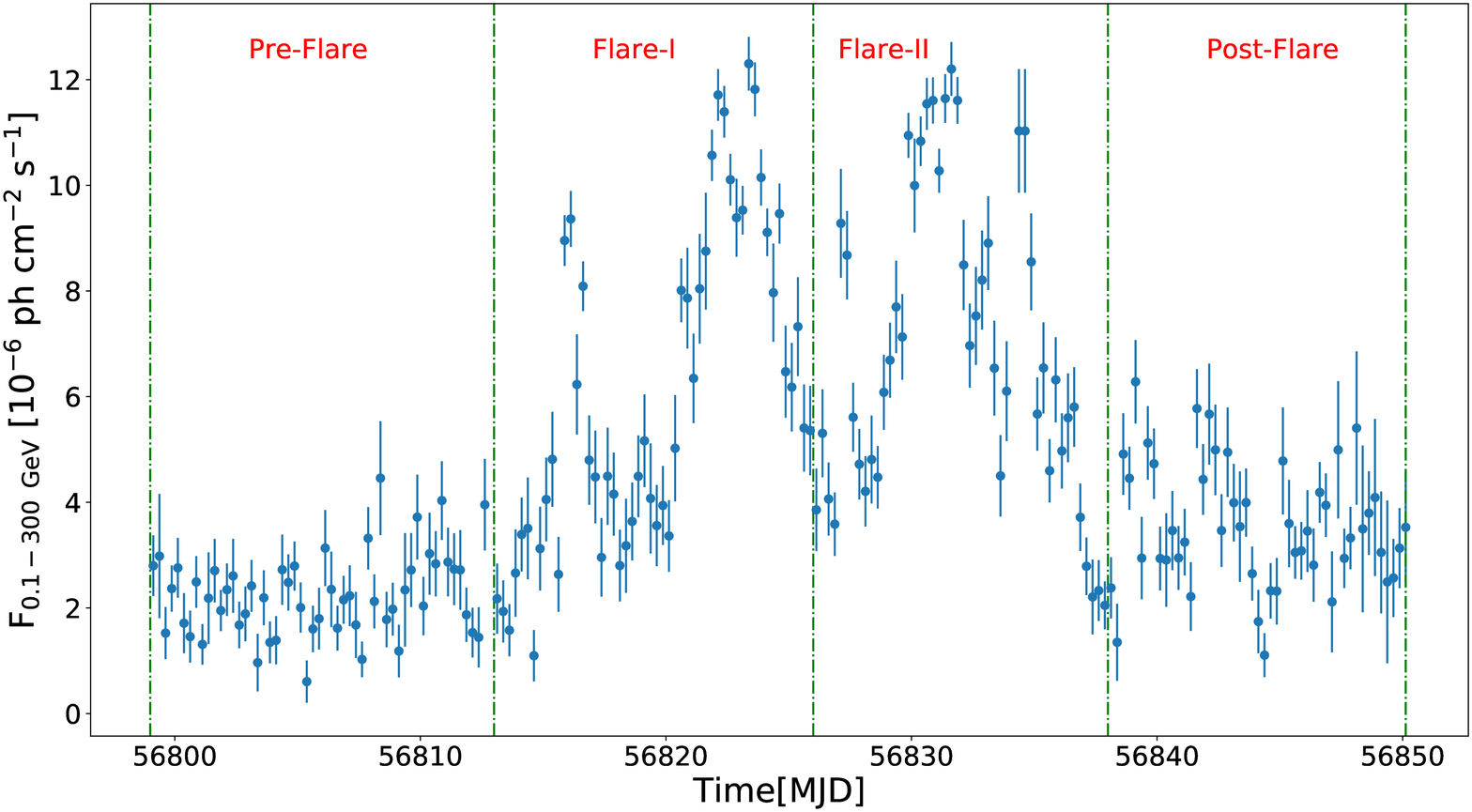}
\caption[optional]{Six-hour binning light curve for Flare-3A. Time durations of all the different periods of activities (shown by broken green line) are: MJD 56799-56813 (Pre-flare), MJD 56813-56826 (Flare-I), MJD 56826-56838 (Flare-II) and MJD 56838-56850 (Post-flare).}

\end{figure*}

\begin{figure*}[h]
\centering

\includegraphics[height=2.6in,width=5.3in]{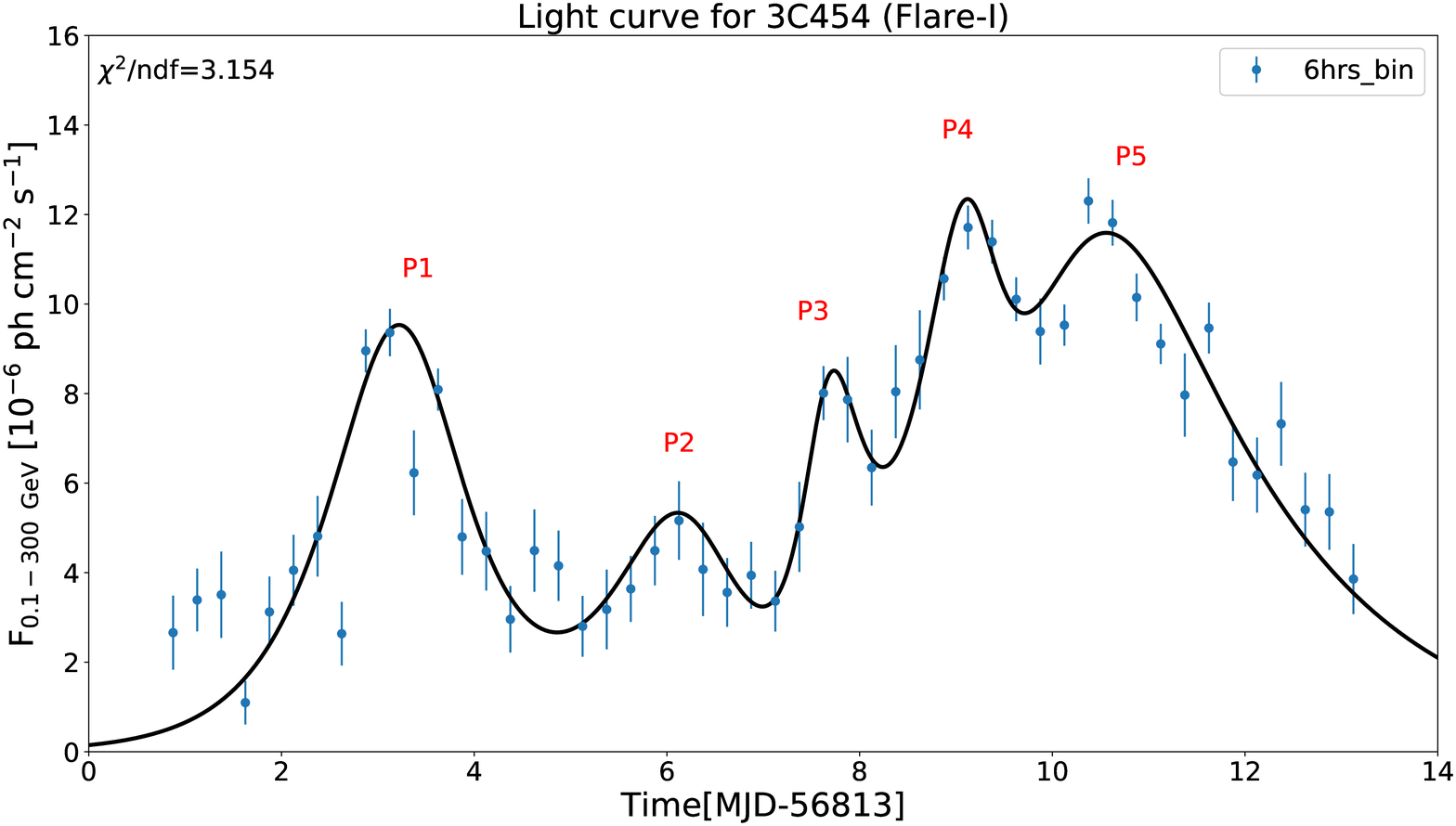}
\caption[optional]{Fitted light curve (fitted by the sum of exponential function) of Flare-3A for Flare-I (MJD 56813-56826) epoch.}

\includegraphics[height=2.5in,width=5.3in]{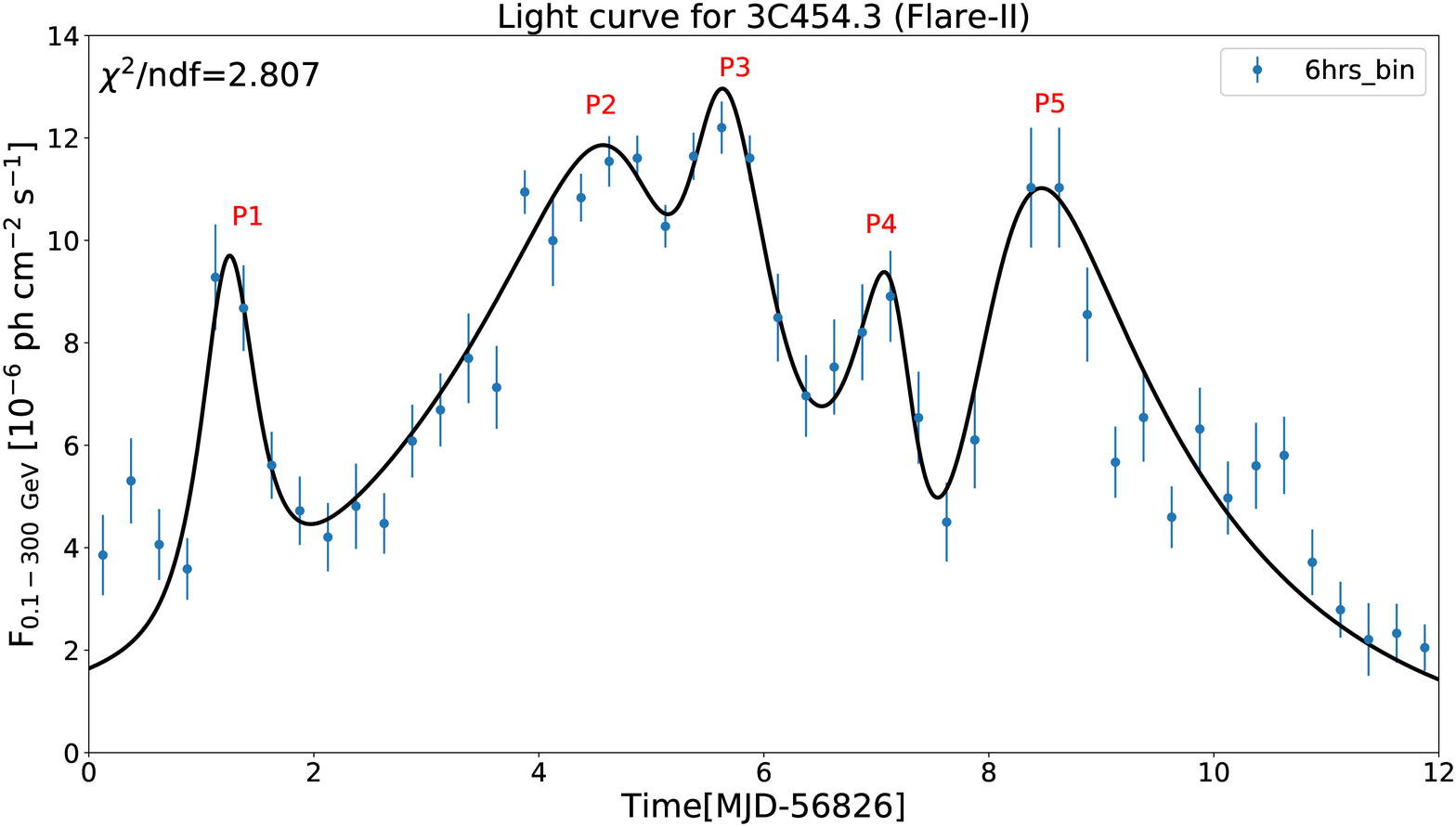}
\caption[optional]{Fitted light curve (fitted by the sum of exponential function) of Flare-3A for Flare-II (MJD 56826-56838) epoch.}

\includegraphics[height=2.6in,width=5.3in]{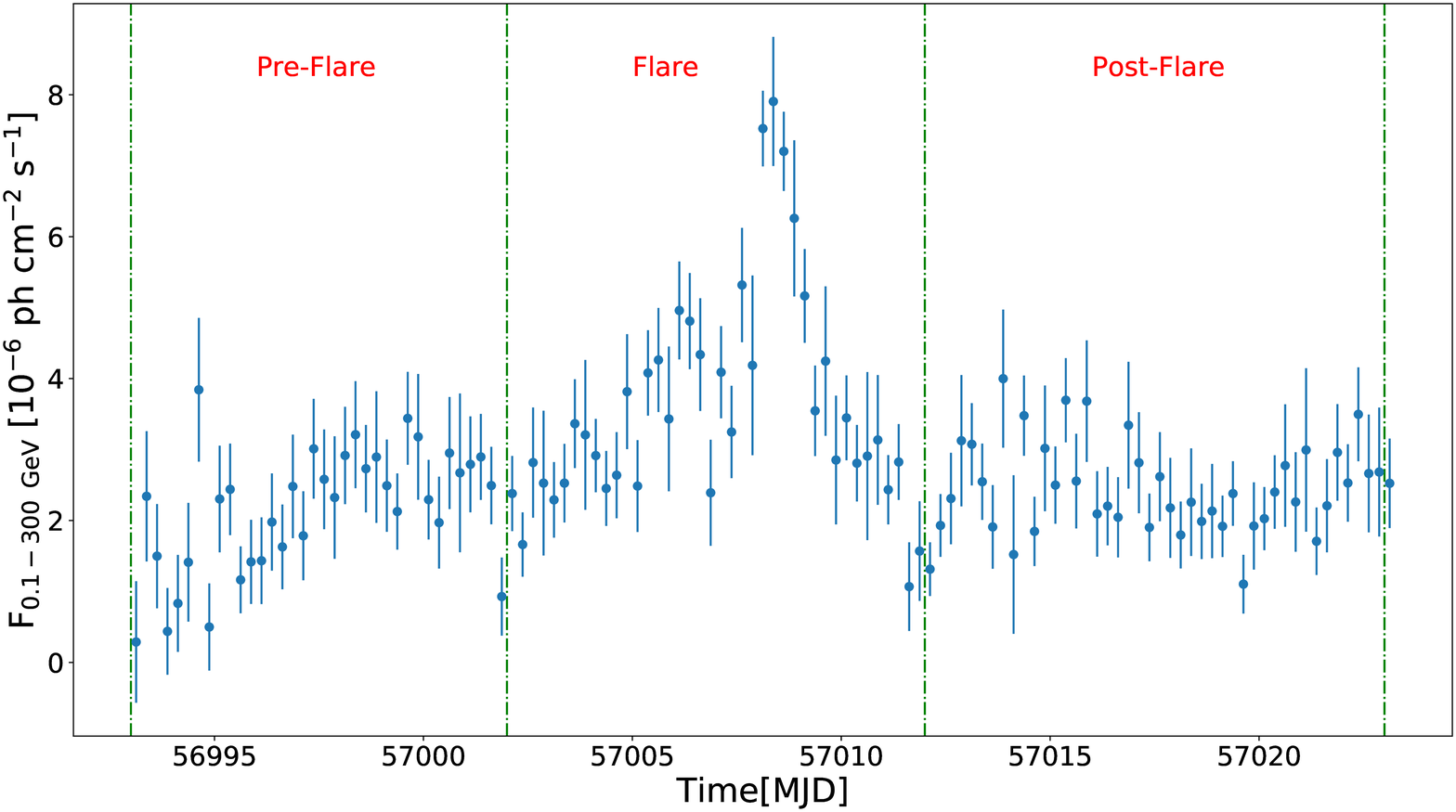}
\caption[optional]{Six-hour binning light curve for Flare-3B. Time durations of all the different periods of activities (shown by broken green line) are: MJD 56993-57002 (Pre-flare), MJD 57002-57012 (Flare) and MJD 57012-57023 (Post-flare).}

\end{figure*}

\begin{figure*}[h]
\centering

\includegraphics[height=2.5in,width=5.3in]{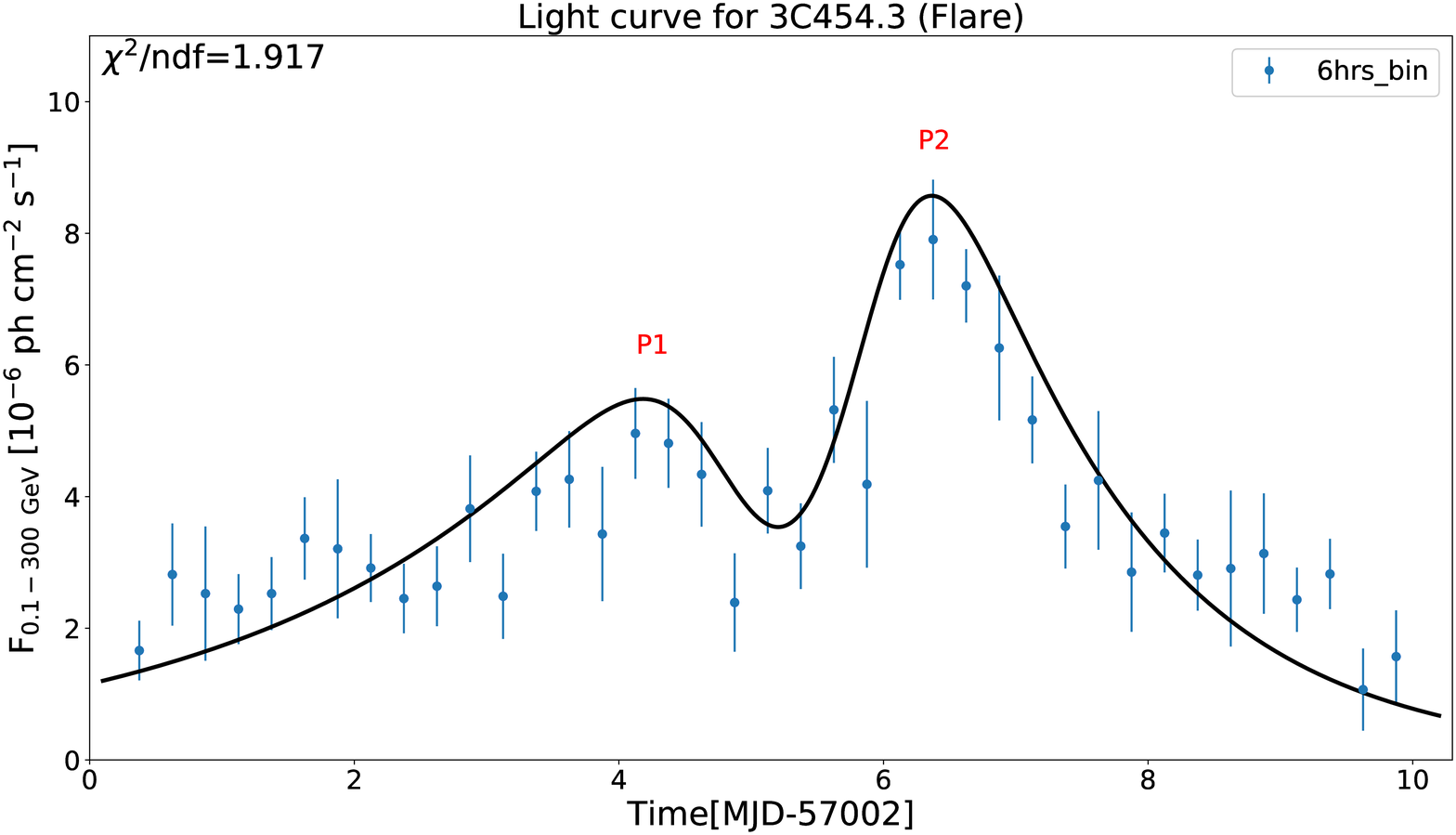}
\caption[optional]{Fitted light curve (fitted by the sum of exponential function) of Flare-3B for Flare (MJD 57002-57012) epoch.}

\includegraphics[height=2.5in,width=5.3in]{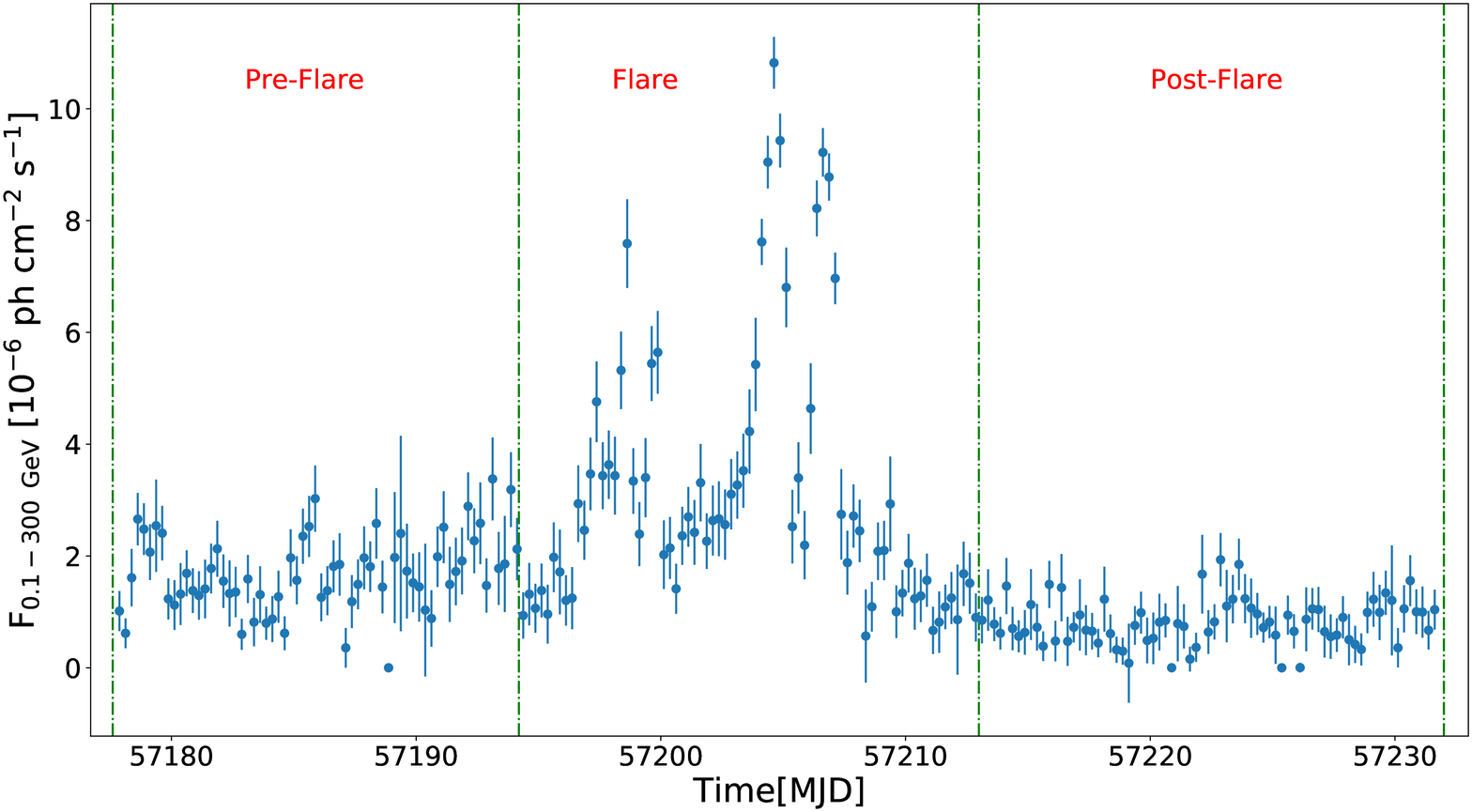}
\caption[optional]{Six-hour binning light curve for Flare-4A.Time durations of all the different periods of activities (shown by broken green line)are: MJD 57160-57190 (Pre-flare),MJD 57194-57213 (Flare) and MJD 57213-57232 (Post-flare).}

\includegraphics[height=2.5in,width=5.3in]{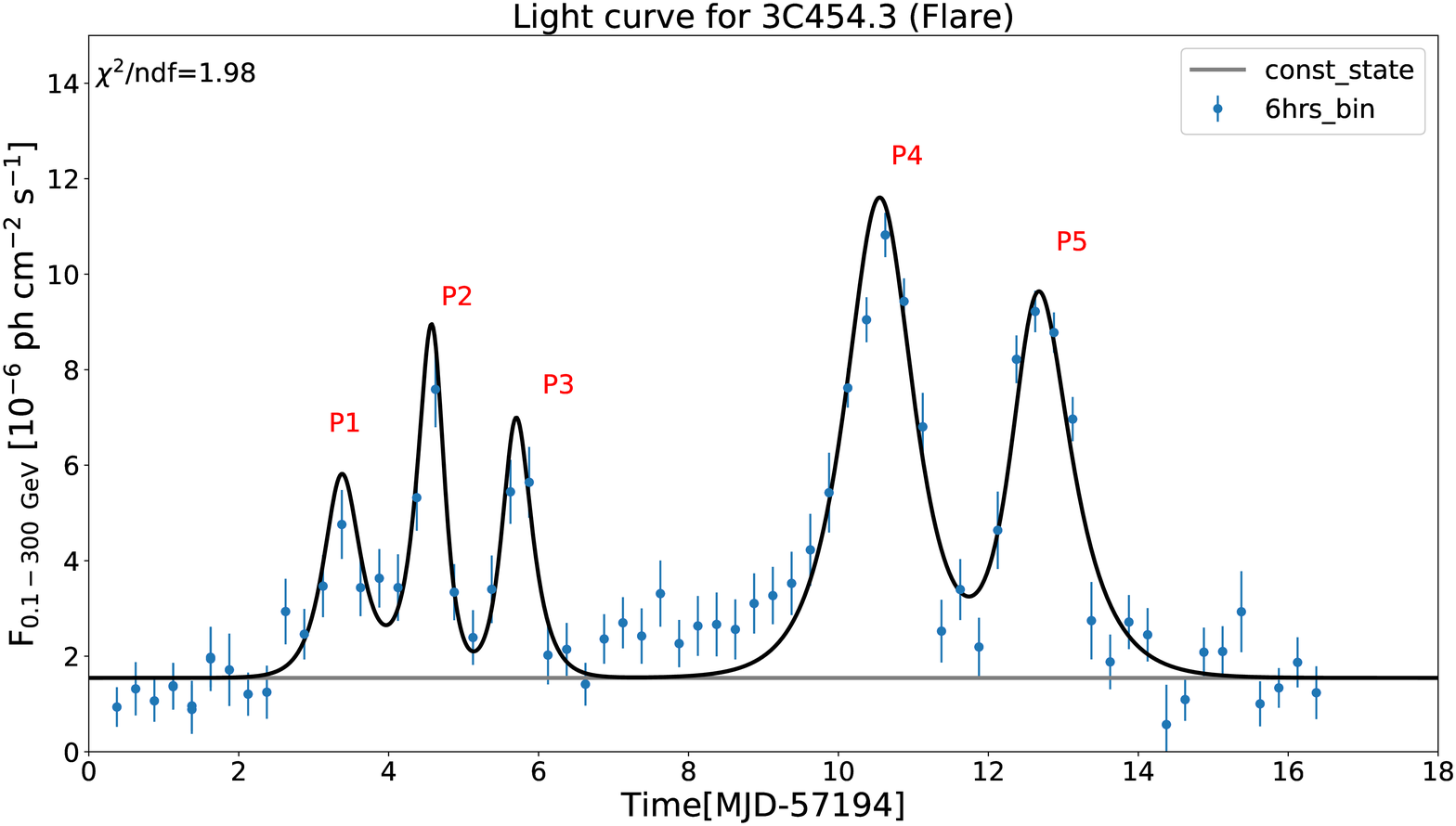}
\caption[optional]{Fitted light curve (fitted by the sum of exponential function) of Flare-4A for Flare (MJD 57194-57213) epoch.}

\end{figure*}

\begin{figure*}[h]
\centering

\end{figure*}

\begin{figure*}[h]
\centering

\includegraphics[height=2.6in,width=5.3in]{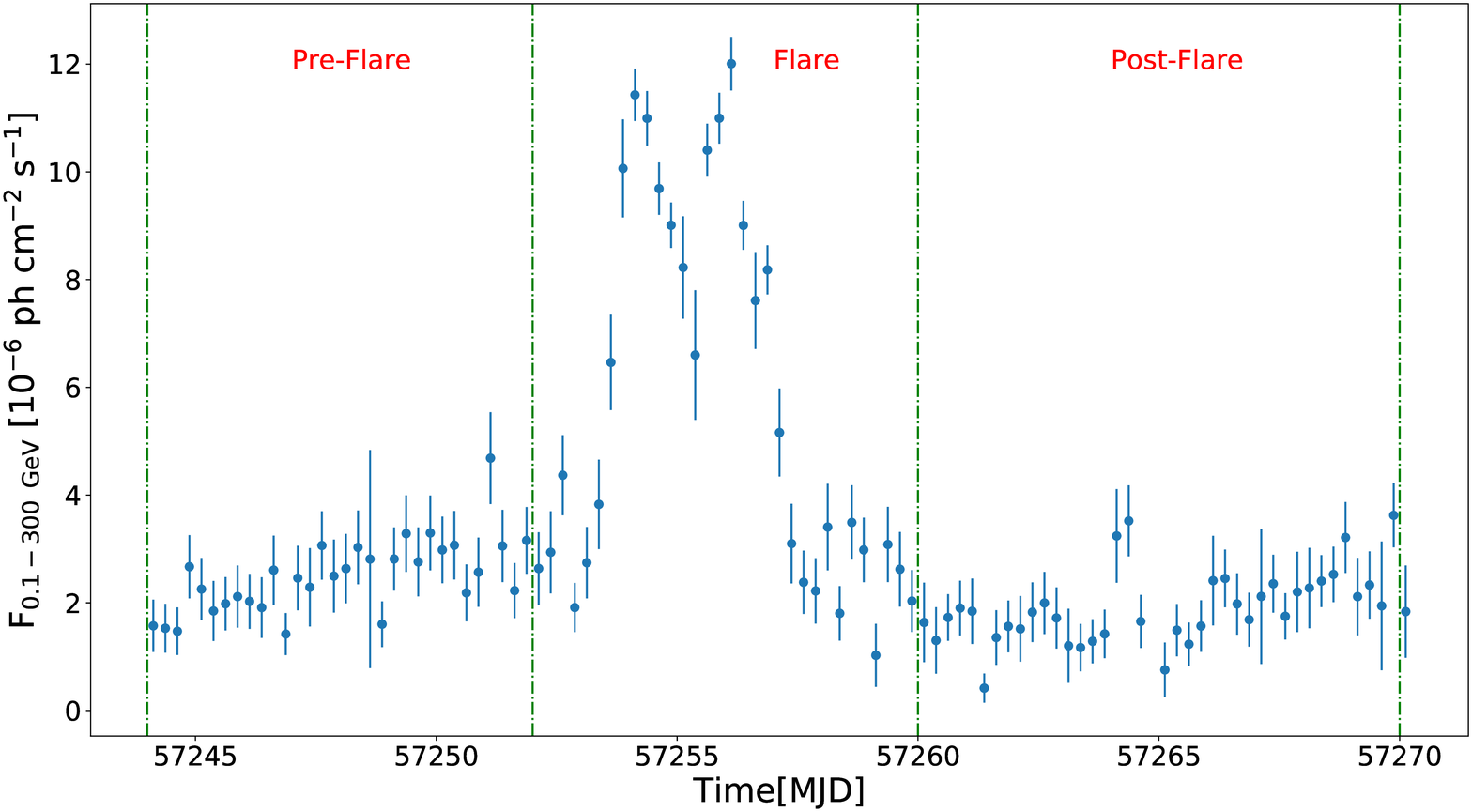}
\caption[optional]{Six-hour binning light curve for Flare-4B. Time durations of all the different periods of activities (shown by broken green line) are: MJD 57244-57252 (Pre-flare), MJD 57251-57260 (Flare) and MJD 57260-57270 (Post-flare).}

\includegraphics[height=2.5in,width=5.3in]{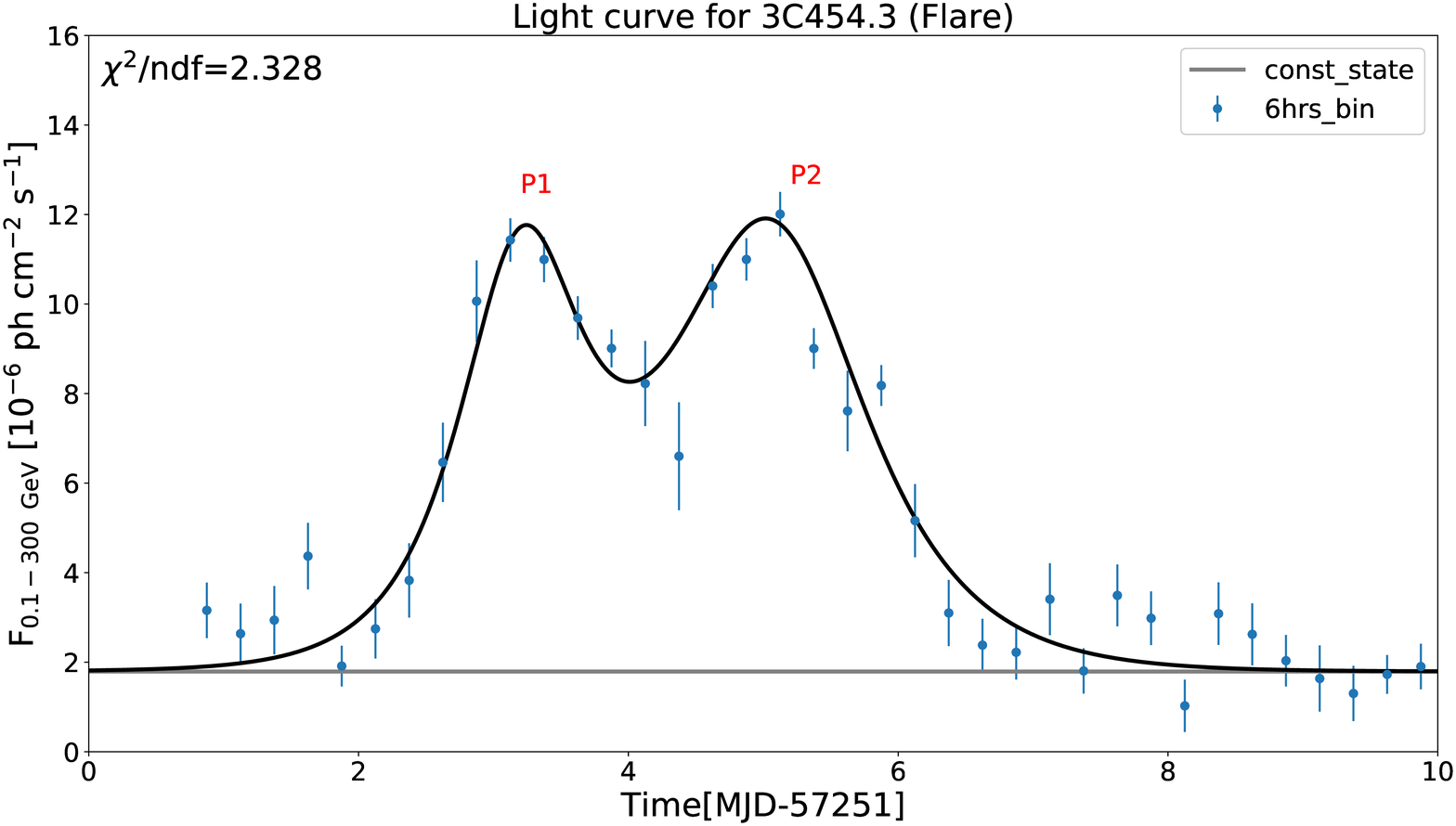}
\caption[optional]{Fitted light curve (fitted by the sum of exponential function) of Flare-4B for Flare (MJD 57251-57260) epoch.}

\includegraphics[height=2.5in,width=5.3in]{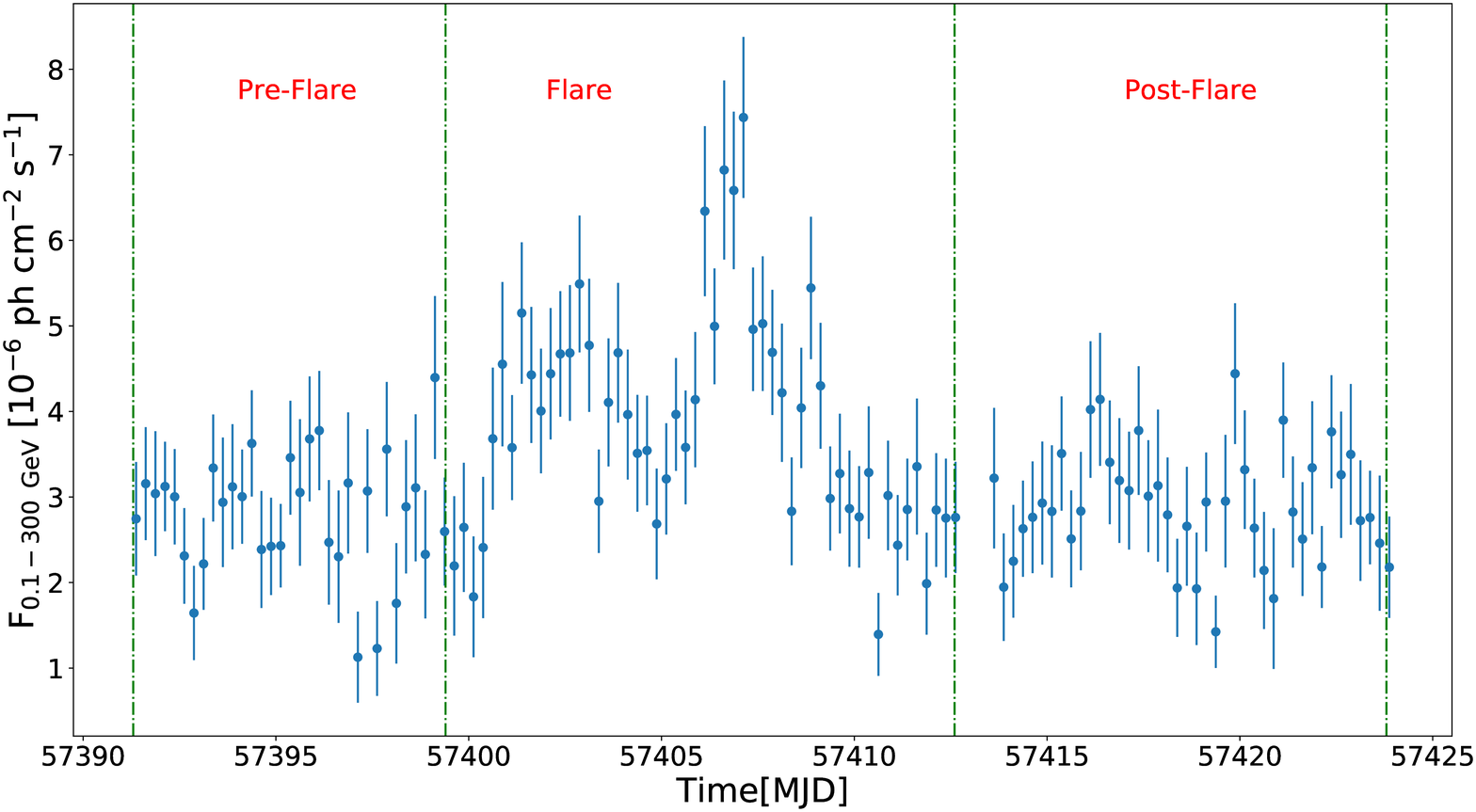}
\caption[optional]{Six-hour binning light curve for Flare-4C. Time durations of all the different periods of activities (shown by broken green line) are: MJD 57391-57399 (Pre-flare), MJD 57399-57413 (Flare) and MJD 57413-57424 (Post-flare).}

\end{figure*}

\begin{figure*}[h]
\centering

\includegraphics[height=2.5in,width=5.3in]{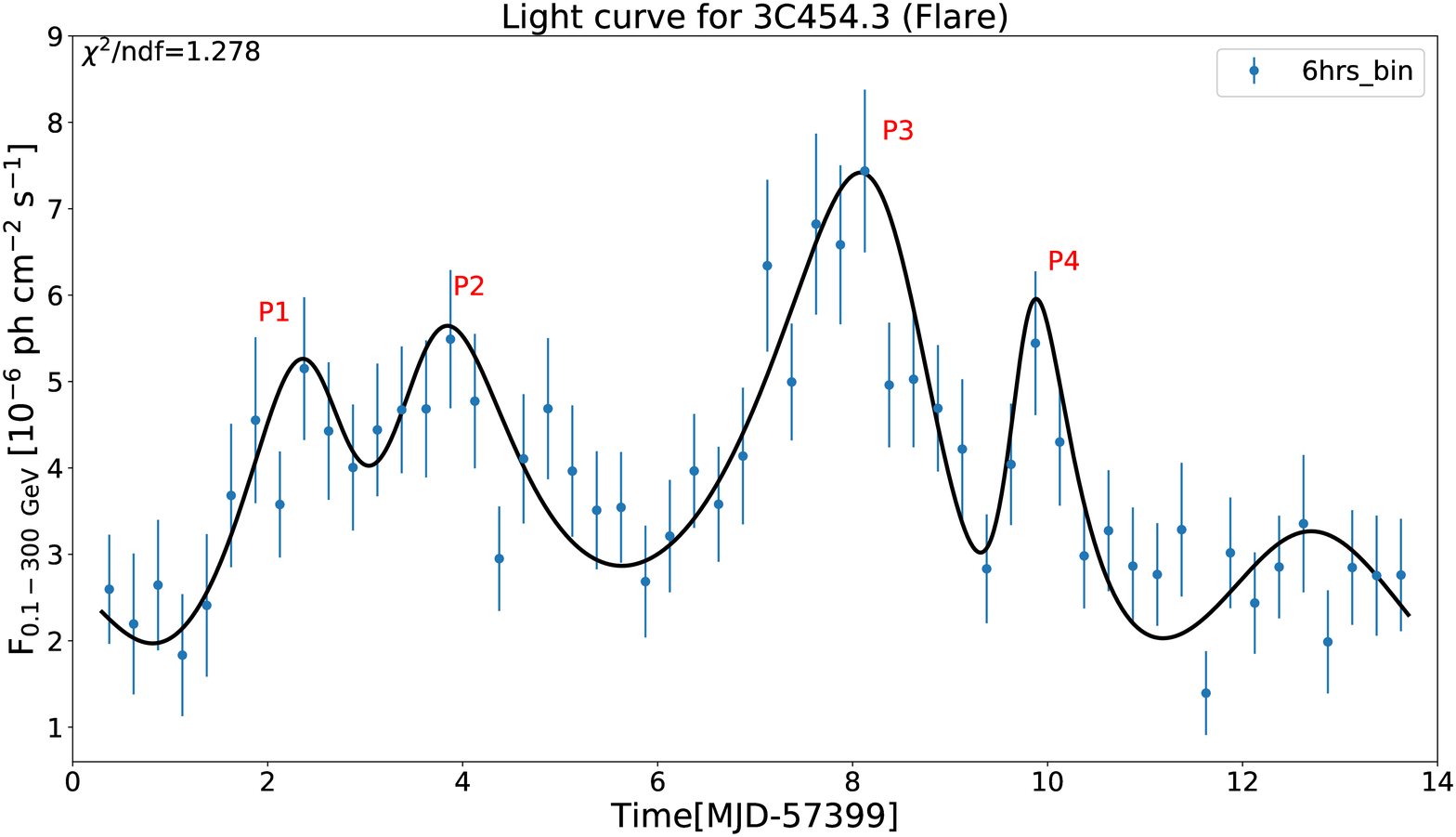}
\caption[optional]{Fitted light curve (fitted by the sum of exponential function) of Flare-4C for Flare (MJD 57399-57413) epoch.}

\includegraphics[height=2.5in,width=5.3in]{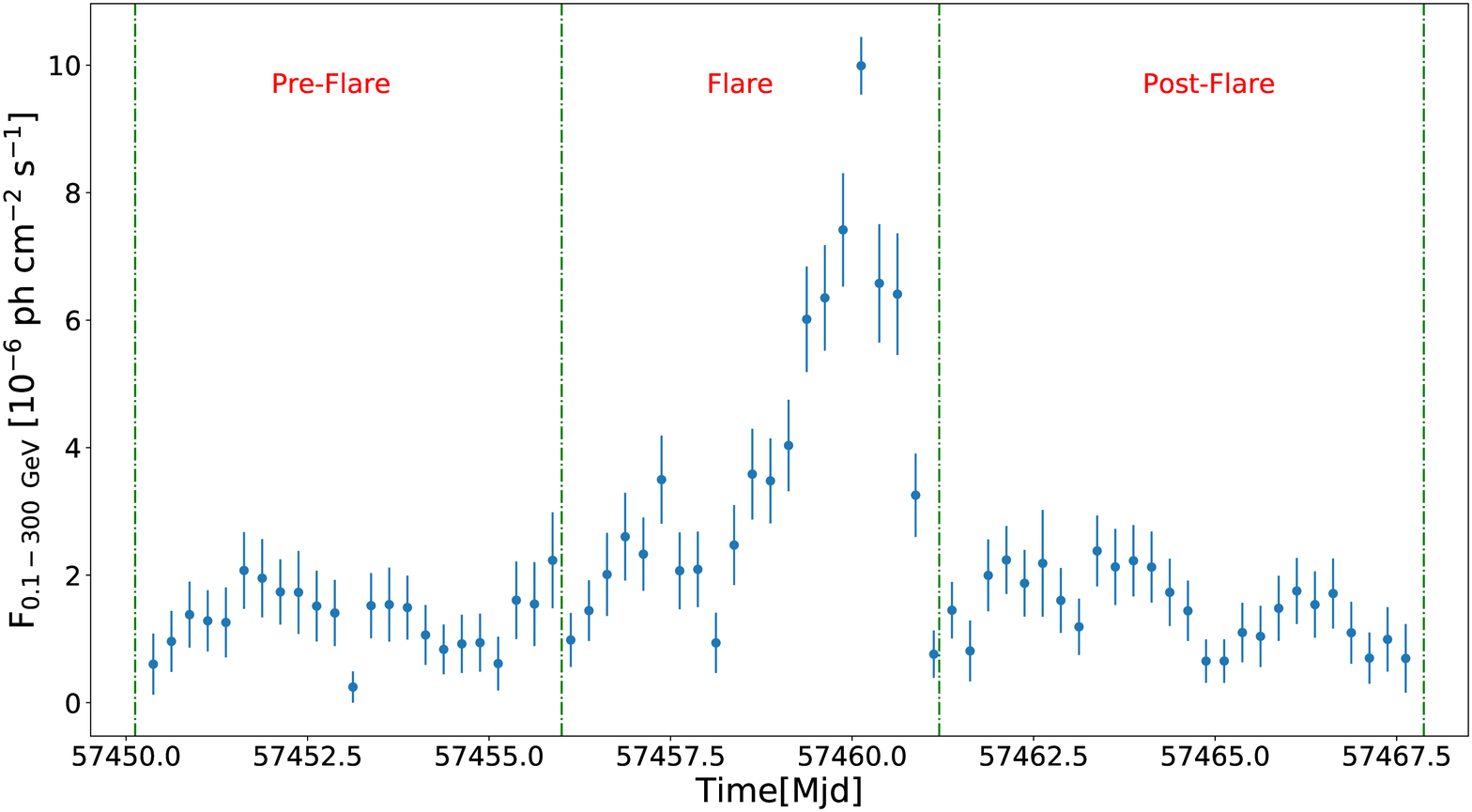}
\caption[optional]{Six-hour binning light curve for Flare-4D. Time durations of all the different periods of activities (shown by broken green line)are: MJD 57450-57456 (Pre-flare), MJD 57454-57461 (Flare) and MJD 57461-57468 (Post-flare).}

\includegraphics[height=2.5in,width=5in]{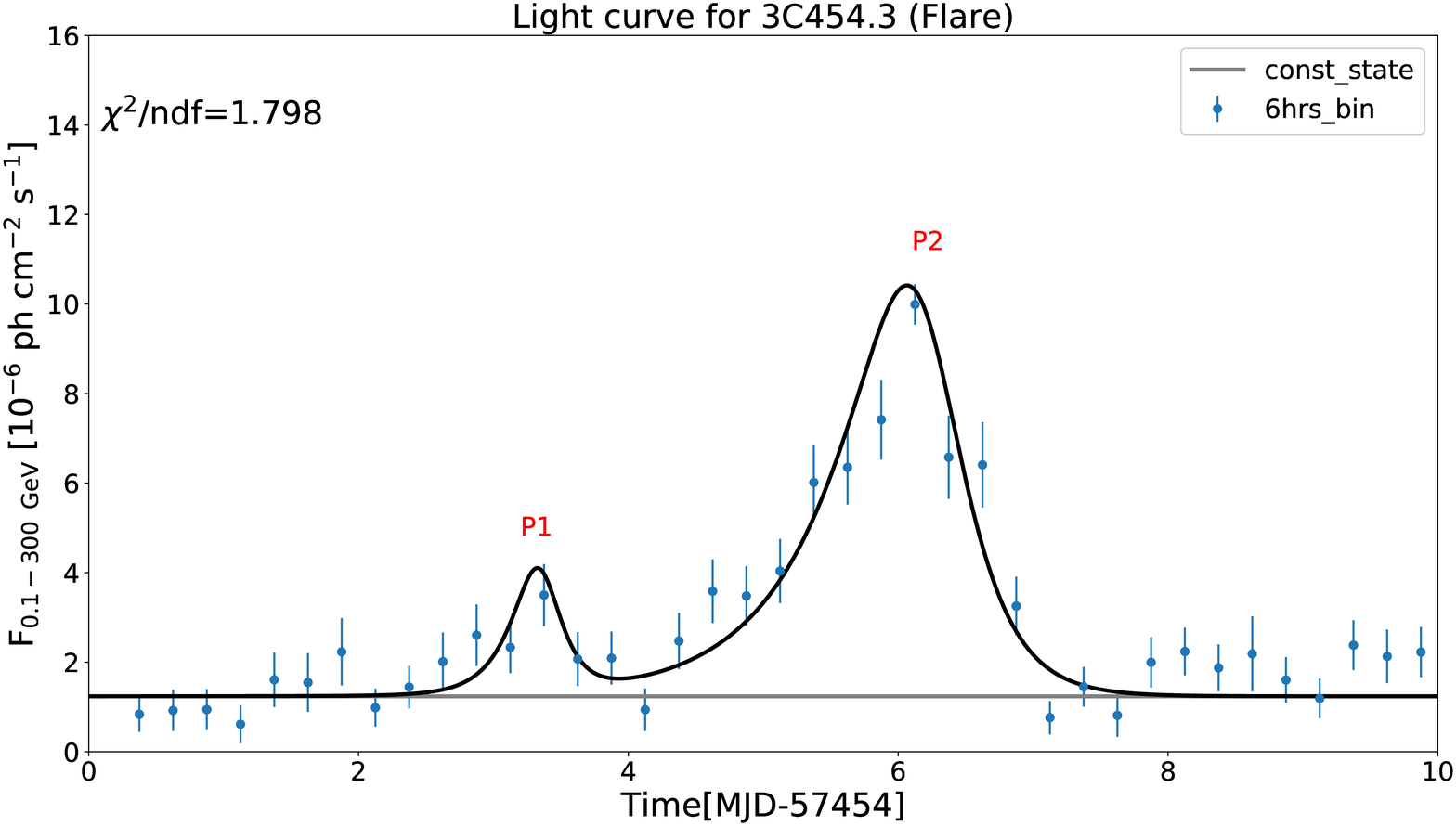}
\caption[optional]{Fitted light curve (fitted by the sum of exponential function) of Flare-4D for Flare (MJD 57454-57461) epoch.}

\end{figure*}



\begin{figure*}[h]
\centering

\includegraphics[height=2.5in,width=5.3in]{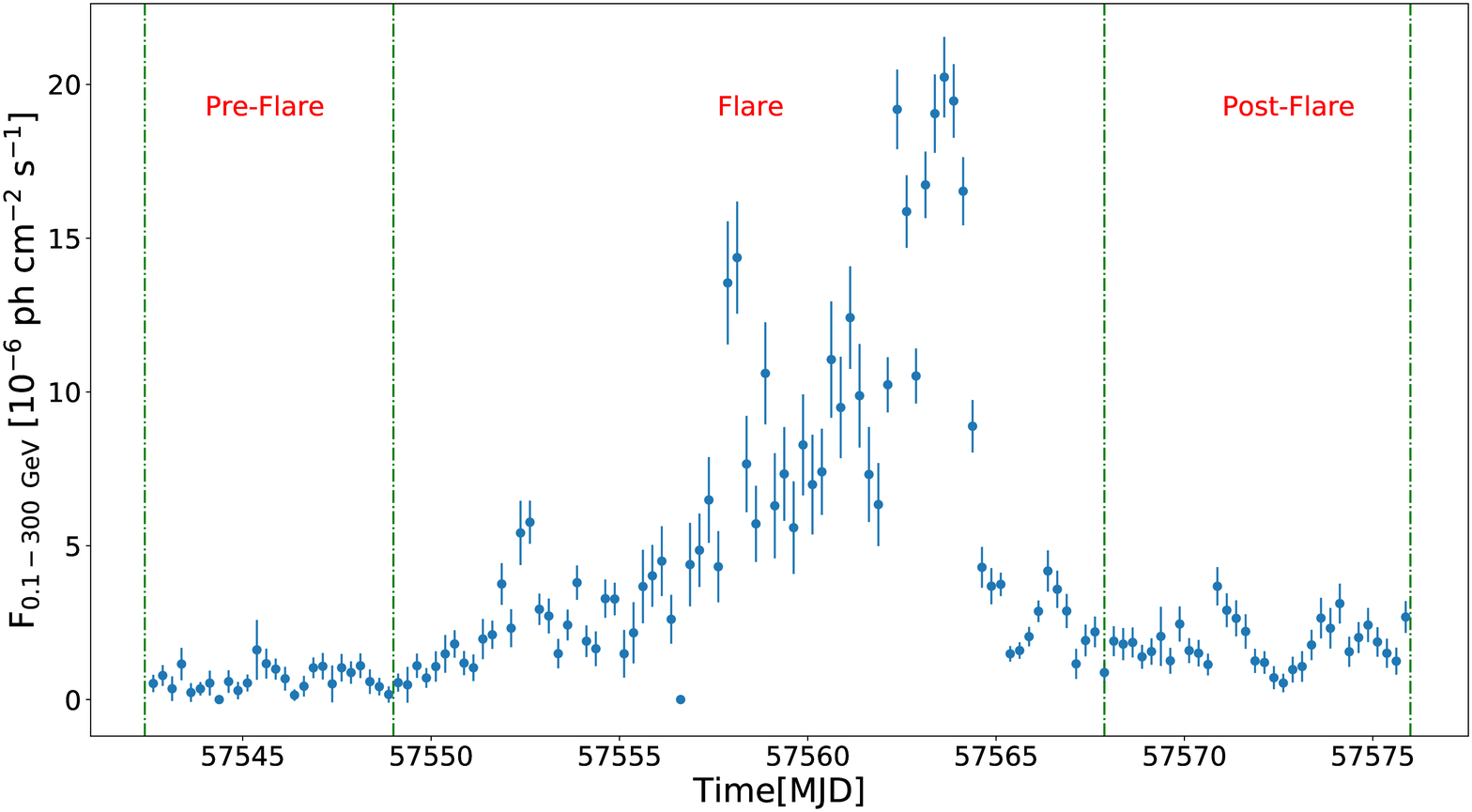}
\caption[optional]{Six-hour binning light curve for Flare-5A. Time durations of all the different periods of activities (shown by broken green line) are: MJD 57542-57549 (Pre-flare), MJD 57549-57568 (Flare) and MJD 57568-57576 (Post-flare).}

\includegraphics[height=2.5in,width=5.3in]{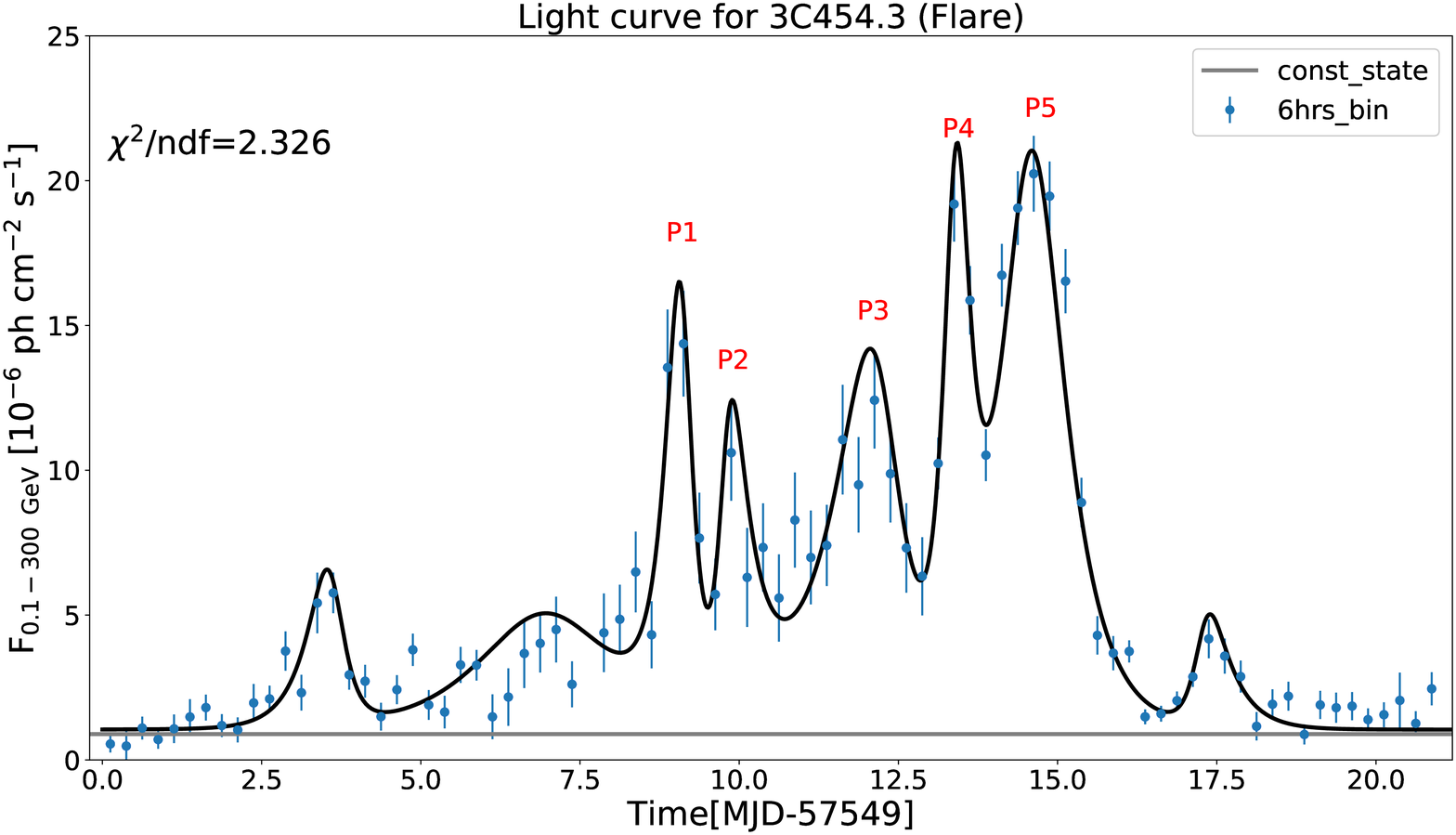}
\caption[optional]{Fitted light curve (fitted by the sum of exponential function) of Flare-5A for Flare (MJD 57549-57568) epoch.}

\includegraphics[height=2.5in,width=5.3in]{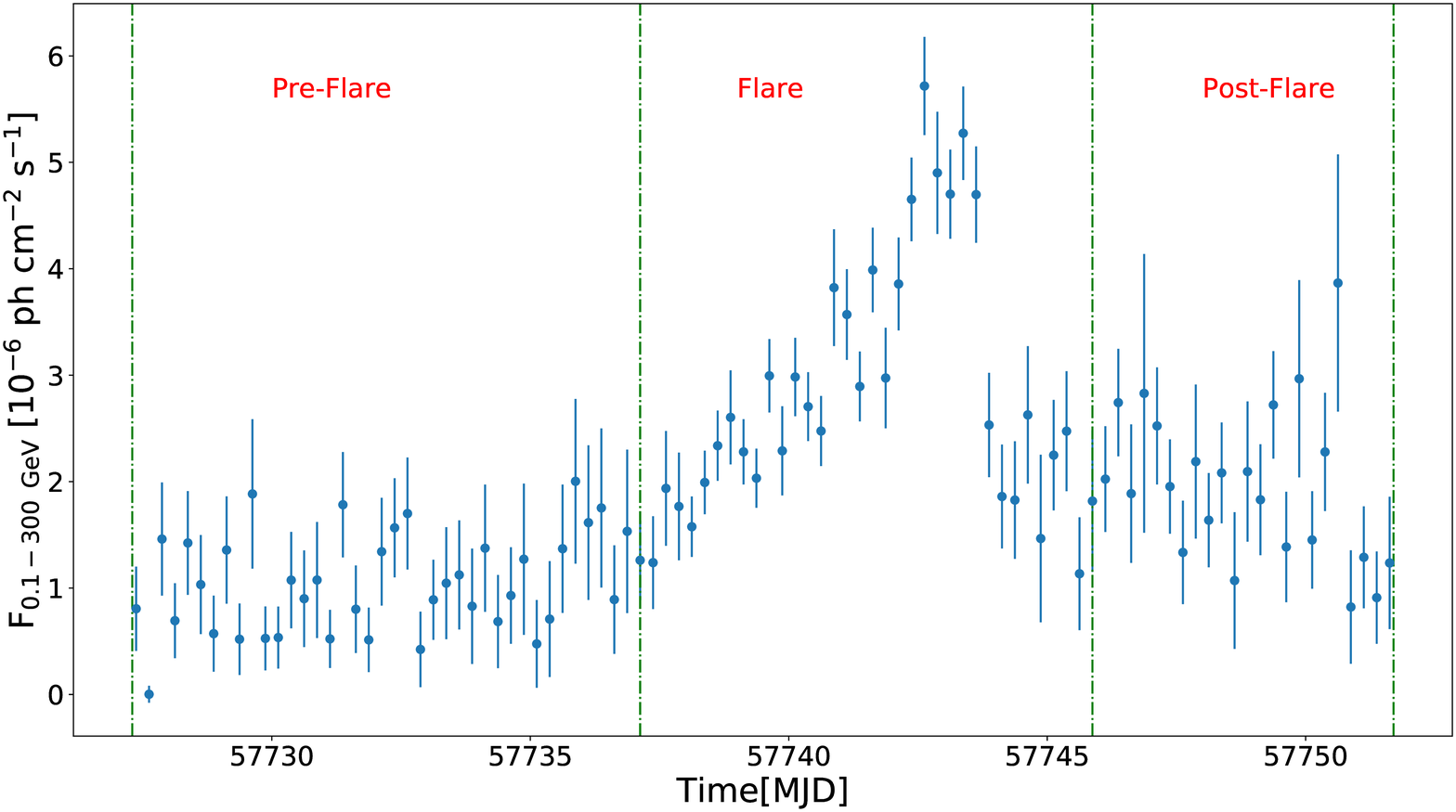}
\caption[optional]{Six-hour binning light curve for Flare-5B.Time durations of all the different periods of activities (shown by broken green line)are: MJD 57727-57737 (Pre-flare),MJD 57737-57746 (Flare) and MJD 57746-57752 (Post-flare).}

\end{figure*}

\begin{figure*}[h]
\centering

\includegraphics[height=2.6in,width=5.3in]{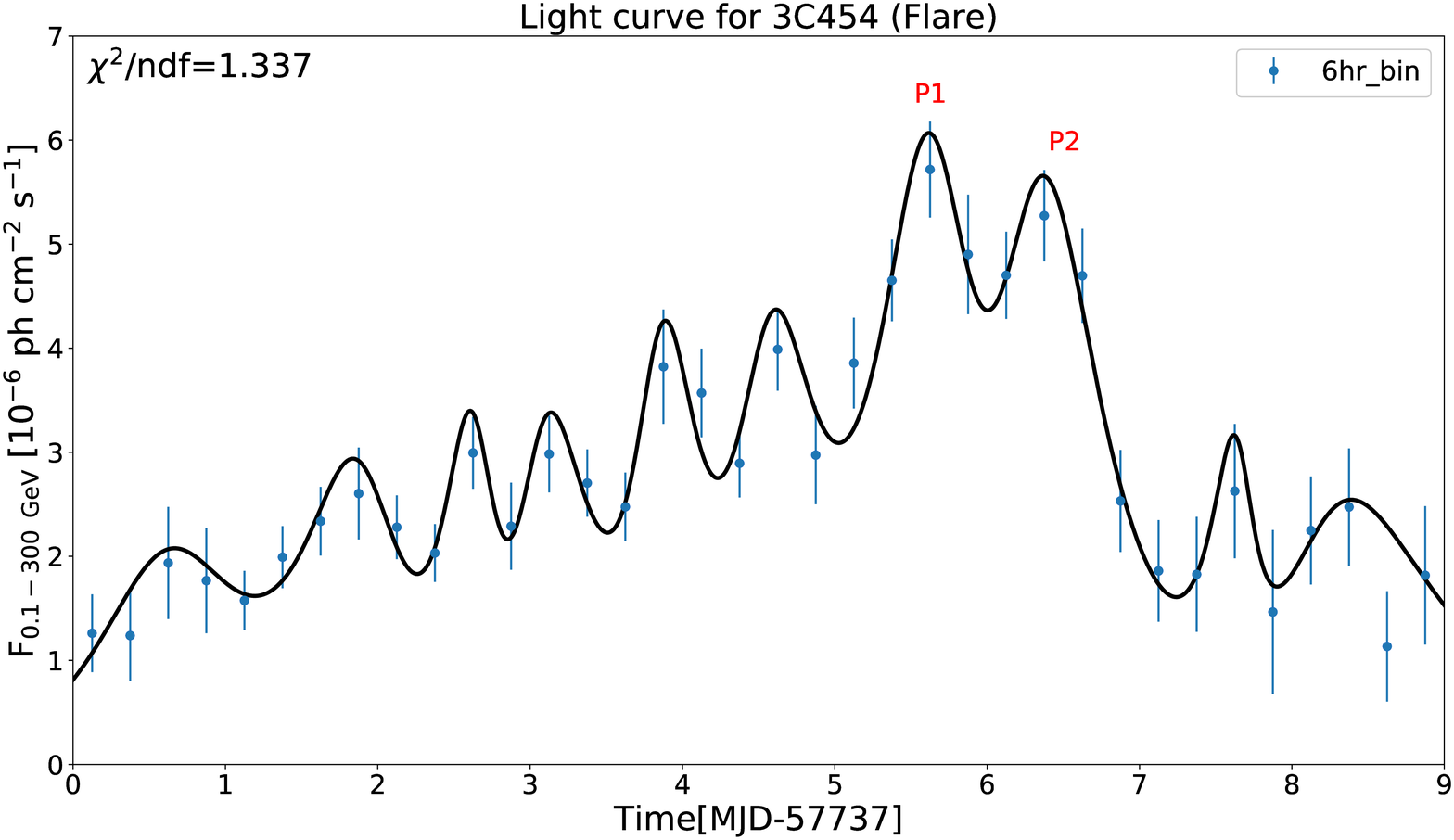}
\caption[optional]{Fitted light curve (fitted by the sum of exponential function) of Flare-5B for Flare (MJD 57737-57746) epoch.}

\includegraphics[height=1.77in,width=2.5in]{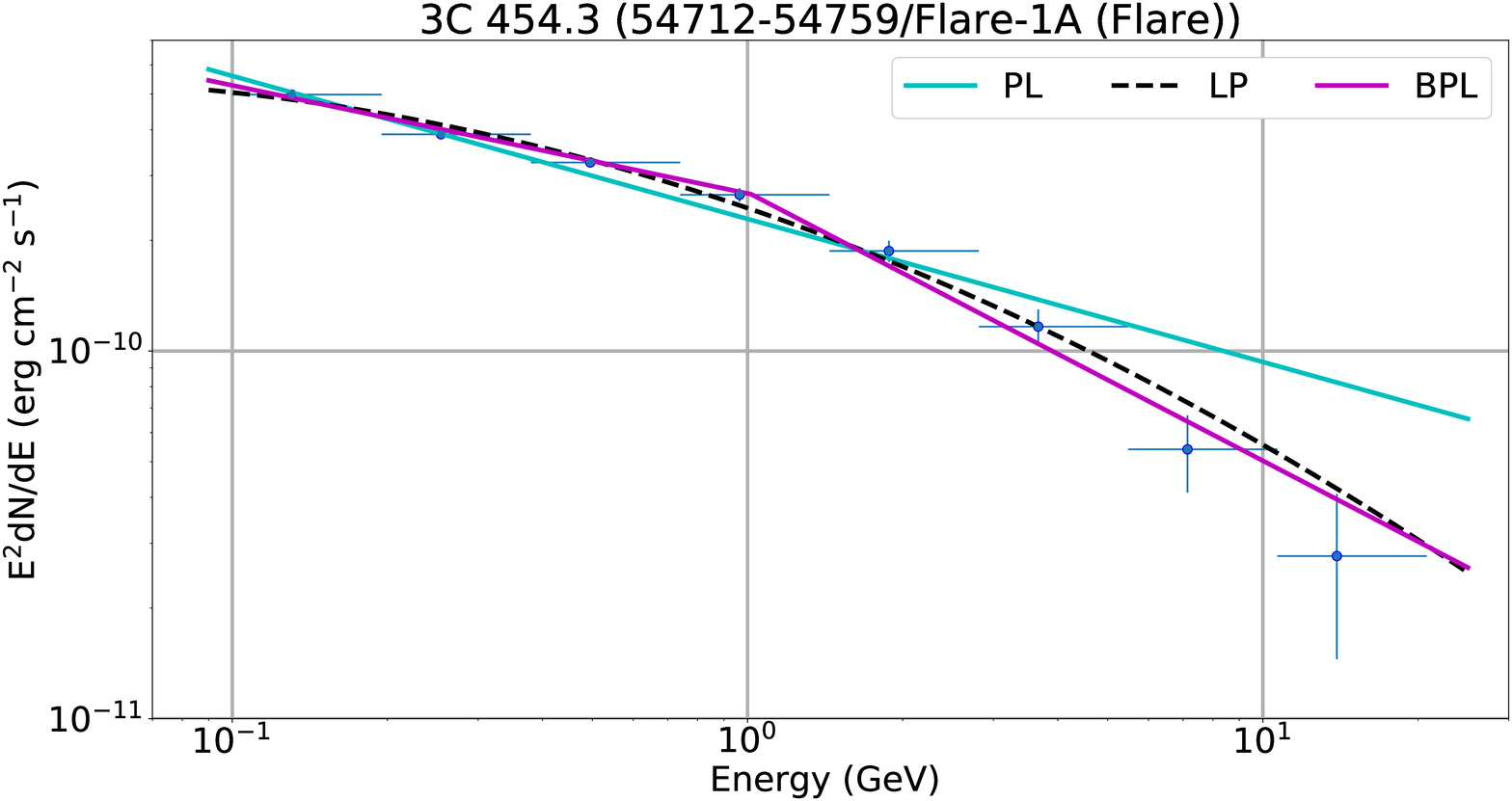}
\includegraphics[height=1.77in,width=2.5in]{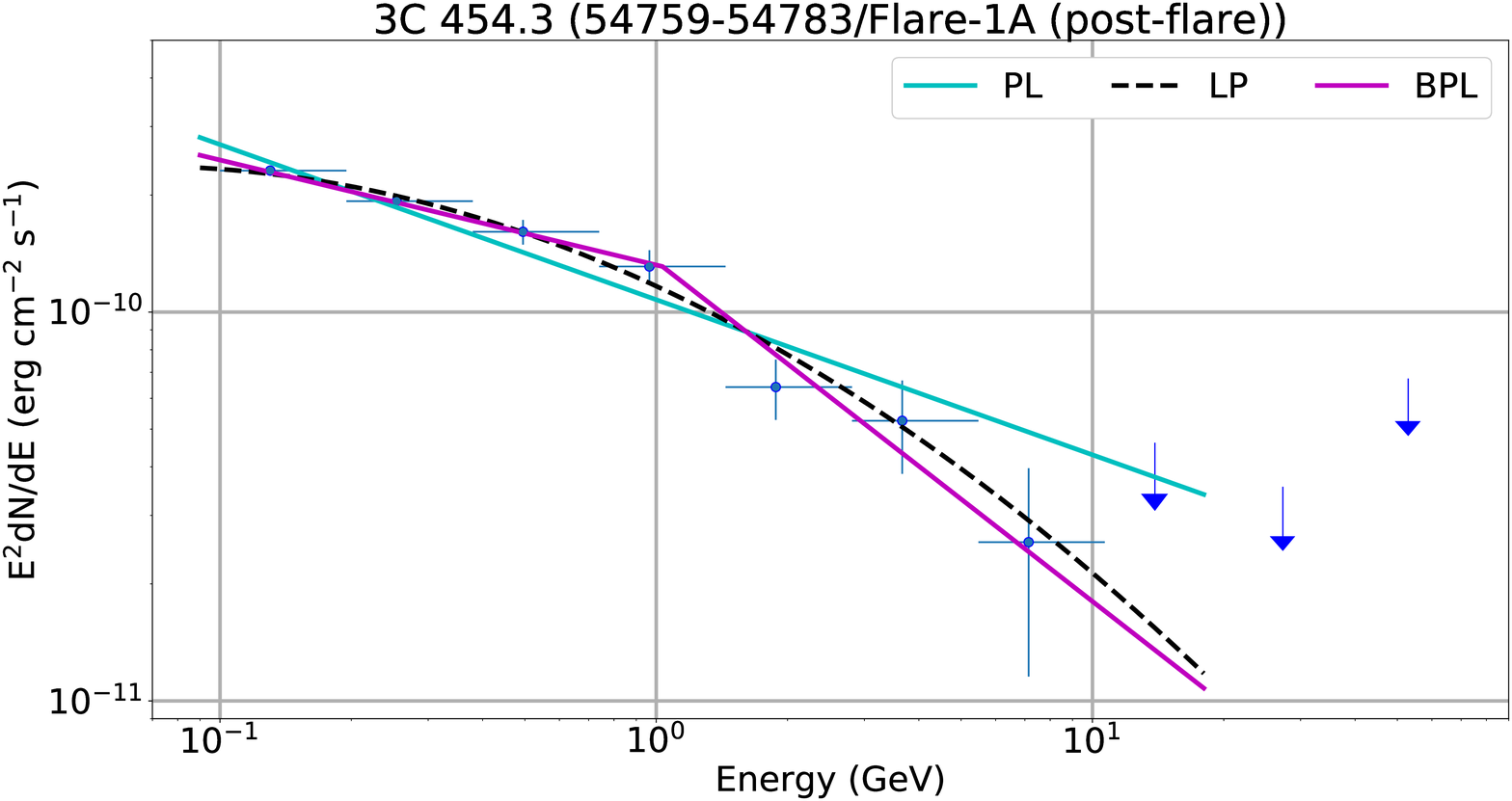}
\caption[optional]{SED of different periods of Flare-1A as given in Figure-2. PL, LP \& BPL describe the Powerlaw, Logparabola and Broken-powerlaw model respectively, which are fitted to data points.}

\includegraphics[height=1.77in,width=2.5in]{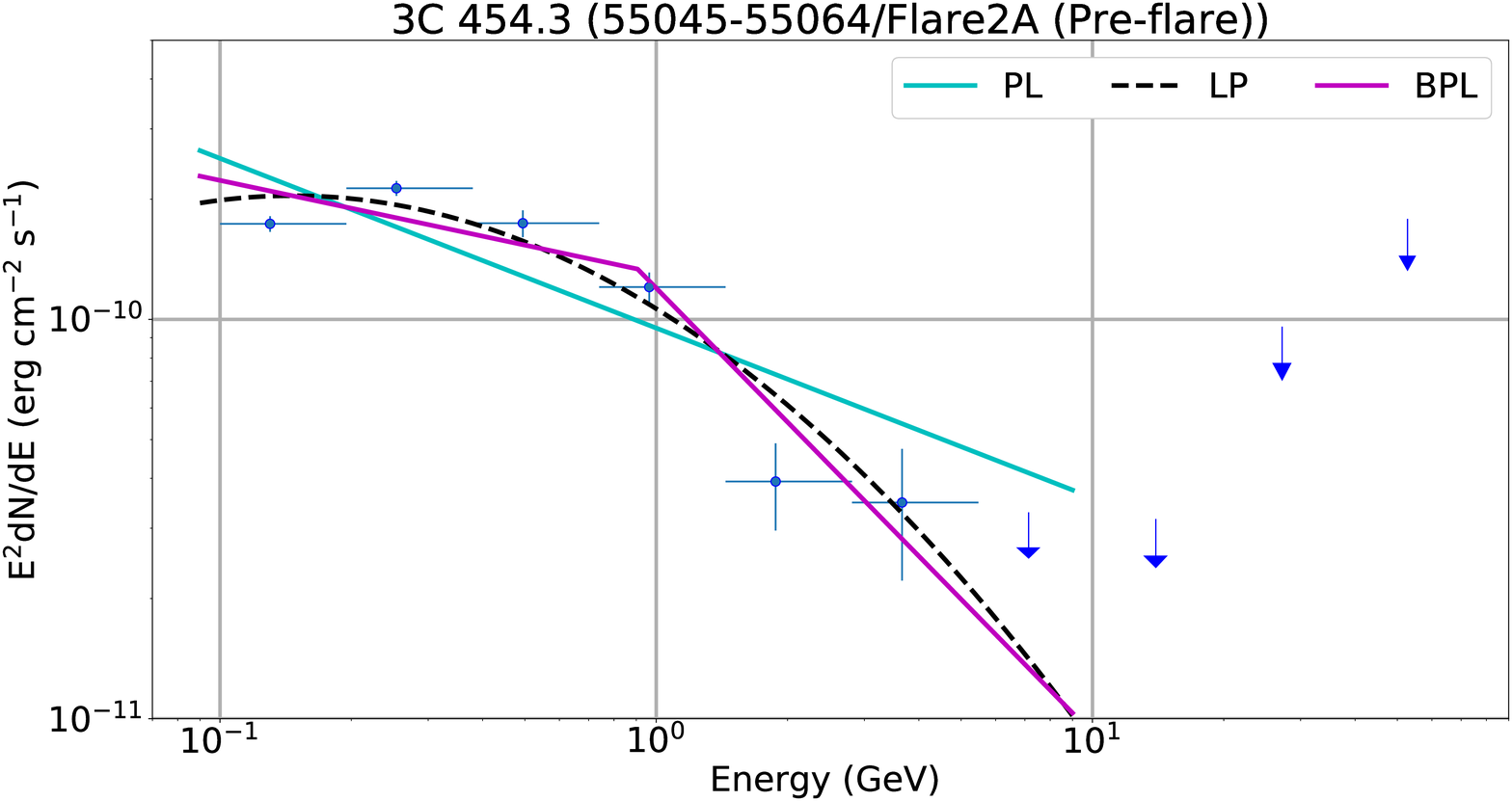}
\includegraphics[height=1.77in,width=2.5in]{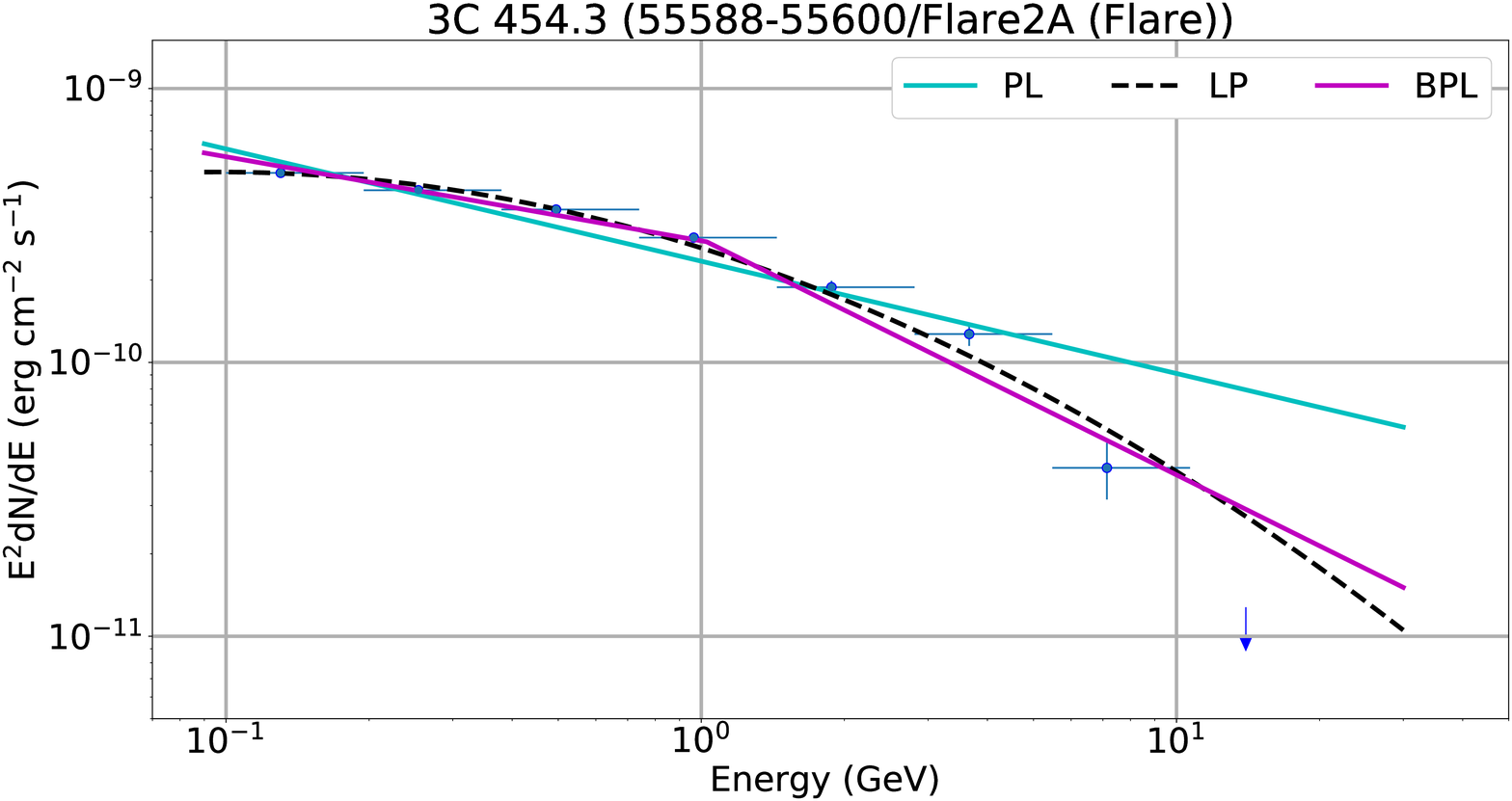}
\caption[optional]{SED of different periods of Flare-2A as given in Figure-4. PL, LP \& BPL describe the Powerlaw, Logparabola and Broken-powerlaw model respectively, which are fitted to data points.}

\includegraphics[height=1.77in,width=2.5in]{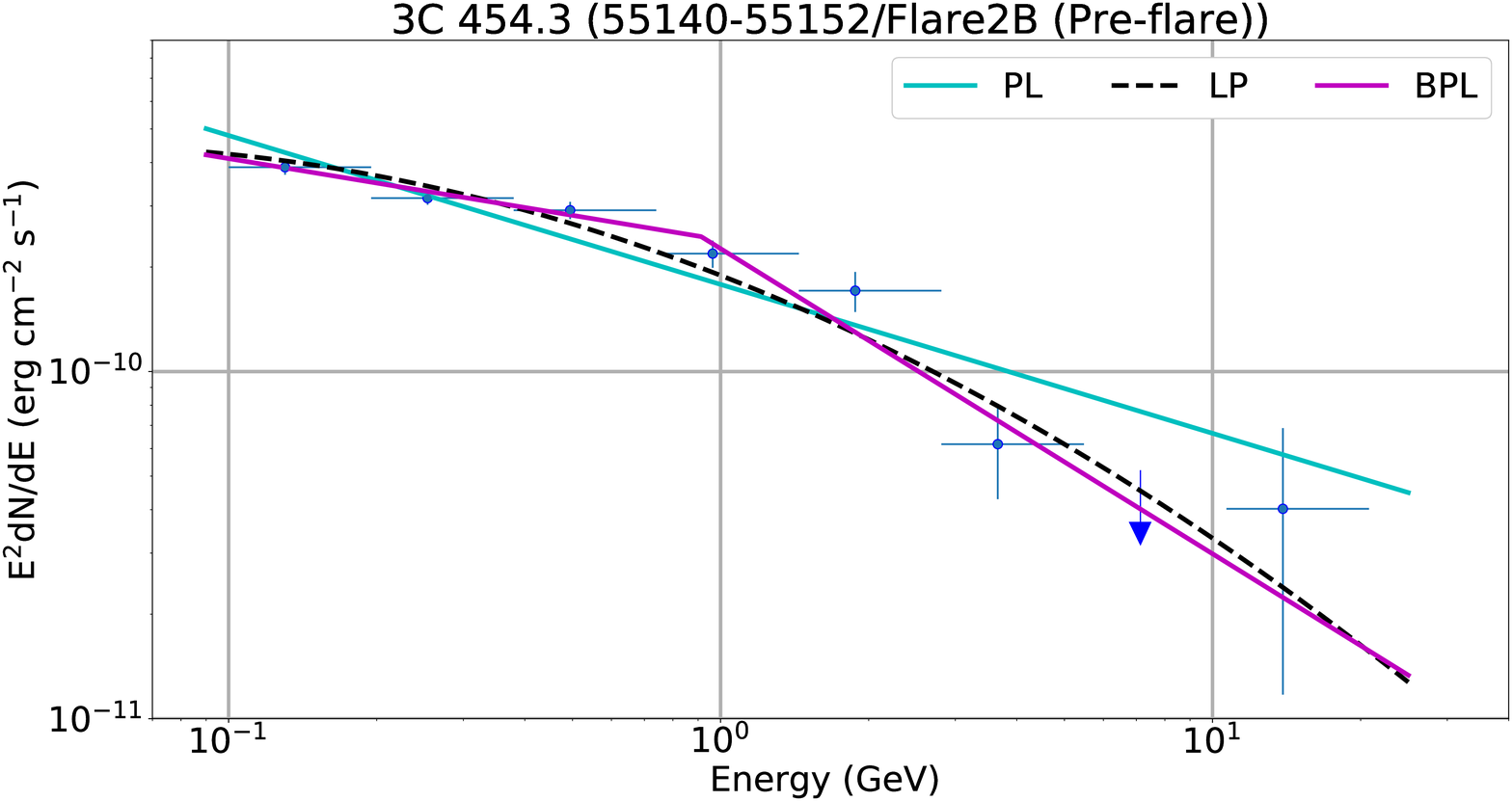}
\includegraphics[height=1.77in,width=2.5in]{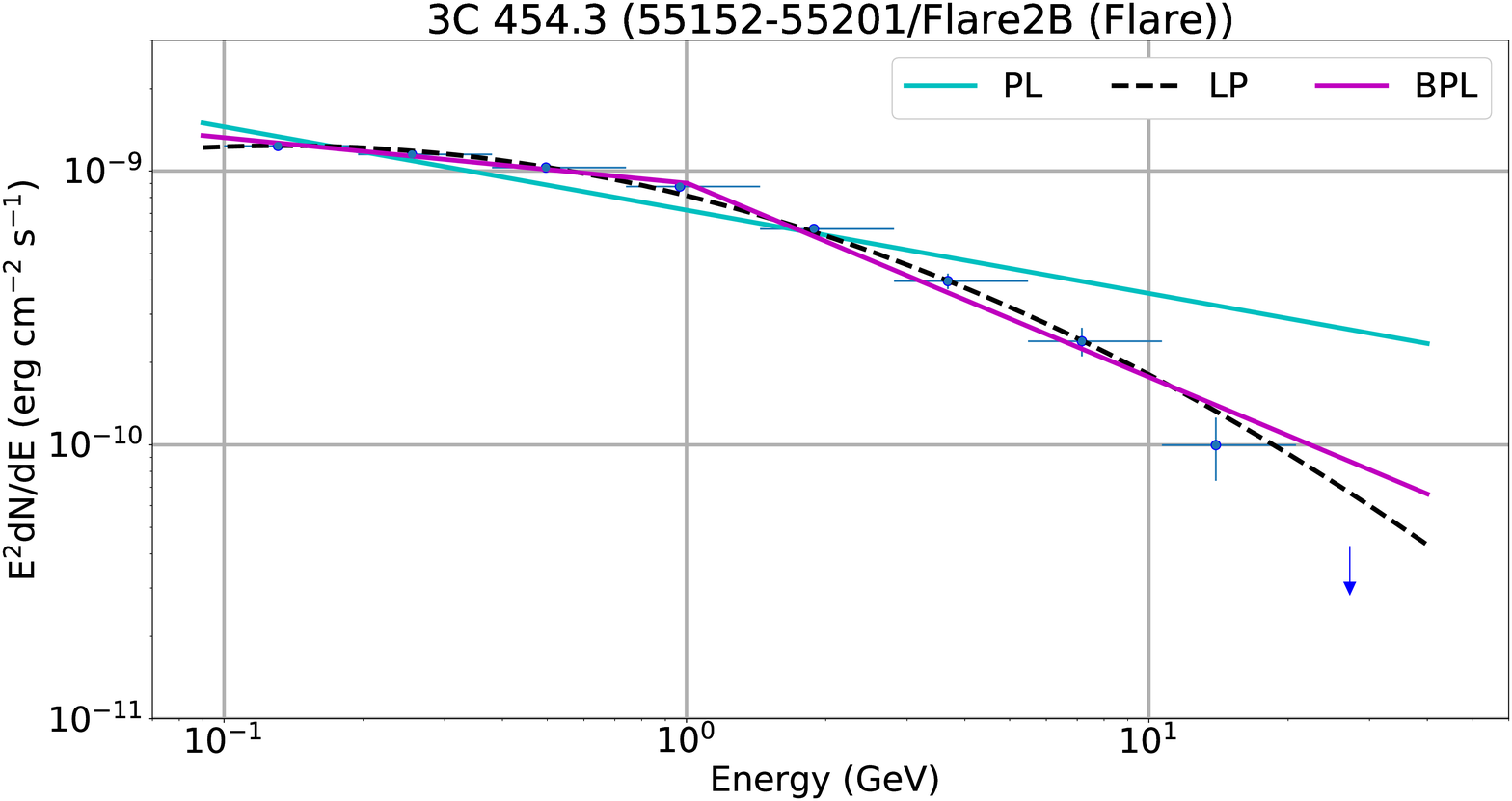}
\caption[optional]{SED of different periods of Flare-2B as given in Figure-6. PL, LP \& BPL describe the Powerlaw, Logparabola and Broken-powerlaw model respectively, which are fitted to data points.}

\end{figure*}

\begin{figure*}[h!]
\centering
\includegraphics[height=1.77in,width=2.5in]{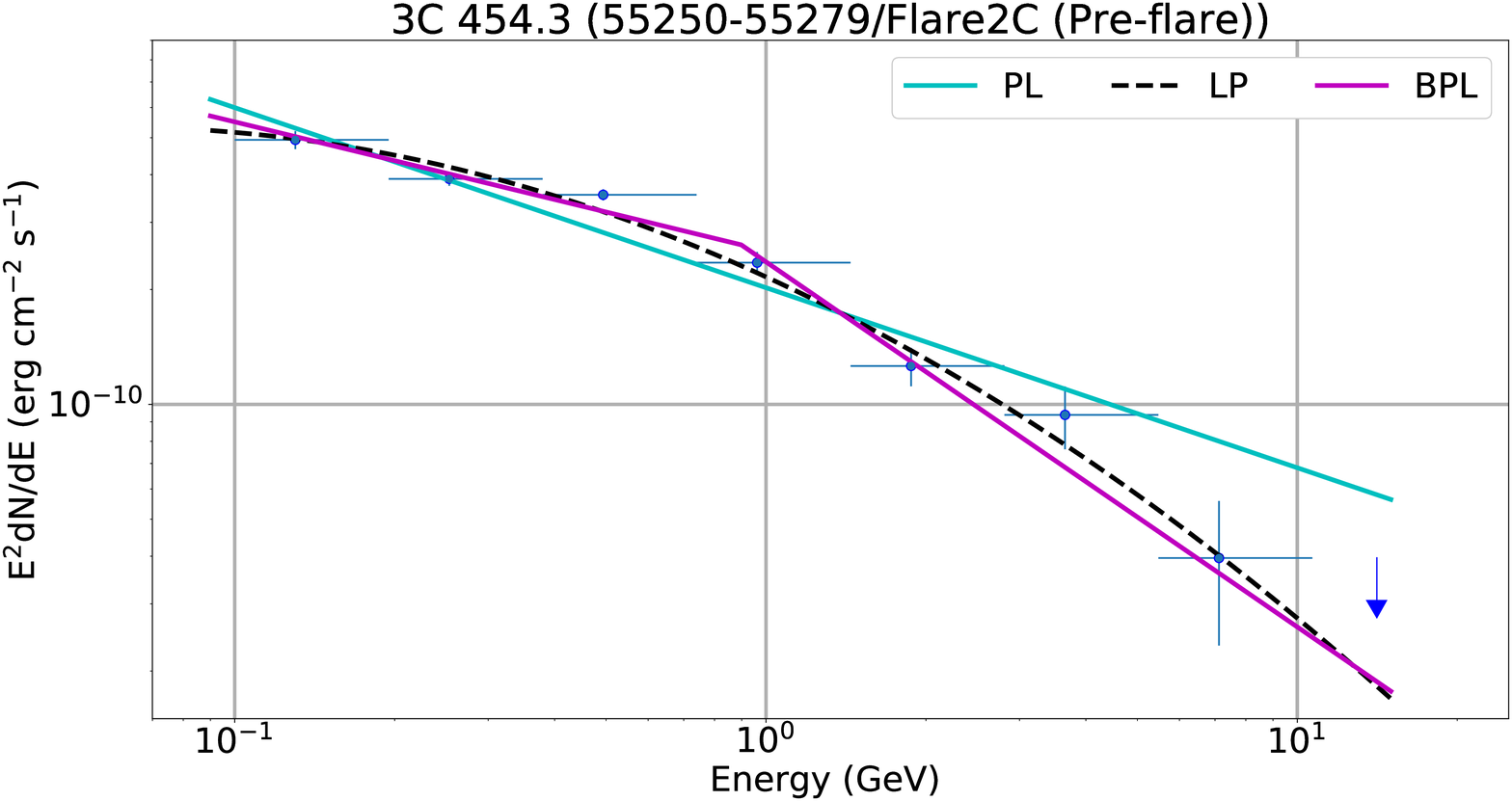}
\includegraphics[height=1.77in,width=2.5in]{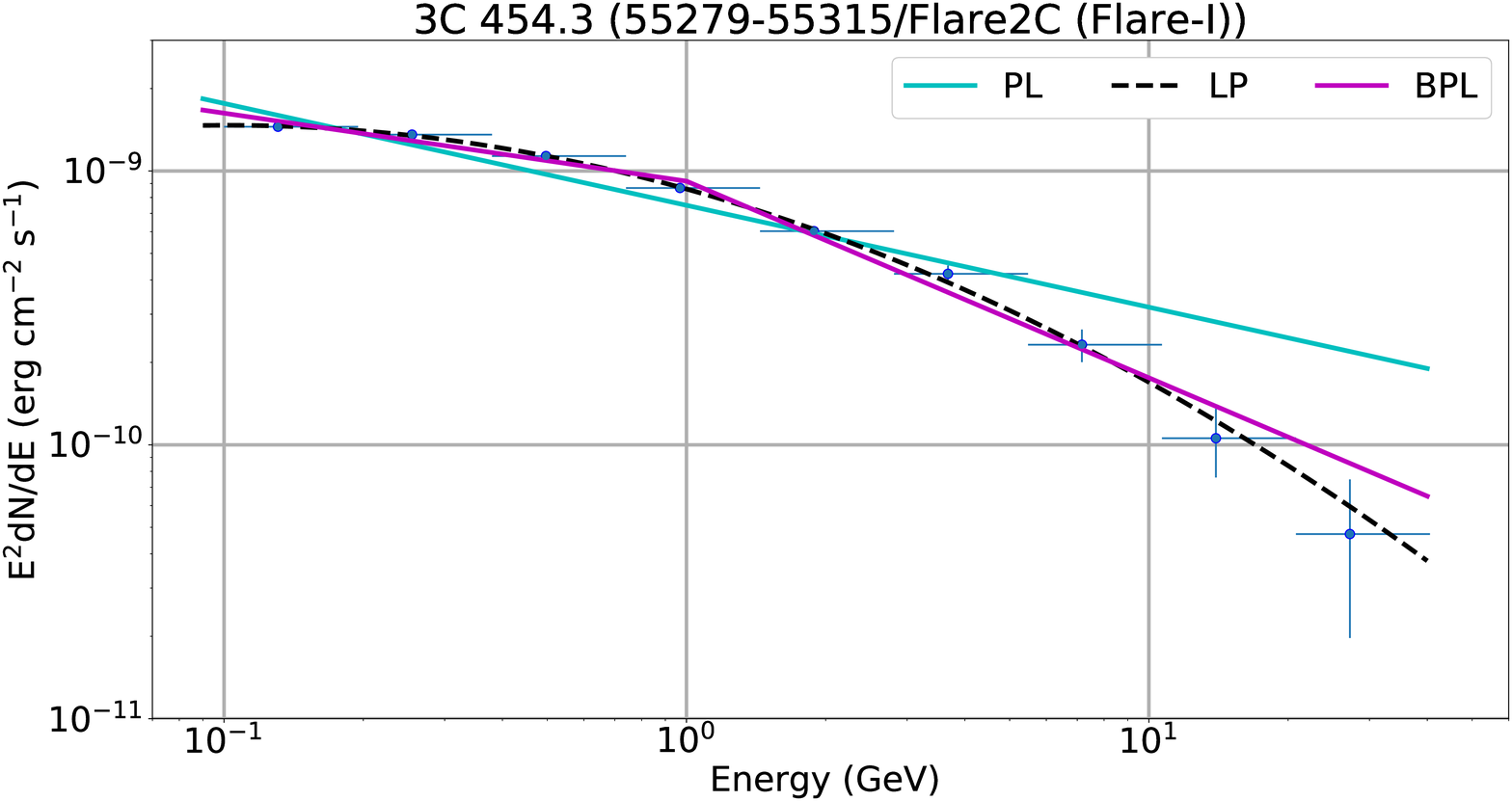}

\includegraphics[height=1.77in,width=2.5in]{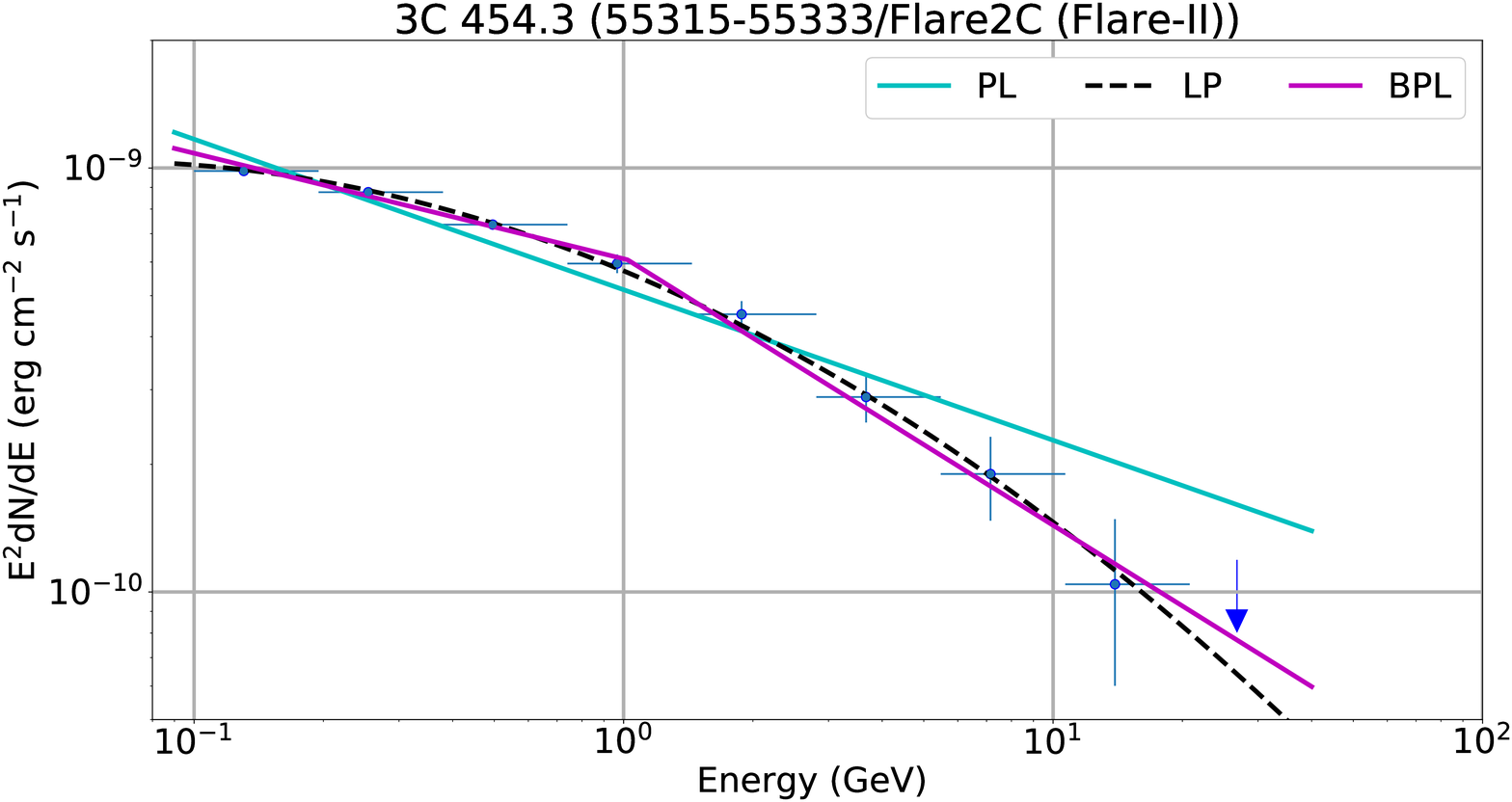}
\includegraphics[height=1.77in,width=2.5in]{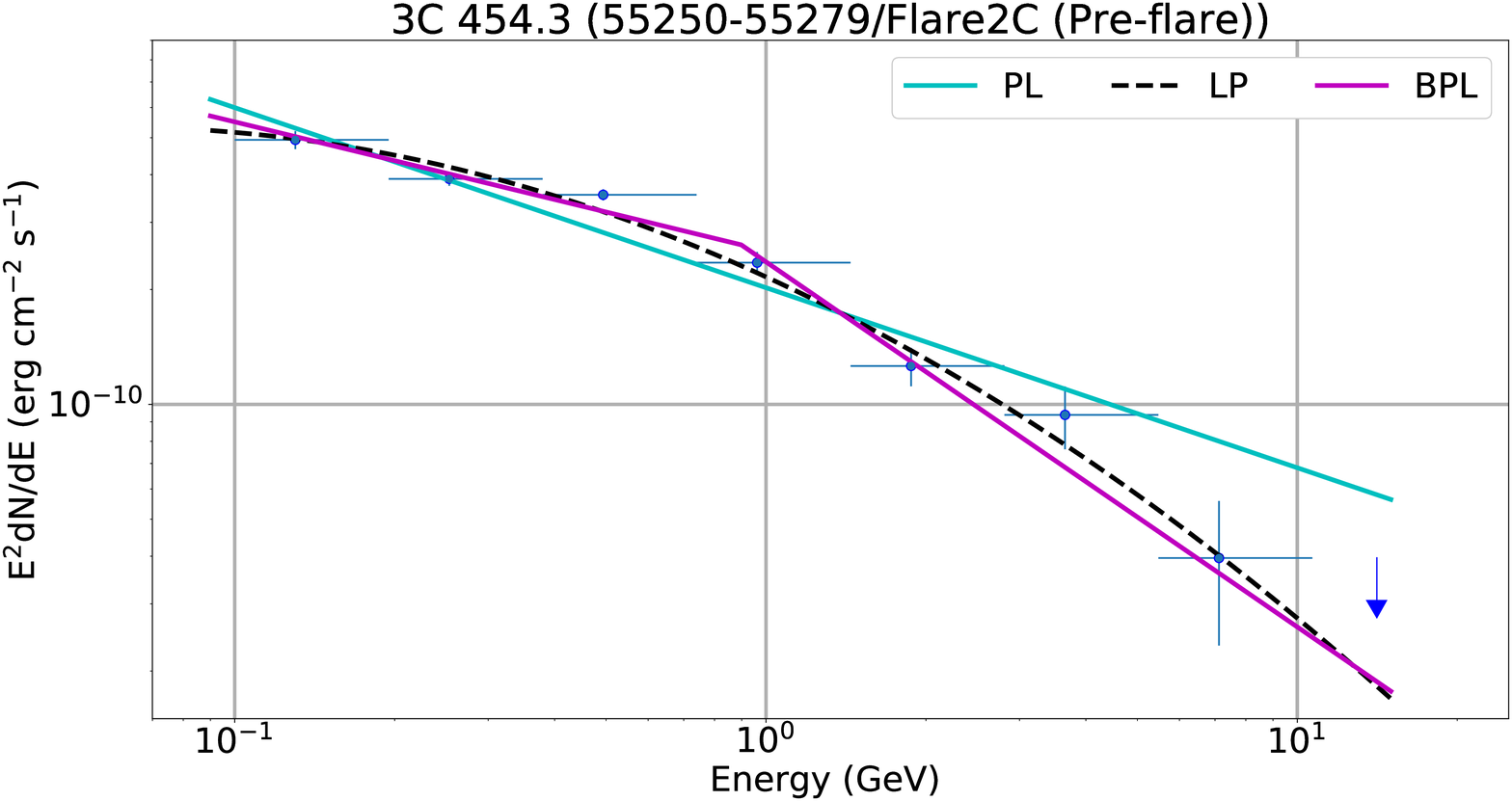}
\caption[optional]{SED of different periods of Flare-2C as given in Figure-9. PL, LP \& BPL describe the Powerlaw, Logparabola and Broken-powerlaw model respectively, which are fitted to data points.}

\includegraphics[height=1.77in,width=2.5in]{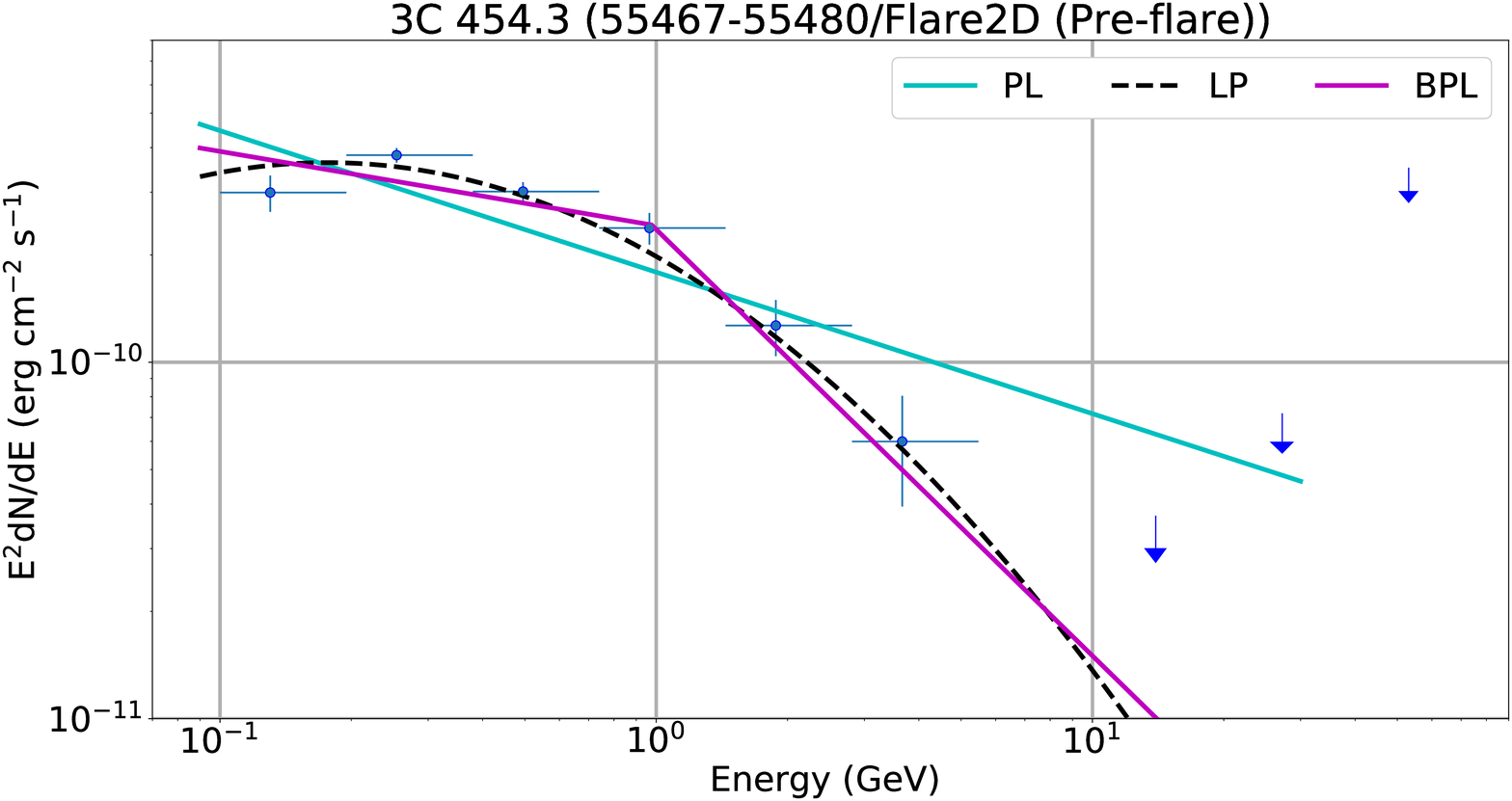}
\includegraphics[height=1.77in,width=2.5in]{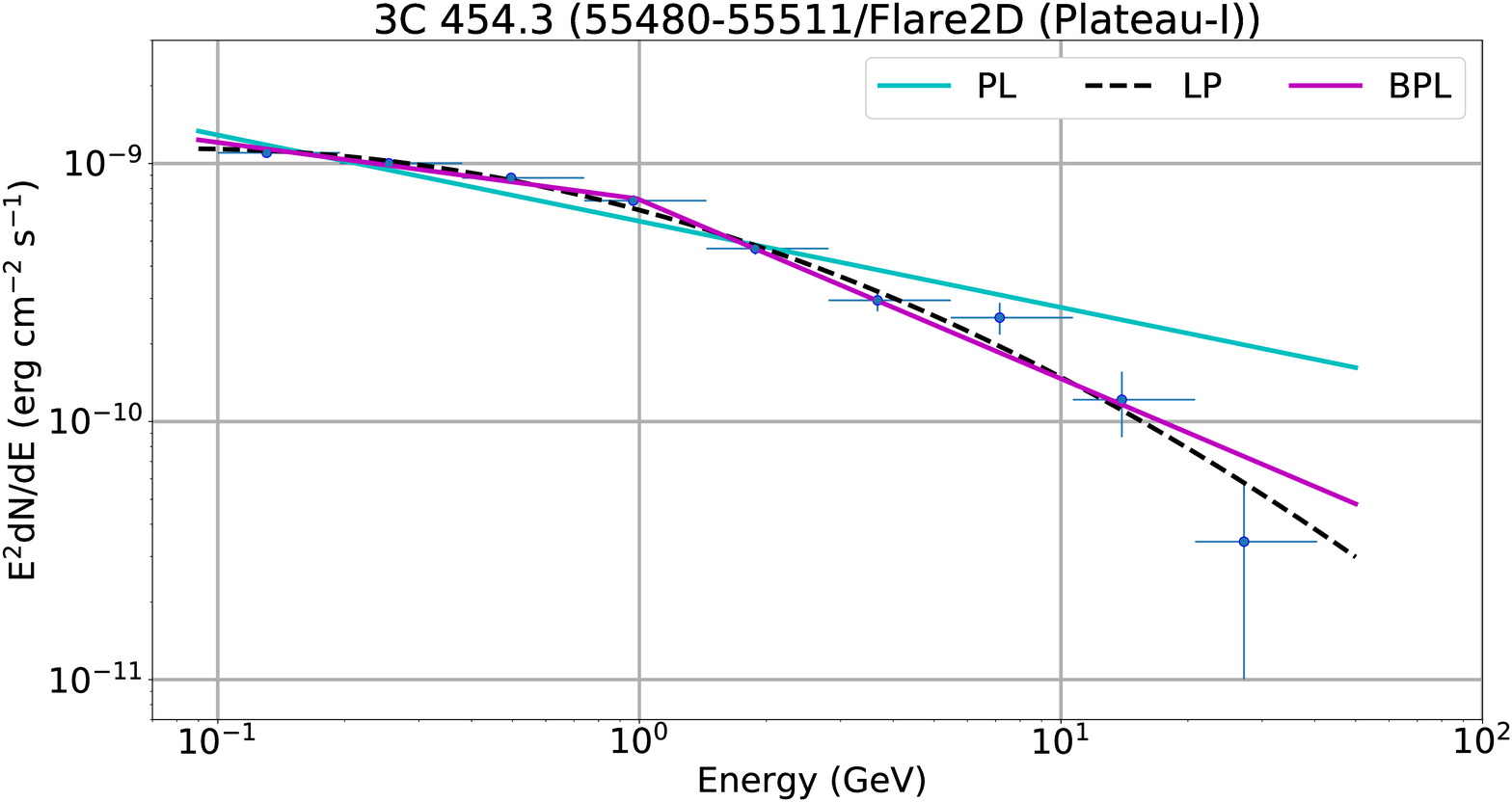}
\includegraphics[height=1.77in,width=2.5in]{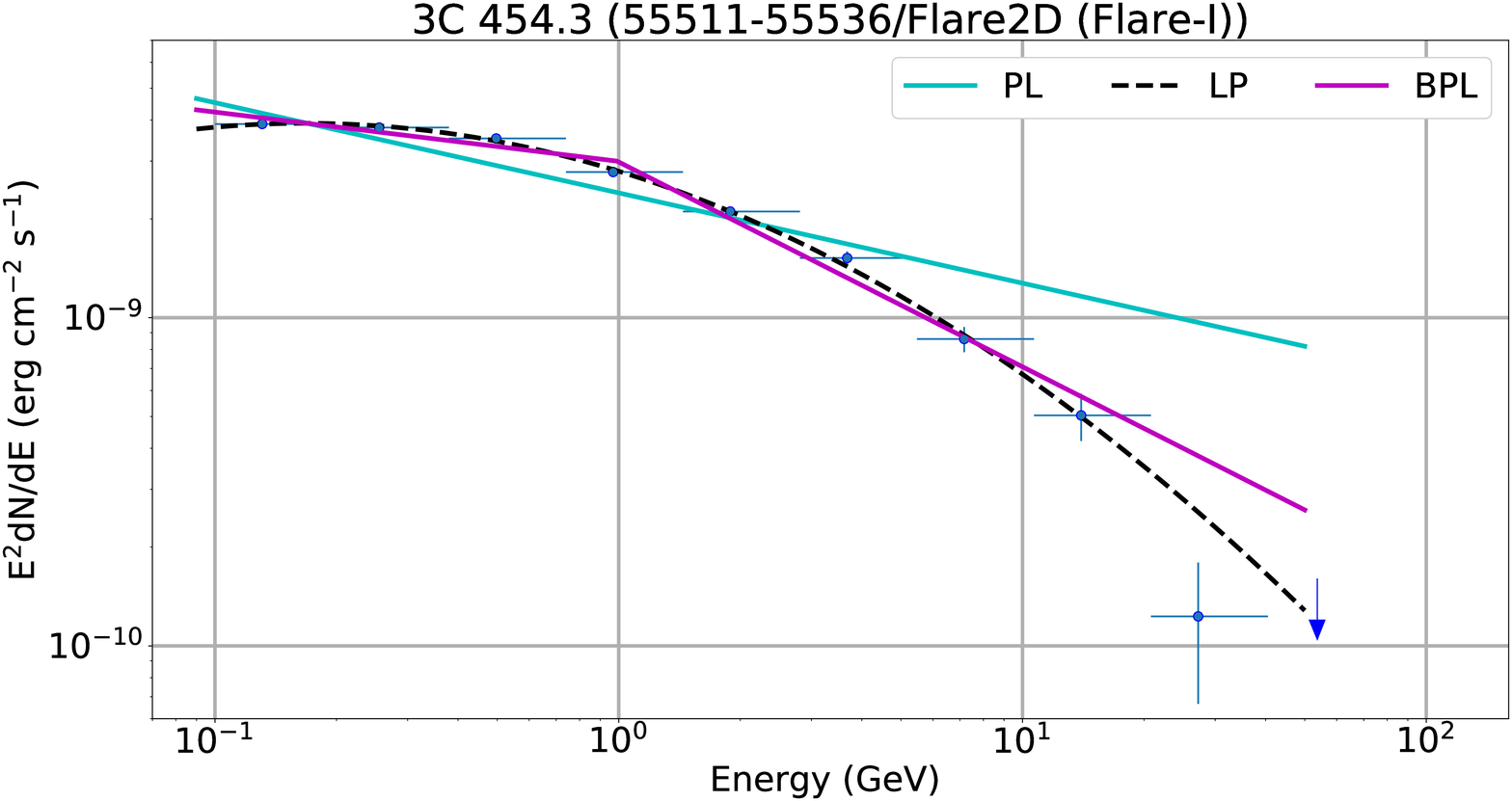}
\includegraphics[height=1.77in,width=2.5in]{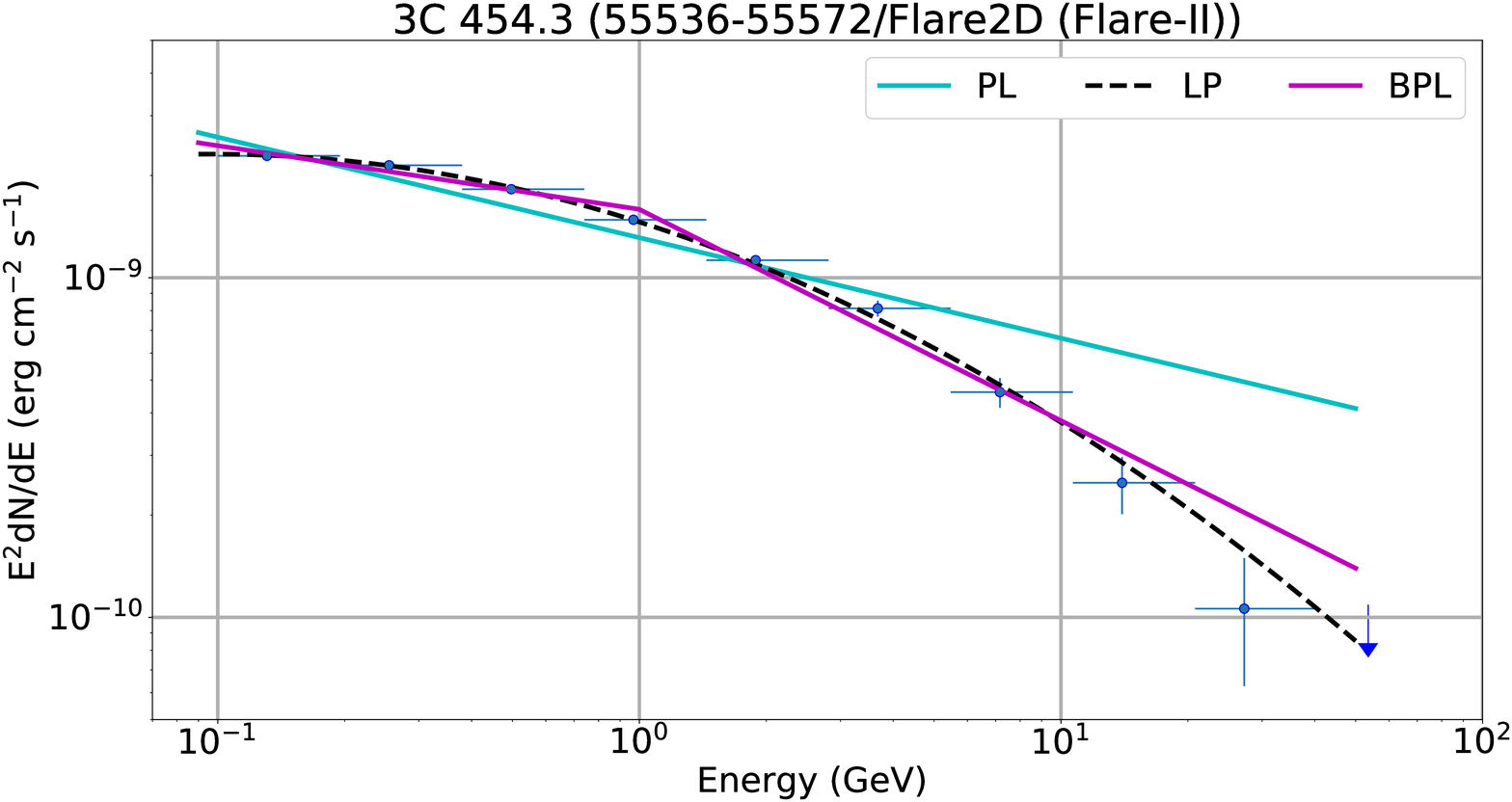}
\includegraphics[height=1.77in,width=2.5in]{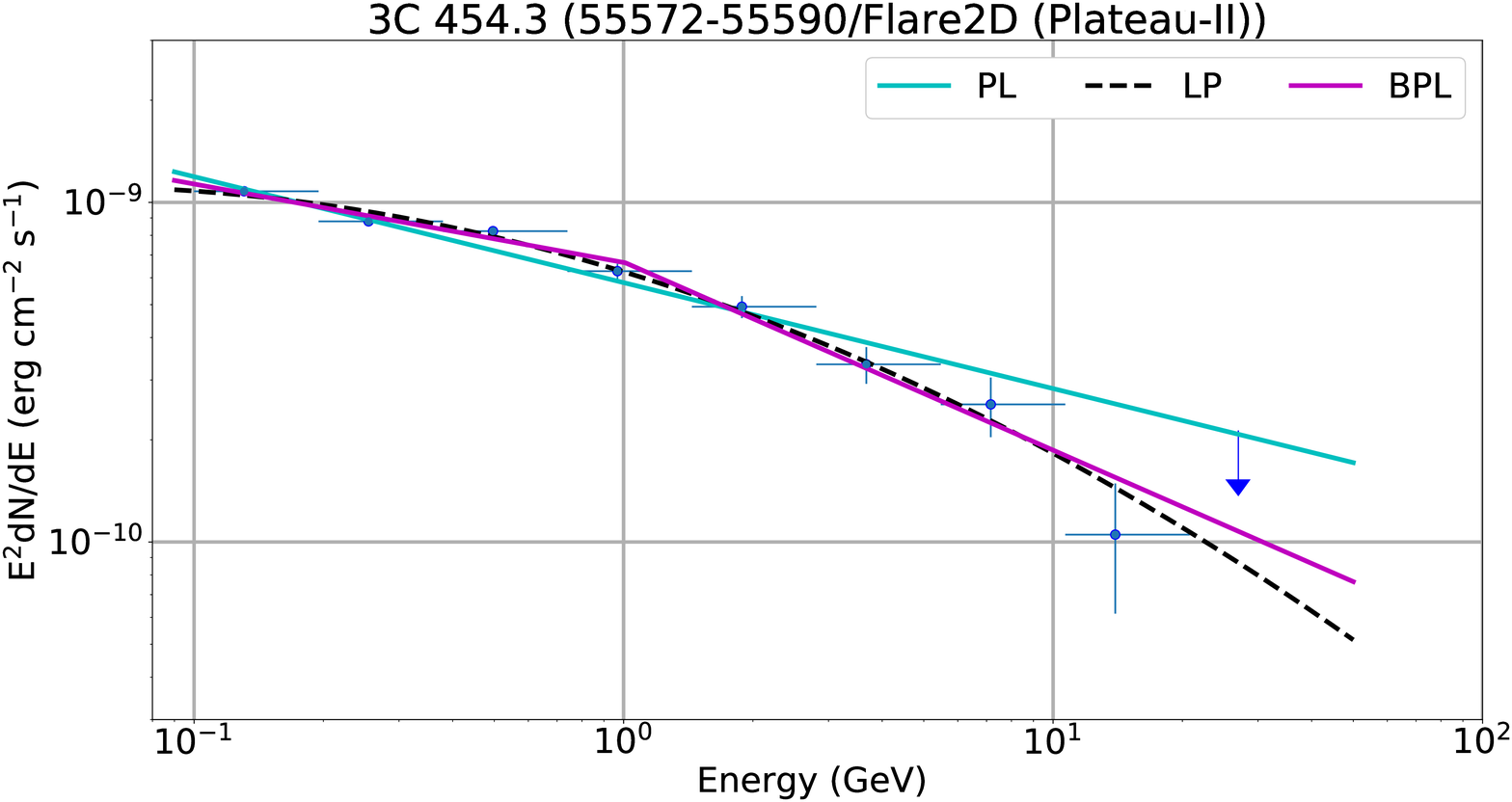}
\includegraphics[height=1.77in,width=2.5in]{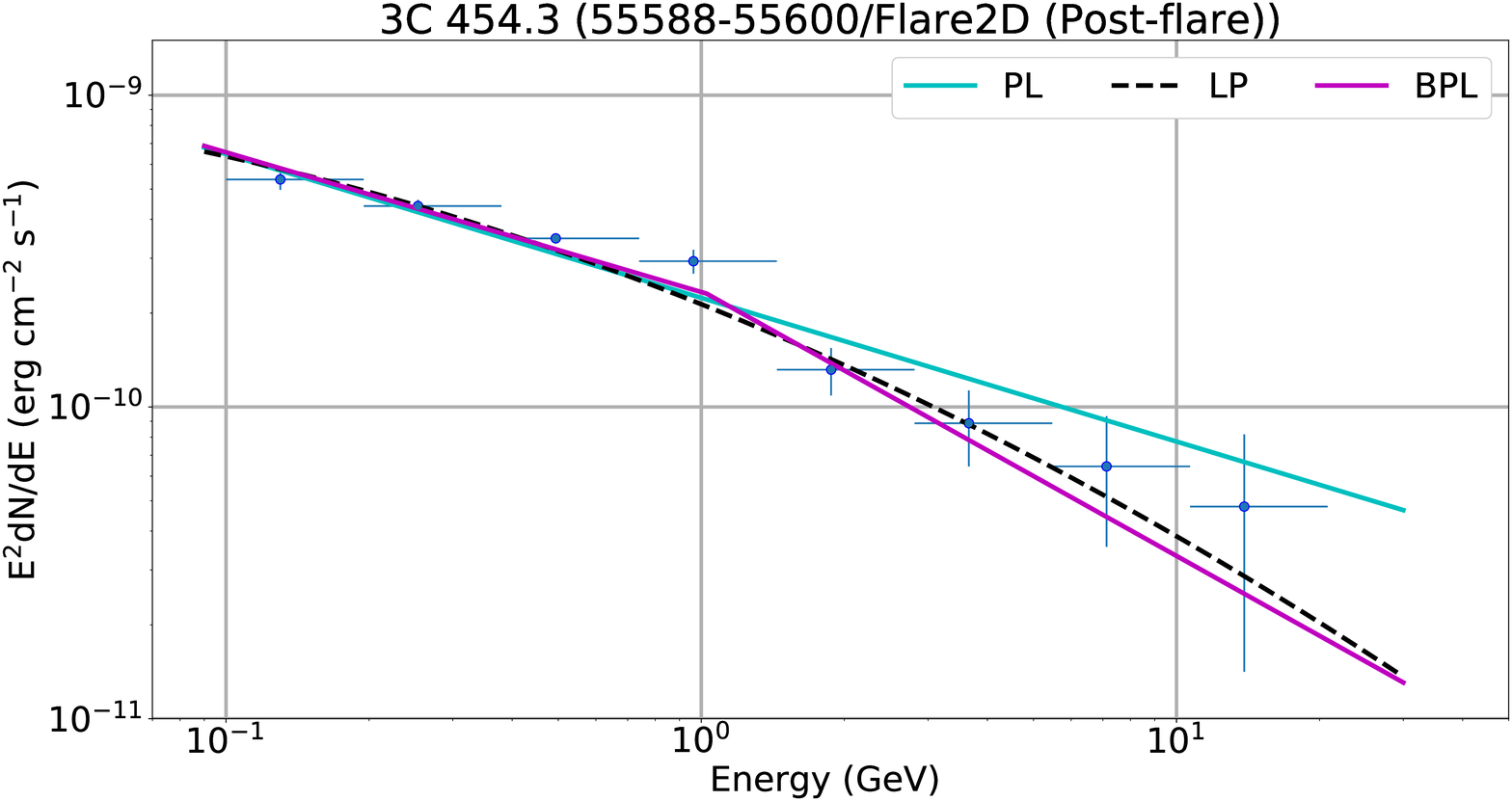}
\caption[optional]{SED of different periods of Flare-2D as given in Figure-12. PL, LP \& BPL describe the Powerlaw, Logparabola and Broken-powerlaw model respectively, which are fitted to data points.}

\end{figure*}

\begin{figure*}[h!]
\centering
\includegraphics[height=1.77in,width=2.5in]{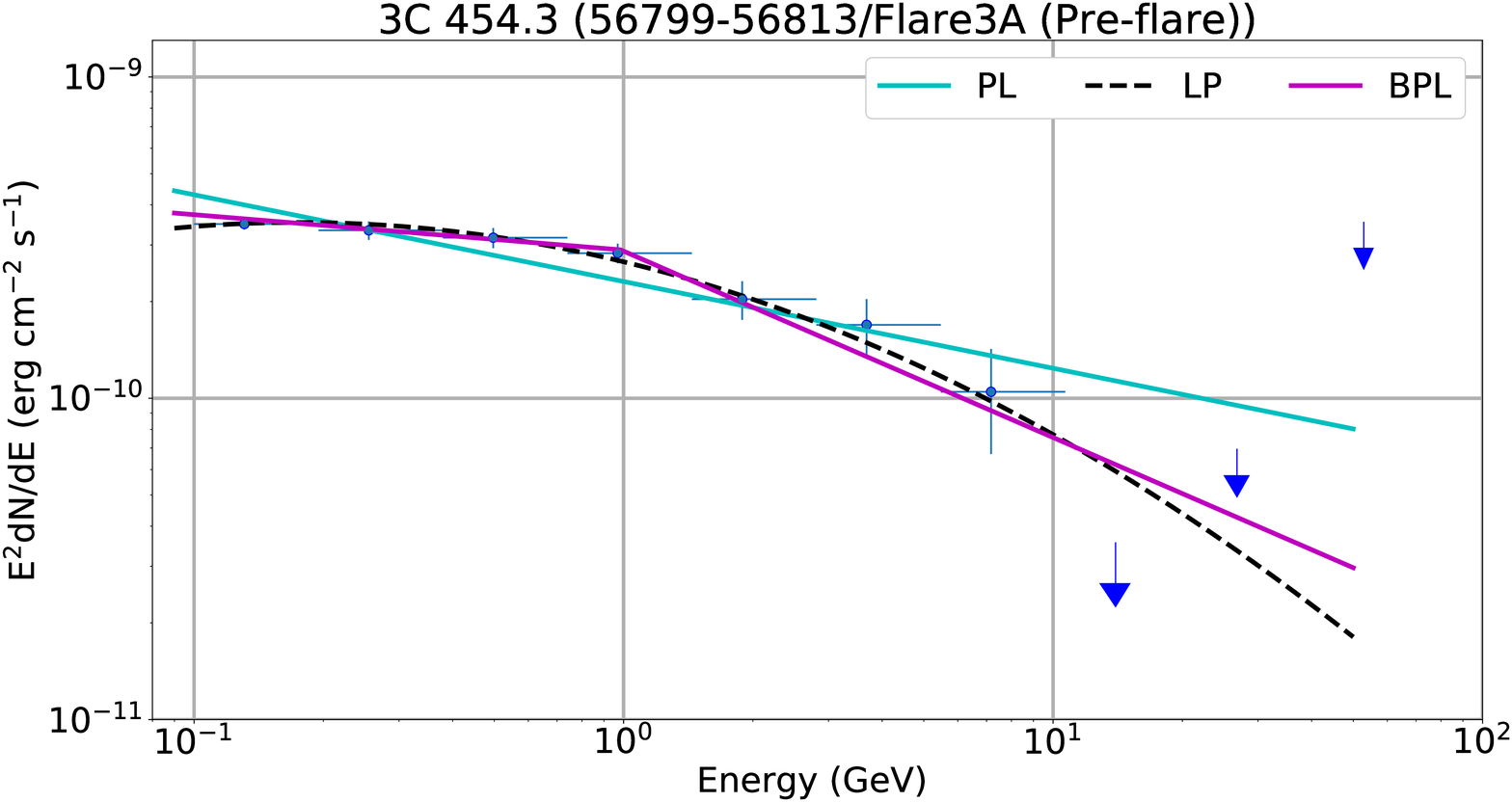}
\includegraphics[height=1.77in,width=2.5in]{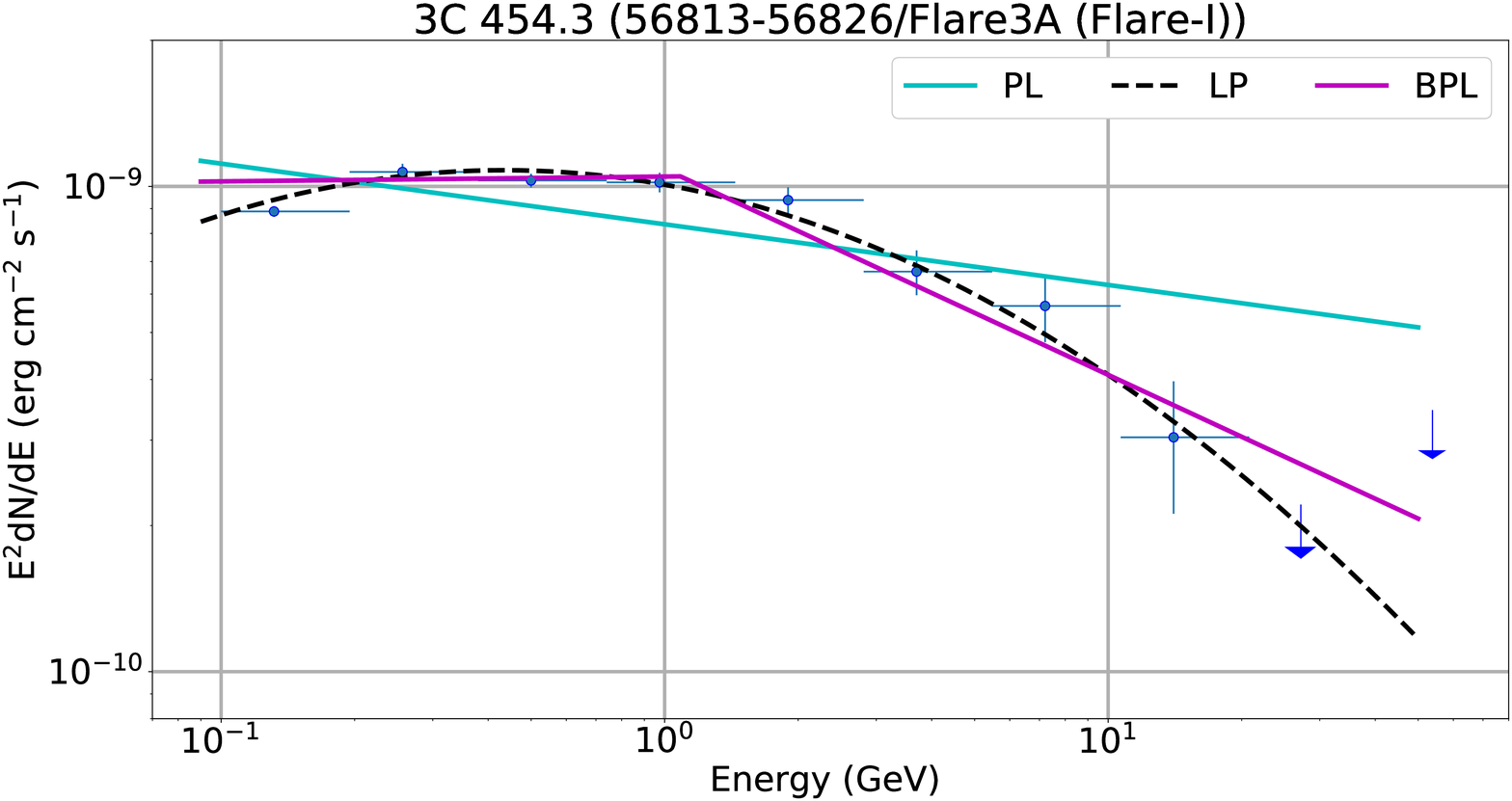}
\includegraphics[height=1.77in,width=2.5in]{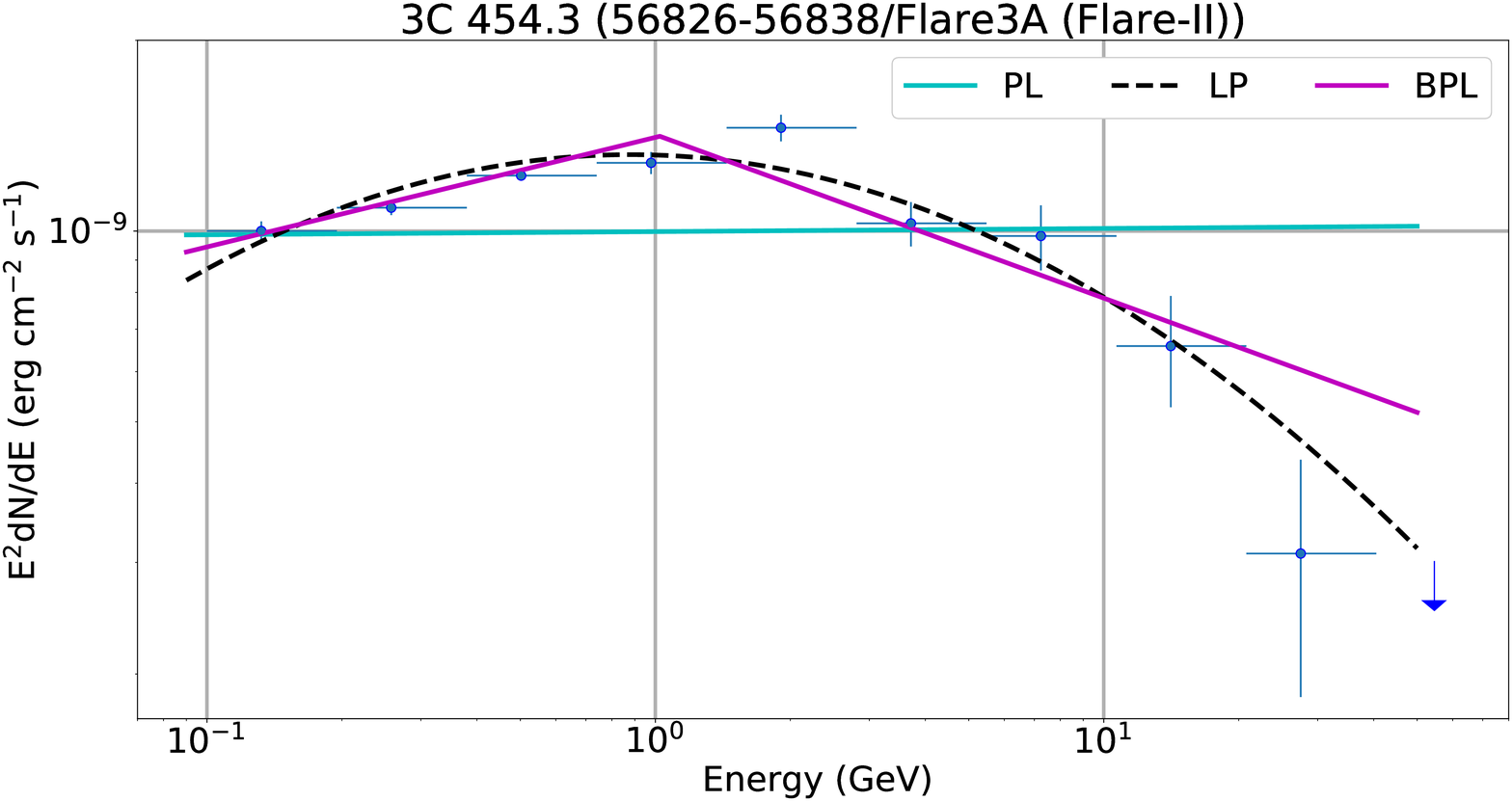}
\includegraphics[height=1.77in,width=2.5in]{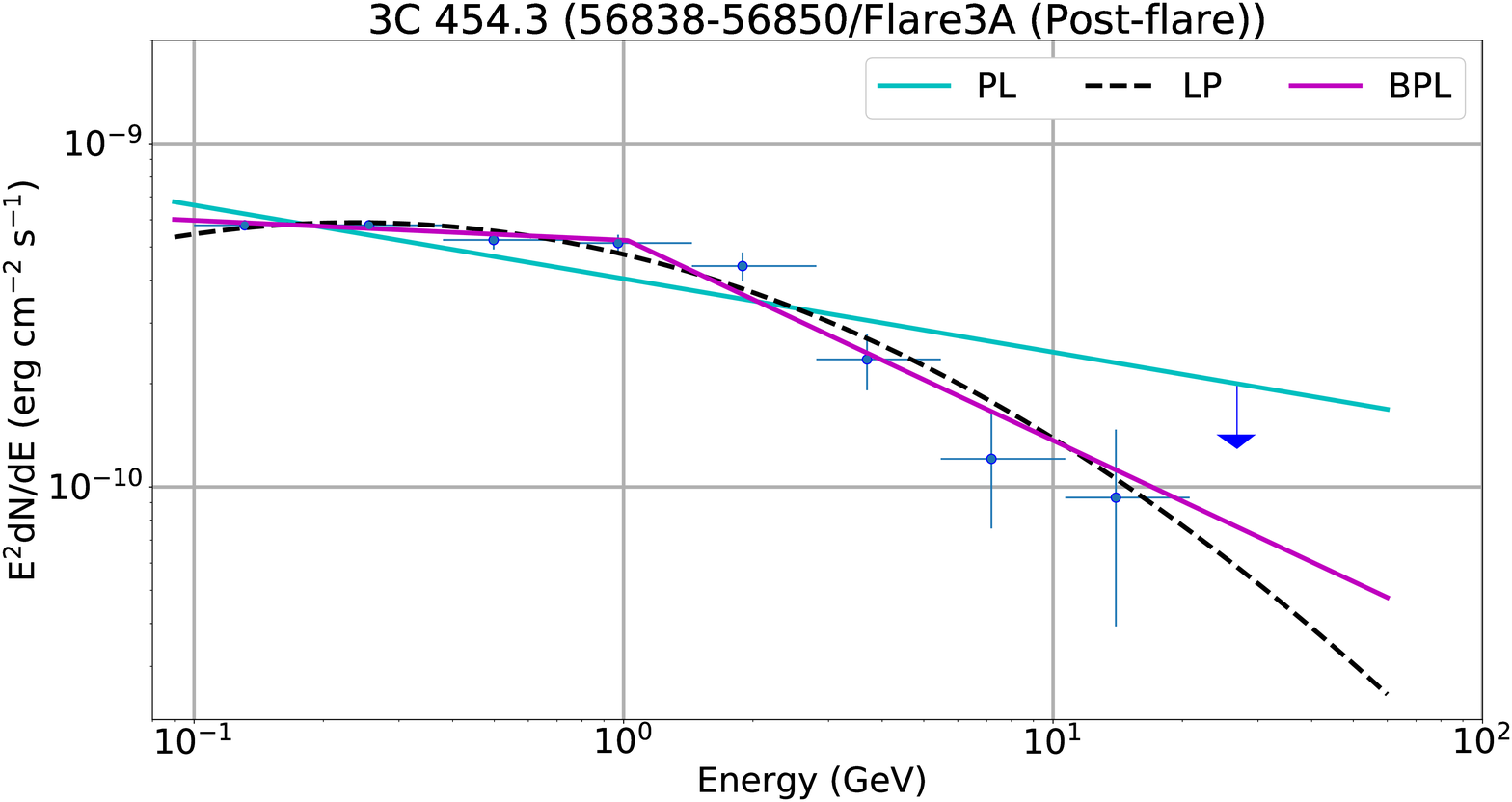}
\caption[optional]{SED of different periods of Flare-3A as given in Figure-15. PL, LP \& BPL describe the Powerlaw, Logparabola and Broken-powerlaw model respectively, which are fitted to data points.}

\includegraphics[height=1.77in,width=2.5in]{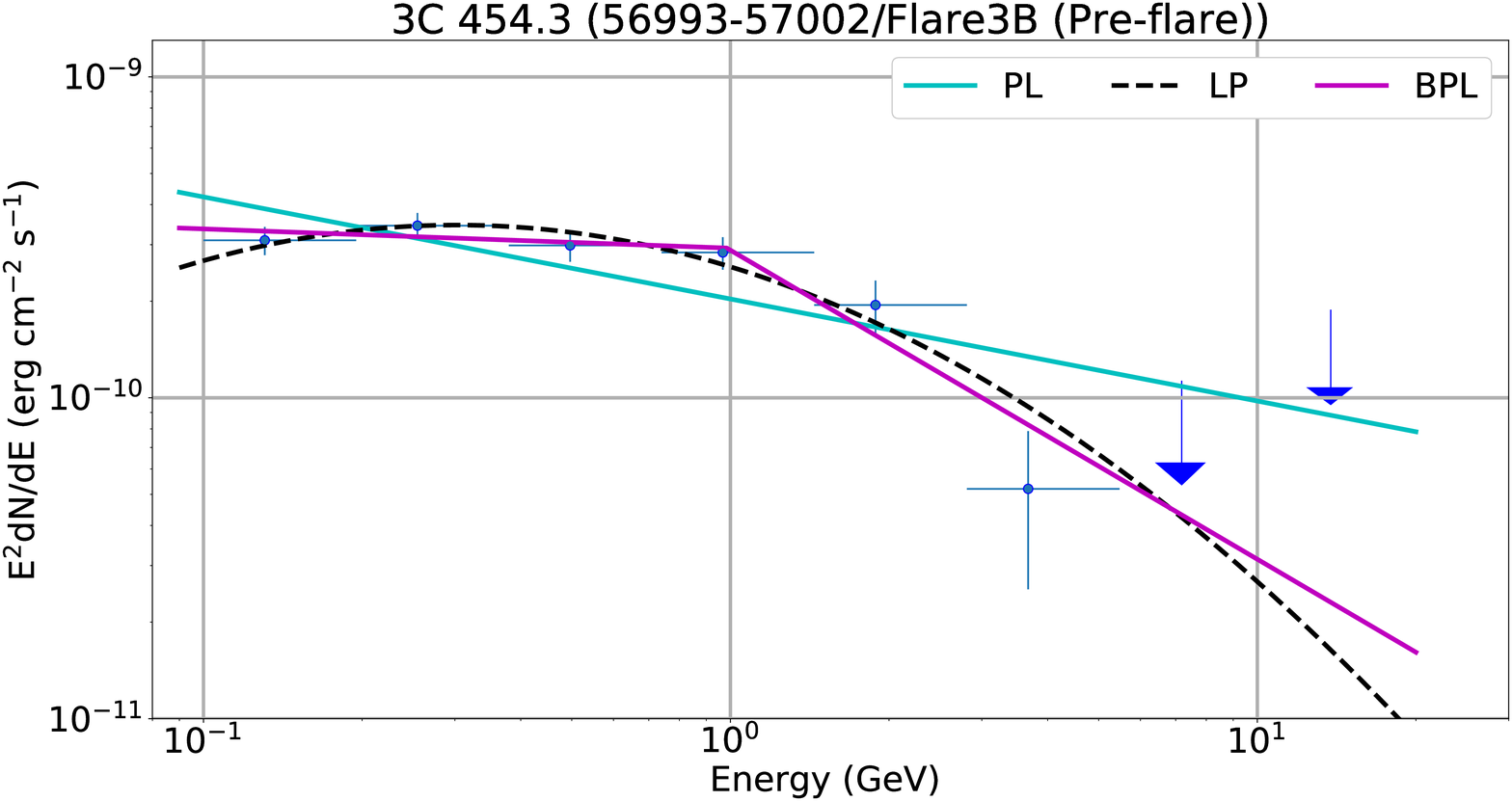}
\includegraphics[height=1.77in,width=2.5in]{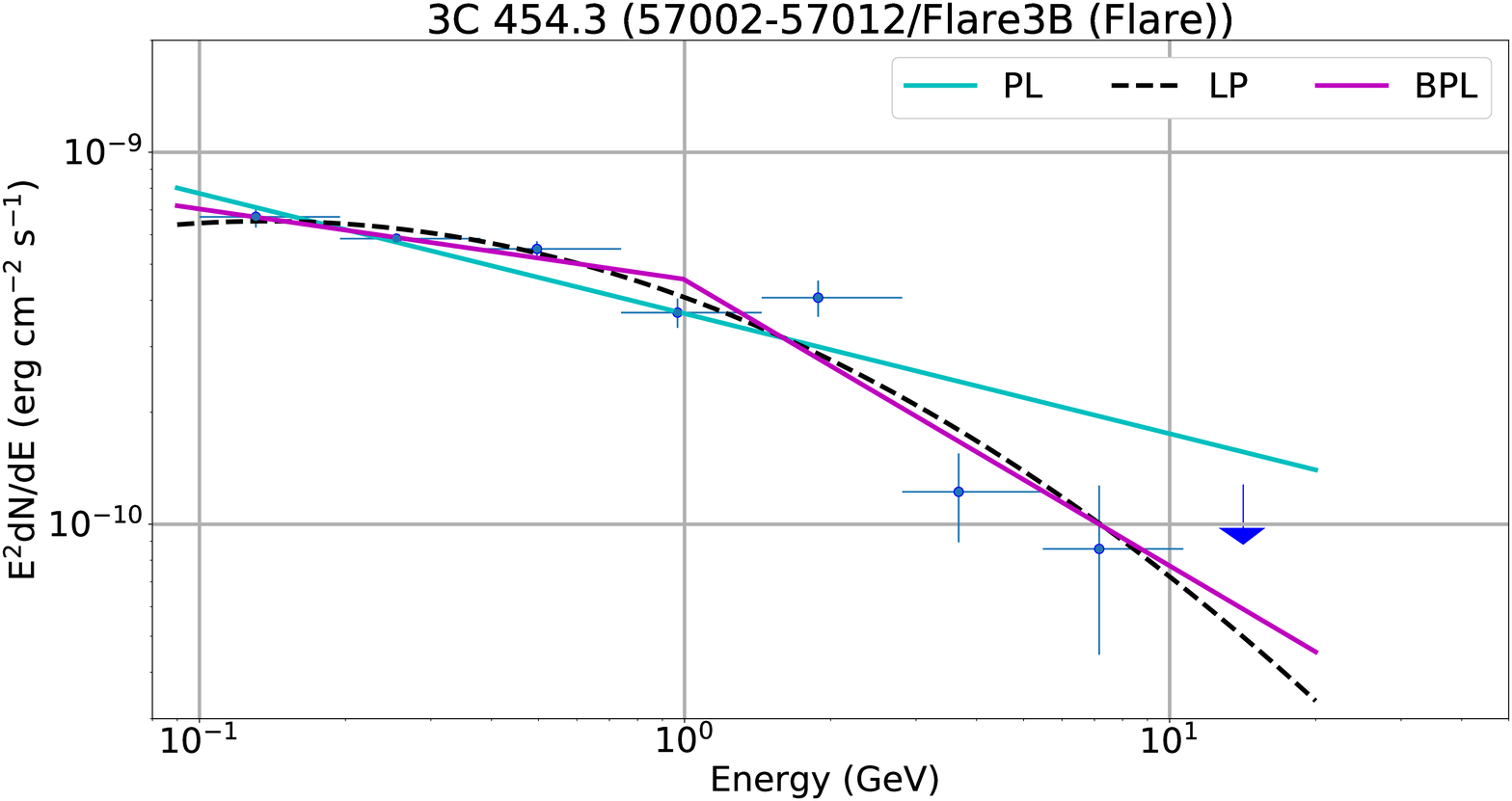}
\includegraphics[height=1.77in,width=2.5in]{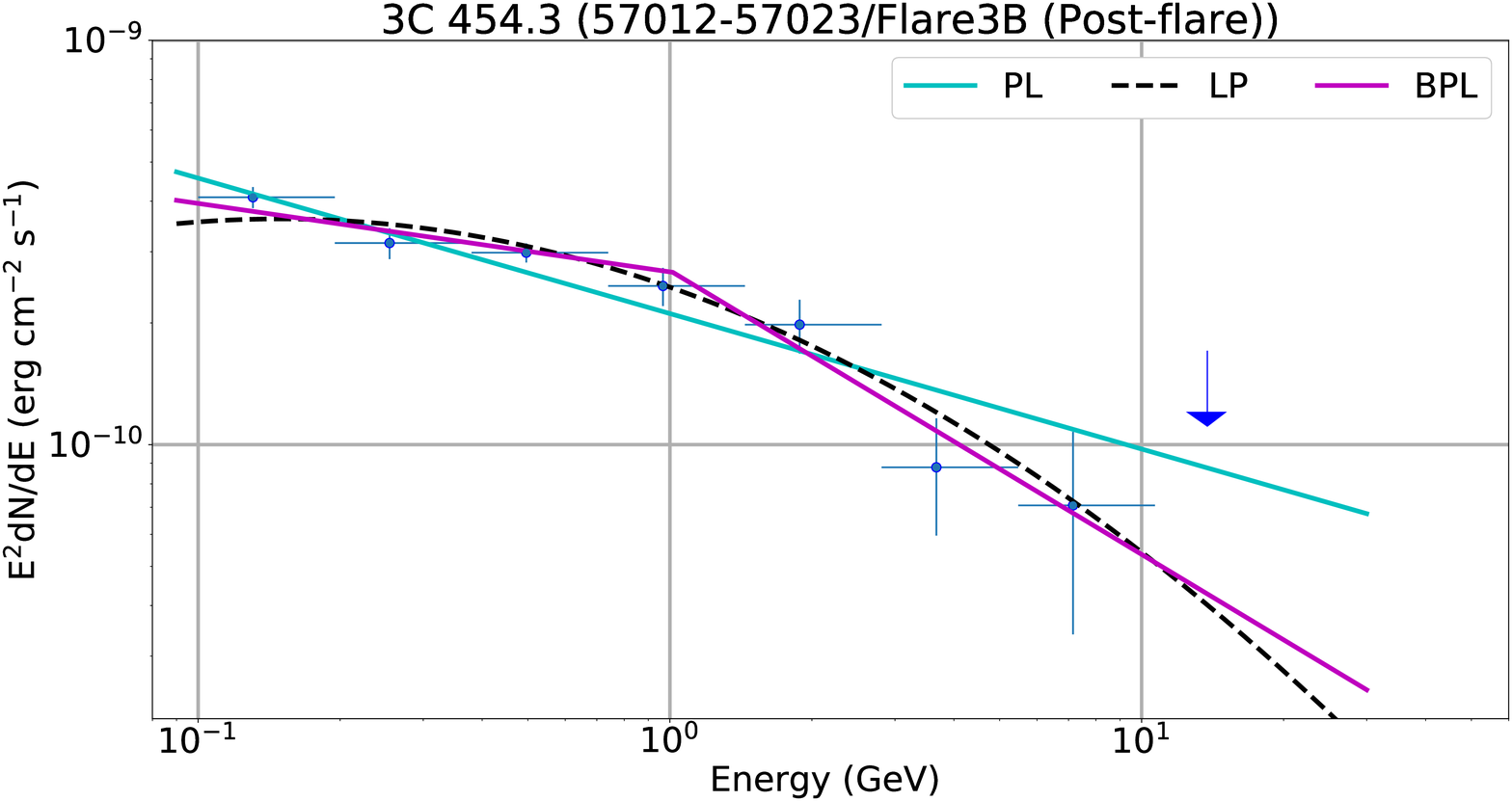}
\caption[optional]{SED of different periods of Flare-3B as given in Figure-18. PL, LP \& BPL describe the Powerlaw, Logparabola and Broken-powerlaw model respectively, which are fitted to data points.}

\includegraphics[height=1.77in,width=2.5in]{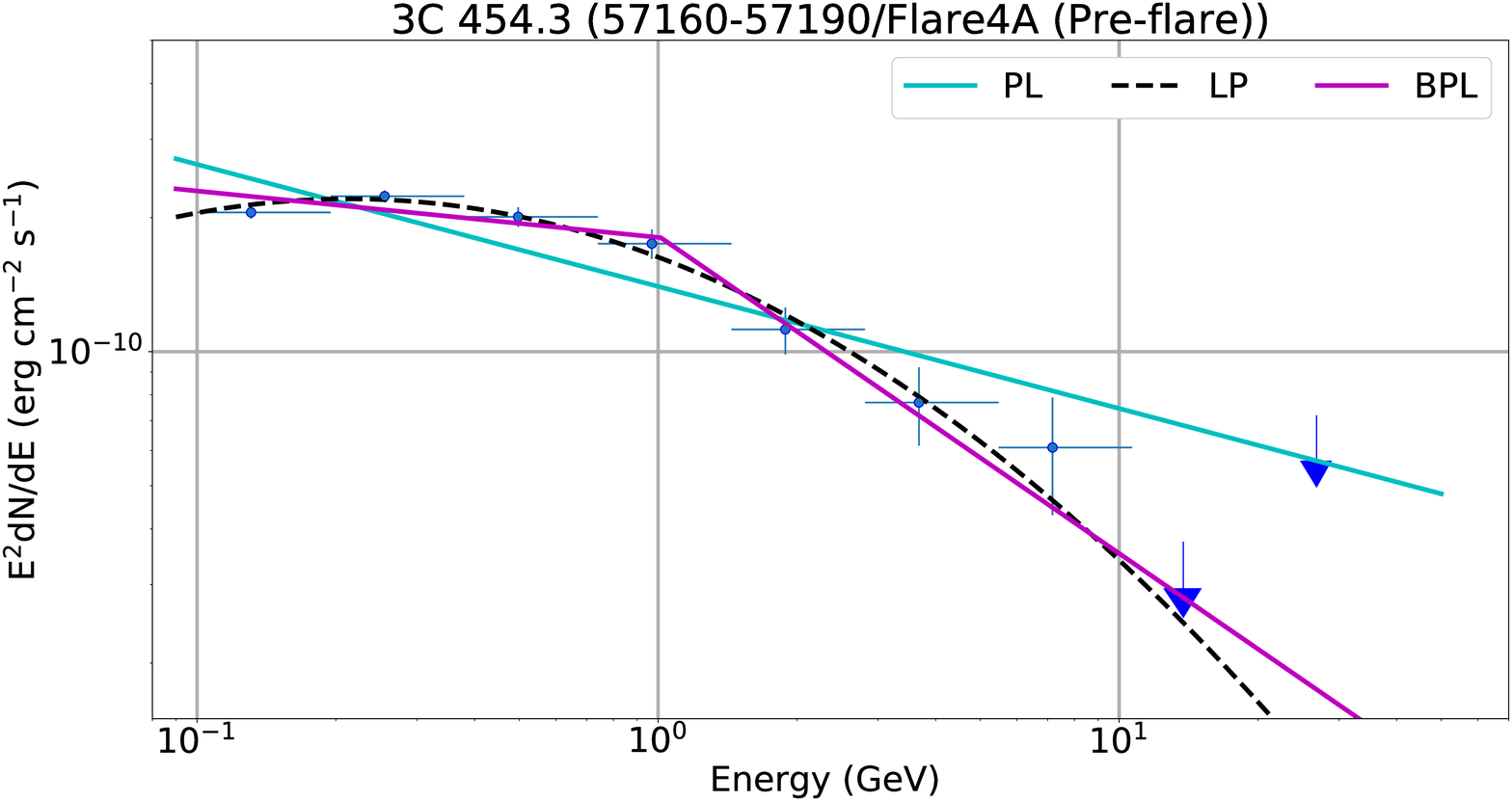}
\end{figure*}

\begin{figure*}[h!]
\centering

\includegraphics[height=1.75in,width=2.5in]{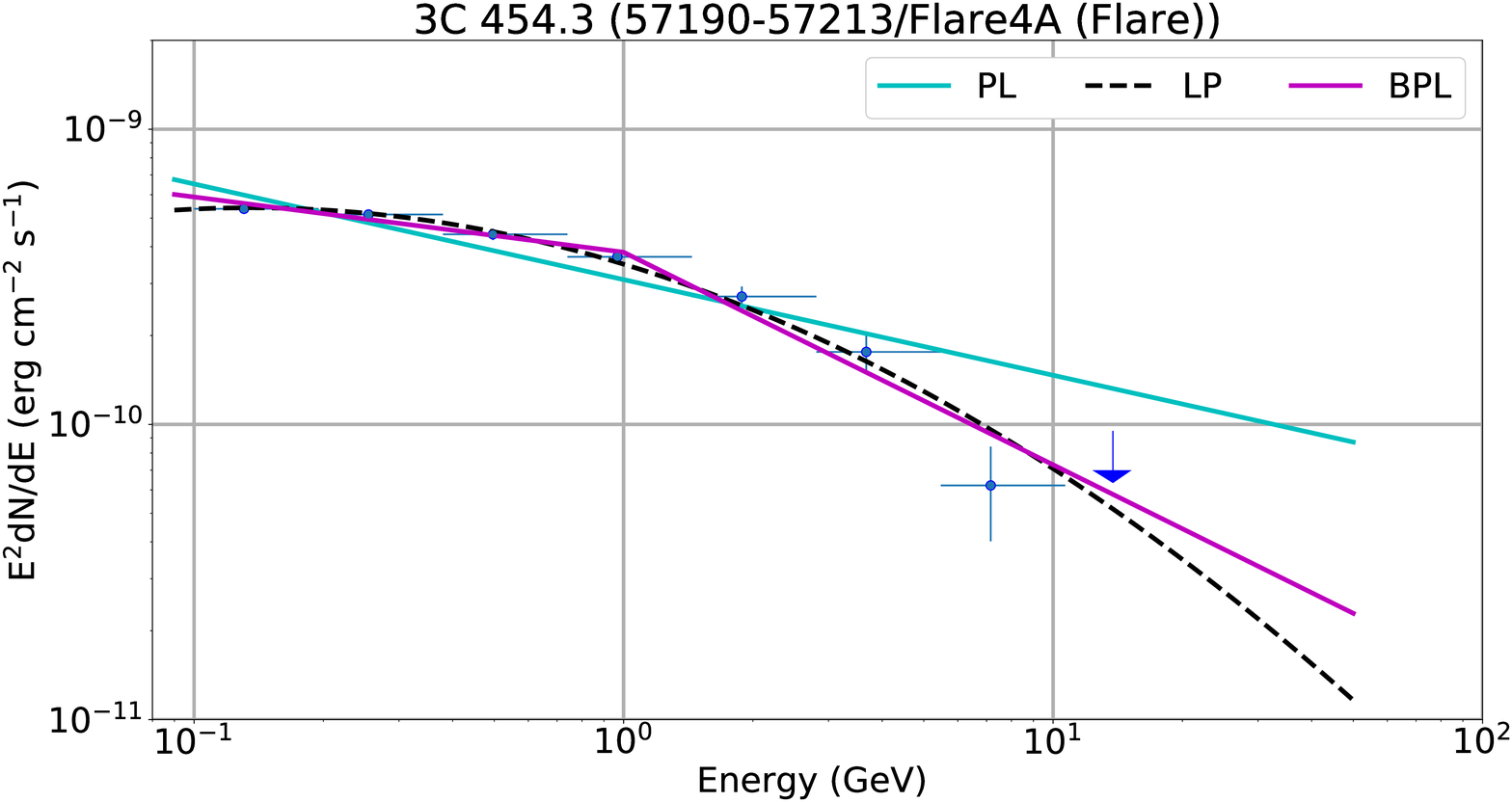}
\includegraphics[height=1.75in,width=2.5in]{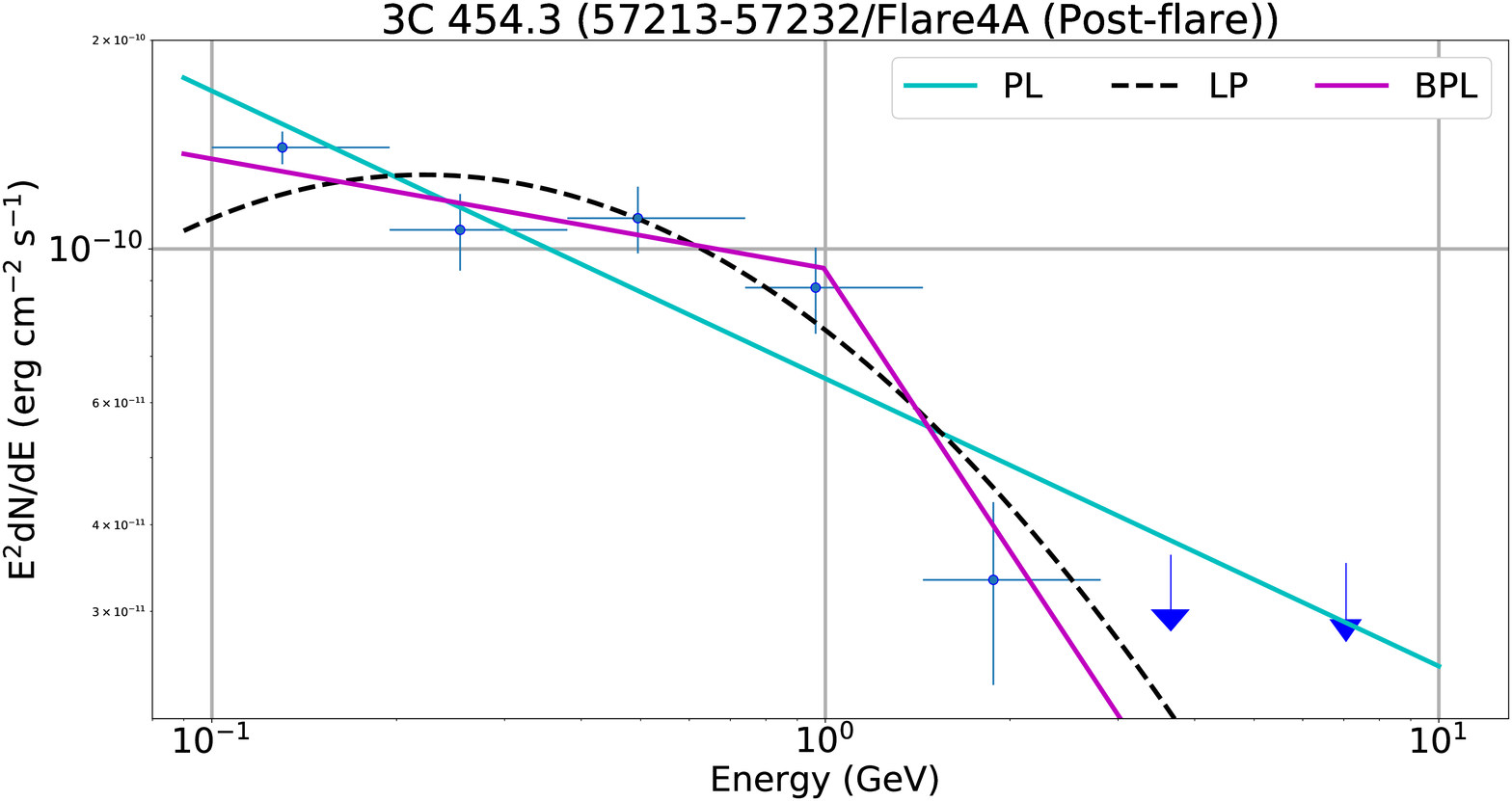}
\caption[optional]{SED of different periods of Flare-4A as given in Figure-20. PL, LP \& BPL describe the Powerlaw, Logparabola and Broken-powerlaw model respectively, which are fitted to data points.}

\includegraphics[height=1.75in,width=2.5in]{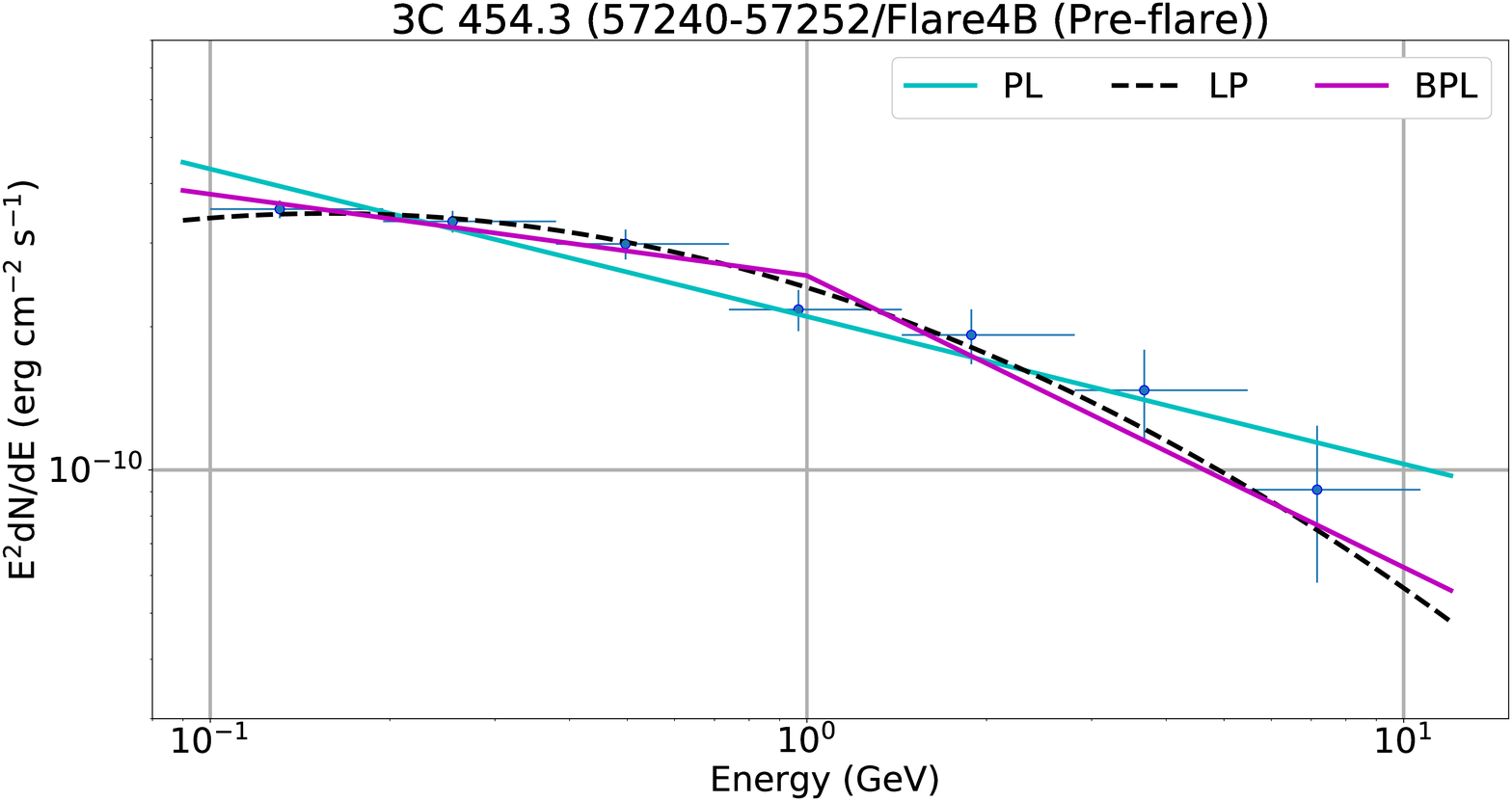}
\includegraphics[height=1.75in,width=2.5in]{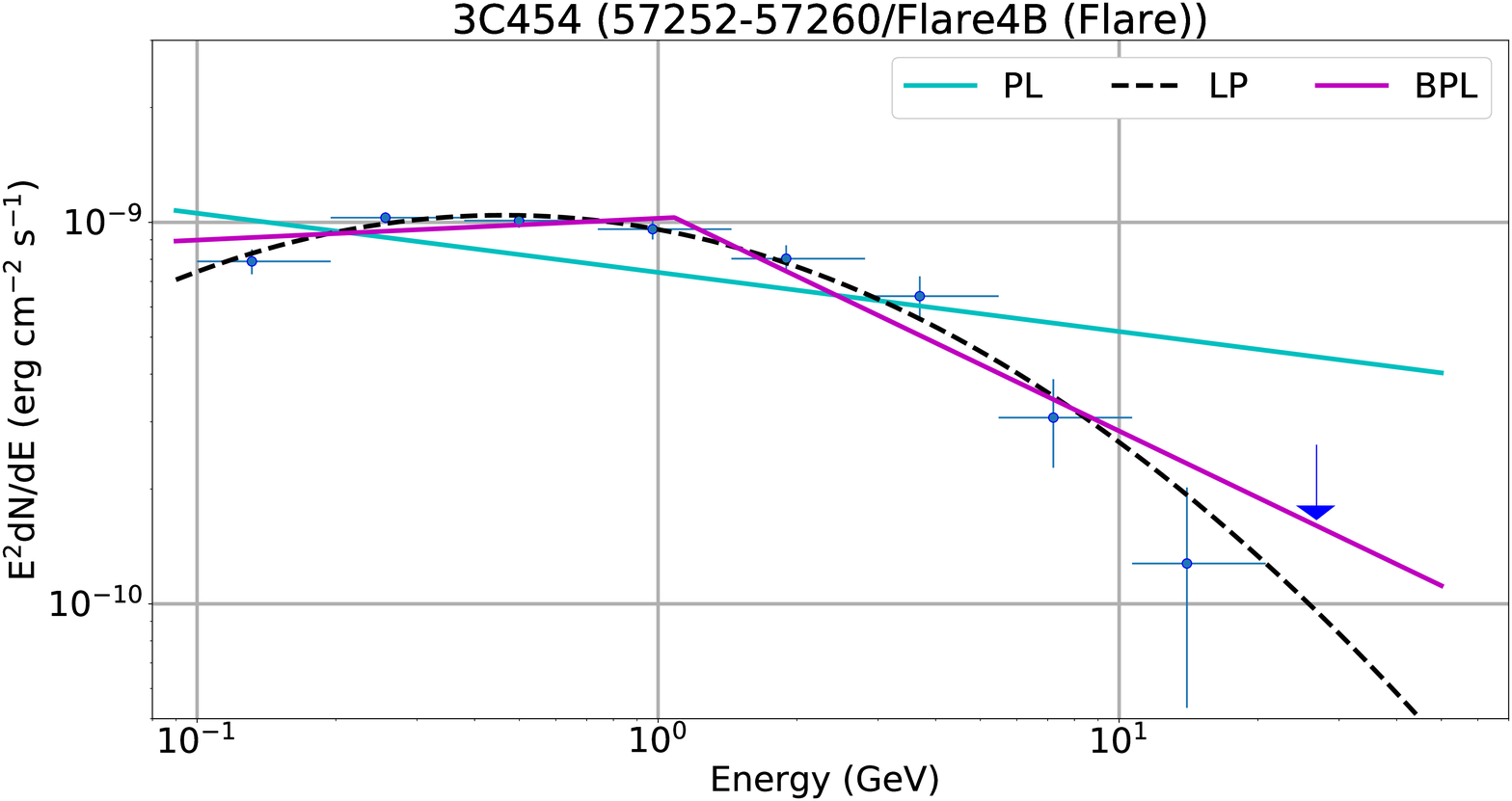}
\includegraphics[height=1.75in,width=2.5in]{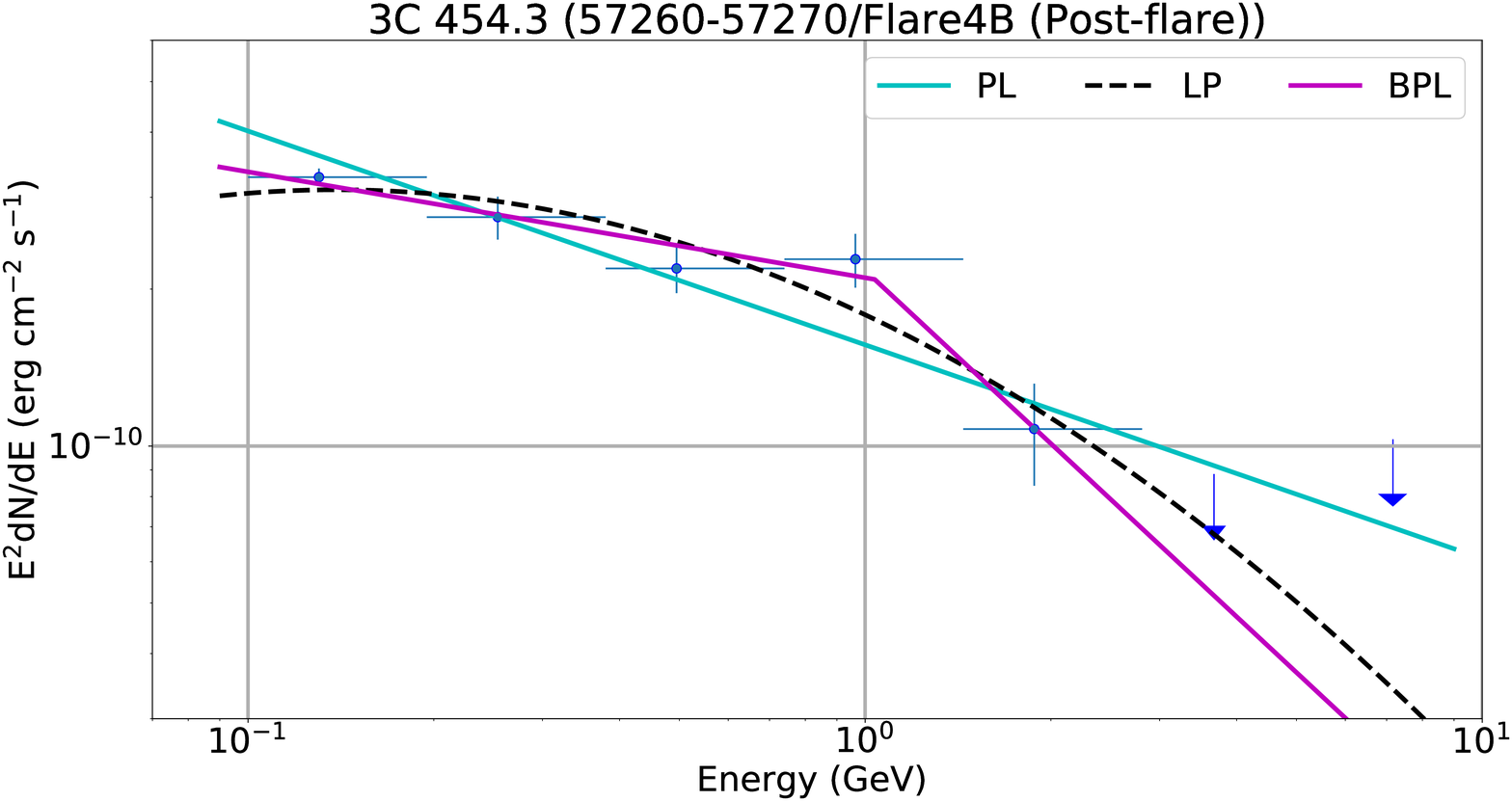}
\caption[optional]{SED of different periods of Flare-4B as given in Figure-22. PL, LP \& BPL describe the Powerlaw, Logparabola and Broken-powerlaw model respectively, which are fitted to data points.}

\includegraphics[height=1.75in,width=2.5in]{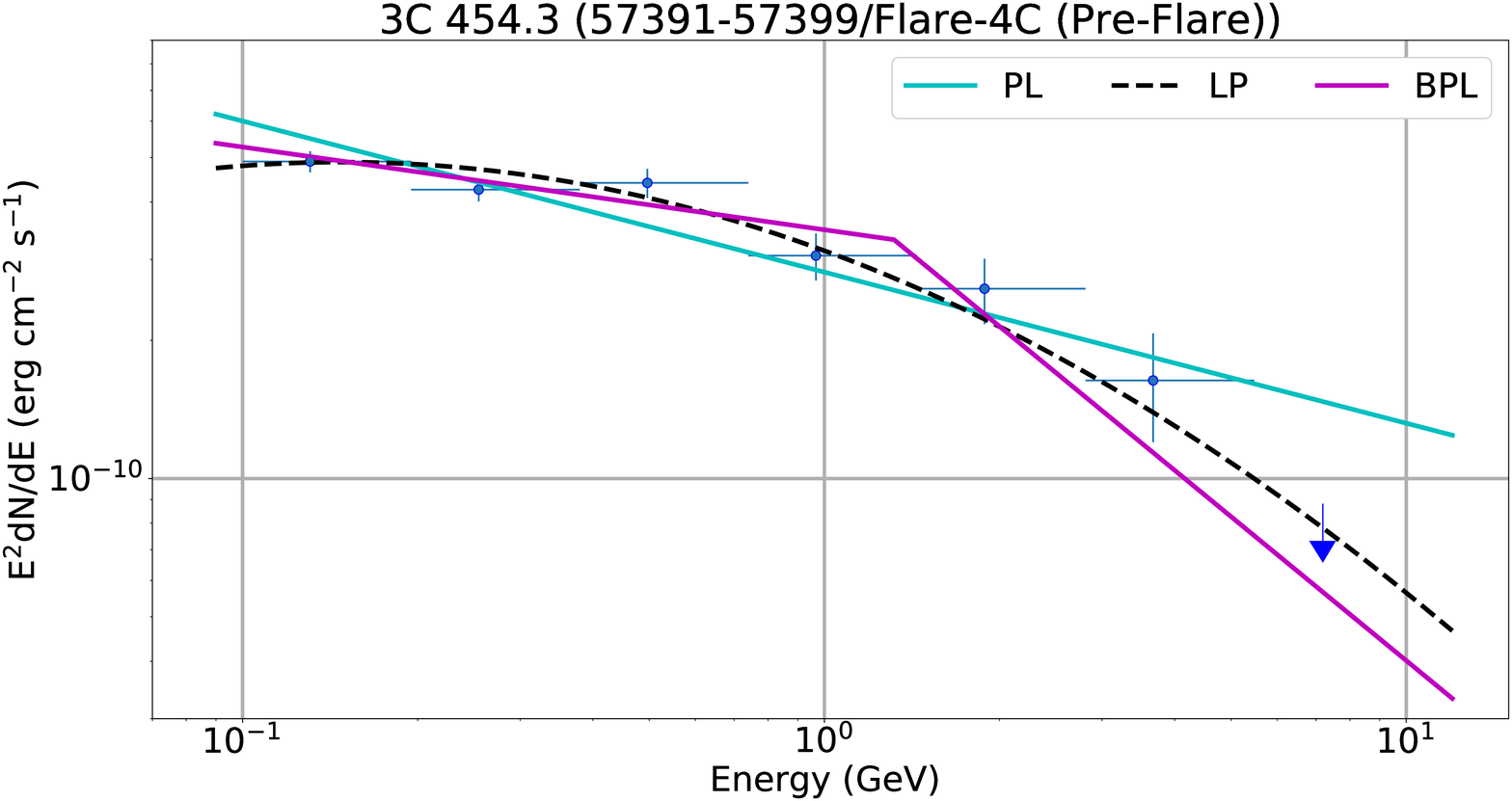}
\includegraphics[height=1.75in,width=2.5in]{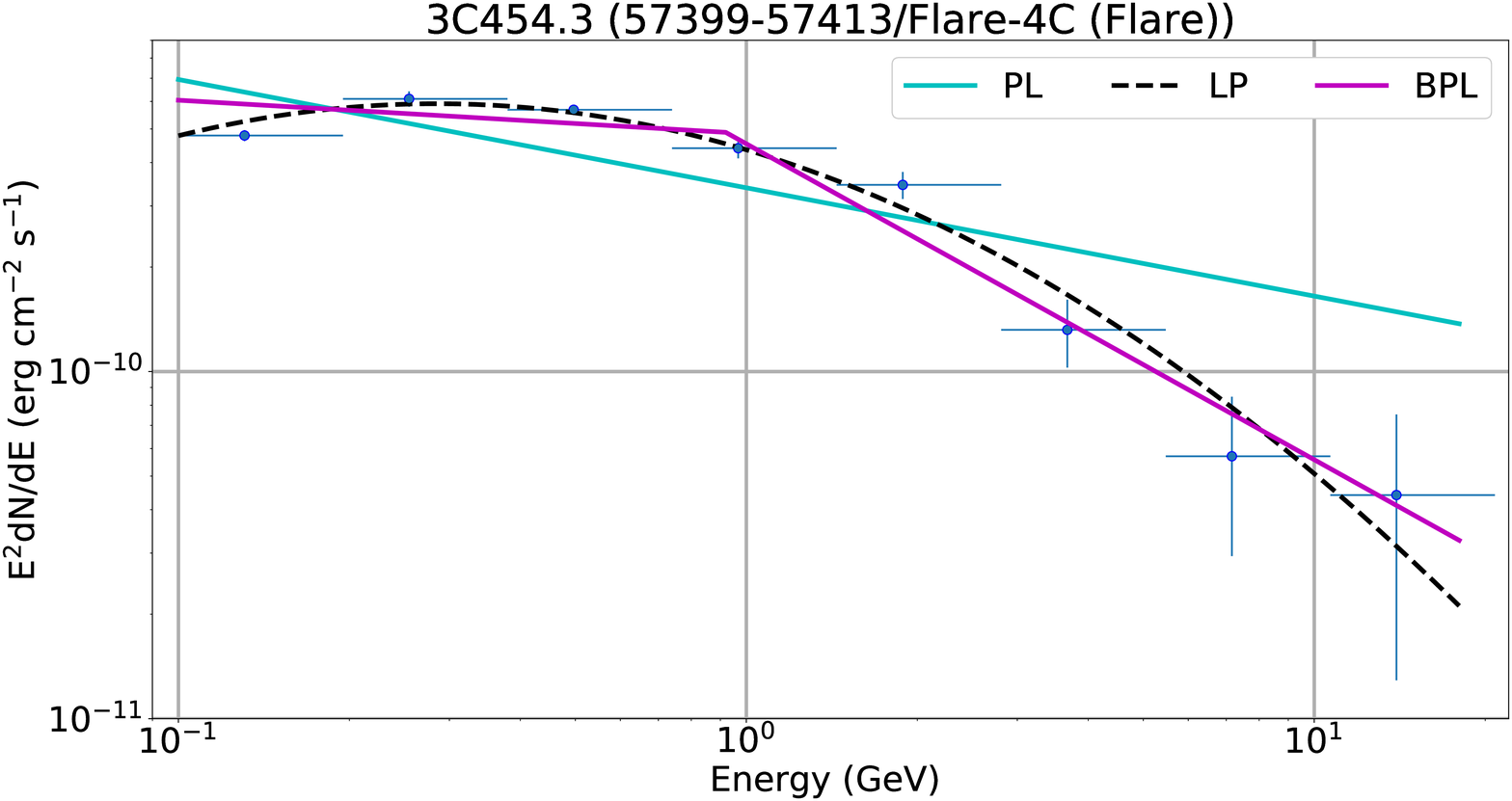}
\includegraphics[height=1.75in,width=2.5in]{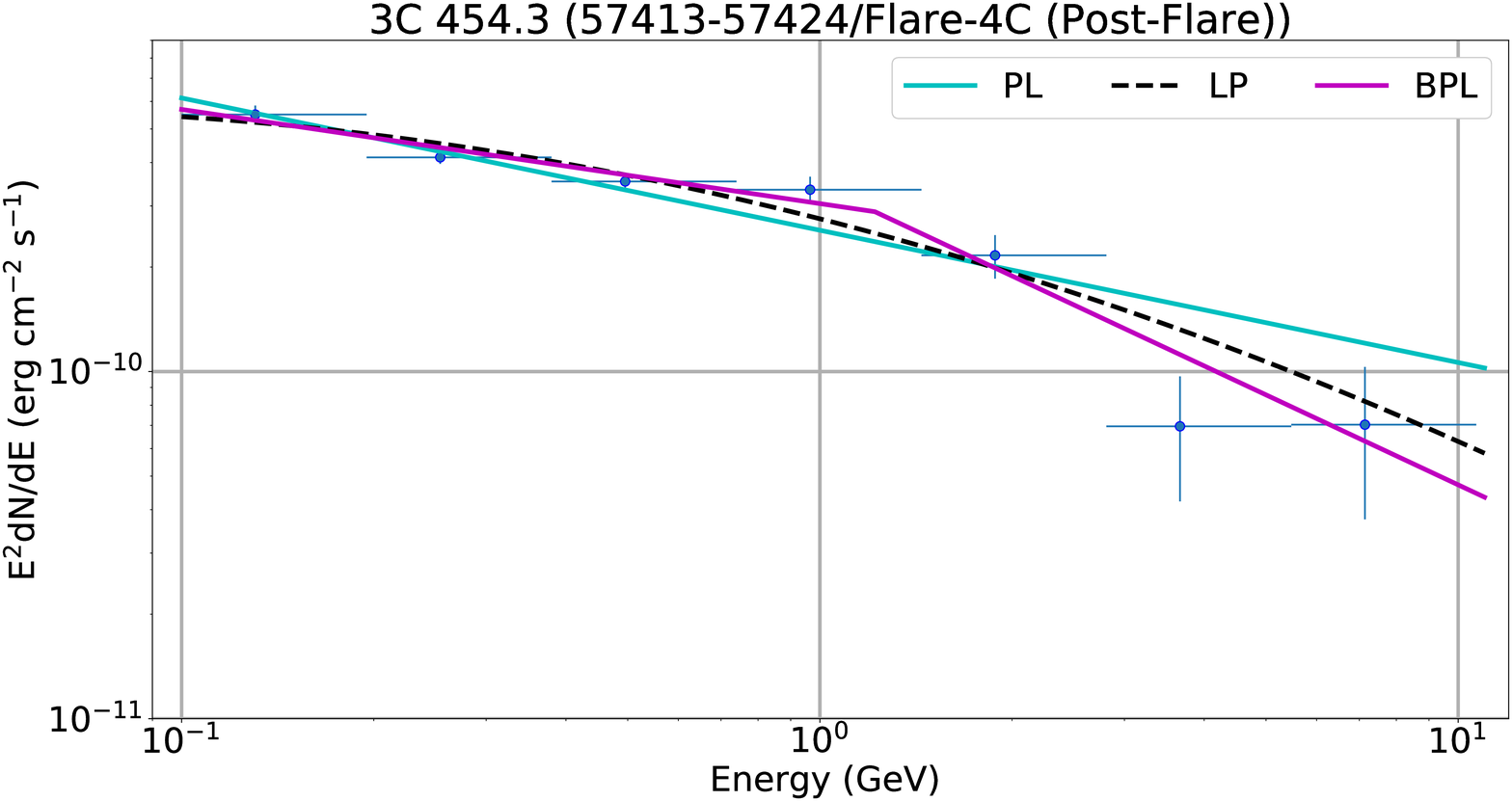}
\caption[optional]{SED of different periods of Flare-4C as given in Figure-24. PL, LP \& BPL describe the Powerlaw, Logparabola and Broken-powerlaw model respectively, which are fitted to data points.}

\end{figure*}

\begin{figure*}[h!]
\centering
\includegraphics[height=1.77in,width=2.5in]{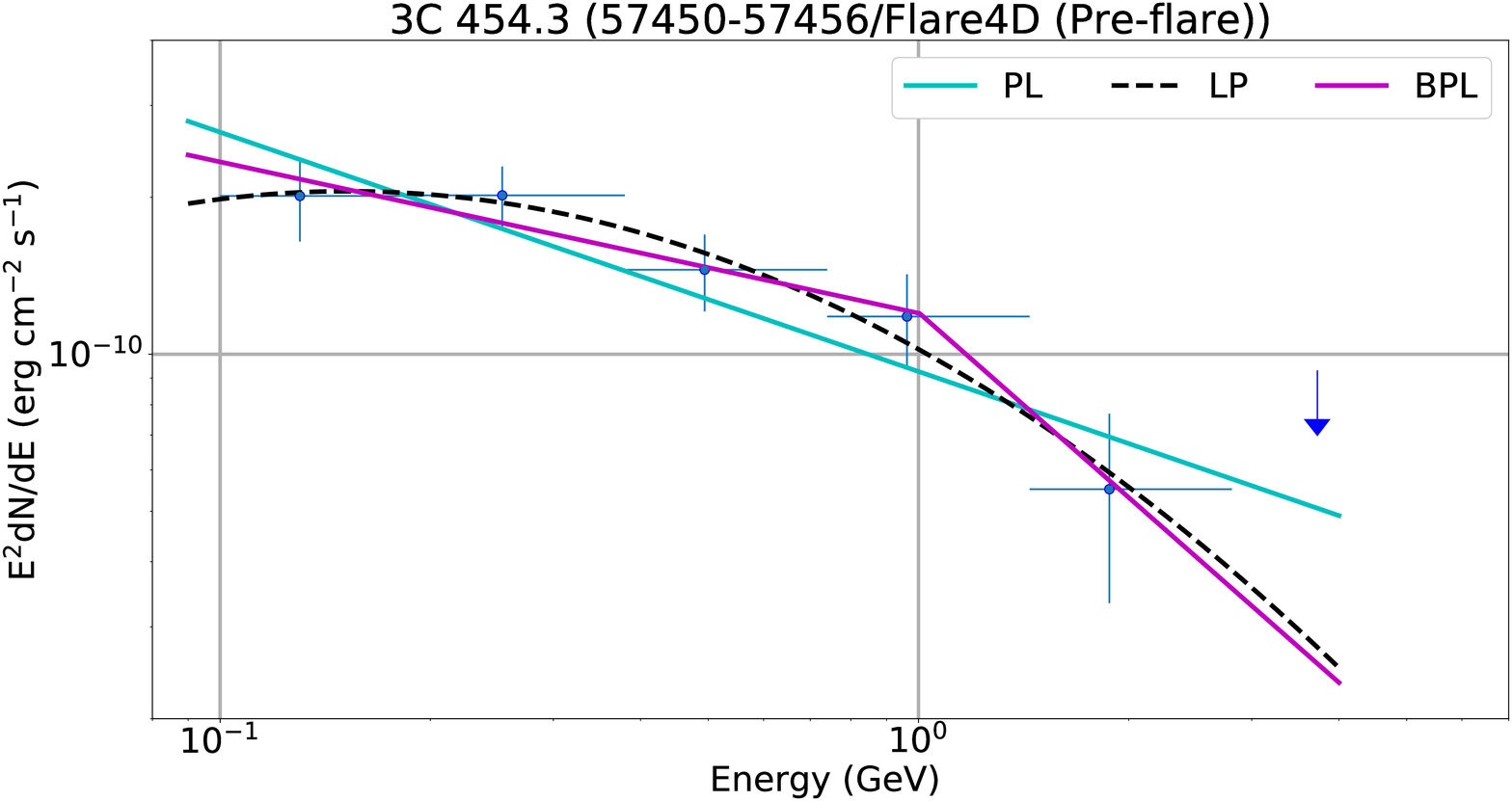}
\includegraphics[height=1.77in,width=2.5in]{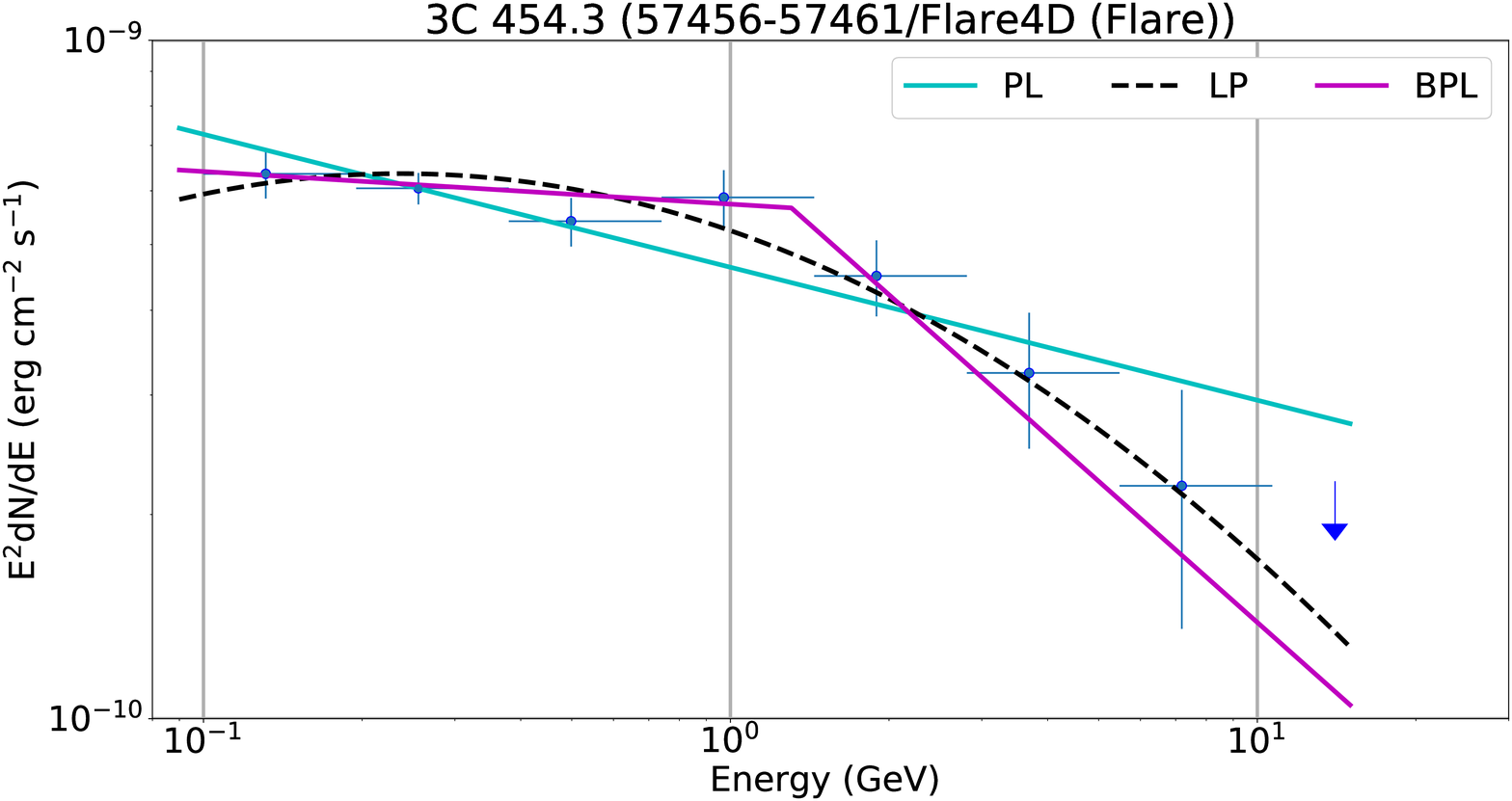}
\includegraphics[height=1.77in,width=2.5in]{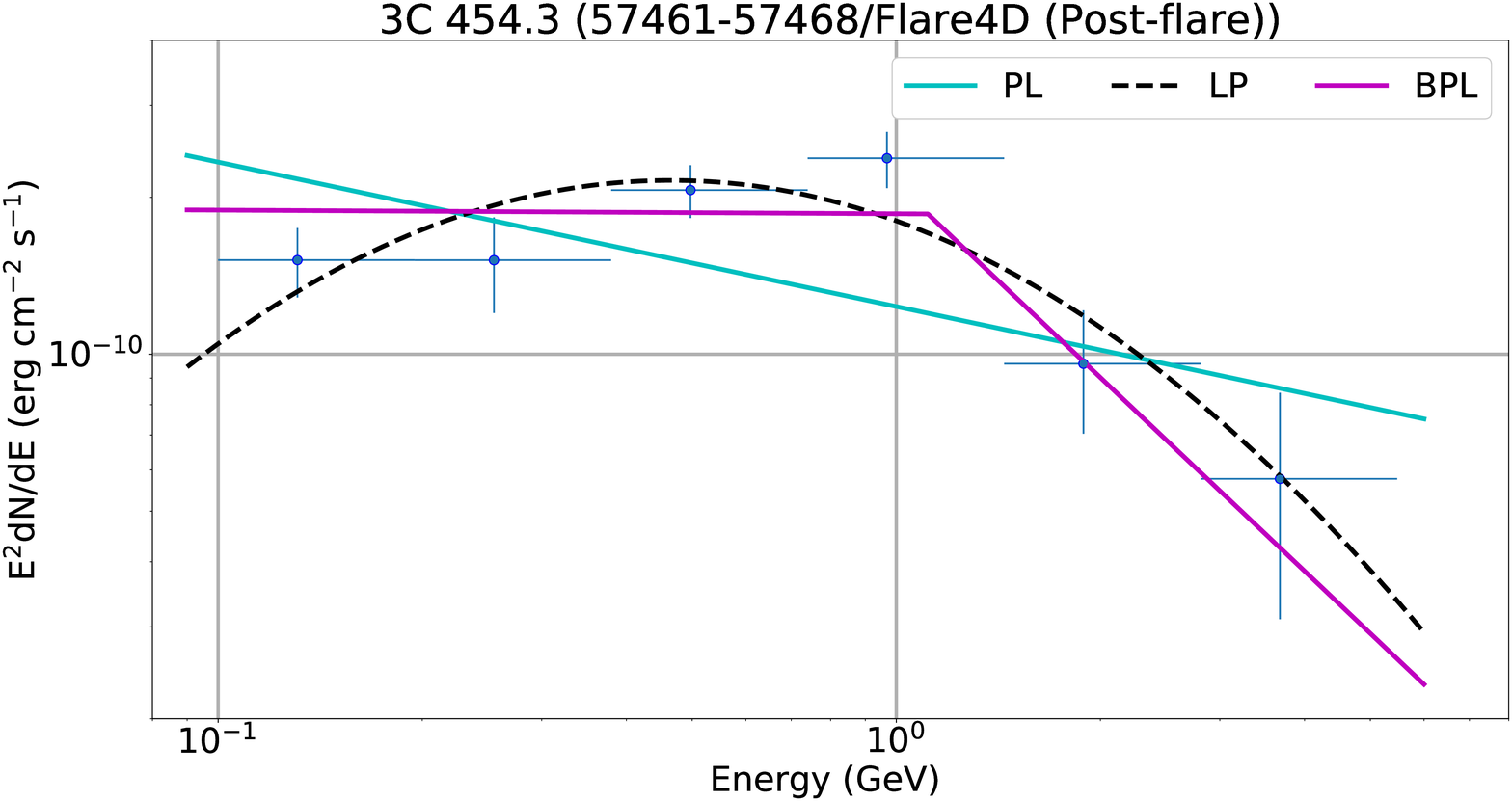}
\caption[optional]{SED of different periods of Flare-4D as given in Figure-26. PL, LP \& BPL describe the Powerlaw, Logparabola and Broken-powerlaw model respectively, which are fitted to data points.}

\includegraphics[height=1.77in,width=2.5in]{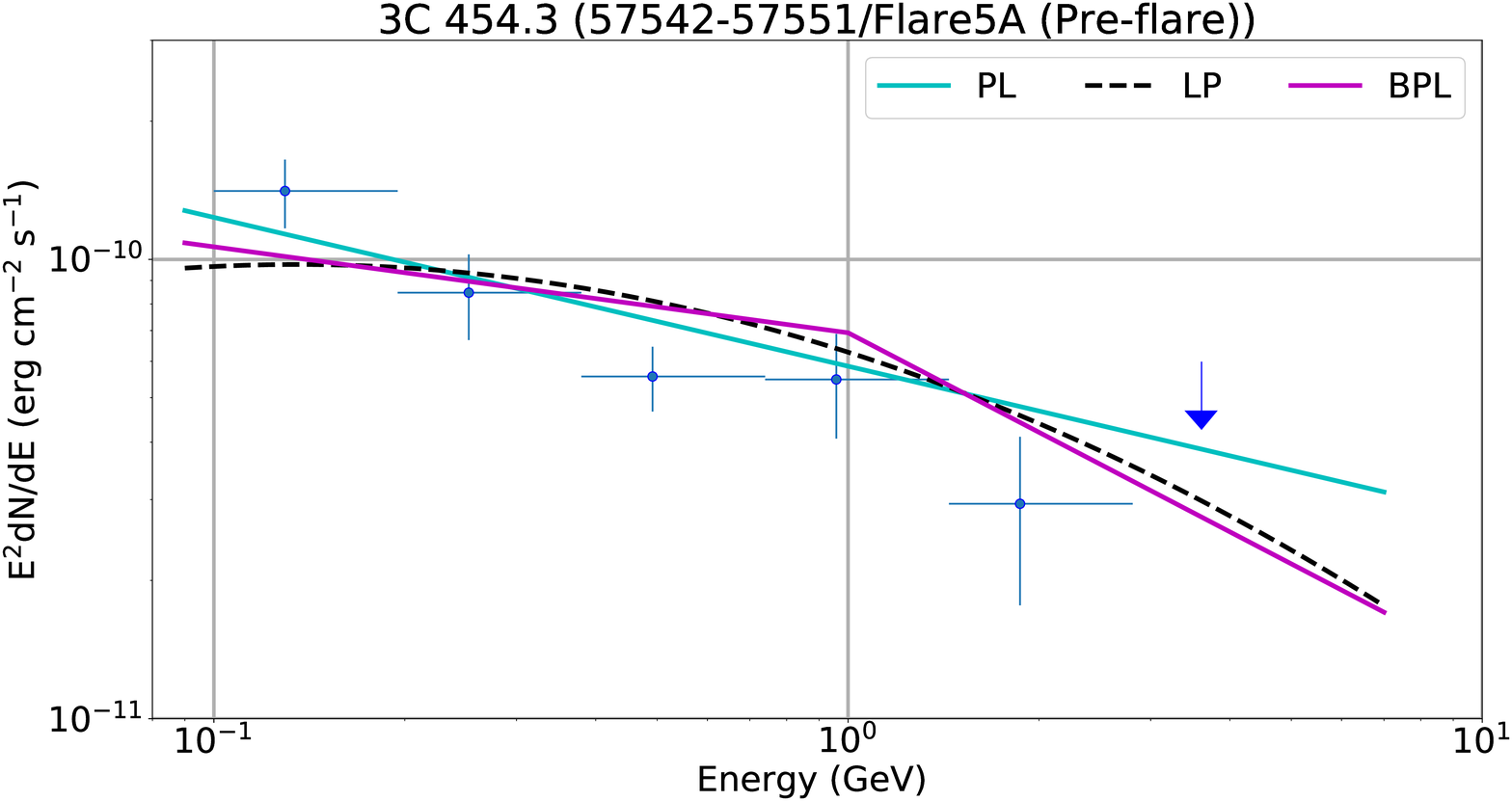}
\includegraphics[height=1.77in,width=2.5in]{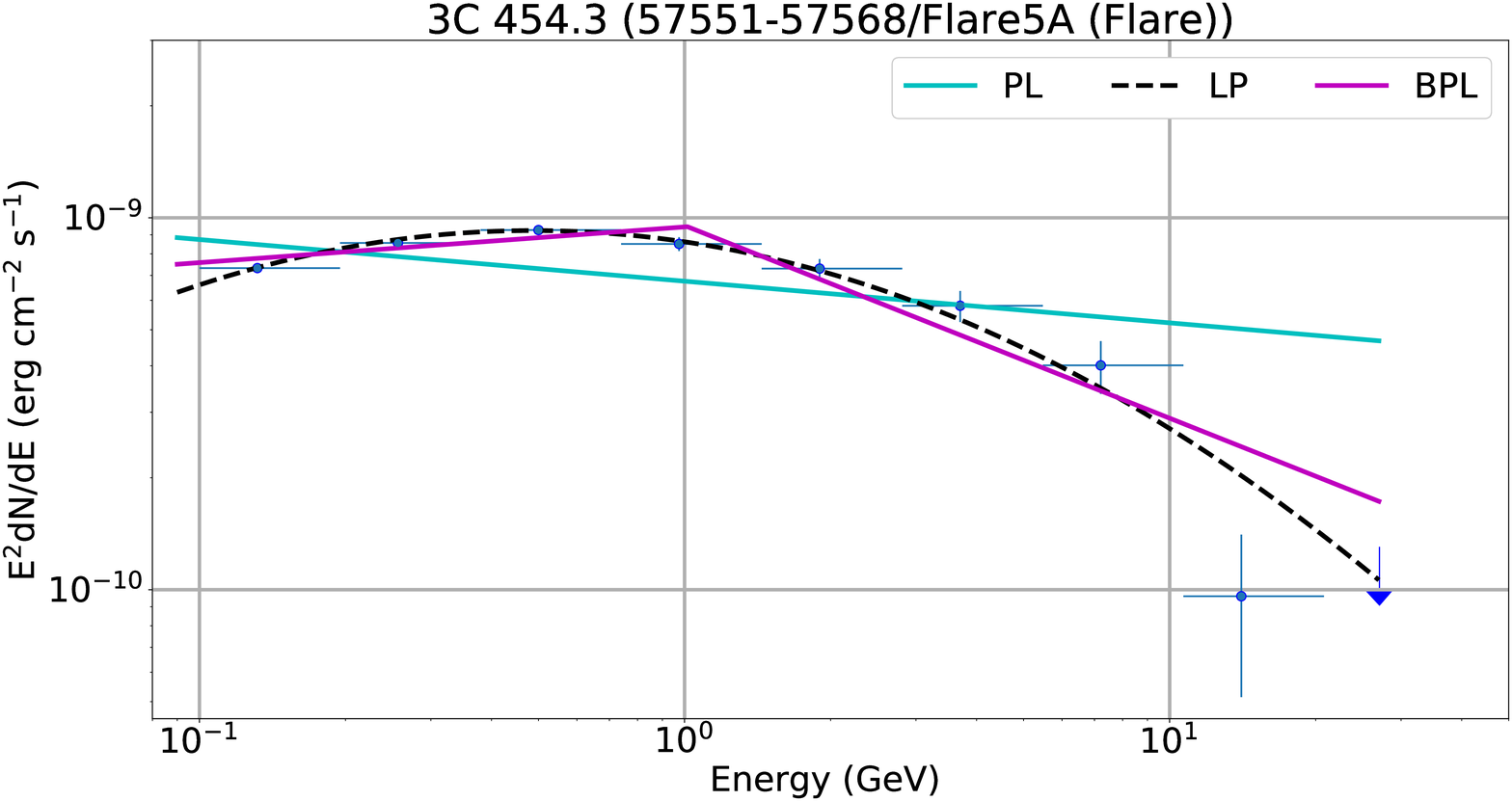}
\includegraphics[height=1.77in,width=2.5in]{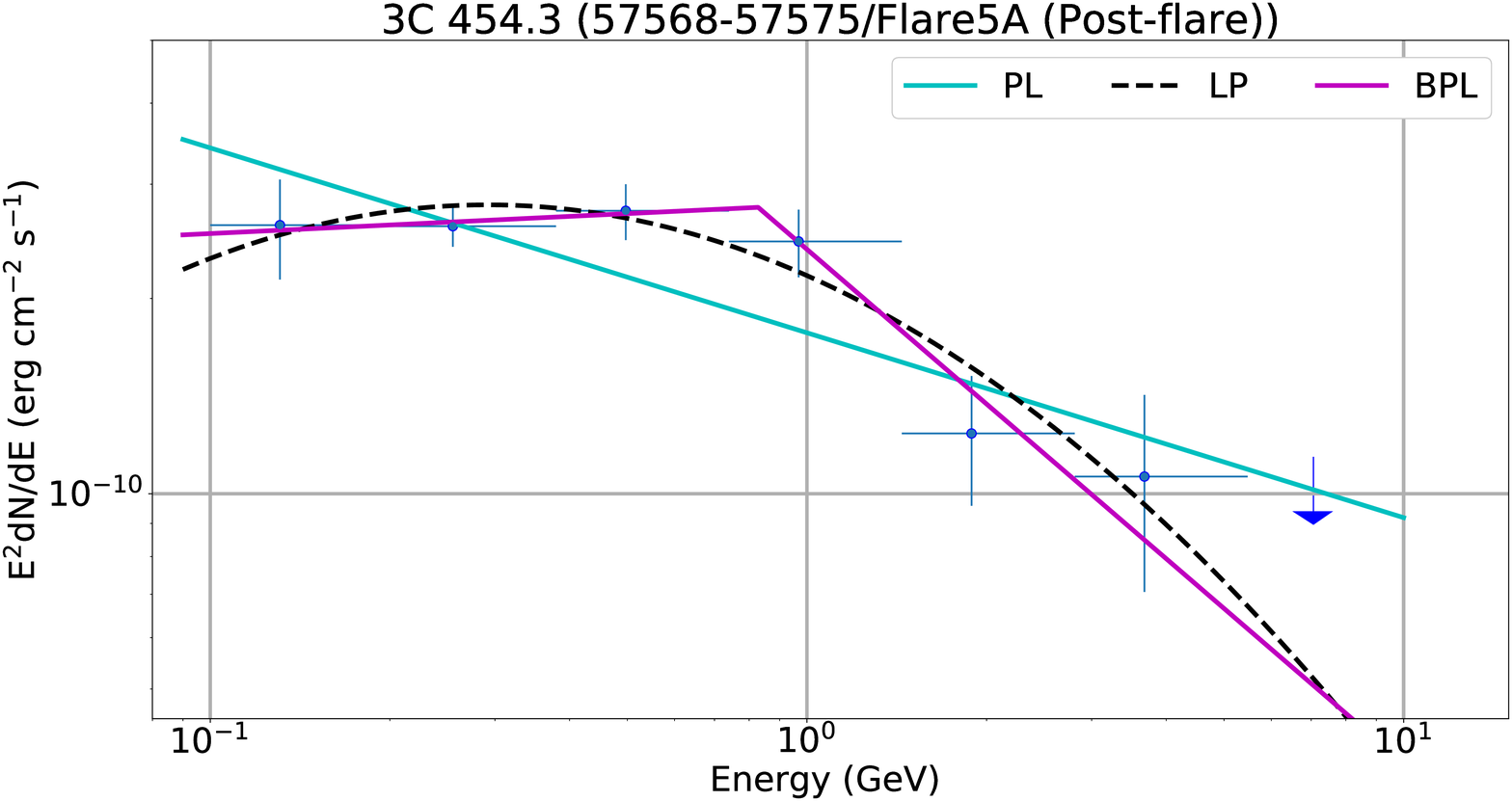}
\caption[optional]{SED of different periods of Flare-5A as given in Figure-28. PL, LP \& BPL describe the Powerlaw, Logparabola and Broken-powerlaw model respectively, which are fitted to data points.}

\includegraphics[height=1.77in,width=2.5in]{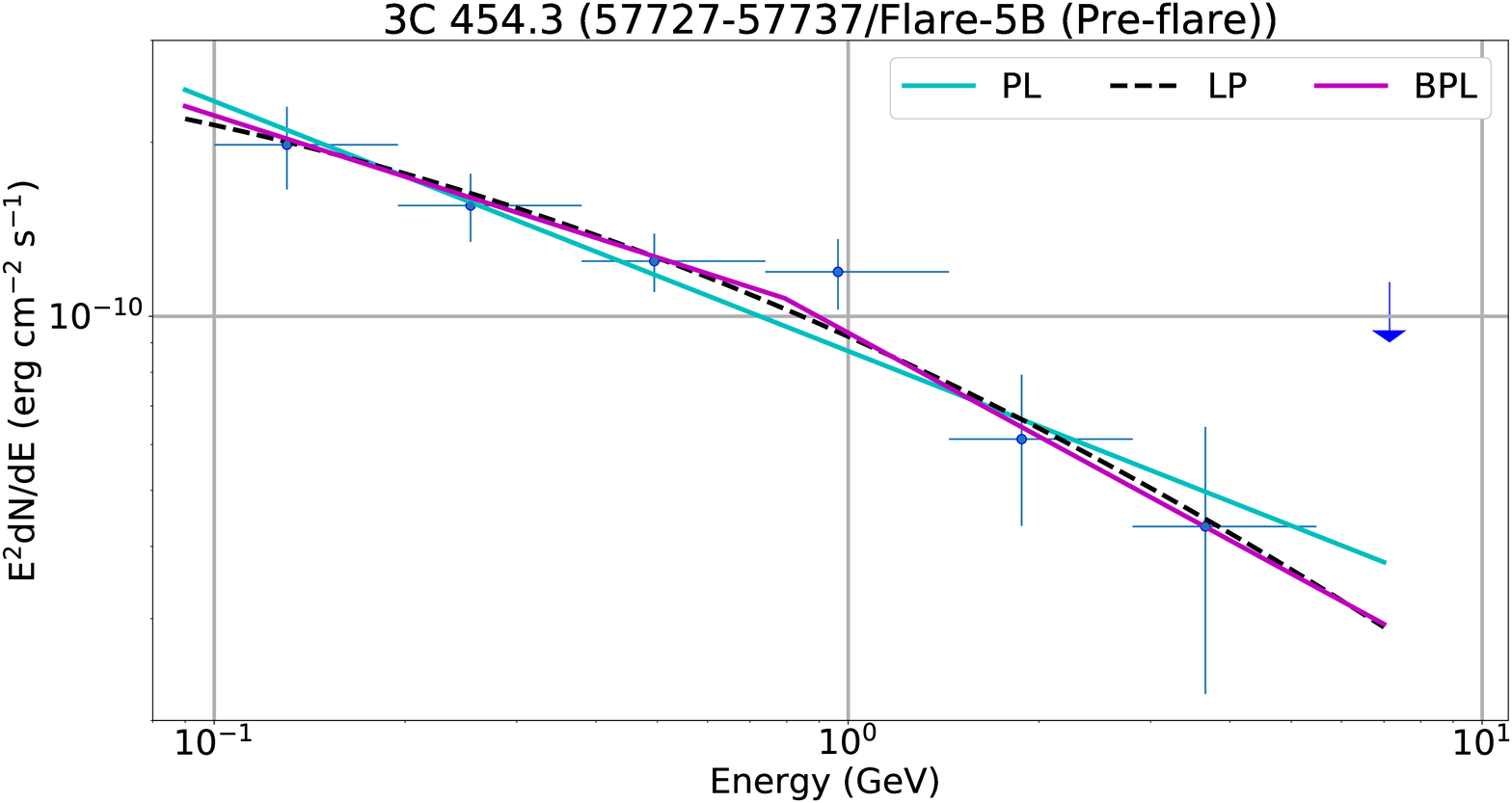}

\end{figure*}

\begin{figure*}[h!]
\centering
\includegraphics[height=1.77in,width=2.5in]{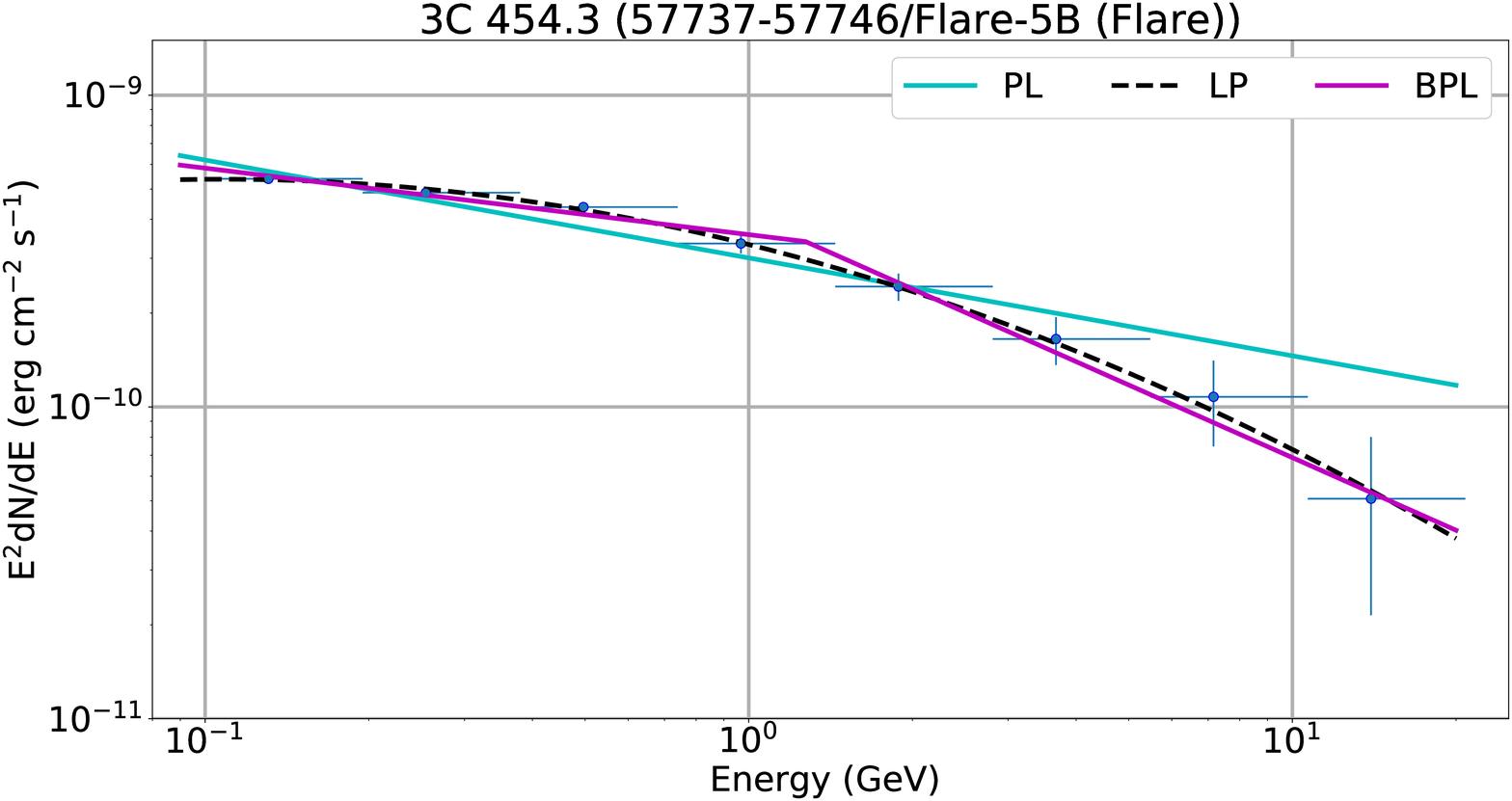}
\includegraphics[height=1.77in,width=2.5in]{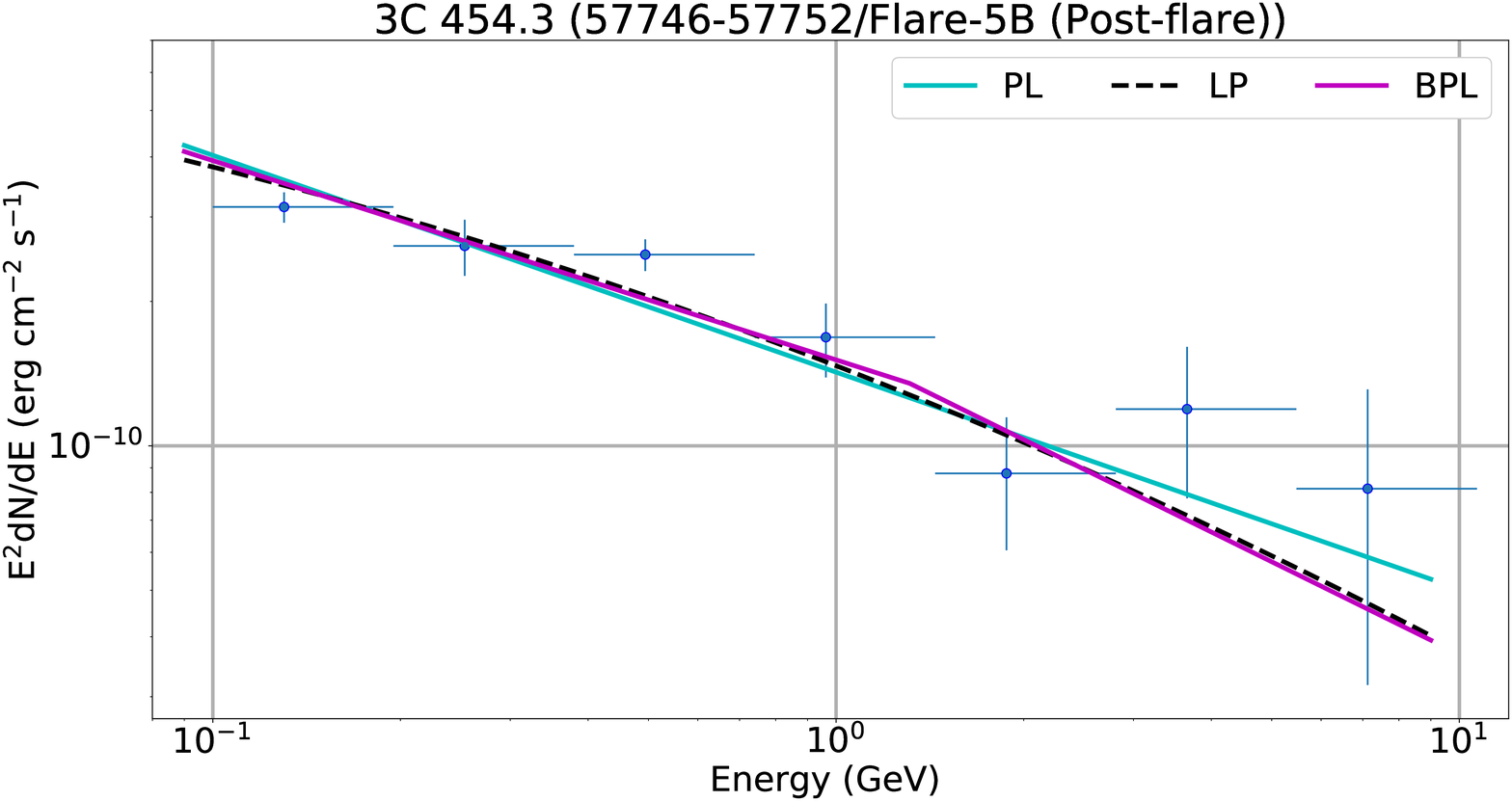}
\caption[optional]{SED of different periods of Flare-5B as given in Figure-30. PL, LP \& BPL describe the Powerlaw, Logparabola and Broken-powerlaw model respectively, which are fitted to data points.}

\end{figure*}

\begin{figure*}[h!]
\centering

\includegraphics[height=4.5in,width=8.2in]{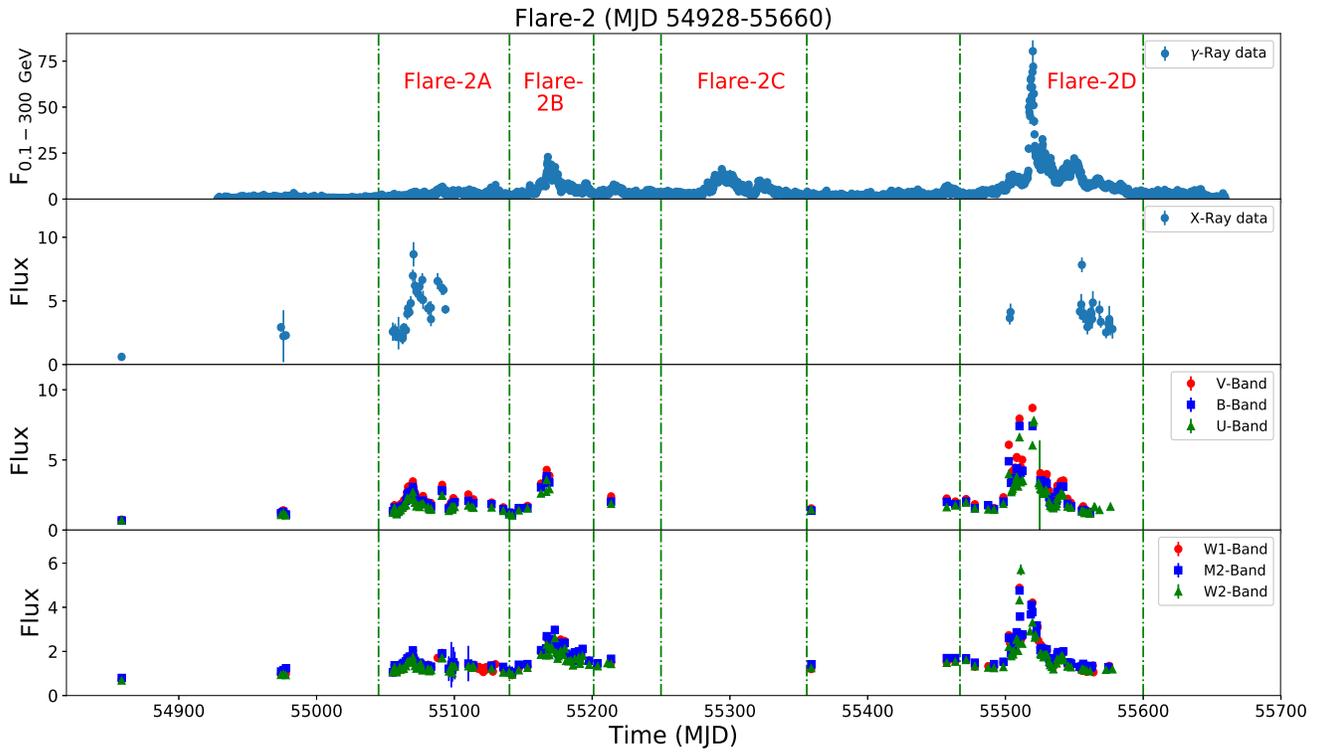}
\centering
\caption[optional]{Multi-wavelength light curve of Flare-2. Four distinctive major Flare have been identified. $\gamma$-ray flux ($F_{0.1-300 GeV}$) is in unit of $10^{-6}$ $ph$ $cm^{-2} s^{-1}$. X-ray, Optical (V, B \& U-Band) \& Ultra-violet (W1, M2 \& W2-Band) fluxes are in unit of $10^{-11}$ $erg$ $cm^{-2}$ $s^{-1}$}

\end{figure*}

\begin{figure*}[h!]
\centering

\includegraphics[height=4.5in,width=8.0in]{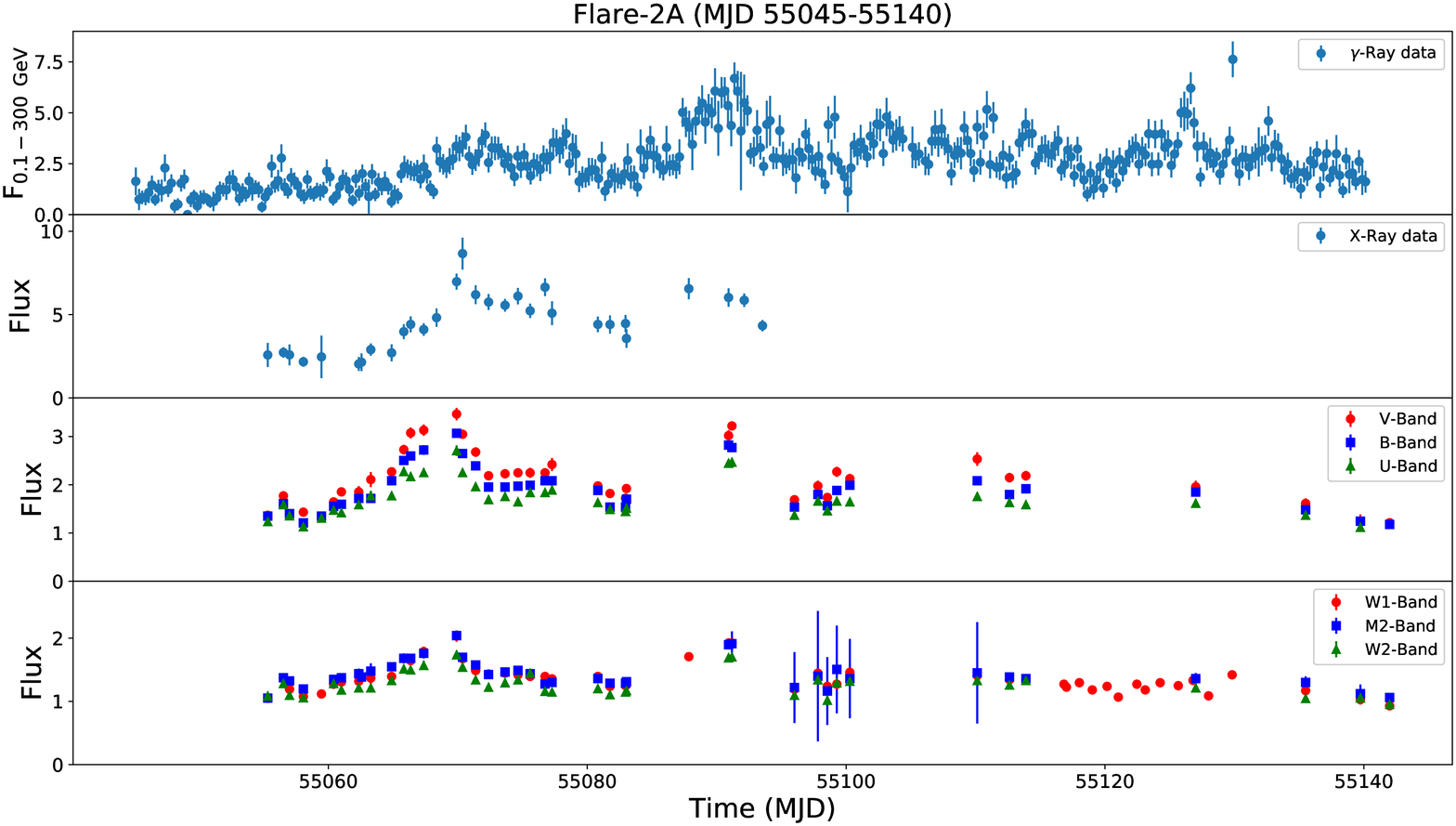}
\caption[optional]{Multi-wavelength light curve of Flare-2A. $\gamma$-ray flux ($F_{0.1-300 GeV}$) is in unit of $10^{-6}$ $ph$ $cm^{-2} s^{-1}$.X-ray, Optical (V, B \& U-Band) \& Ultra-violet (W1, M2 \& W2-Band) fluxes are in unit of $10^{-11}$ $erg$ $cm^{-2}$ $s^{-1}$.}

\includegraphics[height=4.5in,width=8.0in]{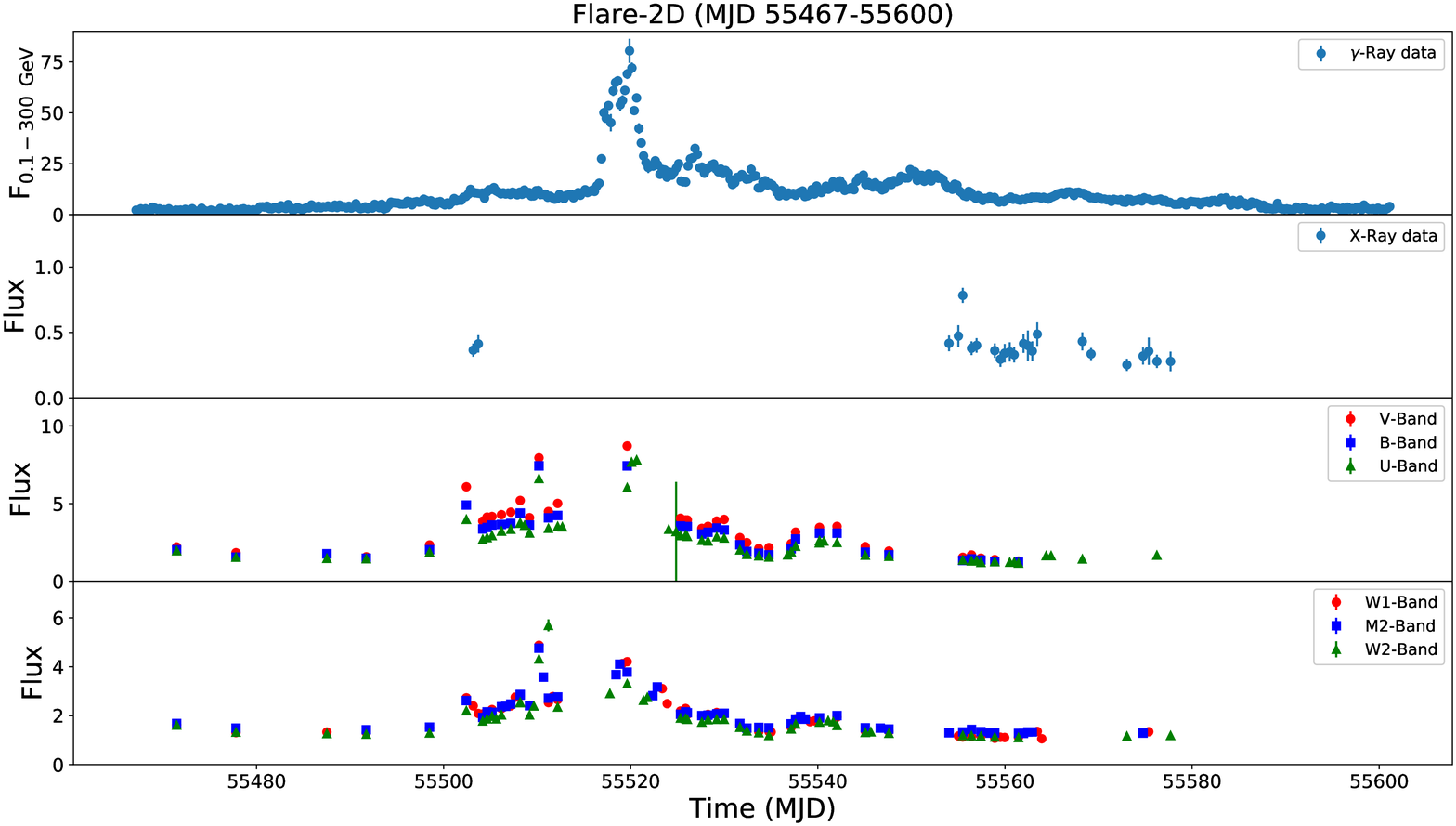}
\caption[optional]{Multi-wavelength light curve of Flare-2D.  $\gamma$-ray flux ($F_{0.1-300 GeV}$) is in unit of $10^{-6}$ $ph$ $cm^{-2} s^{-1}$. X-ray, Optical (V, B \& U-Band) \& Ultra-violet (W1, M2 \& W2-Band) fluxes are in unit of $10^{-11}$ $erg$ $cm^{-2}$ $s^{-1}$.}

\end{figure*}

\begin{figure*}[h!]
\centering

\includegraphics[height=2.7in,width=3.2in]{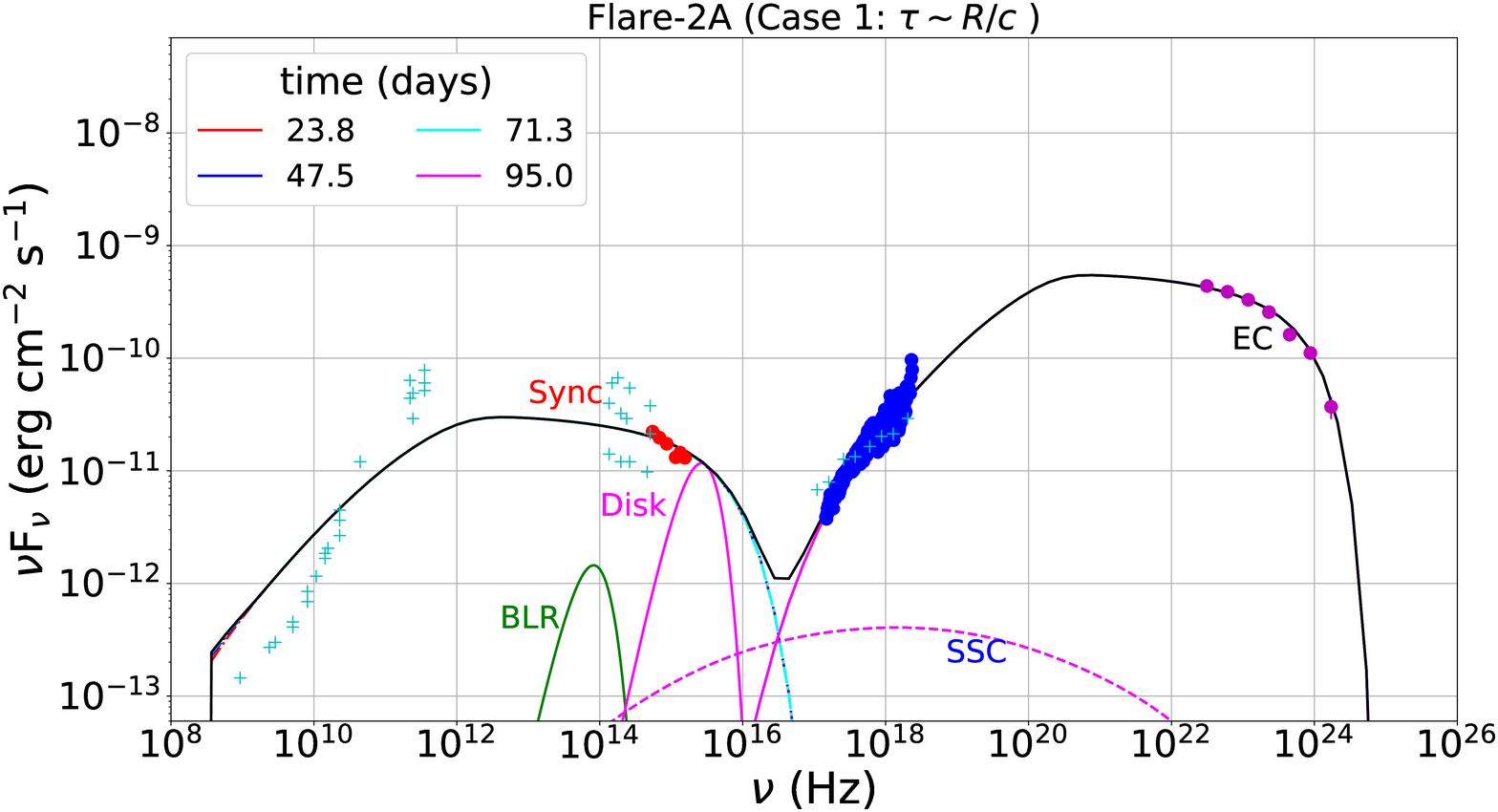}
\includegraphics[height=2.7in,width=3.2in]{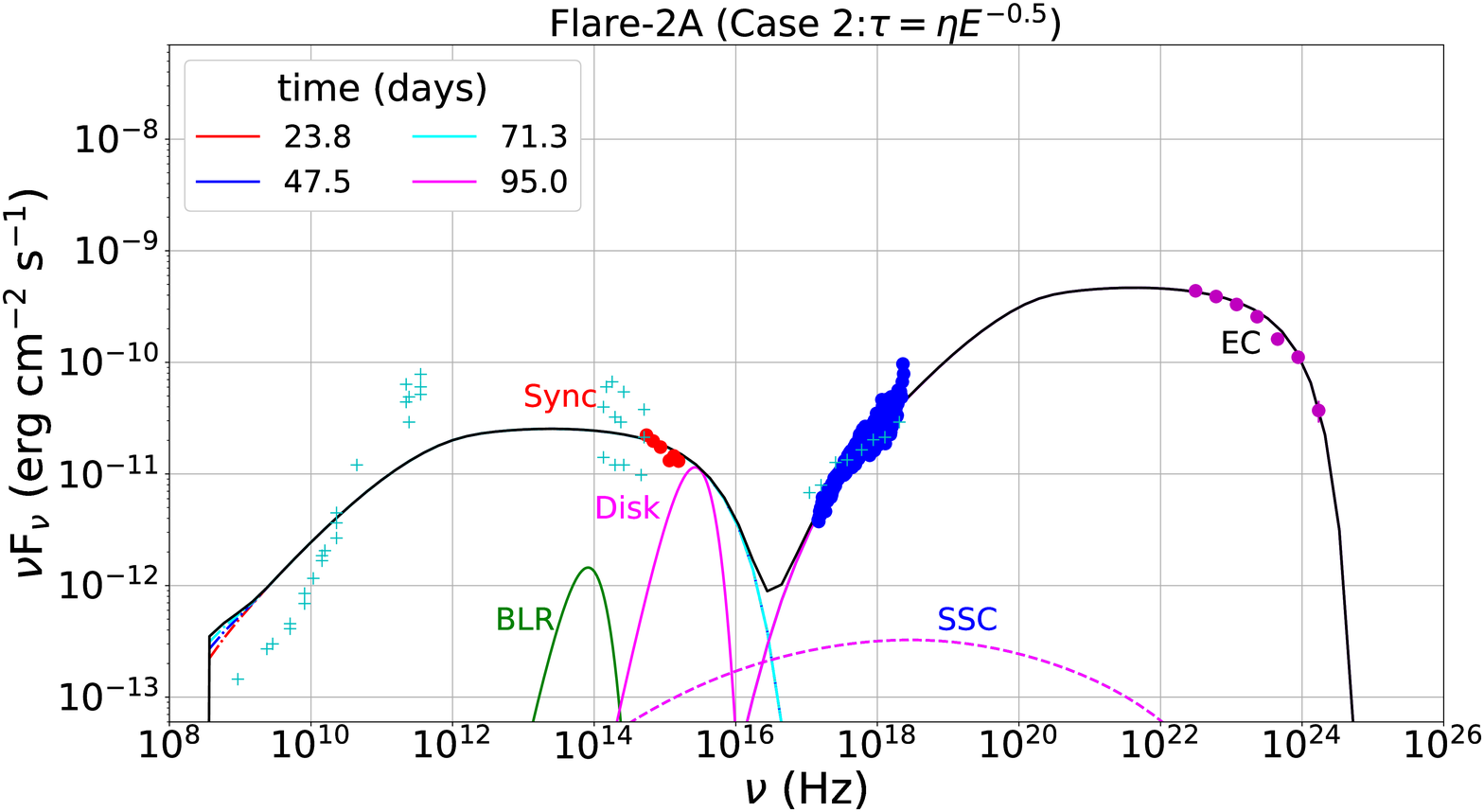}
\caption[optional]{Multiwavelength SED of Flare-2A for two different cases of escape timescale. Our analyzed data is shown in Red, Blue \& purple color. Non-simultaneous data (see text for more details) is shown in cyan plus point, which is taken from \citet{Abdo et al. (2010c)}.}

\includegraphics[height=2.7in,width=3.2in]{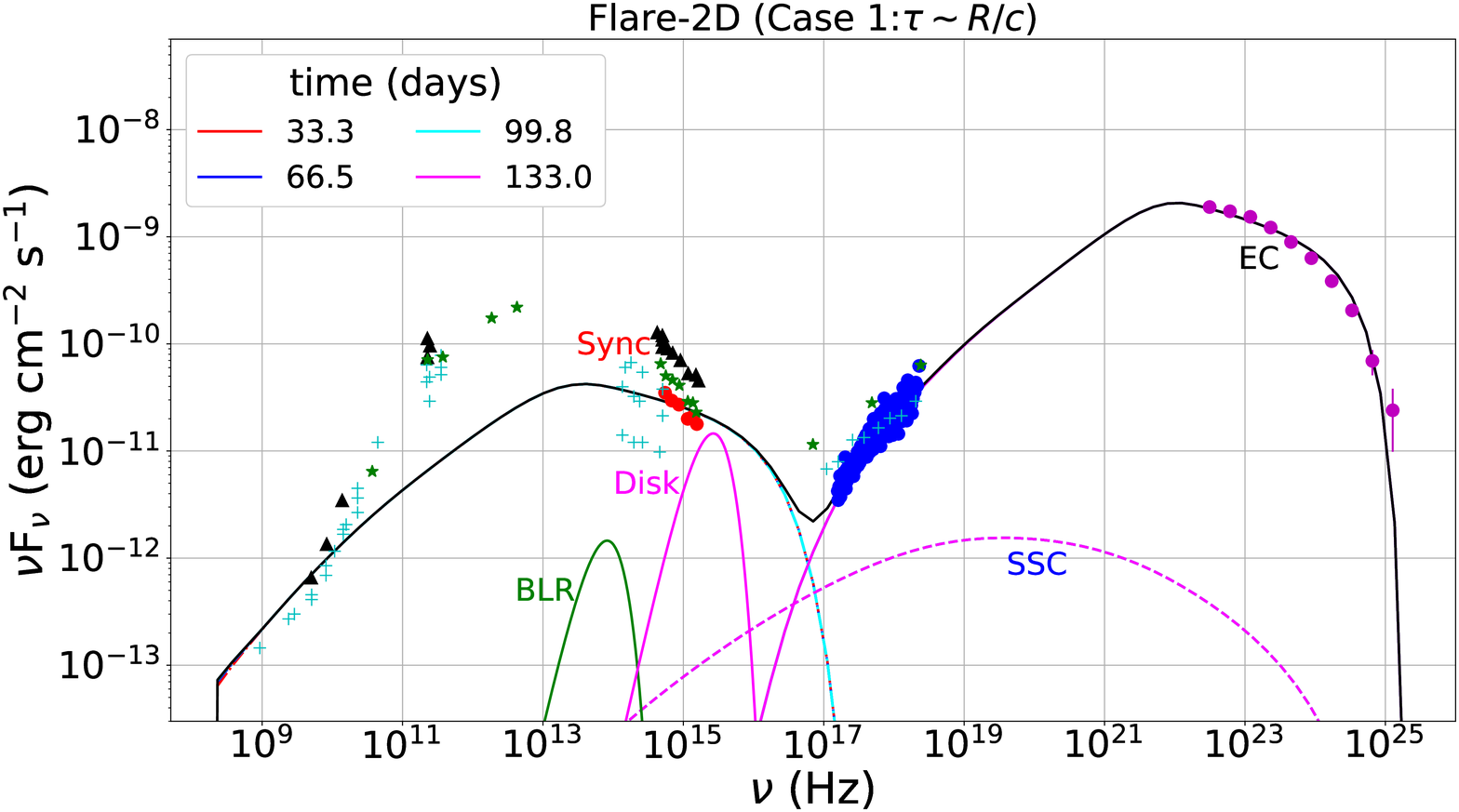}
\includegraphics[height=2.7in,width=3.2in]{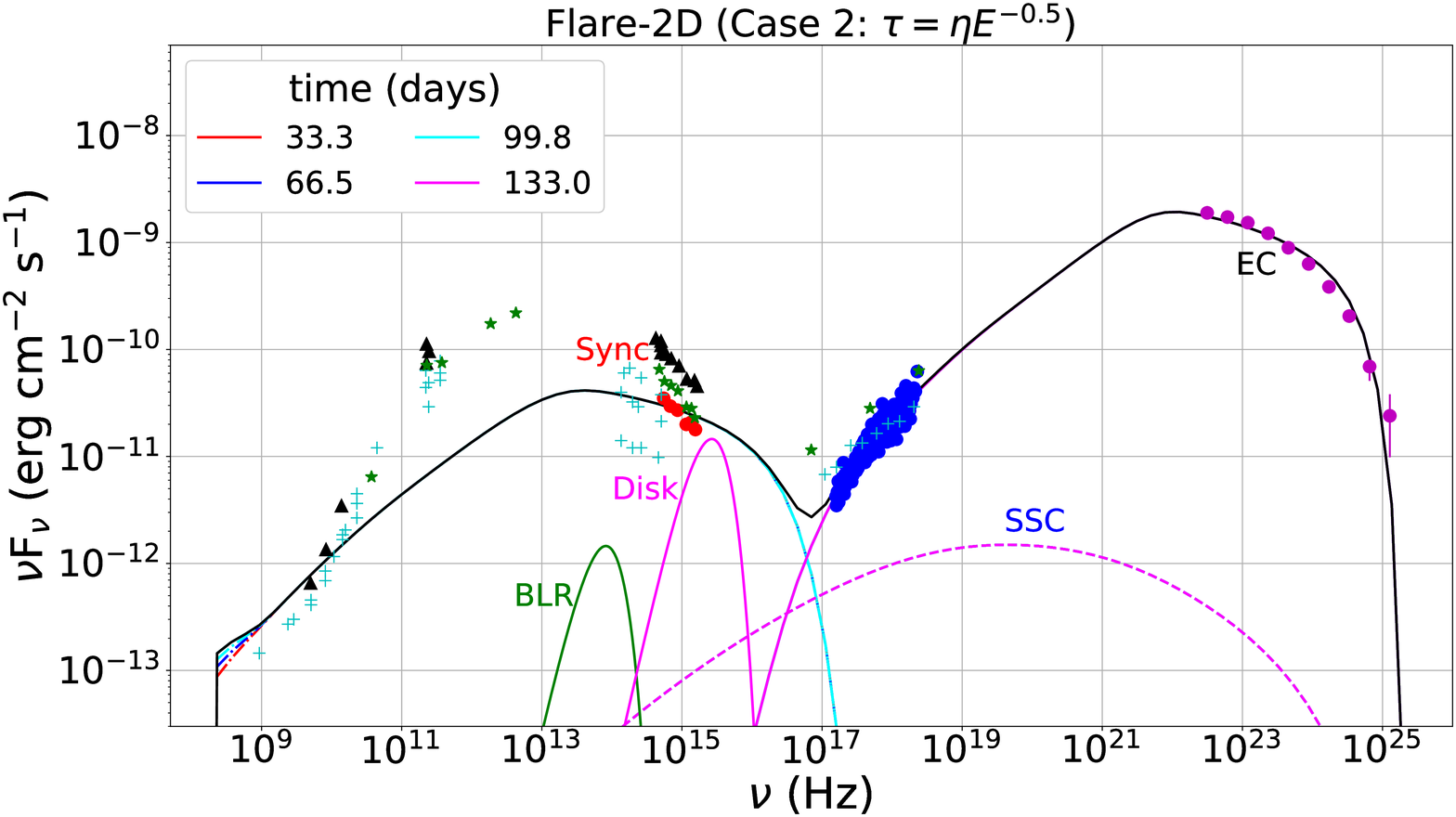}
\centering
\caption[optional]{Multi-wavelength SED of Flare-2D for two different cases of escape timescale. Our analyzed data is shown in Red, Blue \& purple color. Non-simultaneous data (see text for more details) is shown in cyan plus point, which is taken from \citet{Abdo et al. (2010c)}. quasi-simultaneous data is also shown in black triangle \citep{Vercellone et al. (2011)} and green star point \citep{Jorstad et al. (2013)}.}

\end{figure*}

\begin{figure*}[h]
\centering

\includegraphics[height=2.5in,width=5.3in]{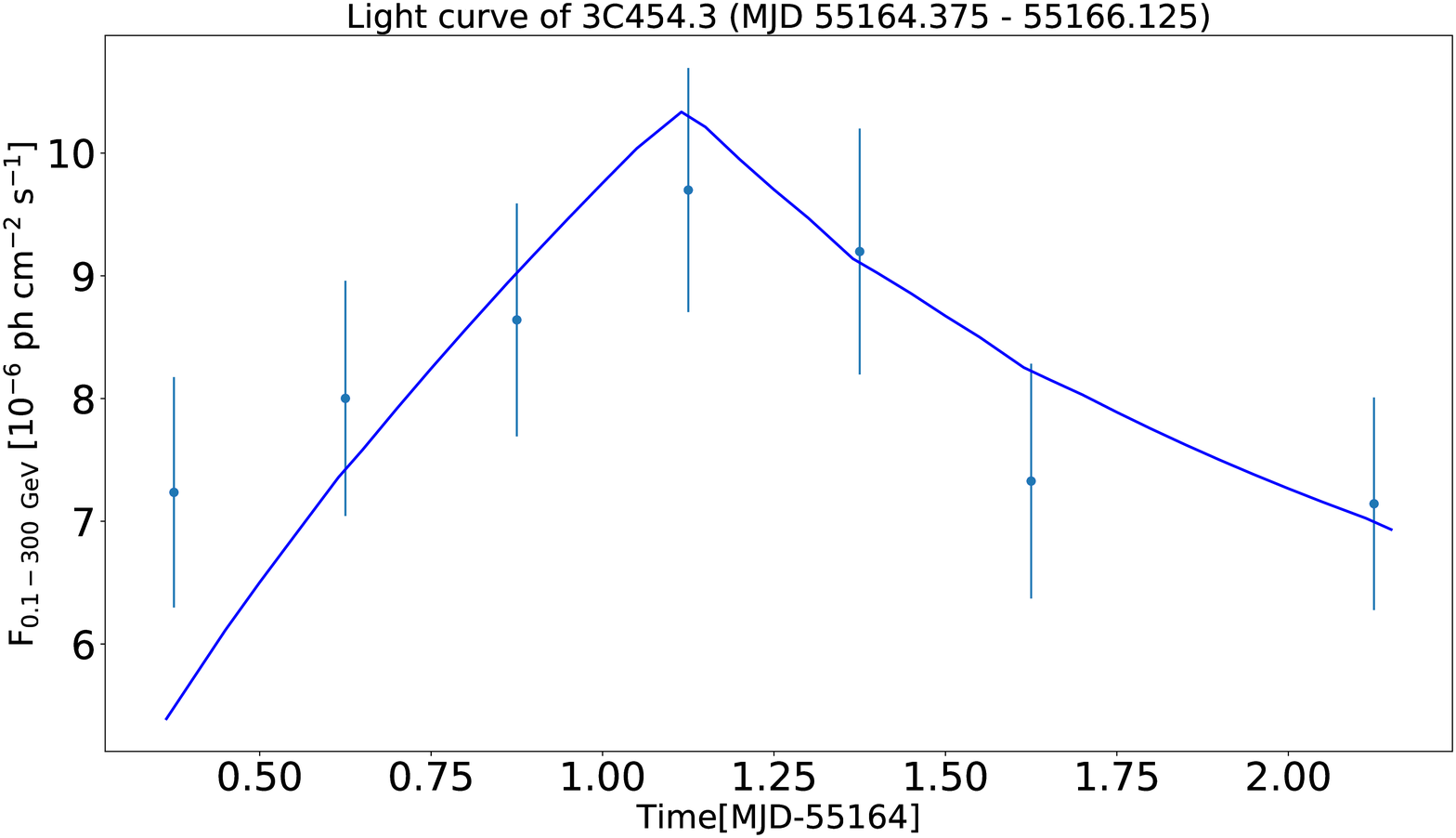}
\caption[optional]{Modelled light curve (by varying Doppler factor) between the data of MJD 55164.375 - 55166.125, which corresponds to P3 peak of 1st part of Flare-2B.}

\includegraphics[height=2.5in,width=5.3in]{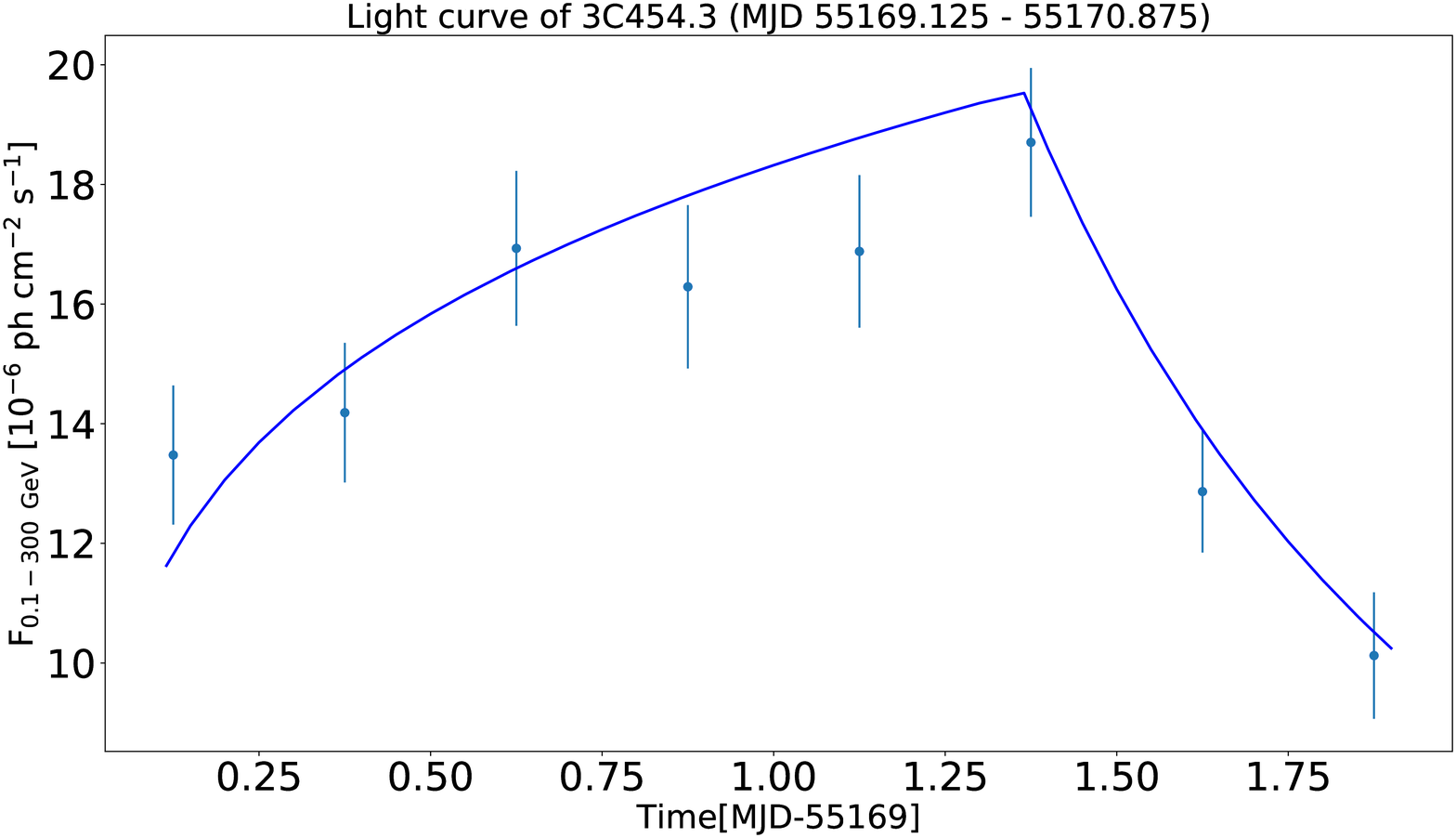}
\caption[optional]{Modelled light curve (by varying Doppler factor) between the data of MJD 55169.125 - 55170.875, which corresponds to P5 peak of 1st part of Flare-2B.}

\includegraphics[height=2.5in,width=5.3in]{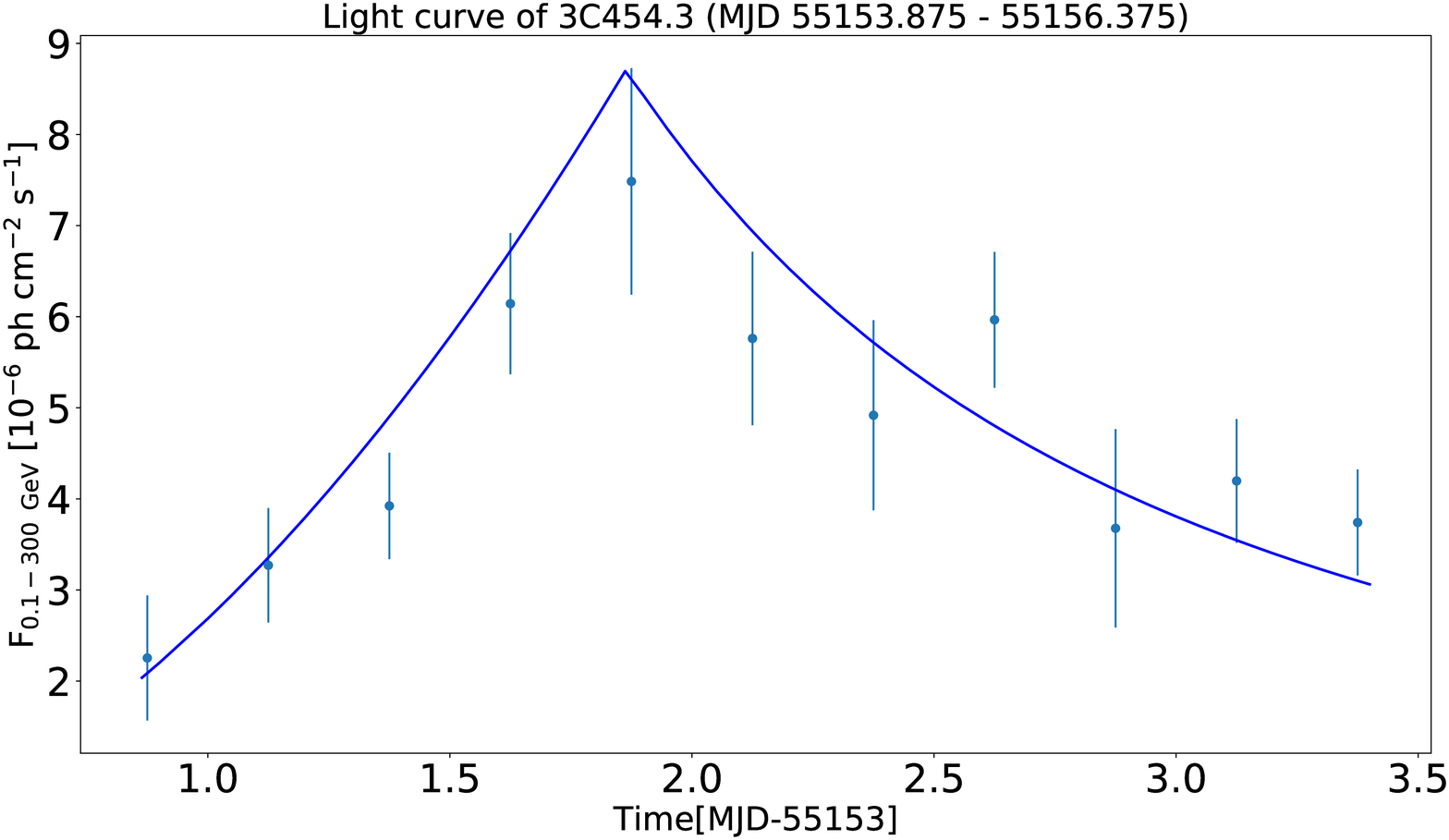}
\caption[optional]{Modelled light curve (by varying Doppler factor) between the data of MJD 55153.875 - 55156.375, which corresponds to P1 peak of 1st part of Flare-2B.}

\end{figure*}

\begin{figure*}[h]
\centering

\includegraphics[height=2.5in,width=5.3in]{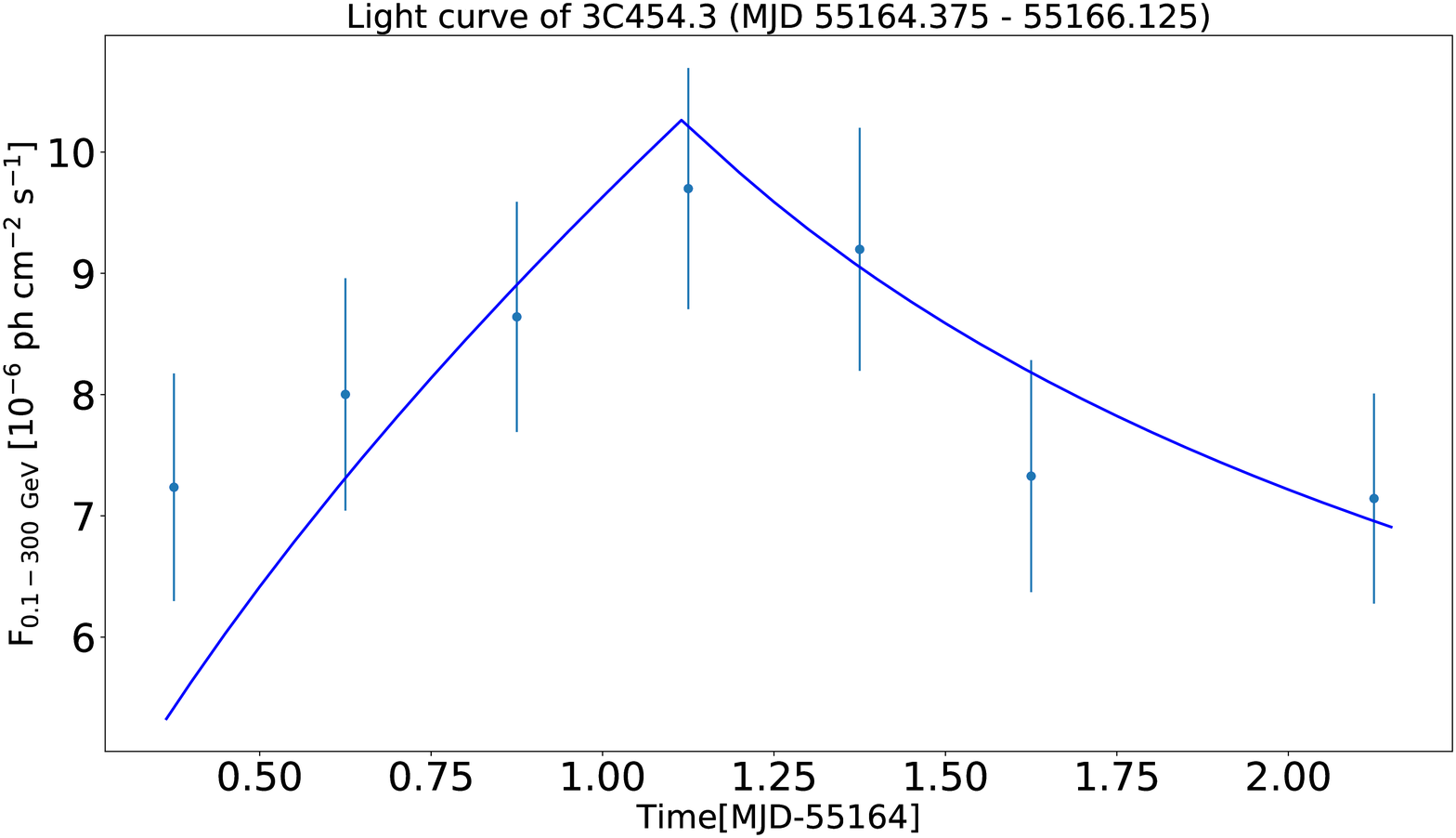}
\caption[optional]{Modelled light curve (by varying normalisation constant of the flux of injected electrons) between the data of MJD 55164.375 - 55166.125, which corresponds to P3 peak of 1st part of Flare-2B.}

\includegraphics[height=2.5in,width=5.3in]{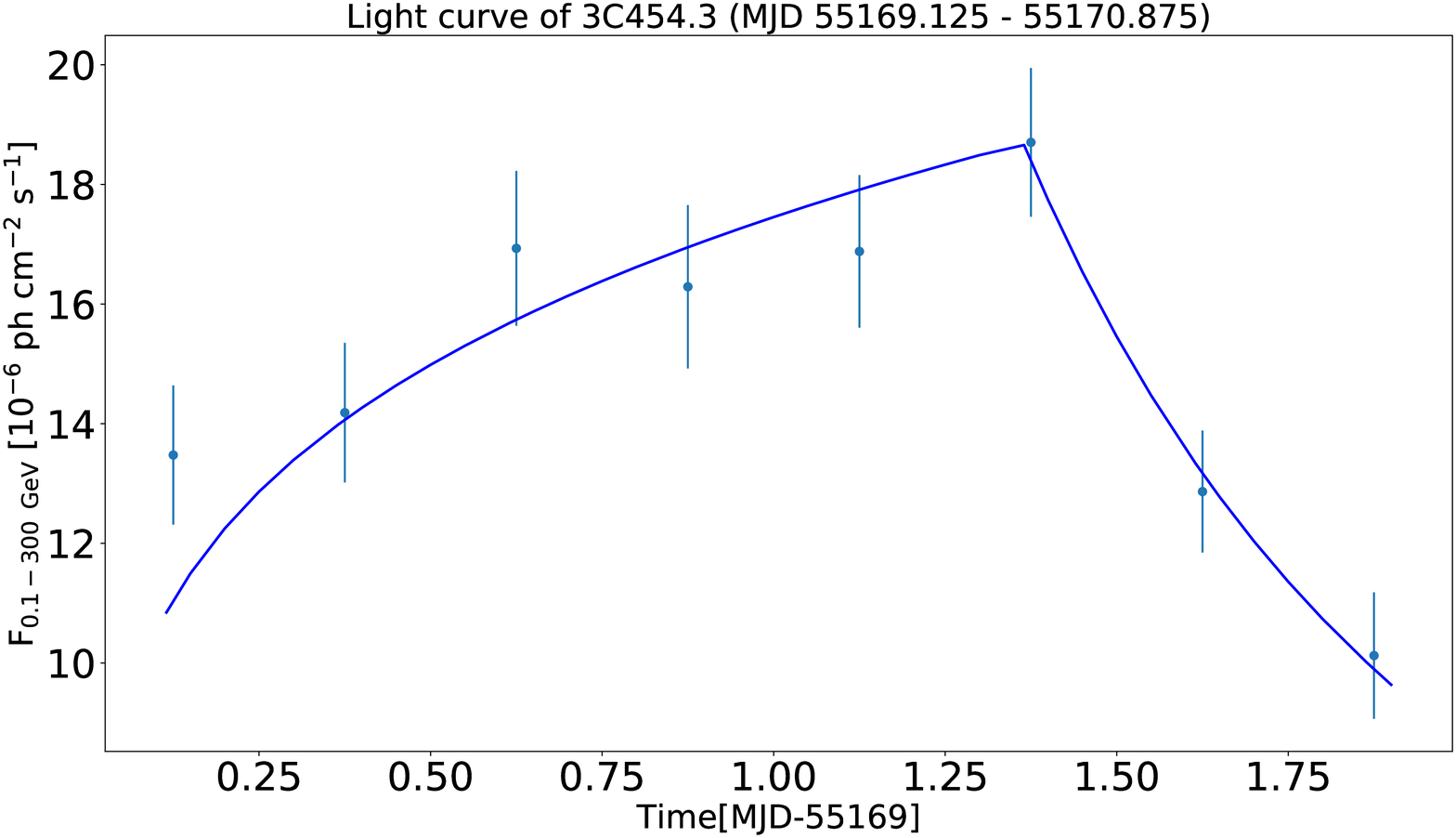}
\caption[optional]{Modelled light curve (by varying normalisation constant of the flux of injected electrons) between the data of MJD 55169.125 - 55170.875, which corresponds to P5 peak of 1st part of Flare-2B.}

\includegraphics[height=2.5in,width=5.3in]{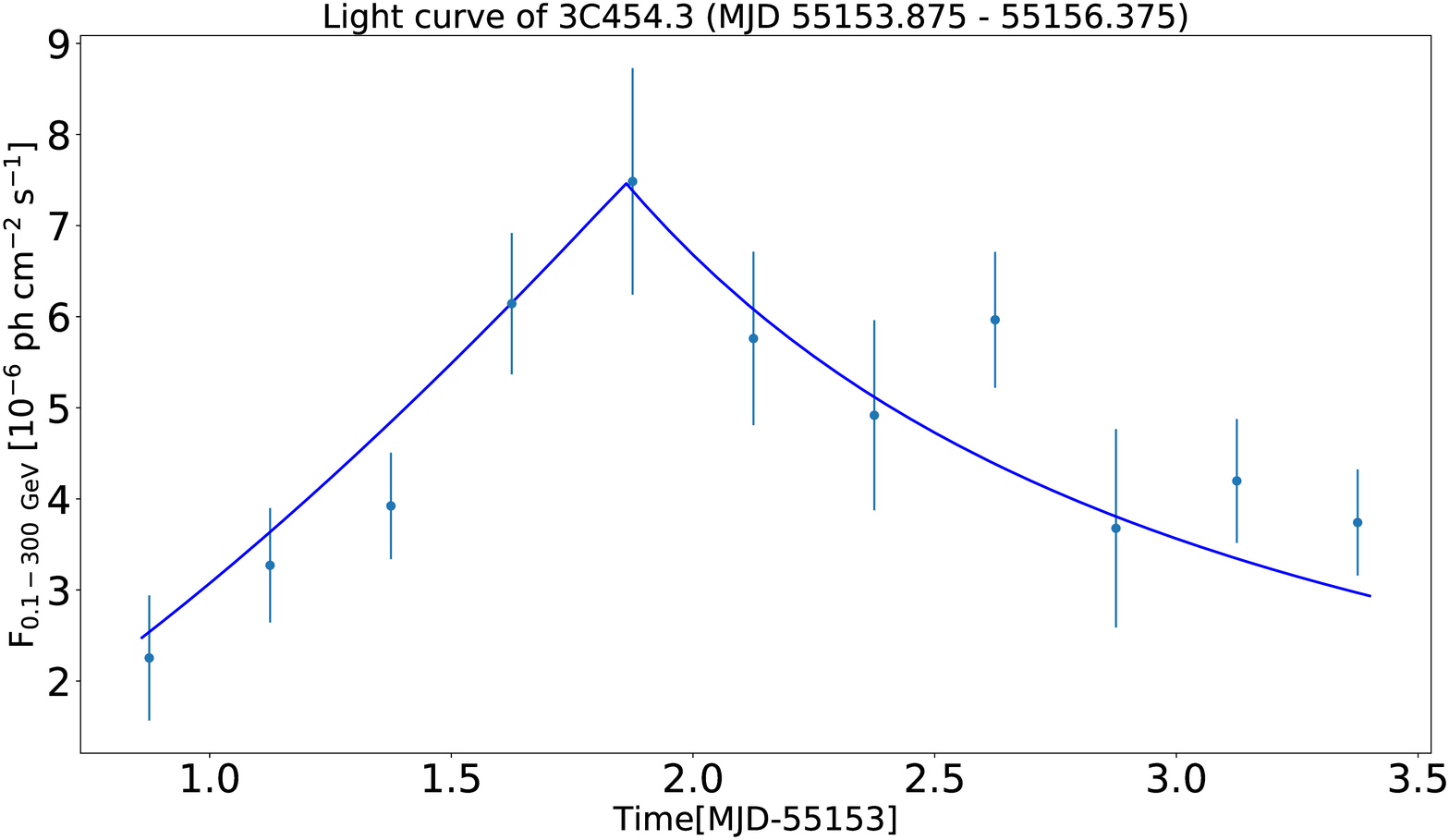}
\caption[optional]{Modelled light curve (by varying normalisation constant of the flux of injected electrons) between the data of MJD 55153.875 - 55156.375, which corresponds to P1 peak of 1st part of Flare-2B.}

\end{figure*}

\begin{table*}[h]
\caption{Rising and Decay time [$T_r$(column 4) and $T_d$ (column 5)] for given peak time [$t_0$ (column 2)] and peak flux [$F_0$ (column 3)] which is calculated by temporal fitting of light curve (Flare-1A) with sum of exponential function. Column 1 represent peak number. Here results are shown for 1 day binning.} 
\centering
\begin{tabular}{ccccc rrrr}   
\\
 & & Flare-1A\\
\hline\hline                        
Peak & $t_0$ & $F_0$ & $T_r$ & $T_d$ \\ [0.8ex]
& [MJD] & [10$^{-6}$ ph cm$^{-2}$ s$^{-1}$] & [hr] & [hr]\\
\hline
P1 &  54719.1 & 5.45$\pm$0.42 &  73.35$\pm$4.21 & 54.79$\pm$14.38 & \\
P2 &  54729.1 & 4.31$\pm$0.35 &  19.95$\pm$5.78 & 57.03$\pm$9.96 & \\
P3 &  54738.1 & 3.66$\pm$0.28 &  15.55$\pm$4.91 & 30.52$\pm$9.86 & \\
\hline                          
\end{tabular}
\label{tab:hresult}
\end{table*}

\begin{table*}[h]
\caption{All the parameters represented here are similar to the parameters of Table 1. Results (1 day binning) are shown for Flare-2A} 
\centering
\begin{tabular}{ccccc rrrr}   
\\
 & & Flare-2A\\
\hline\hline                        
Peak & $t_0$ & $F_0$ & $T_r$ & $T_d$ \\ [0.8ex]
& [MJD] & [10$^{-6}$ ph cm$^{-2}$ s$^{-1}$] & [hr] & [hr]\\
\hline
P1 &  55070.5 & 3.32$\pm$0.29 &  80.87$\pm$8.40 & 100.13$\pm$17.21 & \\
P2 &  55077.5 & 3.31$\pm$0.36 &  16.46$\pm$9.12 & 39.40$\pm$13.49 & \\
P3 &  55091.5 & 5.95$\pm$0.52 &  100.14$\pm$8.79 & 45.80$\pm$7.77 & \\
P2 &  55103.5 & 4.08$\pm$0.36 &  114.94$\pm$16.77 & 56.86$\pm$17.67 & \\
\hline                          
\end{tabular}
\label{tab:hresult}
\end{table*}

\begin{table*}[h]
\caption{All the parameters represented here are similar to the parameters of Table 1. Results (6 hour binning) are shown for Flare-2B} 
\centering
\begin{tabular}{ccccc rrrr}   
\\
 & & 1st part of Flare-2B\\
\hline\hline                        

Peak & $t_0$ & $F_0$ & $T_r$ & $T_d$ \\ [0.8ex]
& [MJD] & [10$^{-6}$ ph cm$^{-2}$ s$^{-1}$] & [hr] & [hr] \\
\hline
P1 &  55154.9 & 7.48$\pm$1.24 &  7.30$\pm$2.38 & 36.74$\pm$3.77 & \\
P2 &  55163.1 & 9.69$\pm$1.41 &  51.81$\pm$3.13 & 23.60$\pm$4.48 & \\
P3 &  55165.1 & 9.69$\pm$0.99 &  7.13$\pm$2.81 & 11.34$\pm$3.91 & \\
P4 &  55167.9 & 22.86$\pm$1.48 &  19.36$\pm$1.51 & 12.85$\pm$1.90 & \\
P5 &  55170.4 & 18.70$\pm$1.24 &  19.46$\pm$2.75 & 9.96$\pm$1.20 & \\
P6 &  55172.1 & 14.56$\pm$1.21 &  13.19$\pm$2.04 & 93.62$\pm$3.90 &\\
\hline                          

 & & 2nd part of Flare-2B\\
\hline\hline                        
Peak & $t_0$ & $F_0$ & $T_r$ & $T_d$ \\ [0.8ex]
& [MJD] & [10$^{-6}$ ph cm$^{-2}$ s$^{-1}$] & [hr] & [hr]\\
\hline
P1 &  55178.4 & 8.73$\pm$1.04 &  11.94$\pm$1.79 & 16.73$\pm$4.63 & \\
P2 &  55180.4 & 7.85$\pm$0.95 &  12.29$\pm$5.32 & 14.71$\pm$5.85 & \\
P3 &  55182.6 & 8.48$\pm$0.98 &  18.86$\pm$6.58 & 9.59$\pm$3.04 & \\
P4 &  55185.1 & 7.68$\pm$0.87 &  27.05$\pm$5.76 & 67.89$\pm$5.56 & \\
P5 &  55195.1 & 8.52$\pm$0.83 &  37.47$\pm$3.88 & 61.49$\pm$3.49 & \\
\hline                          
\end{tabular}
\label{tab:hresult}
\end{table*}

\begin{table*}[h]
\caption{All the parameters represented here are similar to the parameters of Table 1. Results (6 hour binning) are shown for Flare-2C} 
\centering
\begin{tabular}{ccccc rrrr}   
\\
& & Flare-2C\\
\hline
& & Flare-I \\
\hline\hline                        
Peak & $t_0$ & $F_0$ & $T_r$ & $T_d$ \\ [0.8ex]
& [MJD] & [10$^{-6}$ ph cm$^{-2}$ s$^{-1}$] & [hr] & [hr]\\
\hline
P1 &  55289.1 & 13.04$\pm$1.21 &  23.45$\pm$3.83 & 8.09$\pm$1.97 & \\
P2 &  55290.6 & 12.73$\pm$1.43 &  7.37$\pm$1.98 & 18.13$\pm$4.33 & \\
P3 &  55292.4 & 12.77$\pm$0.64 &  12.70$\pm$3.49 & 13.10$\pm$3.04 & \\
P4 &  55294.1 & 16.29$\pm$0.82 &  17.06$\pm$2.98 & 18.77$\pm$2.99 & \\
P5 &  55300.1 & 13.29$\pm$2.66 &  26.06$\pm$7.01 & 32.56$\pm$4.66 & \\
\hline                          

 & & Flare-II\\ 
\hline\hline                        
Peak & $t_0$ & $F_0$ & $T_r$ & $T_d$ \\ [0.8ex]
& [MJD] & [10$^{-6}$ ph cm$^{-2}$ s$^{-1}$] & [hr] & [hr]\\
\hline
P1 &  55320.6 & 9.71$\pm$0.94 &  17.61$\pm$2.88 & 7.93$\pm$2.57 & \\
P2 &  55321.9 & 10.05$\pm$0.95 &  4.71$\pm$1.92 & 8.90$\pm$2.18 & \\
P3 &  55322.6 & 7.79$\pm$0.90 &  9.92$\pm$3.28 & 31.44$\pm$5.41 & \\
P4 &  55327.1 & 9.12$\pm$1.26 &  26.26$\pm$4.90 & 14.72$\pm$2.59 & \\
P5 &  55329.4 & 5.96$\pm$0.78 &  16.87$\pm$4.40 & 59.31$\pm$5.26 & \\
\hline                          
\end{tabular}
\label{tab:hresult}
\end{table*}

\begin{table*}[h]
\caption{All the parameters represented here are similar to the parameters of Table 1. Results (6 hour binning) are shown for Flare-2D} 
\centering
\begin{tabular}{ccccc rrrr}   
\\
& & Flare-2D\\
\hline
& & Flare-I \\
\hline\hline                        
Peak & $t_0$ & $F_0$ & $T_r$ & $T_d$ \\ [0.8ex]
& [MJD] & [10$^{-6}$ ph cm$^{-2}$ s$^{-1}$] & [hr] & [hr]\\
\hline
P1 &  55517.6 & 53.51$\pm$2.08 &  11.52$\pm$1.45 & 8.46$\pm$1.67 & \\
P2 &  55518.6 & 65.66$\pm$2.34 &  8.92$\pm$1.15 & 6.19$\pm$1.40 & \\
P3 &  55519.9 & 80.41$\pm$5.92 &  13.82$\pm$1.57 & 17.51$\pm$1.34 & \\
\hline                          

 & & Flare-II\\ 
\hline\hline                        
Peak & $t_0$ & $F_0$ & $T_r$ & $T_d$ \\ [0.8ex]
& [MJD] & [10$^{-6}$ ph cm$^{-2}$ s$^{-1}$] & [hr] & [hr]\\
\hline
P1 &  55541.6 & 16.12$\pm$1.29 & 82.92$\pm$4.83 & 33.91$\pm$4.07 & \\
P2 &  55544.9 & 19.41$\pm$1.25 &  12.21$\pm$2.50 & 37.04$\pm$8.23 & \\
P3 &  55549.9 & 22.12$\pm$1.39 &  43.56$\pm$6.03 & 25.18$\pm$3.92 & \\
P4 &  55551.9 & 20.02$\pm$1.31 &  10.00$\pm$3.45 & 21.92$\pm$4.61 & \\
P5 &  55563.1 & 13.44$\pm$1.56 &  12.20$\pm$5.34 & 10.20$\pm$4.54 & \\
P6 &  55566.9 & 18.22$\pm$1.53 &  41.75$\pm$5.33 & 57.73$\pm$3.23 &\\
\hline                          
\end{tabular}
\label{tab:hresult}
\end{table*}

\begin{table*}[h]
\caption{All the parameters represented here are similar to the parameters of Table 1. Results (6 hour binning) are shown for Flare-3A} 
\centering
\begin{tabular}{ccccc rrrr}   
\\
& & Flare-3A\\
\hline
& & Flare-I \\
\hline\hline                       
Peak & $t_0$ & $F_0$ & $T_r$ & $T_d$ \\ [0.8ex]
& [MJD] & [10$^{-6}$ ph cm$^{-2}$ s$^{-1}$] & [hr] & [hr]\\
\hline
P1 &  56816.1 & 9.36$\pm$0.52 &  16.04$\pm$1.64 & 14.23$\pm$2.39 & \\
P2 &  56819.1 & 5.16$\pm$0.87 &  16.23$\pm$4.78 & 13.15$\pm$5.48 & \\
P3 &  56820.6 & 8.01$\pm$0.60 &  5.30$\pm$2.87 & 8.45$\pm$4.11 & \\
P4 &  56822.1 & 11.71$\pm$0.48 &  12.35$\pm$4.02 & 6.51$\pm$1.19 & \\
P5 &  56823.4 & 12.30$\pm$0.50 &  16.33$\pm$1.82 & 40.02$\pm$2.63 & \\
\hline                          

 & & Flare-II\\ 
\hline\hline                        
Peak & $t_0$ & $F_0$ & $T_r$ & $T_d$ \\ [0.8ex]
& [MJD] & [10$^{-6}$ ph cm$^{-2}$ s$^{-1}$] & [hr] & [hr]\\
\hline
P1 &  56827.1 & 9.27$\pm$1.03 &  5.17$\pm$1.23 & 4.87$\pm$1.30 & \\
P2 &  56830.9 & 11.60$\pm$0.43 &  50.68$\pm$3.20 & 8.25$\pm$1.05 & \\
P3 &  56831.6 & 12.20$\pm$0.50 &  5.46$\pm$0.78 & 14.07$\pm$3.94 & \\
P4 &  56833.1 & 8.90$\pm$0.88 &  11.37$\pm$5.28 & 3.50$\pm$1.77 & \\
P5 &  56834.4 & 11.02$\pm$1.16 &  7.97$\pm$2.15 & 38.00$\pm$2.28 & \\
\hline                          
\end{tabular}
\label{tab:hresult}
\end{table*}

\begin{table*}[h]
\caption{All the parameters represented here are similar to the parameters of Table 1. Results (6 hour binning) are shown for Flare-3B} 
\centering
\begin{tabular}{cccc rrrr}   
& & Flare-3B \\
\hline\hline                        
Peak & $t_0$ & $F_0$ & $T_r$ & $T_d$ \\ [0.8ex]
& [MJD] & [10$^{-6}$ ph cm$^{-2}$ s$^{-1}$] & [hr] & [hr]\\
\hline
P1 &  57006.1 & 4.95$\pm$0.69 &  58.87$\pm$5.01 & 9.22$\pm$2.72 & \\
P2 &  57008.4 & 7.90$\pm$0.90 &  8.96$\pm$1.54 & 33.04$\pm$2.12 & \\
\hline                          
\end{tabular}
\label{tab:hresult}
\end{table*}

\begin{table*}[h]
\caption{All the parameters represented here are similar to the parameters of Table 1. Results (6 hour binning) are shown for Flare-4A} 
\centering
\begin{tabular}{ccccc rrrr}   
& & Flare-4A\\
\hline\hline                        
Peak & $t_0$ & $F_0$ & $T_r$ & $T_d$ \\ [0.8ex]
& [MJD] & [10$^{-6}$ ph cm$^{-2}$ s$^{-1}$] & [hr] & [hr]\\
\hline
P1 &  57197.4 & 4.75$\pm$0.72 &  5.02$\pm$1.36 & 5.40$\pm$1.53 & \\
P2 &  57198.6 & 7.58$\pm$0.79 &  4.52$\pm$0.86 & 3.16$\pm$0.73 & \\
P3 &  57199.9 & 5.64$\pm$0.74 &  3.65$\pm$1.17 & 4.80$\pm$1.08 & \\
P4 &  57204.6 & 10.82$\pm$0.46 &  9.88$\pm$0.79 & 9.44$\pm$0.88 & \\
P5 &  57206.6 & 9.21$\pm$0.43 &  6.68$\pm$0.99 & 10.59$\pm$0.97 & \\
\hline                          
\end{tabular}
\label{tab:hresult}
\end{table*}

\begin{table*}[h]
\caption{All the parameters represented here are similar to the parameters of Table 1. Results (6 hour binning) are shown for Flare-4B} 
\centering
\begin{tabular}{cccc rrrr}   
& & Flare-4B\\
\hline\hline                        
Peak & $t_0$ & $F_0$ & $T_r$ & $T_d$ \\ [0.8ex]
& [MJD] & [10$^{-6}$ ph cm$^{-2}$ s$^{-1}$] & [hr] & [hr]\\
\hline
P1 &  57254.1 & 11.43$\pm$0.48 &  9.83$\pm$1.58 & 9.55$\pm$1.89 & \\
P2 &  57256.1 & 12.00$\pm$0.49 &  17.35$\pm$2.58 & 14.33$\pm$1.40 & \\
\hline                          
\end{tabular}
\label{tab:hresult}
\end{table*}

\begin{table*}[h]
\caption{All the parameters represented here are similar to the parameters of Table 1. Results (6 hour binning) are shown for Flare-4C} 
\centering
\begin{tabular}{ccccc rrrr}   
& & Flare-4C\\
\hline\hline                        
Peak & $t_0$ & $F_0$ & $T_r$ & $T_d$ \\ [0.8ex]
& [MJD] & [10$^{-6}$ ph cm$^{-2}$ s$^{-1}$] & [hr] & [hr]\\
\hline
P1 &  57401.4 & 5.15$\pm$0.83 &  19.75$\pm$6.27 & 8.96$\pm$2.39 & \\
P2 &  57402.9 & 5.49$\pm$0.80 &  9.56$\pm$2.57 & 24.49$\pm$4.33 & \\
P3 &  57407.1 & 7.44$\pm$0.94 &  32.60$\pm$4.33 & 12.08$\pm$2.10 & \\
P4 &  57408.9 & 5.44$\pm$0.83 &  4.16$\pm$1.71 & 13.26$\pm$3.50 & \\
\hline                          
\end{tabular}
\label{tab:hresult}
\end{table*}

\begin{table*}[h]
\caption{All the parameters represented here are similar to the parameters of Table 1. Results (6 hour binning) are shown for Flare-4D} 
\centering
\begin{tabular}{cccc rrrr}   
& & Flare-4D\\
\hline\hline                        
Peak & $t_0$ & $F_0$ & $T_r$ & $T_d$ \\ [0.8ex]
& [MJD] & [10$^{-6}$ ph cm$^{-2}$ s$^{-1}$] & [hr] & [hr]\\
\hline
P1 &  57457.4 & 3.49$\pm$0.69 & 4.53$\pm$1.78 & 3.26$\pm$1.85 & \\
P2 &  57460.1 & 9.99$\pm$0.45 &  13.64$\pm$1.28 & 6.60$\pm$0.91 & \\
\hline                          
\end{tabular}
\label{tab:hresult}
\end{table*}

\begin{table*}[h]
\caption{All the parameters represented here are similar to the parameters of Table 1. Results (6 hour binning) are shown for Flare-5A} 
\centering
\begin{tabular}{ccccc rrrr}   
& & Flare-5A\\
\hline\hline                        
Peak & $t_0$ & $F_0$ & $T_r$ & $T_d$ \\ [0.8ex]
& [MJD] & [10$^{-6}$ ph cm$^{-2}$ s$^{-1}$] & [hr] & [hr]\\
\hline
P1 &  57558.1 & 14.37$\pm$1.82 &  5.49$\pm$1.47 & 3.56$\pm$1.39 & \\
P2 &  57558.9 & 10.60$\pm$1.66 &  2.83$\pm$1.35 & 8.12$\pm$2.62 & \\
P3 &  57561.1 & 12.41$\pm$1.66 &  15.37$\pm$2.74 & 7.81$\pm$2.20 & \\
P4 &  57562.4 & 19.19$\pm$1.29 &  4.06$\pm$0.83 & 4.87$\pm$1.13 & \\
P5 &  57563.6 & 20.23$\pm$1.30 &  9.79$\pm$1.54 & 11.74$\pm$0.58 & \\
\hline                          
\end{tabular}
\label{tab:hresult}
\end{table*}

\begin{table*}[h]
\caption{All the parameters represented here are similar to the parameters of Table 1. Results (6 hour binning) are shown for Flare-5B} 
\centering
\begin{tabular}{ccccc rrrr}   
& & Flare-5B\\
\hline\hline                        
Peak & $t_0$ & $F_0$ & $T_r$ & $T_d$ \\ [0.8ex]
& [MJD] & [10$^{-6}$ ph cm$^{-2}$ s$^{-1}$] & [hr] & [hr]\\
\hline
P1 &  57742.6 & 5.72$\pm$0.46 &  7.85$\pm$1.91 & 5.49$\pm$1.05 & \\
P2 &  57743.4 & 5.27$\pm$0.44 &  5.10$\pm$1.18 & 10.39$\pm$1.45 & \\
\hline                          
\end{tabular}
\label{tab:hresult}
\end{table*}

\begin{table*}[h]
\caption{Constant flux value for four Sub-structures} 
\centering
\begin{tabular}{ccccc rrrr}   
\hline\hline                        
Sub-structures & Constant-Flux & \\ [0.8ex]
& $F_{0.1-300 Gev}$ \\
& [10$^{-6}$ ph cm$^{-2}$ s$^{-1}$] & \\
\hline
Flare-4A & 1.54$\pm$0.13 & \\
Flare-4B & 1.79$\pm$0.23 & \\
Flare-4D & 1.24$\pm$0.13 & \\
Flare-5A & 0.52$\pm$0.07 & \\
\hline\hline
\end{tabular}
\label{tab:hresult}
\end{table*}

\begin{table*}[h]
\caption{Results of variability time [$t_{var}$ (column 5)] which is calculated by scanning the 6 hour binning $\gamma$-ray light curve for each flare. $\Delta$ $t_{d/h}$ (column 6) is the redshift corrected doubling/halving time. Rise/Decay (column 7) represent the behaviour of the flux in a given time interval[$T_{start}$ (column 1) and $T_{stop}$ (column 2)]. Results are shown here from MJD 54728 - 57207.} 
\centering
\begin{tabular}{ccccc rrrr}   
\hline\hline                        

$T_{start}(t_1)$ & $T_{stop}(t_2)$ & $Flux start[F(t_1)]$ & $Flux stop[F(t_2)]$ & $\tau_{d/h}$ & $\Delta$ $t_{d/h}$ & Rise/Decay \\ [0.8ex]
[MJD] & [MJD] & [10$^{-6}$ ph cm$^{-2}$ s$^{-1}$] & [10$^{-6}$ ph cm$^{-2}$ s$^{-1}$] & [hr] & [hr] \\
\hline
& & & Flare-1A \\
\hline
54728.125 &  54728.375 & 2.37$\pm$0.60 &  5.22$\pm$0.77 & 2.63$\pm$0.98 & 1.41$\pm$0.53 & R \\
54744.625 &  54744.875 & 1.55$\pm$0.34 &  3.10$\pm$0.50 & 3.00$\pm$1.18 & 1.61$\pm$0.63 & R \\
54749.125 &  54749.375 & 2.77$\pm$0.52 &  1.27$\pm$0.32 & -2.67$\pm$1.07 & -1.44$\pm$0.57 & D \\
54756.625 &  54756.875 & 1.78$\pm$0.40 &  0.87$\pm$0.34 & -2.90$\pm$1.83 & -1.56$\pm$0.98 & D \\
\hline                          
& & & Flare-2A \\
\hline
\hline
\end{tabular}

\label{tab:hresult}
\end{table*}

\begin{table*}
\caption{All the parameters represented here are similar to the parameters of Table-15, but result are shown here from MJD 57263 - 57762.} 
\centering
\begin{tabular}{ccccc rrrr}   
\hline\hline
$T_{start}(t_1)$ & $T_{stop}(t_2)$ & $Flux start[F(t_1)]$ & $Flux stop[F(t_2)]$ & $\tau_{d/h}$ & $\Delta$ $t_{d/h}$ & Rise/Decay \\ [0.8ex]
[MJD] & [MJD] & [10$^{-6}$ ph cm$^{-2}$ s$^{-1}$] & [10$^{-6}$ ph cm$^{-2}$ s$^{-1}$] & [hr] & [hr] \\
\hline
& & & Flare-3B \\
\hline
57013.625 & 57013.875 & 1.91$\pm$0.59 & 4.00$\pm$0.97 & 2.81$\pm$1.49 & 1.51$\pm$0.80 & R \\
57013.875 & 57014.125 & 4.00$\pm$0.97 & 1.52$\pm$1.11 & -2.15$\pm$1.70 & -1.16$\pm$0.91 & D \\
57014.125 & 57014.375 & 1.52$\pm$1.11 & 3.48$\pm$0.56 & 2.51$\pm$2.27 & 1.35$\pm$1.22 & R \\
57019.375 & 57019.625 & 2.38$\pm$0.45 & 1.10$\pm$0.41 & -2.69$\pm$1.46 & -1.45$\pm$0.78 & D \\
\hline
& & & Flare-4A \\
\hline
\hline
\end{tabular}
\label{tab:hresult}
\end{table*}

\begin{table*}
\caption{All the parameters represented here are similar to the parameters of Table-15, but result are shown here from MJD 57749 - 57762.} 
\centering
\begin{tabular}{ccccc rrrr}   
\hline\hline
$T_{start}(t_1)$ & $T_{stop}(t_2)$ & $Flux start[F(t_1)]$ & $Flux stop[F(t_2)]$ & $\tau_{d/h}$ & $\Delta$ $t_{d/h}$ & Rise/Decay \\ [0.8ex]
[MJD] & [MJD] & [10$^{-6}$ ph cm$^{-2}$ s$^{-1}$] & [10$^{-6}$ ph cm$^{-2}$ s$^{-1}$] & [hr] & [hr] \\
\hline
& & & Flare-5B \\
\hline
57749.375 & 57749.625 & 2.72$\pm$0.50 & 1.38$\pm$0.52 & -3.06$\pm$1.89 & -1.65$\pm$1.02 & D \\
57749.625 & 57749.875 & 1.38$\pm$0.52 & 2.97$\pm$0.93 & 2.71$\pm$1.73 & 1.46$\pm$0.93 & R \\
57749.875 & 57750.125 & 2.97$\pm$0.93 & 1.45$\pm$0.46 & -2.90$\pm$1.80 & -1.56$\pm$0.97 & D \\
57761.875 & 57762.125 & 2.62$\pm$0.72 & 1.18$\pm$0.42 & -2.61$\pm$1.47 & -1.40$\pm$0.79 & D \\
\hline
\hline
\end{tabular}
\label{tab:hresult}
\end{table*}

\begin{table*}[h]
\caption{Result of SED for Flare-1A fitted with different models (Powerlaw, Logparabola and Broken-powerlaw). Column 1 represents the different periods of activity, column 2 and column 3 to column 4 represent Flux value ($F_0$) and spectral indices for different models respectively. Break energy ($E_{break}$) for Broken-powerlaw model is given in column 5. The goodness of fit (log of Likelihood) is mentioned in column 6. Column 7 represents the difference in the goodness of fit w.r.t. powerlaw model.} 
\centering
\begin{tabular}{ccccc rrrr}   
\hline\hline                        
& & & Powerlaw\\
\hline
Activity & $F_0$ & $\Gamma$ & & & log(Likelihood) \\ [1.5ex]
& [10$^{-6}$ ph cm$^{-2}$ s$^{-1}$] &  \\
\hline
Flare &  2.60$\pm$0.03 & 2.39$\pm$0.01 &  - & - & 162999.70 & - \\
Post-Flare &  1.30$\pm$0.04 & 2.40$\pm$0.03 &  - & - & 57805.94 & - \\
\hline                          
 & & & Logparabola\\ 
\hline
Activity & $F_0$ & $\alpha$ & $\beta$ & - & log(Likelihood) & $\Delta$log(Likelihood) \\ [1.5ex]
& [10$^{-6}$ ph cm$^{-2}$ s$^{-1}$] &  \\
\hline
Flare &  2.50$\pm$0.04 & 2.31$\pm$0.02 &  0.07$\pm$0.01 & - & 162979.78 & -19.92 \\
Post-Flare &  1.30$\pm$0.05 & 2.29$\pm$0.05 &  0.09$\pm$0.03 & - & 57799.58 & -6.36 \\

\hline                          
 & & & Broken-Powerlaw\\ 
\hline
Activity & $F_0$ & $\Gamma_1$ & $\Gamma_2$ & $E_{break}$ & log(Likelihood) & $\Delta$log(Likelihood) \\ [1.5ex]
& [10$^{-6}$ ph cm$^{-2}$ s$^{-1}$] & & & [Gev] \\
\hline
Flare &  2.60$\pm$0.04 & 2.29$\pm$0.02 & 2.73$\pm$0.06 & 1.02$\pm$0.03 & 162960.52 & -39.18 \\
Post-Flare &  1.30$\pm$0.06 & 2.27$\pm$0.09 & 2.87$\pm$0.20 & 1.03$\pm$0.18 & 57797.77 & -8.17 \\

\hline 
\end{tabular}
\label{tab:hresult}
\end{table*}

\begin{table*}[h]
\caption{All the parameters represented here are similar to the parameters of Table-17. Results are shown for Flare-2A} 
\centering
\begin{tabular}{ccccc rrrr}   
\hline\hline                        
& & & Powerlaw\\
\hline
Activity & $F_0$ & $\Gamma$ & & & log(Likelihood) \\ [1.5ex]
& [10$^{-6}$ ph cm$^{-2}$ s$^{-1}$] &  \\
\hline
Pre-Flare &  1.30$\pm$0.05 & 2.42$\pm$0.04 &  - & - & 50345.11 & - \\
Flare &  2.90$\pm$0.006 & 2.41$\pm$0.001 &  - & - & 231037.12 & - \\

\hline                          
 & & & Logparabola\\ 
\hline
Activity & $F_0$ & $\alpha$ & $\beta$ & - & log(Likelihood) & $\Delta$log(Likelihood) \\ [1.5ex]
& [10$^{-6}$ ph cm$^{-2}$ s$^{-1}$] &  \\
\hline
Pre-Flare &  1.20$\pm$0.05 & 2.25$\pm$0.05 &  0.18$\pm$0.04 & - & 50368.45 & 23.34 \\
Flare &  2.70$\pm$0.05 & 2.26$\pm$0.02 &  0.12$\pm$0.01 & - & 230979.32 & -57.80 \\

\hline                          
 & & & Broken-Powerlaw\\ 
\hline
Activity & $F_0$ & $\Gamma_1$ & $\Gamma_2$ & $E_{break}$ & log(Likelihood) & $\Delta$log(Likelihood) \\ [1.5ex]
& [10$^{-6}$ ph cm$^{-2}$ s$^{-1}$] & & & [Gev] \\
\hline
Pre-Flare &  1.20$\pm$0.07 & 2.23$\pm$0.09 & 3.11$\pm$0.24 & 0.91$\pm$0.16 & 50368.09 & 22.98 \\
Flare &  2.90$\pm$0.04 & 2.31$\pm$0.02 & 2.86$\pm$0.05 & 1.03$\pm$0.02 & 231038.29 & 1.17 \\

\hline 
\end{tabular}
\label{tab:hresult}
\end{table*}

\begin{table*}[h]
\caption{All the parameters represented here are similar to the parameters of Table-17. Results are shown for Flare-2B} 
\centering
\begin{tabular}{ccccc rrrr}   
\hline\hline                        
& & & Powerlaw\\
\hline
Activity & $F_0$ & $\Gamma$ & & & log(Likelihood) \\ [1.5ex]
& [10$^{-6}$ ph cm$^{-2}$ s$^{-1}$] &  \\
\hline
Pre-Flare &  2.36$\pm$0.067 & 2.42$\pm$0.03 &  - & - & 43514.22 & - \\
Flare &  7.4$\pm$0.065 & 2.30$\pm$0.008 &  - & - & 238878.89 & - \\

\hline                          
 & & & Logparabola\\ 
\hline
Activity & $F_0$ & $\alpha$ & $\beta$ & - & log(Likelihood) & $\Delta$log(Likelihood) \\ [1.5ex]
& [10$^{-6}$ ph cm$^{-2}$ s$^{-1}$] &  \\
\hline
Pre-Flare &  2.20$\pm$0.06 & 2.34$\pm$0.04 &  0.08$\pm$0.02 & - & 43507.63 & -6.59 \\
Flare &  7.00$\pm$0.08 & 2.16$\pm$0.01 &  0.10$\pm$0.007 & - & 238755.82 & -123.07 \\

\hline                          
 & & & Broken-Powerlaw\\ 
\hline
Activity & $F_0$ & $\Gamma_1$ & $\Gamma_2$ & $E_{break}$ & log(Likelihood) & $\Delta$log(Likelihood) \\ [1.5ex]
& [10$^{-6}$ ph cm$^{-2}$ s$^{-1}$] & & & [Gev] \\
\hline
Pre-Flare &  2.10$\pm$0.09 & 2.23$\pm$0.04 & 2.87$\pm$0.11 & 0.91$\pm$0.04 & 43493.02 & -21.2 \\
Flare &  7.10$\pm$0.07 & 2.16$\pm$0.01 & 2.70$\pm$0.03 & 1.00$\pm$0.02 & 238745.81 & -133.08 \\

\hline 
\end{tabular}
\label{tab:hresult}
\end{table*}

\begin{table*}[h]
\caption{All the parameters represented here are similar to the parameters of Table-17. Results are shown for Flare-2C} 
\centering
\begin{tabular}{ccccc rrrr}   
\hline\hline                        
& & & Powerlaw\\
\hline
Activity & $F_0$ & $\Gamma$ & & & log(Likelihood) \\ [1.5ex]
& [10$^{-6}$ ph cm$^{-2}$ s$^{-1}$] &  \\
\hline
Pre-Flare &  2.70$\pm$0.05 & 2.47$\pm$0.02 &  - & - & 79966.93 & - \\
Flare-I &  8.61$\pm$1.8e-4 & 2.37$\pm$1.38e-5 &  - & - & 198341.69 & - \\
Flare-II &  5.66$\pm$0.09 & 2.35$\pm$0.01 &  - & - & 74663.22 & - \\
Post-Flare &  2.70$\pm$0.06 & 2.40$\pm$0.02 &  - & - & 69581.59 & - \\
\hline                          
 & & & Logparabola\\
\hline 
\end{tabular}
\label{tab:hresult}
\end{table*}

\begin{table*}[h]
\caption{All the parameters represented here are similar to the parameters of Table-17. Results are shown for Flare-2D} 
\centering
\begin{tabular}{ccccc rrrr}   
\hline\hline                        
& & & Powerlaw\\
\hline
Activity & $F_0$ & $\Gamma$ & & & log(Likelihood) \\ [1.5ex]
& [10$^{-6}$ ph cm$^{-2}$ s$^{-1}$] &  \\
\hline
Pre-Flare &  2.98$\pm$0.04 & 2.41$\pm$0.01 &  - & - & 85950.62 & - \\
Platue-I & 6.40$\pm$0.07 & 2.33$\pm$0.01 &  - & - & 145909.56 & - \\
Flare-I &  24.0$\pm$0.16 & 2.27$\pm$0.006 &  - & - & 250597.83 & - \\
Flare-II &  13.0$\pm$0.09 & 2.29$\pm$0.007 &  - & - & 241670.99 & - \\
Platue-II &  5.90$\pm$0.09 & 2.31$\pm$0.01 &  - & - & 70434.36 & - \\
\hline 
\end{tabular}
\label{tab:hresult}
\end{table*}

\begin{table*}[h]
\caption{All the parameters represented here are similar to the parameters of Table-17. Results are shown for Flare-3A} 
\centering
\begin{tabular}{ccccc rrrr}   
\hline\hline                        
& & & Powerlaw\\
\hline
Activity & $F_0$ & $\Gamma$ & & & log(Likelihood) \\ [1.5ex]
& [10$^{-6}$ ph cm$^{-2}$ s$^{-1}$] &  \\
\hline
Pre-Flare &  2.25$\pm$0.02 & 2.27$\pm$0.005 &  - & - & 36569.38 & - \\
Flare-I &  6.80$\pm$0.13 & 2.12$\pm$0.01 &  - & - & 54645.77 & - \\
Flare-II &  7.20$\pm$0.12 & 1.99$\pm$0.01 &  - & - & 63621.22 & - \\
Post-Flare &  3.65$\pm$0.004 & 2.21$\pm$0.0006 &  - & - & 36381.77 & - \\
\hline                          
 & & & Logparabola\\
\hline 
\end{tabular}
\label{tab:hresult}
\end{table*}

\begin{table*}[h]
\caption{All the parameters represented here are similar to the parameters of Table-17. Results are shown for Flare-3B} 
\centering
\begin{tabular}{ccccc rrrr}   
\hline\hline                        
& & & Powerlaw\\
\hline
Activity & $F_0$ & $\Gamma$ & & & log(Likelihood) \\ [1.5ex]
& [10$^{-6}$ ph cm$^{-2}$ s$^{-1}$] &  \\
\hline
Pre-Flare &  2.30$\pm$0.0006 & 2.31$\pm$0.0001 &  - & - & 20940.21 & - \\
Flare &  3.81$\pm$0.08 & 2.32$\pm$0.01 &  - & - & 34862.33 & - \\
Post-Flare &  2.32$\pm$0.01 & 2.33$\pm$0.005 &  - & - & 32768.08 & - \\
\hline                          
 & & & Logparabola\\ 
\hline
Activity & $F_0$ & $\alpha$ & $\beta$ & - & log(Likelihood) & $\Delta$log(Likelihood) \\
\hline 
\end{tabular}
\label{tab:hresult}
\end{table*}

\begin{table*}[h]
\caption{All the parameters represented here are similar to the parameters of Table-17. Results are shown for Flare-4A} 
\centering
\begin{tabular}{ccccc rrrr}   
\hline\hline                        
& & & Powerlaw\\
\hline
Activity & $F_0$ & $\Gamma$ & & & log(Likelihood) \\ [1.5ex]
& [10$^{-6}$ ph cm$^{-2}$ s$^{-1}$] &  \\
\hline
Pre-Flare &  1.38$\pm$0.01 & 2.27$\pm$0.006 &  - & - & 75077.06 & - \\
Flare &  3.26$\pm$0.004 & 2.32$\pm$0.0009 &  - & - & 81537.39 & - \\
Post-Flare &  0.78$\pm$0.01 & 2.41$\pm$0.01 &  - & - & 40656.04 & - \\
\hline                          
 & & & Logparabola\\ 
\hline
Activity & $F_0$ & $\alpha$ & $\beta$ & - & log(Likelihood) & $\Delta$log(Likelihood) \\
\hline 
\end{tabular}
\label{tab:hresult}
\end{table*}

\begin{table*}[h]
\caption{All the parameters represented here are similar to the parameters of Table-17. Results are shown for Flare-4B} 
\centering
\begin{tabular}{ccccc rrrr}   
\hline\hline                        
& & & Powerlaw\\
\hline
Activity & $F_0$ & $\Gamma$ & & & log(Likelihood) \\ [1.5ex]
& [10$^{-6}$ ph cm$^{-2}$ s$^{-1}$] &  \\
\hline
Pre-Flare &  2.13$\pm$0.005 & 2.31$\pm$0.001 &  - & - & 39002.40 & - \\
Flare &  6.51$\pm$0.01 & 2.15$\pm$0.001 &  - & - & 41130.59 & - \\
Post-Flare &  1.87$\pm$0.01 & 2.41$\pm$0.004 &  - & - & 25474.42 & - \\
\hline                          
 & & & Logparabola\\ 
\hline
Activity & $F_0$ & $\alpha$ & $\beta$ & - & log(Likelihood) & $\Delta$log(Likelihood) \\
\hline 
\end{tabular}
\label{tab:hresult}
\end{table*}

\begin{table*}[h]
\caption{All the parameters represented here are similar to the parameters of Table-17. Results are shown for Flare-4C} 
\centering
\begin{tabular}{ccccc rrrr}   
\hline\hline                        
& & & Powerlaw\\
\hline
Activity & $F_0$ & $\Gamma$ & & & log(Likelihood) \\ [1.5ex]
& [10$^{-6}$ ph cm$^{-2}$ s$^{-1}$] &  \\
\hline
Pre-Flare &  2.90$\pm$0.12 & 2.33$\pm$0.39 &  - & - & 21642.13 & - \\
Flare &  3.90$\pm$0.10 & 2.31$\pm$0.02 &  - & - & 54879.34 & - \\
Post-Flare &  2.90$\pm$0.10 & 2.38$\pm$0.03 &  - & - & 40934.47 & - \\
\hline                          
 & & & Logparabola\\ 
\hline
Activity & $F_0$ & $\alpha$ & $\beta$ & - & log(Likelihood) & $\Delta$log(Likelihood) \\
\hline 
\end{tabular}
\label{tab:hresult}
\end{table*}

\begin{table*}[h]
\caption{All the parameters represented here are similar to the parameters of Table-17. Results are shown for Flare-4D} 
\centering
\begin{tabular}{ccccc rrrr}   
\hline\hline                        
& & & Powerlaw\\
\hline
Activity & $F_0$ & $\Gamma$ & & & log(Likelihood) \\ [1.5ex]
& [10$^{-6}$ ph cm$^{-2}$ s$^{-1}$] &  \\
\hline
Pre-Flare &  1.20$\pm$0.10 & 2.45$\pm$0.07 &  - & - & 17622.70 & - \\
Flare &  4.00$\pm$0.16 & 2.19$\pm$0.03 &  - & - & 19422.69 & - \\
Post-Flare &  1.40$\pm$0.09 & 2.27$\pm$0.05 &  - & - & 23233.62 & - \\
\hline                          
 & & & Logparabola\\ 
\hline
Activity & $F_0$ & $\alpha$ & $\beta$ & - & log(Likelihood) & $\Delta$log(Likelihood) \\
\hline 
\end{tabular}
\label{tab:hresult}
\end{table*}

\begin{table*}[h]
\caption{All the parameters represented here are similar to the parameters of Table-17. Results are shown for Flare-5A} 
\centering
\begin{tabular}{ccccc rrrr}   
\hline\hline                        
& & & Powerlaw\\
\hline
Activity & $F_0$ & $\Gamma$ & & & log(Likelihood) \\ [1.5ex]
& [10$^{-6}$ ph cm$^{-2}$ s$^{-1}$] &  \\
\hline
Pre-Flare &  0.69$\pm$0.01 & 2.60$\pm$0.01 &  - & - & 25761.05 & - \\
Flare &  5.49$\pm$0.006 & 2.11$\pm$0.0005 &  - & - & 80606.46 & - \\
Post-Flare &  1.78$\pm$0.004 & 2.28$\pm$0.001 &  - & - & 25145.68 & - \\
\hline                          
 & & & Logparabola\\ 
\hline
Activity & $F_0$ & $\alpha$ & $\beta$ & - & log(Likelihood) & $\Delta$log(Likelihood) \\
\hline 
\end{tabular}
\label{tab:hresult}
\end{table*}

\begin{table*}[h]
\caption{All the parameters represented here are similar to the parameters of Table-17. Results are shown for Flare-5B} 
\centering
\begin{tabular}{ccccc rrrr}   
\hline\hline                        
& & & Powerlaw\\
\hline
Activity & $F_0$ & $\Gamma$ & & & log(Likelihood) \\ [1.5ex]
& [10$^{-6}$ ph cm$^{-2}$ s$^{-1}$] &  \\
\hline
Pre-Flare &  1.10$\pm$0.08 & 2.43$\pm$0.06 &  - & - & 42871.64 & - \\
Flare &  3.10$\pm$0.07 & 2.31$\pm$0.02 &  - & - & 113517.61 & - \\
Post-Flare &  2.00$\pm$0.10 & 2.45$\pm$0.05 &  - & - & 41430.56 & - \\
\hline                          
 & & & Logparabola\\ 
\hline
Activity & $F_0$ & $\alpha$ & $\beta$ & - & log(Likelihood) & $\Delta$log(Likelihood) \\
\hline
\end{tabular}
\label{tab:hresult}
\end{table*}

\begin{table*}[h]
\caption{Results of reduced-$\chi^2$ value (column 2) for different spectral models (Powerlaw, Logparabola, Broken-powerlaw). Column 1 represents different flares activity.} 
\centering
\begin{tabular}{c|cccc rrrr}   
\hline\hline                        
Activity & & Reduced-$\chi^2$ \\ [0.8ex]
\hline
Flare-1A & Powerlaw & Log-parabola & Broken-Powerlaw \\
\hline
Flare & 12.38 & 3.98 & 1.72 \\
\hline
Flare-2A \\
\hline
Flare & 49.06 & 4.53 & 12.19 \\
\hline
Flare-2B \\
\hline
Flare & 41.92 & 1.64 & 2.00 \\
\hline
Flare-2C \\
\hline
Flare-I & 1796.62 & 0.37 & 122.61 \\
Flare-II & 11.05 & 0.19 & 1.32 \\
\hline
Flare-2D \\
\hline
Flare-I & 129.60 & 1.76 & 16.99 \\
Flare-II & 43.70 & 0.62 & 6.23 \\
\hline
Flare-3A \\
\hline
Flare-I & 15.13 & 1.18 & 5.64 \\
Flare-II & 24.86 & 2.82 & 3.02 \\
\hline                          
Flare-3B \\
\hline
Flare & 5.65 & 4.06 & 2.60 \\
\hline
Flare-4A \\
\hline
Flare & 8.07 & 0.60 & 1.33 \\
\hline
Flare-4B \\
\hline
Flare & 12.04 & 0.54 & 2.15 \\
\hline
Flare-4C \\
\hline
Flare & 241.71 & 2.80 & 31.62 \\
\hline
Flare-4D \\
\hline
Flare & 1.09 & 0.61 & 0.31 \\
\hline
Flare-5A \\
\hline
Flare & 47.71 & 1.05 & 6.60 \\
\hline
Flare-5B \\
\hline
Flare & 4.48 & 0.14 & 0.61 \\
\hline\hline
\end{tabular}
\label{tab:hresult}
\end{table*}

\begin{table*}
\caption{Results of multi-wavelength SED modelling which is shown in Figure-48 \& Figure-49. 1st column represents the study of different cases (see text for more details). Time duration of the Flares is given in last column. } 
\centering
\begin{tabular}{ccccc rrrr}   
\hline\hline
Different Cases & Parameters & Symbol & values & Time duration  \\
\hline
& Flare-2A \\
\hline
& Spectral index of injected electron spectrum (LP) & $\alpha$ & 2.00  \\
& Magnetic field in emission region & B & 3.80 G  \\
& Temperature of BLR region & $T^\prime_{BLR}$ & 2.0$\times10^{4}$ K & 95 days  \\
& Phonton density of BLR region & $U^\prime_{BLR}$ & 5.63 erg/$cm^3$  \\
& Temperature of Disk & $T^\prime_{Disk}$ & 1.0$\times10^{6}$ K  \\
& Phonton density of Disk & $U^\prime_{Disk}$ & 1.48$\times10^{-5}$ erg/$cm^3$  \\
\hline\hline

\end{tabular}
\label{tab:hresult}
\end{table*}

\begin{table*}[h]
\caption{Best fitted values of the parameters when Doppler factor is varying according to equation $\ref{eq:15}$ to model the light curve of different types of flare peaks. Last column represents the range of values of the Doppler factor ($\delta$) for each peak.} 
\centering
\begin{tabular}{c|cccc rrrr}   
\hline\hline                        
Type of flare peak & & & Best fitted parameter value \\ [0.8ex]
\hline
& $t_c$ & $a_1$ & $a_2$ & k & Range of $\delta$\\
\hline
$(T_r \sim T_d)$ & 1.116 & 0.195 & -0.205 & 47.36 & 42.60 - 47.20 \\
\hline
$(T_r > T_d)$ & 1.364 & 0.07 & -0.65 & 58.48 & 48.50 - 57.65 \\ 
\hline
$(T_r < T_d)$ & 1.864 & 0.63 & -0.58 & 30.39 & 28.60 - 43.20 \\
\hline\hline
\end{tabular}
\label{tab:hresult}
\end{table*}

\begin{table*}[h]
\caption{Best fitted values of the parameters when normalisation constant ($l_0$) is varying according to equation $\ref{eq:16}$ to model the light curve of different types of flare peaks. 5th \& 6th column represent the range of values of the normalisation constant ($l_0$) and injected power in electrons ($P_e$) for each peak respectively.} 
\centering
\begin{tabular}{c|cccc rrrr}   
\hline\hline                        
Type of flare peak & & & Best fitted parameter value \\ [0.8ex]
\hline
& $t_c$ & $a_3$ & $a_4$ & k & Range of $l_0 (\times10^{50})$ & Range of $P_e(\times10^{46} erg/s)$ \\
\hline
$(T_r \sim T_d)$ & 1.116 & 0.586 & -0.605 & 20.23 & 14.83 - 20.25 & 2.39 - 3.27 \\
\hline
$(T_r > T_d)$ & 1.364 & 0.22 & -2.00 & 36.69 & 21.00 - 38.95 & 3.39 - 6.27 \\
\hline
$(T_r < T_d)$ & 1.861 & 1.43 & -1.55 & 6.18 & 4.30 - 14.90 & 0.693 - 2.4 \\
\hline\hline
\end{tabular}
\label{tab:hresult}
\end{table*}

\bibliographystyle{plain}
{}

\end{document}